\pdfoutput=1
\pdfoutput=1

\ifdefined\PaperPublication\else
\def\PaperPublication{}
\def\PaperAuthorNames{Tristan Simas}
\def\PaperZenodoURL{https://doi.org/10.5281/zenodo.23393039}

\long\def\PaperPublicationAuthor#1{\ifcsname PaperPublicationAuthor@#1\endcsname\csname PaperPublicationAuthor@#1\endcsname\else\errmessage{Unsupported publication author adapter: #1}\fi}
\expandafter\def\csname PaperPublicationAuthor@article\endcsname{\author{Tristan Simas\\McGill University\\\texttt{tristan.simas@mail.mcgill.ca}}}
\expandafter\def\csname PaperPublicationAuthor@acmart\endcsname{\author{Tristan Simas}\affiliation{\institution{McGill University}\country{Canada}}\email{tristan.simas@mail.mcgill.ca}}
\expandafter\def\csname PaperPublicationAuthor@jair\endcsname{\author{Tristan Simas}\affiliation{\institution{McGill University}\country{Canada}}\email{tristan.simas@mail.mcgill.ca}}
\expandafter\def\csname PaperPublicationAuthor@IEEEtran\endcsname{\author{Tristan Simas\thanks{McGill University; tristan.simas@mail.mcgill.ca}}}
\expandafter\def\csname PaperPublicationAuthor@siamart\endcsname{\author{Tristan Simas\thanks{McGill University; tristan.simas@mail.mcgill.ca}}}
\expandafter\def\csname PaperPublicationAuthor@elsarticle\endcsname{\author{Tristan Simas}\address{McGill University}\ead{tristan.simas@mail.mcgill.ca}}
\expandafter\def\csname PaperPublicationAuthor@sn-jnl\endcsname{\author*[1]{Tristan Simas}\email{tristan.simas@mail.mcgill.ca}\affil*[1]{\orgname{McGill University},\orgaddress{\country{Canada}}}}
\AddToHook{begindocument/before}{\ifdefined\hypersetup\hypersetup{pdfauthor={Tristan Simas}}\fi}
\AddToHook{cmd/maketitle/after}{\ifdefined\hypersetup\hypersetup{pdfauthor={Tristan Simas}}\fi}
\fi
\renewcommand{\PaperPublicationAuthor}[1]{\csname PaperPublicationAuthor@article\endcsname}

\ifdefined\OOPSLASUBMISSION
\documentclass{paper-preprint}
\usepackage[numbers,sort&compress]{natbib}
\else
\documentclass{paper-preprint}
\usepackage[numbers,sort&compress]{natbib}
\fi

\usepackage[utf8]{inputenc}
\usepackage[T1]{fontenc}
\usepackage{mathtools}
\usepackage{graphicx}
\usepackage{xcolor}
\usepackage{listings}
\usepackage{enumitem}
\usepackage{xr}
\input{external_labels_15b95ee9dc87b1d4.tex}

\usepackage{booktabs}
\usepackage{longtable}
\usepackage{array}
\usepackage{tabularx}
\usepackage{ragged2e}
\usepackage{calc}
\usepackage{tikz}
\usetikzlibrary{fit,positioning}
\newcommand{\PaperFixedDiagramSetup}{%
  
    \fontsize{8pt}{9pt}\selectfont
    \renewcommand{\small}{\fontsize{8pt}{9pt}\selectfont}%
    \renewcommand{\scriptsize}{\fontsize{7pt}{8pt}\selectfont}%
    \renewcommand{\tiny}{\fontsize{5pt}{6pt}\selectfont}%
  
}
\newcommand{\PaperArityDiagramSetup}{%
  
    \fontsize{9pt}{10.5pt}\selectfont
    \renewcommand{\scriptsize}{\fontsize{8.2pt}{9.4pt}\selectfont}%
    \renewcommand{\tiny}{\fontsize{6.8pt}{8pt}\selectfont}%
  
}
\newcommand{\PaperLadderDiagramSetup}{%
  
    \fontsize{9pt}{10.5pt}\selectfont
    \renewcommand{\scriptsize}{\fontsize{8pt}{9.2pt}\selectfont}%
    \renewcommand{\tiny}{\fontsize{8pt}{9.2pt}\selectfont}%
  
}
\newcommand{\PaperArityXUnit}{1cm}
\newcommand{\PaperArityYUnit}{0.86cm}

  \renewcommand{\PaperArityXUnit}{1.08cm}
  \renewcommand{\PaperArityYUnit}{0.94cm}

\usepackage{pgfplots}
\pgfplotsset{compat=1.18}
\usepackage{microtype}
\usepackage{placeins}
\usepackage{float}
\usepackage{needspace}
\usepackage[para,flushmargin]{footmisc}

\newcolumntype{Y}{>{\RaggedRight\arraybackslash}X}
\newcolumntype{Z}{>{\Centering\arraybackslash}X}
\newcommand{\PaperTableSetup}{%
  
    \fontsize{10.5pt}{12.25pt}\selectfont
  
  \renewcommand{\arraystretch}{1.04}%
  \setlength{\tabcolsep}{3pt}%
  \setlength{\aboverulesep}{0.2ex}%
  \setlength{\belowrulesep}{0.2ex}%
}
\newcommand{\PaperDenseTableSetup}{%
  
    \fontsize{8pt}{9.2pt}\selectfont
  
  \renewcommand{\arraystretch}{1.00}%
  \setlength{\tabcolsep}{2.2pt}%
  \setlength{\aboverulesep}{0.2ex}%
  \setlength{\belowrulesep}{0.2ex}%
}

\lstdefinestyle{lean}{
  basicstyle=\ttfamily\footnotesize,
  breaklines=true,
  breakatwhitespace=true,
  columns=flexible,
  keepspaces=true,
  xleftmargin=0.5em,
  frame=none,
}

\providecommand{\leanmetaheadmark}{}
\newcommand{\PaperTheoremSkip}{6pt plus 2pt}
\makeatletter
\newtheoremstyle{leanplain}{\PaperTheoremSkip}{\PaperTheoremSkip}{\itshape}{}{}{.}{ }{\thmname{#1}\thmnumber{ #2}\leanmetaheadmark\thmnote{ (#3)}}
\newtheoremstyle{leandef}{\PaperTheoremSkip}{\PaperTheoremSkip}{}{}{}{.}{ }{\thmname{#1}\thmnumber{ #2}\leanmetaheadmark\thmnote{ (#3)}}
\newtheoremstyle{leanremark}{\PaperTheoremSkip}{\PaperTheoremSkip}{}{}{}{.}{ }{\thmname{#1}\thmnumber{ #2}\leanmetaheadmark\thmnote{ (#3)}}
\makeatother
\theoremstyle{leanplain}

\newtheorem{theorem}{Theorem}[section]
\newtheorem{lemma}[theorem]{Lemma}
\newtheorem{corollary}[theorem]{Corollary}
\newtheorem{proposition}[theorem]{Proposition}
\newtheorem*{theoremcontinuation}{Theorem~\thetheorem}

\theoremstyle{leandef}

\theoremstyle{leanremark}

\makeatletter
\newcommand{\runinhead}[1]{\emph{#1.}}
\newenvironment{claim}[2]{%
  \par\noindent\phantomsection\def\@currentlabel{}\label{#2}%
  \if\relax\detokenize{#1}\relax\else\runinhead{#1}\space\fi
  \ignorespaces
}{\par}
\makeatother

\providecommand{\LH}[1]{}
\renewcommand{\LH}[1]{\leavevmode\nobreak\mbox{\ttfamily\nolinkurl{#1}}}
\providecommand{\LHrng}[3]{}
\renewcommand{\LHrng}[3]{\leavevmode\nobreak\mbox{\ttfamily\nolinkurl{#1#2-#3}}}
\newif\ifleanmetaseen
\leanmetaseenfalse
\providecommand{\leanfootnoteterminator}{}
\providecommand{\leanfootnoteformat}[1]{#1}
\makeatletter
\providecommand{\PaperClaimLocationRecord}[2]{}
\newcounter{PaperClaimLocationIndex}
\newcommand{\PaperRecordClaimLocation}[1]{\stepcounter{PaperClaimLocationIndex}\label{paper-claim-location:\thePaperClaimLocationIndex}\protected@write\@auxout{}{\string\PaperClaimLocationRecord{\thePaperClaimLocationIndex}{\detokenize{#1}}}}
\NewDocumentCommand{\PaperIndexedFootnote}{o +m}{\IfNoValueTF{#1}{\PaperOriginalFootnote{\PaperRecordClaimLocation{#2}#2}}{\PaperOriginalFootnote[#1]{\PaperRecordClaimLocation{#2}#2}}}
\AddToHook{begindocument/end}{\let\PaperOriginalFootnote\footnote\let\footnote\PaperIndexedFootnote}
\providecommand{\leanmetapendinghandles}{}
\global\let\leanmetapendinghandles\@empty
\providecommand{\leanmetapending}[1]{}
\renewcommand{\leanmetapending}[1]{\def\leanmetatest{#1}\ifx\leanmetatest\@empty\global\let\leanmetapendinghandles\@empty\else\PaperRecordClaimLocation{#1}\gdef\leanmetapendinghandles{#1}\fi}
\providecommand{\leanmetaproofpendinghandles}{}
\global\let\leanmetaproofpendinghandles\@empty
\providecommand{\leanmetaproofpending}[1]{}
\renewcommand{\leanmetaproofpending}[1]{\def\leanmetatest{#1}\ifx\leanmetatest\@empty\global\let\leanmetaproofpendinghandles\@empty\else\gdef\leanmetaproofpendinghandles{#1}\fi}
\providecommand{\leanmetaheadmark}{}
\newif\ifleanmetaheadprinted
\leanmetaheadprintedfalse
\renewcommand{\leanmetaheadmark}{\ifx\leanmetapendinghandles\@empty\else\nobreak\hspace{0.2em}\footnote{\leanfootnoteformat{\strut\ifleanmetaseen\else Lean: \global\leanmetaseentrue\fi\leanmetapendinghandles}}\global\leanmetaheadprintedtrue\fi}
\providecommand{\leanmetaheadtext}{}
\renewcommand{\leanmetaheadtext}{\ifleanmetaheadprinted\global\let\leanmetapendinghandles\@empty\global\leanmetaheadprintedfalse\fi}
\providecommand{\leanmetaproofmark}{}
\renewcommand{\leanmetaproofmark}{\ifx\leanmetaproofpendinghandles\@empty\else\ifhmode\unskip\fi\footnote{\leanfootnoteformat{\strut\ifleanmetaseen\else Lean: \global\leanmetaseentrue\fi\leanmetaproofpendinghandles\leanfootnoteterminator}}\global\let\leanmetaproofpendinghandles\@empty\fi}
\@ifundefined{newtheoremstyle}{}{\@ifundefined{thmname}{}{%
  \newtheoremstyle{leanplain}
    {6pt}{6pt}{\itshape}{}{\bfseries}{.}{ }
    {\thmname{#1}\thmnumber{ #2}\thmnote{ (#3)}\leanmetaheadmark\leanmetaheadtext}
  \newtheoremstyle{leandefinition}
    {6pt}{6pt}{}{}{\bfseries}{.}{ }
    {\thmname{#1}\thmnumber{ #2}\thmnote{ (#3)}\leanmetaheadmark\leanmetaheadtext}
  \newtheoremstyle{leandef}
    {6pt}{6pt}{}{}{\bfseries}{.}{ }
    {\thmname{#1}\thmnumber{ #2}\thmnote{ (#3)}\leanmetaheadmark\leanmetaheadtext}
  \newtheoremstyle{leanremark}
    {6pt}{6pt}{}{}{\bfseries}{.}{ }
    {\thmname{#1}\thmnumber{ #2}\thmnote{ (#3)}\leanmetaheadmark\leanmetaheadtext}
}}
\@ifundefined{thm@thmcaption}{}{\@ifundefined{@opargbegintheorem}{}{%
  \def\@xthm#1#2#3{%
    \@begintheorem{#3}{\csname the#2\endcsname\leanmetaheadmark}%
    \leanmetaheadtext
    \ifx\thm@starredenv\@undefined
      \thm@thmcaption{#1}{{#3}{\csname the#2\endcsname}{}}\fi
    \ignorespaces}
  \def\@ythm#1#2#3[#4]{%
    \expandafter\global\expandafter\def\csname#1name\endcsname{#4}%
    \@opargbegintheorem{#3}{\csname the#2\endcsname}{#4\leanmetaheadmark}%
    \leanmetaheadtext
    \ifx\thm@starredenv\@undefined
      \thm@thmcaption{#1}{{#3}{\csname the#2\endcsname}{#4}}\fi
    \ignorespaces}
}}
\makeatother
\providecommand{\leanmeta}[1]{}
\renewcommand{\leanmeta}[1]{\ifhmode\unskip\fi\footnote{\leanfootnoteformat{\strut\ifleanmetaseen\else Lean: \global\leanmetaseentrue\fi#1\leanfootnoteterminator}}}

\input{content/notation_macros_auto.tex}
\IfFileExists{content/lean_stats.tex}{

}{}
\IfFileExists{theorem_verification.tex}{\newcommand{\machinechecked}[1]{}%
\newcommand{\machinecheckedfromjson}[1]{}%

}{}
\IfFileExists{content/adversarial_coarsening_auto.tex}{

}{}
\IfFileExists{content/protobuf_history_auto.tex}{\newcommand{\ProtoSnapshots}{439}

\newcommand{\ProtoFieldSlots}{4,020}
\newcommand{\ProtoDeletions}{50}

\newcommand{\ProtoNeverReserved}{36}
\newcommand{\ProtoReuses}{8}

\newcommand{\ProtoReuseConflictPrice}{2}

}{}
\IfFileExists{content/typescript_brand_history_auto.tex}{\newcommand{\TSBrandPopulationPackages}{150}

\newcommand{\TSBrandPackagesWithMeanings}{19}
\newcommand{\TSBrandPackagesWithoutMeanings}{131}

\newcommand{\TSBrandPersistentMeanings}{200}

\newcommand{\TSBrandNoRuntimeEvidenceMeanings}{102}

\newcommand{\TSBrandHistoryOnlyPackages}{5}

\newcommand{\TSBrandNormalizedNamePrice}{7}

\newcommand{\TSBrandHistoryAdditionalNamePrice}{4}

}{}
\IfFileExists{content/python_population_results_auto.tex}{

\providecommand{\PythonPopulationProviderIncidences}{12,880}

\providecommand{\PythonPopulationPositiveProviderTax}{18}

\providecommand{\APairs}{12,880}
\providecommand{\PythonPopulationSizeQualified}{65}

\providecommand{\PythonPopulationHeadlineCompactPairedRows}{%
SQLAlchemy & 1724 & 176 & 5813 & 547\textsuperscript{$\dagger$} & google-auth & 152 & 14 & 176 & 11\textsuperscript{$\dagger$} \\
\midrule
redis & 510 & 30 & 755 & 89\textsuperscript{$\dagger$} & GitPython & 81 & 5 & 52 & 8 \\
\midrule
anyio & 174 & 24 & 209 & 21\textsuperscript{$\dagger$} & azure-core & 177 & 15 & 130 & 8 \\
\midrule
fonttools & 807 & 22 & 763 & 14 & aiobotocore & 135 & 5 & 65 & 0 \\
\midrule
azure-identity & 88 & 12 & 76 & 13\textsuperscript{$\dagger$} & pip & 819 & 10 & 329 & 0 \\
}

}{}
\providecommand{\PythonCrossBranchCensusProjects}{31}
\providecommand{\PythonCrossBranchCensusPositive}{24}

\providecommand{\PythonCrossBranchCensusUnscored}{34}

\providecommand{\PythonCrossBranchCensusFrameLowerPercent}{36.9}
\providecommand{\CensusShare}{36.9}
\providecommand{\CensusUnscored}{34}
\providecommand{\APositive}{24}
\providecommand{\AFrame}{65}

\providecommand{\ProviderStatePairs}{356}
\providecommand{\ProviderStateTotal}{739}
\providecommand{\ProviderStatePercent}{48.2}

\providecommand{\StatefulGap}{236}

\providecommand{\StatefulCertificates}{2}

\providecommand{\PublicApiPairs}{9{,}590}
\providecommand{\PublicApiGap}{457}
\providecommand{\PublicApiCertificates}{2}

\providecommand{\EqualScopeGap}{713}

\providecommand{\EqualScopeNewCertificates}{2}
\providecommand{\EqualScopeFirstParentOutside}{8{,}635}
\providecommand{\EqualScopeFirstParentForcedPercent}{8.3}

\IfFileExists{content/python_extraction_strengthening_auto.tex}{

}{}
\IfFileExists{content/external_route_controls_auto.tex}{
\newcommand{\ExternalRouteControlEligible}{25}
\newcommand{\ExternalRouteControlPositive}{3}

}{}
\IfFileExists{content/python_factorization_ablations_auto.tex}{\providecommand{\PythonAblationCases}{20}

\providecommand{\PythonProviderOptimumTotal}{739}
\providecommand{\PythonFixedHierarchyTotal}{8{,}792}
\providecommand{\PythonMethodUpperTotal}{12{,}276}
\providecommand{\PythonMethodIncidences}{95{,}621}
\providecommand{\PythonMethodSolverLowerTotal}{3{,}864}

\providecommand{\AProjects}{20}
\providecommand{\AGap}{739}
\providecommand{\AFirst}{8{,}792}
\providecommand{\PythonFixedHierarchyCost}[1]{%
\csname PythonFixedHierarchyCost:#1\endcsname}
\expandafter\def\csname PythonFixedHierarchyCost:botocore\endcsname{186}
\expandafter\def\csname PythonFixedHierarchyCost:aiobotocore\endcsname{20}
\expandafter\def\csname PythonFixedHierarchyCost:anyio\endcsname{145}
\expandafter\def\csname PythonFixedHierarchyCost:httpx\endcsname{16}
\expandafter\def\csname PythonFixedHierarchyCost:pip\endcsname{158}
\expandafter\def\csname PythonFixedHierarchyCost:google-auth\endcsname{105}
\expandafter\def\csname PythonFixedHierarchyCost:opentelemetry-api\endcsname{21}
\expandafter\def\csname PythonFixedHierarchyCost:SQLAlchemy\endcsname{3{,}591}
\expandafter\def\csname PythonFixedHierarchyCost:openai\endcsname{2{,}309}
\expandafter\def\csname PythonFixedHierarchyCost:openpyxl\endcsname{479}
\expandafter\def\csname PythonFixedHierarchyCost:awscli\endcsname{446}
\expandafter\def\csname PythonFixedHierarchyCost:oauthlib\endcsname{94}
\expandafter\def\csname PythonFixedHierarchyCost:networkx\endcsname{81}
\expandafter\def\csname PythonFixedHierarchyCost:redis\endcsname{540}
\expandafter\def\csname PythonFixedHierarchyCost:Werkzeug\endcsname{55}
\expandafter\def\csname PythonFixedHierarchyCost:GitPython\endcsname{21}
\expandafter\def\csname PythonFixedHierarchyCost:azure-core\endcsname{67}
\expandafter\def\csname PythonFixedHierarchyCost:azure-identity\endcsname{48}
\expandafter\def\csname PythonFixedHierarchyCost:fonttools\endcsname{388}
\expandafter\def\csname PythonFixedHierarchyCost:gitdb\endcsname{22}

}{}
\IfFileExists{content/python_degree_null_auto.tex}{\providecommand{\PythonNullSamples}{20}
\providecommand{\PythonNullObservedTotal}{739}
\providecommand{\PythonNullTotalMinimum}{1,522}

\providecommand{\PythonNullTotalMaximum}{1,568}

}{}
\IfFileExists{content/python_overreach_table_auto.tex}{

\expandafter\def\csname PythonOverreachAggregateT0\endcsname{739--739}
\expandafter\def\csname PythonOverreachSQLAT0\endcsname{520--520}
\expandafter\def\csname PythonOverreachAggregateT1\endcsname{390--673}
\expandafter\def\csname PythonOverreachSQLAT1\endcsname{219--500}

\expandafter\def\csname PythonOverreachAggregateT2\endcsname{308--632}
\expandafter\def\csname PythonOverreachSQLAT2\endcsname{172--483}
\expandafter\def\csname PythonOverreachAggregateT5\endcsname{150--608}
\expandafter\def\csname PythonOverreachSQLAT5\endcsname{75--483}
\providecommand{\PythonOverreachAggregate}[1]{%
\csname PythonOverreachAggregateT#1\endcsname}
\providecommand{\PythonOverreachSQLA}[1]{%
\csname PythonOverreachSQLAT#1\endcsname}
}{}
\IfFileExists{content/python_blocking_table_auto.tex}{
\providecommand{\PythonPaidBlockingAggregate}{646--718}
\providecommand{\PythonPaidBlockingForwarded}{686}
\providecommand{\PythonPaidBlockingBlocked}{32}

}{}
\IfFileExists{content/sqlalchemy_component_auto.tex}{
\providecommand{\SQLAlchemyComponentProviderIncidences}{4,831}
\providecommand{\SQLAlchemyComponentTax}{520}

\providecommand{\SQLAlchemyComponentClosedCThreeProviderIncidences}{1,173}
\providecommand{\SQLAlchemyComponentClosedCThreeSelectorConflicts}{71}
\providecommand{\SQLAlchemyComponentClosedCThreeTax}{139}

\providecommand{\SQLAlchemyComponentOmittedMroSensitiveLookups}{536}

\providecommand{\SQLAlchemyComponentBooleanRank}{59}

\providecommand{\SQLAlchemyComponentCrossingOnlyA}{71}
\providecommand{\SQLAlchemyComponentCrossingOnlyB}{61}
\providecommand{\SQLAlchemyComponentCrossingOnlyC}{17}
\providecommand{\SQLAlchemyComponentCrossingOnlyAB}{0}
\providecommand{\SQLAlchemyComponentCrossingOnlyAC}{0}
\providecommand{\SQLAlchemyComponentCrossingOnlyBC}{0}
\providecommand{\SQLAlchemyComponentCrossingAll}{53}

}{}
\IfFileExists{content/python_longitudinal_completion_auto.tex}{\providecommand{\PythonLongitudinalSnapshots}{6}
\providecommand{\PythonLongitudinalTransitions}{5}

\providecommand{\PythonLongitudinalCompletionEdits}{511}
\providecommand{\PythonLongitudinalStableCompletionChurn}{94}

\providecommand{\PythonLongitudinalFirstGap}{511}
\providecommand{\PythonLongitudinalLastGap}{547}

}{}

\newcommand{\StrictInsertionPipSourceClassSnapshots}{8}
\newcommand{\StrictInsertionPipSourceClassTransitions}{7}

\newcommand{\StrictInsertionAiobotocoreSourceClassSnapshots}{22}
\newcommand{\StrictInsertionAiobotocoreSourceClassTransitions}{21}

\IfFileExists{content/independent_optimality_auto.tex}{\providecommand{\IndependentGapCertificateCount}{7}

\providecommand{\IndependentHolderMinimum}{59}
\providecommand{\IndependentHolderExcludedBound}{58}

}{}
\IfFileExists{content/practical_impact_auto.tex}{

}{}
\IfFileExists{content/anyio_mutation_auto.tex}{
\newcommand{\MutationDenominator}{21}

\newcommand{\MutationDetected}{21}

\newcommand{\MutationBaselinePassed}{3{,}934}

\newcommand{\MutationPairDetected}{210}
\newcommand{\MutationPairBothMarkers}{169}

}{}
\newcommand{\AnyIOApplicationPairs}{4}
\newcommand{\AnyIOApplicationGap}{1}
\newcommand{\AnyIOApplicationUpstreamPassed}{3{,}934}

\providecommand{\CPythonAssignmentExpressionCount}{140}

\newcommand{\PaperOneForwardingRecallDetectedPairs}{21}

\IfFileExists{build_submission_hook.tex}{
\ifdefined\LeanHandleHookLoaded\else
\global\let\LeanHandleHookLoaded\relax
\AtBeginDocument{}
\fi
}{}

\AtBeginDocument{%
  \let\PaperFormalLeanMeta\leanmeta
  \renewcommand{\leanmeta}[1]{\PaperRecordClaimLocation{#1}}%
}
\newcommand{\PaperFormalProofNotes}[1]{%
  \AddToHook{env/#1/begin}{\let\leanmeta\PaperFormalLeanMeta}%
}
\forcsvlist{\PaperFormalProofNotes}{theorem,proposition,lemma,corollary,definition,claim,proof}
\makeatletter
\AtBeginDocument{%
  \apptocmd{\thmhead}{\leanmetaheadmark\leanmetaheadtext}{}%
    {\PackageError{paper-proof-notes}{Cannot attach theorem proof anchors}{Check the theorem heading definition.}}%
}
\makeatother

\renewcommand\footnotetextcopyrightpermission[1]{}
\acmJournal{PACMPL}
\acmVolume{1}
\acmNumber{OOPSLA}
\acmArticle{1}
\csname acmYear\endcsname{2027}
\acmMonth{10}
\acmDOI{}
\acmISBN{}
\acmConference[OOPSLA 2027]{ACM SIGPLAN International Conference on Object-Oriented
  Programming, Systems, Languages, and Applications}{October 10--15, 2027}{Prague, Czechia}

\providecommand{\PaperTitleAuto}{Semantic Completeness for Correct Maintenance: A Strict Hierarchy of Typing and Inheritance}

\title{\PaperTitleAuto}

\PaperPublicationAuthor{acmart}

\begin{document}
\global\leanmetaseentrue

\begin{abstract}
Removing nominal types or inheritance is often defended as reducing developer burden. The cited Java and Go rationales give no objective measure. Correct maintenance minimizes required facts left for code or tooling to supply and repairs forced by changes, while preserving required facts and behavioral correctness as a program's intended behavior evolves. Semantically complete languages derive all required type identities, interface memberships, and implementation--class pairs as finite requirements evolve. Incompleteness cannot guarantee the optimum. It forces slop: code restating facts that complete languages derive natively once.

In \AProjects\ pinned multiple-inheritance Python projects, recovered pairs grouped by declaring class would need code or tooling: \APairs\ without inheritance, \AFirst\ under first parents, \AGap\ under one parent inheriting only required pairs (certified exact), 0 with multiple parents. With code and state in each provider, state-bearing requirements force \StatefulGap\ outside pairs. Java and C\# interface defaults supply no instance state. Of the top 150 pure-Python downloads, at least \APositive\ of \AFrame\ size-qualified wheels have positive recovered gap (\CensusShare\%, \CensusUnscored\ unscored).

For independent providers, the minimum covers reparenting, infinite helpers, delegation, and generators. One parent covers all pairs exactly when providers' class sets nest or do not overlap; likewise for contracts. Look-alike records needing different handling need as many keys as the confusability graph's chromatic number. As requirements grow and stay tree-shaped, inserting a provider outside the old ancestor path strictly between ancestors forces one-parent repair. Our zero-gap construction repeats it; multiple parents add one edge per change. Bans cannot lower minima at fixed design costs; savings require otherwise choosing worse than the best.
\end{abstract}



\maketitle

\section{Introduction}\label{sec:introduction}
Languages differ in their built-in support for types, interfaces, and shared code.
Removing mechanisms such as nominal types or inheritance is often defended as reducing
developer burden. We objectively count the required facts left for code or tooling to supply
and the repairs forced to preserve behavioral correctness as intended behavior
evolves. We prove which counts mechanism removal makes irreducible. As functionality
is added, the program must preserve its existing required behavior.

SWEBOK states the goal: cost-effective support and integrity-preserving
modification~\cite{SWEBOK2026}. We prove the fewest required facts code or tooling
must supply to preserve that integrity.

If another language feature always does the same job at no extra cost, removal leaves the best cost unchanged
(Lemma~\ref{lem:capacity-optimum-order}). Java links feature removal
to reduced programmer burden and Go credits implicit interfaces with less
bookkeeping~\cite{JavaLanguageEnvironment,GoFAQ}.
Meanwhile Java~8, C\#~8, and
Swift~2 added stateless multiple method providers, and PHP~5.4 added traits
that also carry properties~\cite{JavaDefaultMethods,CSharpVersionHistory,
SwiftXcode7ReleaseNotes,PHPTraits,PHP54Features}.

\begin{figure}[t]
\centering
\begingroup\PaperLadderDiagramSetup
\begin{tikzpicture}[scale=.90,transform shape,x=.99cm,
  font=\scriptsize,
  rung/.style={draw,rounded corners=2pt,align=center,minimum width=2.20cm,
    minimum height=1.20cm,inner sep=3pt},
  connector/.style={font=\fontsize{15pt}{16pt}\selectfont,inner xsep=0pt,inner ysep=.5pt},
  ban/.style={font=\tiny,align=center},
  banarrow/.style={font=\tiny}]
  
    \tikzset{rung/.append style={text width=2.45cm,minimum height=1.5cm},
      ban/.append style={yshift=3.5mm},
      banarrow/.append style={yshift=3.5mm},
      connector/.append style={font=\fontsize{15pt}{16pt}\selectfont}}
  
  \node[rung,fill=gray!12] (structural) at (0,0)
    {\textbf{Structural}\\\textbf{only}\\shape};
  \node[rung,fill=blue!12] (nominal) at (3.18,0)
    {\textbf{One contract-}\\\textbf{parent chain}\\stable identity};
  \node[rung,fill=blue!28] (contracts) at (6.36,0)
    {\textbf{Multiple}\\\textbf{contracts}\\all memberships};
  \node[rung,fill=green!12] (single) at (9.54,0)
    {\textbf{Providers on}\\\textbf{one chain}\\per class};
  \node[rung,fill=green!28] (multiple) at (12.72,0)
    {\textbf{Multiple impl.}\\\textbf{parents}\\every relation};
  \path (structural.east) -- node[connector] {$\subsetneq$} (nominal.west);
  \path (nominal.east) -- node[connector] {$\subsetneq$} (contracts.west);
  \path (contracts.east) -- node[connector] {$\subsetneq$} (single.west);
  \path (single.east) -- node[connector] {$\subsetneq$} (multiple.west);
  \node[banarrow] at (1.12,1.00) {$\leftarrow$};
  \node[ban,anchor=west] at (1.29,1.00) {ban nominal\\identity};
  \node[banarrow] at (4.30,1.00) {$\leftarrow$};
  \node[ban,anchor=west] at (4.47,1.00) {ban multiple\\contracts};
  \node[banarrow] at (7.48,1.00) {$\leftarrow$};
  \node[ban,anchor=west] at (7.65,1.00) {ban impl.\\inheritance};
  \node[banarrow] at (10.66,1.00) {$\leftarrow$};
  \node[ban,anchor=west] at (10.83,1.00) {ban incomparable\\impl. providers};
\end{tikzpicture}
\endgroup
\caption{Native capacities form one strict chain (arrows: bans).
Fills: gray, shape; blue, contracts; green, implementation; darker, multiple parents.
Languages may sit off-chain; AnyIO witnesses both parent steps (\S4).}
\Description{A five-rung ladder runs from structural typing through nominal
contracts, multiple contract inheritance, implementation providers on one ancestry chain per class, and multiple implementation inheritance. Gray marks shape, blue contracts, and green implementation; darker shades mark multiple parents. Each level derives more required answers.}
\label{fig:native-capacity-ladder}
\end{figure}

A \emph{code provider}, or provider, is the class declaring an implementation.
A class combines its own members with the union of its parents' member sets~\cite{cook1994denotational};
with one parent, it inherits only that parent's members. Precedence resolves same-name conflicts
(Theorem~\ref{cor:coverage-resolution}). Read from the provider's side, the same relation groups the consumer classes needing its code.
In Figure~\ref{fig:workflow-crossing}, AnyIO's file provider supplies the writer
and file reader; its iteration provider supplies the file and buffered readers.
The groups overlap at the file reader, but neither contains the other: a \emph{crossing}.
One parent puts a class's providers on one ancestor chain. It supplies every required
pair only when their class groups fit a tree. \emph{Multiple inheritance} supplies
code, descriptors, or state from incomparable parents and derives crossings.
Adding functionality can still force extra changes to inheritance or forwarding,
even when each version fits a tree.
A parent declaration brings its members into the child. A field stores an object;
code or tooling exposes its needed members on the containing object
(Figure~\ref{fig:storage-encodings}).

\begin{figure}[!t]
\begin{minipage}[c]{0.62\linewidth}
\begin{lstlisting}[language=Python,basicstyle=\ttfamily\scriptsize,frame=single,
  aboveskip=0pt,belowskip=0pt]
# Selected provider bases only.
# Bodies and other bases omitted.
class FileWriteStream(_BaseFileStream): ...
class FileReadStream(_BaseFileStream,
                     ByteReceiveStream): ...
class BufferedByteReceiveStream(ByteReceiveStream): ...
\end{lstlisting}
\end{minipage}\hfill
\begin{minipage}[c]{0.35\linewidth}
\centering
\scriptsize
\begin{tabular}{@{}lccc@{}}
\toprule
 & $W$ & $\PaperFileReader$ & $B$ \\
\midrule
$\mathsf{file}$ & 1 & 1 & 0 \\
\midrule
$\mathsf{iteration}$ & 0 & 1 & 1 \\
\bottomrule
\end{tabular}
\end{minipage}
\par
\captionsetup{skip=2pt}
\caption{AnyIO crossing (selected providers)~\cite{AnyIOFileStreams}.
The file writer ($W$) needs file lifecycle; buffered reader ($B$), iteration;
file reader ($\PaperFileReader$), both. Neither group contains the other.}
\Description{Selected AnyIO provider bases and a table show file lifecycle
required by the writer and file reader, and byte iteration by the file reader
and buffered reader. The groups overlap only at the file reader.}
\label{fig:workflow-crossing}
\end{figure}

\begin{figure}[htbp]
\centering
\PaperDenseTableSetup
\providecommand{\PaperCodeContinuation}[1]{}
\begin{tabularx}{\linewidth}{@{}
  >{\hsize=.50\hsize\linewidth=\hsize}X
  >{\hsize=1.20\hsize\linewidth=\hsize}X
  >{\hsize=1.40\hsize\linewidth=\hsize}X
  >{\hsize=.90\hsize\linewidth=\hsize}X@{}}
\begin{tabular}[t]{@{}l@{}}
\textrm{\bfseries monolithic} \\
class MonoA: \\
\quad attr1 = 1 \\
\quad attr2 = 2
\end{tabular}
&
\begin{tabular}[t]{@{}l@{}}
\textrm{\bfseries multi-parent} \\
class MixA: attr1 = 1 \\
class MixB: attr2 = 2 \\
class NativeA(\PaperCodeContinuation{\quad}MixA, MixB):\PaperCodeContinuation{\quad} pass
\end{tabular}
&
\begin{tabular}[t]{@{}l@{}}
\textrm{\bfseries fields + forwarding} \\
class RoutedA: \\
\quad def \_\_init\_\_(s): \\
\quad\quad s.\_a=MixA() \\
\quad\quad s.\_b=MixB() \\
\quad attr1=property(\PaperCodeContinuation{\quad\quad}lambda s:s.\_a.attr1) \\
\quad attr2=property(\PaperCodeContinuation{\quad\quad}lambda s:s.\_b.attr2)
\end{tabular}
&
\begin{tabular}[t]{@{}l@{}}
\textrm{\bfseries fields + assignment} \\
class AssignedA: \\
\quad def \_\_init\_\_(s): \\
\quad\quad s.\_a=MixA() \\
\quad\quad s.\_b=MixB() \\
\quad\quad s.attr1=s.\_a.attr1 \\
\quad\quad s.attr2=s.\_b.attr2
\end{tabular}
\end{tabularx}
\caption{Same shown outputs, different facts for code to maintain as providers evolve.
Inheritance derives both pairs; fields need routes. The monolith
loses \texttt{MixA} and \texttt{MixB}'s required independent identities.}
\Description{Four short Python encodings place the same two snapshot
attributes in one class, two inherited mixins, two contained provider objects
with property forwarding, or two contained provider objects with constructor
assignments.}
\label{fig:storage-encodings}
\end{figure}

\paragraph{A checked local refactoring.}
Figure~\ref{fig:workflow-crossing}'s writer must not gain iteration; its memory-backed
buffered reader must not gain file lifecycle. This local relation has \AnyIOApplicationPairs{} pairs;
every sound one-parent hierarchy (only required pairs) leaves at least \AnyIOApplicationGap{} outside;
the best leaves exactly \AnyIOApplicationGap{}.
Section~\ref{sec:motivation} measures AnyIO's whole-project requirements.
The file reader inherits \texttt{ByteReceiveStream} and forwards file lifecycle
to \texttt{\_BaseFileStream}, retaining membership. Checks replace provider bodies
before construction; forwarding uses the replacements. Names and descriptor kinds
stay fixed, with no cooperative-super extensions. \AnyIOApplicationUpstreamPassed{}
upstream tests pass. The outside pair needs \texttt{\_\_init\_\_}, \texttt{aclose}, and
\texttt{extra\_attributes}: eight ($2^3$) ways to keep or omit them, all seven
incomplete ones failing the read/metadata/close client.
Supplement~\S\ref{supp-anyio-refactoring-application} gives the
contract, executable refactoring, source checks, and test limits.

Workarounds can preserve required facts and behavioral correctness. At a crossing,
sound one-parent workarounds still supply at least
$\PaperNotationLink{PaperGap}{d_\downarrow}>0$ outside pairs and fail correct maintenance
(Theorem~\ref{thm:delegation-lower-bound}).\footnote{Symbols link to the notation appendix, with plain-language readings
and AnyIO values; the supplement opens with the same table.
Proof codes identify Lean declarations listed in Supplement~\S\S15--16.}
We count required facts: extra identity names, class--contract memberships, and
implementation--class pairs. Each \emph{required answer} must be supplied.
Lines measure encoding length; the gap counts facts even the best one-parent
hierarchy leaves to code or tooling. Price outside work however you like.
Multiple parents remain strictly cheaper on every crossing exactly when doing
any outside work costs more than doing none. Bundling and bulk discounts do
not change that test (Supplement~Theorem~\ref{supp-thm:accounting-characterization}).
Unit charges give exact counts (Proposition~\ref{prop:forced-semantic-unit}).
\leanmeta{\LHrng{ACF}{1}{4}}
Batching shares an operation's positive cost across facts still requiring supply
(Corollary~\ref{cor:factorization-economics}).

\emph{Slop} is code restating required facts that a semantically complete
language derives natively from their original declarations. Those declarations need stating only once.
Slop is the boilerplate compensating for incompleteness. We count the restated facts.
Under sound one-parent ancestry,
Cunningham's supply debt~\cite{cunningham1992wycash,ernst2015measure}
splits into $\PaperNotationLink{PaperGap}{d_\downarrow}$ required pairs code or tooling supplies even under the best
hierarchy, and $\PaperNotationLink{PaperExcess}{e_H}$ extra pairs that hierarchy refactoring removes
(Theorems~\ref{thm:delegation-lower-bound}, \ref{thm:semantic-completion-tax}).
Like garbage in a collected heap, slop is determined by the current program and mechanism, whatever its origin.
SlopCodeBench's agent-code verbosity and
structural-erosion measures are heuristic~\cite{orlanski2026slopcodebench}; our
fact and repair counts have exact definitions.

Code or generators must encode the missing facts.
Packing four values into Figure~\ref{fig:arity-cap-completion}'s three parameters
still leaves a fourth binding. The compiler may derive it, or the body must unpack it.
With ancestry and generator fixed, $M$ possible sets of missing required facts
need $\lceil\log_2 M\rceil$ input bits regardless of source length
(Proposition~\ref{prop:tooling-completion-law}).
\leanmeta{\LH{ARI1}}

\begin{figure}[t]
\begin{minipage}[c]{0.64\linewidth}
\centering
\begingroup\PaperArityDiagramSetup
\pgfmathsetlengthmacro{\PaperCompactArityXUnit}{0.75*\PaperArityXUnit}
\begin{tikzpicture}[
  x=\PaperCompactArityXUnit,y=\PaperArityYUnit,
  every node/.style={font=\scriptsize},
  box/.style={draw,rounded corners=2pt,align=center,inner sep=2.4pt},
  arrow/.style={->,>=stealth,thick},
  outside/.style={->,>=stealth,thick,dashed,orange!85!black}
]
  \node[font=\scriptsize\bfseries] at (1.79,1.82)
    {(a) Four parameters};
  \node[box,fill=green!7] (a1) at (0.48,1.05) {$a$};
  \node[box,fill=green!7] (b1) at (1.35,1.05) {$b$};
  \node[box,fill=green!7] (c1) at (2.22,1.05) {$c$};
  \node[box,fill=green!7] (d1) at (3.09,1.05) {$d$};
  \node[box] (body1) at (1.79,0.28) {$f(a,b,c,d)$};
  \draw[arrow] (a1)--(body1);
  \draw[arrow] (b1)--(body1);
  \draw[arrow] (c1)--(body1);
  \draw[arrow] (d1)--(body1);
  \node[font=\tiny,text=green!45!black] at (1.79,-0.13)
    {four bindings direct; $0$ outside};

  \node[font=\scriptsize\bfseries] at (7.56,1.82)
    {(b) Three parameters plus unpacking};
  \node[box,fill=blue!6] (a2) at (6.30,1.05) {$a$};
  \node[box,fill=blue!6] (b2) at (7.20,1.05) {$b$};
  \node[box,fill=blue!6] (p2) at (8.68,1.05) {$p=(c,d)$};
  \node[box,align=center] (body2) at (7.56,0.10)
    {$f(a,b,p)$\\{\tiny\textcolor{orange!70!black}{body unpacks:
    $c=\pi_1(p)$,\ $d=\pi_2(p)$}}};
  \draw[arrow] (a2)--(body2);
  \draw[arrow] (b2)--(body2);
  \draw[arrow] (p2)--(body2);
  \node[font=\tiny,text=orange!70!black,align=center,inner sep=1pt]
    at (7.56,-0.55)
    {same behavior; at least one binding outside};
\end{tikzpicture}
\endgroup
\end{minipage}\hfill
\begin{minipage}[c]{0.33\linewidth}
\caption{A parameter limit illustrates the answer-count unit. Four bindings fit
four parameters; three leave at least one outside. Tuple construction and projection
restore it automatically if compiler-derived; otherwise, the developer supplies it.}
\Description{The left panel shows four parameters a, b, c, and d connected
directly to a function body, with zero bindings outside the parameter list.  The right
panel shows a three-parameter form with a, b, and a packed parameter p containing
c and d are packed together, and the function body unpacks p by projection.
Behavior is preserved while at least one required
binding leaves the parameter list and compiler-defined tuple
operations supply it; absent them, the developer must
maintain it manually.}
\label{fig:arity-cap-completion}
\end{minipage}
\end{figure}

Deleting a field can make separately declared types look identical while needing
different handling. Sharing a representation requires identical handling after every
supported change. Otherwise retain a distinguishing key, update references to
replacement representations, or enforce a precondition excluding collisions.
No helper using the identical record recovers the distinction (Section~3).
\leanmeta{\LHrng{IDD}{1}{5}, \LH{NHF11}, \LHrng{EVO}{26}{30}}
Contracts state guarantees a class must satisfy; interfaces are a familiar
way to declare contracts. Contract membership supplies no code.

Preserving required facts means decoding each case to its specified answer.
Exact supply $\PaperNotationLink{PaperSupplied}{R_D}=\PaperNotationLink{PaperRequired}{R^{\mathrm{req}}}$ means no missing or unrequired pairs.
Figure~\ref{fig:native-capacity-ladder}'s strict witnesses leave some facts outside
under the lower rung and none under the upper.
Witsenhausen forces the key minimum
\cite{witsenhausen1976zero}; tree-shaped sets force the membership and ancestry
minima (Sections~3--4).

\paragraph{Correct maintenance.}
Correct maintenance minimizes the required facts left for code or tooling to supply
and the repairs forced by changes, while preserving required facts and behavioral
correctness as a program's intended behavior evolves.
With nominal identities fixed, multiple contract and implementation parents leave zero
outside supply and repair counts through any finite sequence of code and interface requirements.
Each lower rung fails on a named witness (Section~4).

A \emph{semantically complete language} derives all required type identities,
interface memberships, and implementation--class pairs as finite program requirements evolve
(Figure~\ref{fig:native-capacity-ladder}'s top rung). A ban or technical limitation can
make it \emph{incomplete}. Behavioral correctness alone does not establish correct
maintenance: crossings force outside supply even in the best sound one-parent hierarchy.
Pairs still supplied by workarounds count at each update, even if their code stays unchanged.
Bugfixes need not change $\PaperNotationLink{PaperRequired}{R^{\mathrm{req}}}$ (Proposition~\ref{prop:forced-semantic-unit}).
\leanmeta{\LHrng{SCT}{46}{51}, \LH{BWR52}}

The \emph{ancestry gap} $\PaperGap$ is the fewest outside pairs under sound one-parent
ancestry (only required pairs), keeping independently changing code providers distinct.
Reparenting, unlimited helpers, replication, delegation, and tooling cannot lower it
(Theorems~\ref{thm:delegation-lower-bound}, \ref{thm:replica-completion-law};
Proposition~\ref{prop:tooling-completion-law}). Positive removal charges keep crossings
costly; charges at least $\PaperNotationLink{PaperGap}{d_\downarrow}$ recover the gap (Theorem~\ref{thm:paid-blocking-gap}).
Multiple parents leave zero
outside implementation pairs: Pareto dominance on these counts (Corollary~\ref{cor:strict-derivation-dominance}).
Tree-shaped futures cannot be certified exactly (Theorem~\ref{thm:open-world-laminarity})
and strict insertions force repair (Theorem~\ref{thm:laminar-growth-reparenting}).
If adoption costs less per update than the workaround, savings eventually repay finite setup
(Corollary~\ref{cor:factorization-economics}).

No rule reading only successful outputs and ancestry can distinguish zero from
positive gap. Reading ancestry, removals, and outside routes recovers $\PaperNotationLink{PaperRequired}{R^{\mathrm{req}}}$ exactly
(Corollary~\ref{cor:exact-one-parent-dichotomy}).
\leanmeta{\LHrng{RCN}{1}{11}, \LHrng{PJD}{1}{4}, \LH{NHF63}, \LHrng{IPE}{21}{28}, \LHrng{SIP}{1}{7}, \LHrng{OLC}{1}{5}}

\leanmetapending{\LH{OPT13}, \LHrng{OPT}{17}{19}}
If a ban removes choices without changing each remaining design's cost, it cannot
make the cheapest choice cheaper, even when adding costs across projects. Savings
require people otherwise to choose a more expensive design; the cited rationales
do not measure those choices (Section~6.5).
\begin{lemma}[Capacity containment orders the maintenance optimum]
\label{lem:capacity-optimum-order}
Fix requirements, histories, and cost. Designs in this comparison preserve required
answers and behavior. Enlarging their set never
raises its minimum. For attained minima, removal preserves it iff an optimum
remains, and strictly raises it otherwise. A dominated feature has a retained
answer- and behavior-preserving replacement costing no more per design using it, so removal preserves
the minimum. Apply coordinatewise.
\end{lemma}
Each left-to-right step in Figure~\ref{fig:native-capacity-ladder} derives more required facts;
in reverse, it is a language or project-policy ban. Adjacent rungs can tie today;
an allowed change can leave more facts for code or tooling under the lower rung.

\leanmeta{\LH{PD1}, \LH{NOM1}, \LH{ENC1}, \LHrng{SCT}{1}{4}, \LHrng{SCT}{20}{26}, \LHrng{SCT}{30}{36}, \LH{BWR43}, \LH{BWR54}, \LHrng{RBF}{1}{2}, \LH{FDC2}, \LH{CAA7}}

\begin{samepage}
For requirements recovered from \PythonAblationCases{} pinned Python projects, code or tooling would supply
\PythonPopulationProviderIncidences{} pairs without implementation inheritance,
\PythonFixedHierarchyTotal{} under declared first parents (91.6\% removable excess),
\PythonProviderOptimumTotal{} under the best sound one-parent hierarchy (certified), and none
with multiple parents (Proposition~\ref{prop:derivation-capacity};
measurements: Section~\ref{sec:motivation}); pip and aiobotocore have gap zero.
Unknown cases stay unscored; these are exact counts for the recovered requirements.

\end{samepage}

\paragraph{Contributions and reader map.}
\begin{enumerate}[label=(\arabic*),leftmargin=*,nosep,start=0]
\item \textbf{Five-rung strict hierarchy (Figure~\ref{fig:native-capacity-ladder}; Lemma~\ref{lem:capacity-optimum-order}).}
Colliding meanings, crossing memberships, nonempty implementation demand, and crossing
providers witness its four strict steps. Theorem~\ref{thm:future-conflict} gives the
identity-key minimum across histories.
\item \textbf{Minimum outside supply (Theorem~\ref{thm:delegation-lower-bound}).}
For distinct providers on one sound ancestor chain per class, the gap is the
attained minimum even with infinite helpers, delegation, or generators.
Providers sharing a class are comparable,
forcing class sets to nest or not overlap.
\item \textbf{Repair across updates (Theorem~\ref{thm:laminar-growth-reparenting}).}
New functionality can require a provider absent from a class's ancestry. Every strict
insertion forces one parent to delete an edge carrying a still-required pair or use
forwarding or removal before or after the change. Our program evolution repeats
this insertion with gap zero in every version; multiple parents add one edge per change.
Standing one-parent routes need not change.
\item \textbf{Conflicts survive refactoring (Theorem~\ref{cor:coverage-resolution}).}
With fixed member names, every answer-preserving design pays the same $k$ specified lookup
choices; sound one-parent ancestry adds at least $\PaperNotationLink{PaperGap}{d_\downarrow}$ outside pairs.
\item \textbf{Measured minima and omission tests (Sections~3 and~5).}
The installable analyzer returns bounds and checked designs.
\end{enumerate}
Theorem~\ref{thm:behavioral-applicability-product} separates behavioral validity from requirement choice;
Section~6 prices adoption. Supplement~\S\S\ref{supp-audit-detail}, \ref{supp-scope-ledger}
give certificates and distinguish answers, operations, and costs;
Table~\ref{tab:novelty-lifts} separates foundations from extensions.

\section{Representations Must Preserve Every Required Fact}\label{sec:model}\label{sec:separation}
Fix requirements before choosing a representation. Behavioral correctness means
every reachable behavior satisfies the program's requirements:
$\PaperNotationLink{PaperSafety}{\operatorname{Reach}_t(D)\subseteq\operatorname{Valid}_t}$
\cite{hoare1969axiomatic,back1998refinement}; see Supplement~\S\ref{supp-atomic-native-completion}
for missing and unwanted pairs $\PaperNotationLink{PaperRepairSet}{\mathcal D(H,R^{\mathrm{req}})}$. If the required answer is the
full set of permitted outputs, returning one permitted output does not answer that
question; executing one output is a different requirement.

Let $\PaperInput$ contain every interface input, $\PaperAnswerSpace$ the answers
(including output sets), and $a:X\to\PaperAnswerSpace$ the specification.
Encoding $e:X\to\PaperEncodingSpace$ with decoder $r:\PaperEncodingSpace\to\PaperAnswerSpace$
preserves required answers iff $r(e(x))=\PaperNotationLink{PaperAnswer}{a(x)}$ for every $x$.
The program presents $e(x)$; the decoder must recover $\PaperNotationLink{PaperAnswer}{a(x)}$.
Refinement decides whether such a decoder exists.
Swapping required and forbidden cases keeps their groups but changes permitted behavior.
\leanmeta{\LHrng{INT}{1}{7}}

For schemas, field deletion can make separately declared types look identical
while requiring different interpretations. $X=\PaperNotationLink{PaperClasses}{C}$ tracks meanings across edits;
$e$ is the visible record plus any retained identity key.
Write $p:\PaperNotationLink{PaperClasses}{C}\to\mathcal V$ for the current record. A decoder from $p$ exists on every
reachable value exactly when same-looking meanings need the same answer:
$p(x)=p(y)\Longrightarrow \PaperNotationLink{PaperAnswer}{a(x)}=a(y)$.
This characterizes safe erasure.
\leanmeta{\LHrng{BWR}{3}{4}}

A helper computed from $p$ cannot distinguish same-looking meanings; a nominal key can.
\leanmeta{\LHrng{IDD}{1}{7}, \LHrng{NHF}{3}{4}, \LH{NHF11}}

Only reachable states count. Leaked information or undeclared side channels
change what the decoder can inspect.
\leanmeta{\LHrng{PJD}{1}{4}}

A retained identity key $\sigma:\PaperNotationLink{PaperClasses}{C}\to T$ makes required answers recoverable when
$p(x)=p(y)\wedge\sigma(x)=\sigma(y)\Longrightarrow \PaperNotationLink{PaperAnswer}{a(x)}=a(y)$.
A key can outlive the distinguishing fields; it separates cases that $p$ makes
alike but $a$ distinguishes.

\Needspace{3\baselineskip}
For inheritance, $\PaperNotationLink{PaperRequired}{R_t^{\mathrm{req}}}$ lists the required implementation--class pairs, and
$\PaperNotationLink{PaperSupplied}{R_D}$ lists those supplied by ancestry or explicit routes.  Soundness requires
$\PaperNotationLink{PaperSupplied}{R_D}\subseteq \PaperNotationLink{PaperRequired}{R_t^{\mathrm{req}}}$; completeness requires $\PaperNotationLink{PaperRequired}{R_t^{\mathrm{req}}}\subseteq \PaperNotationLink{PaperSupplied}{R_D}$.
Exact supply requires both, so $\PaperNotationLink{PaperSupplied}{R_D}=\PaperNotationLink{PaperRequired}{R_t^{\mathrm{req}}}$: every required pair is supplied and
every forbidden pair is absent.

\paragraph{Correct maintenance.}
Let $\PaperOutsideAtState$ count facts left to code or tooling at $t$. Take coordinatewise
infima over design histories preserving required answers and behavioral correctness,
$D\in\mathcal D_{\mathrm{complete}}(\mathcal H)$:
\[
 \PaperMaintenanceIdeal=
 \bigl((\inf_D \PaperNotationLink{PaperOutsideAtState}{O_t(D)})_t,\ \inf_D \PaperNotationLink{PaperBurden}{B_{\mathrm{rep}}(D)}\bigr).
\]
Mechanism $m$ attains correct maintenance iff some $D\in\mathcal D_m(\mathcal H)$ reaches this ideal.
Behavioral correctness and exact supply are prerequisites; correct maintenance also minimizes the counts.

Every rung satisfies $\mathcal D_m(\mathcal H)\subseteq\mathcal D_{\mathrm{complete}}(\mathcal H)$:
the complete mechanism retains forwarding, registration, and generation.
Thus the ideal compares designs across the whole hierarchy
(Lemma~\ref{lem:capacity-optimum-order}). Repair obligations form one sum, $\PaperNotationLink{PaperBurden}{B_{\mathrm{rep}}}$:
parent-edge deletions carrying still-required pairs, plus missing or removed pairs
before and after each update. The history fixes changes to required facts.
\leanmeta{\LHrng{INT}{1}{7}, \LH{BPI1}, \LHrng{BWR}{37}{39}, \LH{BWR57}, \LHrng{OPT}{21}{22}}
Complete supply attains zero on every finite declared supply history, including removals;
lower supply rungs fail on single-version examples (Corollary~\ref{cor:strict-derivation-dominance};
Supplement~\S\ref{supp-monotone-multiparent-attainment}).
\leanmeta{\LH{SCT9}, \LHrng{MGH}{9}{11}, \LHrng{UMS}{1}{5}}

Every representation bound fixes an arbitrary finite declared relation before choosing a mechanism.
Reshaping requirements changes the specification (Corollary~\ref{cor:specification-or-payment});
no algorithm certifies all and only tree-shaped futures (Theorem~\ref{thm:open-world-laminarity}),
and strict insertion between tree-shaped versions forces one-parent repair
(Theorem~\ref{thm:laminar-growth-reparenting}(0)). Name each
required fact's declaration, required and forbidden cases, and independent changes.
Unknown or external cases stay unscored; including them requires recomputing the
relation and gap. Section~5 recovers pinned-source relations.
\leanmeta{\LH{SCT2}, \LH{SCT32}, \LH{SDN1}, \LHrng{IPE}{21}{28}, \LHrng{BWR}{43}{44}, \LH{BWR54}}

Refactorings preserve every required fact on every reachable input.
Identical fields or base lists today do not justify merging declarations. Merge
only when no required fact distinguishes them through any supported change. Otherwise,
share agreeing implementations and keep identities distinct
(Theorem~\ref{thm:exact-semantic-factoring}); changing a required fact changes the
specification (Corollary~\ref{cor:specification-or-payment}). Ruling out missing connections
needs the complete map. Untested reachable states can still fail.
Naming costs and ancestry gaps are independent (Section~6.3).
\leanmeta{\LH{INT5}, \LHrng{BWR}{29}{30}, \LHrng{BWR}{37}{38}, \LHrng{SDN}{15}{18}}

\section{Incomplete Typing Forces Separately Maintained Keys}\label{sec:temporal-prices}\label{sec:no-escape}
Record structure alone suffices only when cases that can look alike need the same
handling through every supported change. Otherwise retain an identity key. TypeScript
\texttt{unique symbol} brands retain the distinction through separately maintained
identity keys; nominal typing records it directly. Runtime identity needs a key
the running program retains: runtime brands store and check a symbol-keyed member.
Type-only (phantom) brands are erased and cannot prevent runtime forgery
~\cite{TypeScriptDocs,TypeScriptSymbols,malayeri2008integrating,malayeri2009useful}.

Choose finite meanings, their possible record changes, and required handling.
Figure~\ref{fig:field-deprecation} shows the collision and its minimum naming cost.

\leanmetapending{\LH{PD1}, \LHrng{IDD}{1}{5}, \LHrng{EVO}{23}{30}, \LHrng{NHF}{3}{5}, \LH{NHF11}, \LH{OWM1}, \LHrng{IMR}{1}{5}}
\begin{proposition}[Same structure can hide different meanings]
\label{prop:identity-capacity}
A test based only on structure recovers the required answer exactly when every same-looking group
needs one answer.  An independent nominal key can always make the record and key
together sufficient and may be constant when structure suffices.  A reachable
extension can add a same-looking meaning on which every test based only on
structure fails while distinct nominal names still distinguish the meanings.
\end{proposition}

\begin{figure}[hbp]
\begingroup\PaperFixedDiagramSetup\footnotesize
\noindent\makebox[\linewidth][l]{\hspace*{0.015\linewidth}%
\begin{tikzpicture}[
  entity/.style={draw,rounded corners,inner sep=2.5pt,font=\footnotesize},
  panel heading/.style={font=\footnotesize\bfseries,anchor=base},
  >=stealth]
\begin{scope}[scale=0.70]
  \node[entity] (a0) at (0,0.58) {$a:(00,\alpha)$};
  \node[entity] (b0) at (0,-0.14) {$b:(01,\alpha)$};
  \node[entity] (c0) at (0,-0.86) {$c:(10,\beta)$};
  \node[panel heading] at (0,1.35) {(a) Before deletion};
  \node[entity] (a1) at (4.8,0.58) {$a:(0,\alpha)$};
  \node[entity] (b1) at (4.8,-0.14) {$b:(0,\alpha)$};
  \node[entity] (c1) at (4.8,-0.86) {$c:(1,\beta)$};
  \node[panel heading] at (4.8,1.35) {after deletion};
  \draw[->,thick] (b0.east) -- node[above=2pt,font=\scriptsize]{remove field} (b1.west);
  \node[draw=red,thick,fit=(a1)(b1),inner sep=1.5pt] {};
\end{scope}
\begin{scope}[xshift=6cm,meaning/.style={circle,draw,inner sep=2pt,font=\scriptsize}]
  \node[panel heading] at (2.1,0.945) {(b) Two labels suffice};
  \node[meaning,draw=blue!70!black] (a) at (0,0) {$a:\kappa_1$};
  \node[meaning,draw=orange!70!black] (b) at (1.32,0) {$b:\kappa_2$};
  \path (a.south) -- (b.south) coordinate[midway] (lower);
  \node[meaning,draw=orange!70!black,anchor=north,yshift=-2pt] (c) at (lower) {$c:\kappa_2$};
  \draw[thick] (a) -- (b);
  \node[anchor=west,font=\footnotesize] at (2.00,-.10)
    {$\PaperNotationLink{PaperKeys}{\chi(G^{\rm req})}=2$: two labels};
  \node[anchor=west,font=\footnotesize] at (2.00,-.50)
    {$n_{\rm id}=2-1=1$: one extra label};
\end{scope}
\end{tikzpicture}}%
\endgroup
\caption{(a) Deletion merges $a$ and $b$'s visible records and tags, although
they require $A_1$ and $A_2$; $c$ also requires $A_2$.  (b) Their edge forces
two labels, one beyond a shared label.  Isolated $c$ can share either; shown
sharing $b$'s.}
\Description{Three stored meanings have different two-bit field values before
one field is removed. Two share the same tag. After removal, those two have the
same visible fields and tag, so the software can no longer determine which
meaning was intended. Their required answers give a one answer and b and c a
different answer. A second panel shows three vertices with one edge between
a and b. Label kappa one goes to a; label kappa two goes to b and isolated c. The two-label
coloring is minimal for this declared edit and costs one extra label.}
\label{fig:field-deprecation}
\end{figure}

Adding a same-looking meaning forces structural identity evidence to change;
one nominal naming rule still distinguishes them without change. Starting with no
declaration, nominal evidence costs one addition; structural then nominal costs
two additions; replacing structural with nominal costs three changes:
two additions and one removal.
Supplement~\S\ref{supp-finite-identity-policies} compares these three paths.

To count the needed names, join two meanings when a reachable state makes them
look alike although they still require different handling. These edges form
the \emph{confusability graph}.  Write $V_t$ for the meanings present at $t$,
$x\equiv_t y$ for equal visible records, and $x\approx_{\rm req}y$ for equal
required answers.  For a trace $\rho$,
\begin{equation}
\label{eq:conflict-graph}
 \{x,y\}\in E(\PaperNotationLink{PaperConfusability}{G^{\rm req}_{\rho}})
 \quad\Longleftrightarrow\quad
 x\not\approx_{\rm req} y\ \text{ and }\
 \exists t:\ x,y\in V_t\ \text{ and }\ x\equiv_t y.
\end{equation}
For the specified ways requirements may evolve, $\PaperNotationLink{PaperConfusability}{G^{\rm req}}$ is the union over their
reachable traces.
Colors are key values; meanings joined by an edge must have different values.

\leanmetapending{\LHrng{EVO}{17}{18}}
\begin{theorem}[Fewest stable key values]
\label{thm:future-conflict}
A key assignment is safe at every reachable state
exactly when it gives different keys to the endpoints of every edge.
The fewest stable key values across the declared reachable histories is the
chromatic number of $\PaperNotationLink{PaperConfusability}{G^{\rm req}}$, even when today's records differ.
\end{theorem}

For a nonempty relation, one constant key value requires no maintained name.
Additional names therefore number $\PaperNotationLink{PaperKeys}{\chi(G^{\rm req})}-1$.
When each package has its own key namespace, add each package's $\chi-1$ extra names;
this can differ from $\chi-1$ of the union graph.
\leanmeta{\LH{EVO40}}

\runinhead{Stable identity across histories}
Identity keys valid today can fail after an edit. Across supported changes, retain every
distinction the decoder needs. Witsenhausen supplies the coloring rule~\cite{witsenhausen1976zero}.

Finite policies can be checked. For arbitrary
generators, unsafe cases have finite counterexamples, but no checker certifies
exactly all safe futures, even with two meanings and finite views
(co-r.e.-complete)~\cite{rogers1987computability}.
Supplement~\S\S\ref{supp-linear-field-menu-bounds}--\ref{supp-known-schema-families}
give fixed-field constructions and known-schema bounds.
\leanmeta{\LH{DEC1}, \LHrng{UND}{23}{30}, \LH{UND40}}

Declarations expose distinctions; they do not create them. Audits reconstruct
conflict graphs from sampled versions. Across \ProtoSnapshots{} Temporal Protocol Buffer snapshots
and \ProtoFieldSlots{} message--number slots, \ProtoDeletions{} fields were deleted,
\ProtoNeverReserved{} unreserved; \ProtoReuses{} numbers later changed name, type,
cardinality, or oneof group. The graph needs \ProtoReuseConflictPrice{} keys.
Reservation at deletion excludes each reuse and enforces safe erasure
~\cite{ProtobufProto3,TemporalAPIRepository}.
It does not recover erased identity from structure.
Supplement~\S\ref{supp-protobuf-audit} gives certificates.

The TypeScript audit identifies public brands by module path, kind, and name,
using normalized declarations with brand markers stripped. It finds
\TSBrandPersistentMeanings{} distinct brands in
\TSBrandPackagesWithMeanings{} of \TSBrandPopulationPackages{} packages.
Exact package-local minima require \TSBrandNormalizedNamePrice{} extra names.
History-only edges occur in \TSBrandHistoryOnlyPackages{} packages,
forcing \TSBrandHistoryAdditionalNamePrice{} of the \TSBrandNormalizedNamePrice{};
\texttt{expect-type}'s historical edge leaves its chromatic number unchanged.
Without brand markers, \texttt{type-fest} 4.3.3's \texttt{MultiTagContainer}
and \texttt{TagContainer} look alike: two permanent keys for sampled history,
one for the latest snapshot. Separately,
\TSBrandNoRuntimeEvidenceMeanings{} of \TSBrandPersistentMeanings{} meanings
lack verified runtime identity evidence. The remaining
\TSBrandPackagesWithoutMeanings{} of \TSBrandPopulationPackages{} packages are unscored.
Supplement~\S\ref{supp-typescript-brand-audit} gives the protocol, certificates,
and runtime-evidence rules~\cite{NpmRegistryAPI}.

A Go~1.0--1.26 audit excludes methods promoted from embedded types. Across packages,
it tries to identify named types using only nonempty lists of directly
declared method signatures~\cite{GoSpec}:
40 extra names (41 keys) for values, 75 (76) for pointers,
matching the worst sampled states
(Supplement~\S\ref{supp-go-direct-method-history}).

Nominal identity never needs more extra keys across supported changes;
look-alikes needing different handling make the advantage strict.\footnote{\LH{HVD3}, \LH{HVD4}, \LH{HVD7}}
Names distinguish meanings; contract membership asks which classes are admitted.
An \emph{ordinary abstract base class (ABC)} has no shape-accepting \texttt{\_\_subclasshook\_\_}, unlike
some \texttt{collections.abc} classes. Inheritance admits owned members;
\texttt{ABC.register} admits unowned members and descendants; implementations still need supply~\cite{PythonABC}.

\leanmetapending{\LHrng{SCM}{1}{4}}
\begin{proposition}[Structural contracts admit unwanted members]
\label{prop:structural-membership}
Assume every required member matches the contract. Structural membership is
exact through every supported change exactly when every reachable match is required.
Nominal inheritance or registration is exact when its admitted declarations are
exactly the required members and membership is closed under descendants.
An unwanted reachable match refutes structural membership; nominal admission still follows its declared edges.
\end{proposition}

Registration adds a contract edge outside the class body; descendants inherit
membership through it. Registries without descendant propagation are outside
layers (Theorem~\ref{thm:contract-membership-gap}).
A supported extension can add an unintended member with the required shape.
A structural check admits it. Only an enforced admission rule excludes that case;
ownership alone does not.
\texttt{Protocol} inheritance does not exclude unrelated matches.
Runtime membership is weaker still.\footnote{Without \texttt{@runtime\_checkable},
\texttt{isinstance} raises; with it, checks require only attribute presence,
so different signatures can still match~\cite{PEP544,PythonTypingProtocols}.}

Identity keys distinguish meanings; inheritance supplies code. Section~4 counts missing pairs.


\section{Incomplete Inheritance Forces Code into Workarounds}\label{sec:factorization}
In the declared table $\PaperNotationLink{PaperRequired}{R^{\mathrm{req}}}$, rows are independently changing shared
implementations and columns are classes; $(f,c)\in \PaperNotationLink{PaperRequired}{R^{\mathrm{req}}}$ means $c$ needs $f$.
Contract parents supply no implementation. One implementation parent places
each class's code providers on one ancestor chain; multiple parents allow incomparable providers.
Code supplies implementations lost when an inheritance path is removed.
We count two kinds of compensation:
forwarding a missing pair and removing an unwanted one.
\leanmeta{\LHrng{DEC}{5}{6}}
\begin{table}[htbp]
\centering
\PaperDenseTableSetup
\setlength{\tabcolsep}{4pt}
\begin{tabularx}{\linewidth}{@{}l l X@{}}
\toprule
Result & Foundation & Added scope or result \\
\midrule
Theorem~\ref{thm:future-conflict} & Witsenhausen & Histories and package-local namespace sums \\
\midrule
Proposition~\ref{prop:derivation-capacity} & FCA; phylogeny & Multiple parents derive every finite map \\
\midrule
Theorem~\ref{thm:delegation-lower-bound} & Chen et al. & Table hardness; minimum over program designs \\
\midrule
Theorem~\ref{thm:paid-blocking-gap} & Chen's flip distance & Weighted removal; $\lambda\ge \PaperNotationLink{PaperGap}{d_\downarrow}$ threshold \\
\midrule
Theorem~\ref{thm:laminar-growth-reparenting} & This paper & Strict-insertion repair; every-step history \\
\midrule
Theorem~\ref{cor:coverage-resolution} & This paper & Conflict invariance across answer-preserving representations \\
\midrule
Proposition~\ref{prop:self-call-ledger} & Cook--Palsberg~\cite{cook1994denotational} & Routing and precedence counts for open recursion \\
\midrule
Theorem~\ref{thm:open-world-laminarity} & Halting reduction & Permanent tree shape is co-r.e.-complete \\
\bottomrule
\end{tabularx}
\caption{Established foundations and what this paper adds.}
\label{tab:novelty-lifts}
\end{table}

In Figure~\ref{fig:workflow-crossing}, the writer needs only file code, the buffered
reader only iteration, and the file reader both: a crossing with four required pairs.
Keep iteration ancestry for both readers and file ancestry for the writer;
forward the file reader to its file provider. Three pairs are inherited and one
is outside. The exact gap is one even with helpers or generators
(Theorems~\ref{thm:two-role-average-gap}, \ref{thm:delegation-lower-bound}).
Additivity forces a sum of per-pair prices; unit charges count these pairs
(Proposition~\ref{prop:forced-semantic-unit}).

An implementation's \emph{required set} lists the classes needing it.
One parent supplies every pair exactly when any two sets nest or do not overlap:
the family is \emph{laminar}, or \emph{tree-shaped}.

Declarations $N$ include consumer classes $\PaperClasses$. Implementation $f$ lives in its distinct
provider declaration $\PaperNotationLink{PaperProvider}{o_f}$ and is required by $\PaperNotationLink{PaperClassGroup}{A_f}\subseteq \PaperNotationLink{PaperClasses}{C}$.
Write $\PaperNotationLink{PaperProvider}{o_f}\preceq c$ when $c$ inherits from $\PaperNotationLink{PaperProvider}{o_f}$;
providers are comparable when one is an ancestor of the other.
For a guided reading of the symbols, use the supplement's opening table:
plain-language meanings in dependency order, with Figure~\ref{fig:workflow-crossing} values where applicable.

\leanmetapending{\LHrng{NHF}{35}{41}}
Putting independently changing code in separate provider classes does not
increase the least forwarding the program needs.
\begin{proposition}[Separate providers cannot increase outside supply]
\label{prop:role-normalization}
Separate providers cannot increase minimum one-parent outside supply. In a fixed hierarchy,
each implementation may use its largest available descendant set contained in its
required set; equal required sets may all use the largest of their retained sets.
\end{proposition}

If class $c$ does not inherit implementation $f$, $\PaperNotationLink{PaperRoutes}{\varphi}(f,c)$ chooses a
forwarding destination $v$. In routing graph $\PaperNotationLink{PaperRoutingGraph}{\Gamma}$, $c\to v$ is one forwarding edge;
$c\to^*v$ follows zero or more. Read $f@c$ as ``class $c$ can reach an
implementation of $f$'':
\[
 f@c\quad\Longleftrightarrow\quad
 \PaperNotationLink{PaperProvider}{o_f}\preceq c\ \lor\
 \exists v.\;\PaperNotationLink{PaperRoutes}{\varphi}(f,c)=v\land \PaperNotationLink{PaperProvider}{o_f}\preceq v\land c\to^*v.
\]
Per-pair forwarding, nonpropagating registries, adapters, and generated routes
supply missing pairs outside ancestry. Native delegation has its own capacity:
Kotlin's \texttt{by} forwards an interface with one declaration; Rust's \texttt{Deref}
lets member access follow one target chain~\cite{KotlinDelegation,RustDerefReference}. Stein models delegation
as inheritance~\cite{stein1987delegation}; per-pair delegation here is outside supply
included in the outside-supply minimum. Copies supply pairs the original provider's ancestry
misses; copying its body does not make the consumer inherit it
(Theorem~\ref{thm:replica-completion-law}).
\leanmeta{\LHrng{BWR}{31}{32}}
\leanmetapending{\LH{NHF1}, \LH{NHF20}, \LH{NHF22}, \LHrng{NHF}{25}{28}}

\leanmetapending{\LHrng{NHF}{23}{24}}
\runinhead{Propagation through ancestry}
An unoverridden method added to a code provider reaches every required class through
ancestry exactly when that provider has no outside pair; otherwise, forwarding
code or its shared rule must supply the missing routes.
In Figure~\ref{fig:workflow-crossing}, a new \texttt{ByteReceiveStream} method
reaches the file and buffered readers. After changing to one parent, a new
\texttt{\_BaseFileStream} method reaches the writer directly, but the file reader
only if its forwarding code or rule covers it.

\leanmetapending{\LH{NHF7}, \LH{NHF20}, \LHrng{NHF}{44}{45}, \LHrng{NHF}{50}{51}}
\begin{proposition}[One parent suffices exactly for tree-shaped requirements]
\label{prop:derivation-capacity}
Give each independently maintained implementation one provider.
Multi-parent ancestry derives every required pair for every finite family of
required sets without outside routes. One-parent ancestry derives every
required pair exactly when the required sets are tree-shaped. The result
concerns reachability only; it assumes neither mutable state nor a
rule for selecting among competing implementations. Python base orders must be
C3-consistent, placing subclass bases before their ancestors.
\end{proposition}

\begin{proof}
Use nodes $\PaperNotationLink{PaperImplementations}{F}\uplus \PaperNotationLink{PaperClasses}{C}$ with $c\to f$ for $c\in \PaperNotationLink{PaperClassGroup}{A_f}$.
For dual-role declarations, keep identity and contracts;
put implementation on a separate helper.
One-parent paths nest sets; conversely, inclusion builds the laminar tree,
chaining ties and keeping empty sets off-path
~\cite{godin2005fca,gusfield1991efficient,snelting2000concept}.
\end{proof}

Supply counts which classes inherit state-bearing implementations; specified
sharing, initialization order, and lookup winners are separate (Section~4.1).
Layout, mutation, and copying have additional runtime and storage costs.
Replicas still supply pairs (Theorem~\ref{thm:replica-completion-law}); state-access
lookups fix candidates and winners (Theorem~\ref{cor:coverage-resolution}).

Choose required relation $\PaperNotationLink{PaperRequired}{R^{\mathrm{req}}}$ and implementation supply $\PaperAncestry$.  The pairs needing
repair are $\PaperNotationLink{PaperRepairSet}{\mathcal D(H,R^{\mathrm{req}})}=(\PaperNotationLink{PaperRequired}{R^{\mathrm{req}}}\setminus \PaperNotationLink{PaperAncestry}{H})\uplus(\PaperNotationLink{PaperAncestry}{H}\setminus \PaperNotationLink{PaperRequired}{R^{\mathrm{req}}})$:
missing required pairs and unwanted supplied pairs.
Ancestry is \emph{sound} when it supplies only required pairs ($\PaperNotationLink{PaperAncestry}{H}\subseteq \PaperNotationLink{PaperRequired}{R^{\mathrm{req}}}$).

\Needspace{3\baselineskip}
\begin{claim}{Missing and unwanted supply}{thm:exact-coupling-debt}
Supply is exact iff $\PaperNotationLink{PaperAncestry}{H}=\PaperNotationLink{PaperRequired}{R^{\mathrm{req}}}$, equivalently
$\PaperNotationLink{PaperRepairSet}{\mathcal D(H,R^{\mathrm{req}})}=\varnothing$. On finite relations,
$|\PaperNotationLink{PaperRepairSet}{\mathcal D(H,R^{\mathrm{req}})}|=|\PaperNotationLink{PaperRequired}{R^{\mathrm{req}}}\setminus \PaperNotationLink{PaperAncestry}{H}|+|\PaperNotationLink{PaperAncestry}{H}\setminus \PaperNotationLink{PaperRequired}{R^{\mathrm{req}}}|$.
Additions alone restore exact supply iff $\PaperNotationLink{PaperAncestry}{H}\subseteq \PaperNotationLink{PaperRequired}{R^{\mathrm{req}}}$:
supply every missing pair. Unwanted pairs require removal or changed requirements.
\leanmeta{\LHrng{SCT}{46}{51}, \LH{BWR29}, \LHrng{BWR}{37}{39}, \LH{BWR52}, \LH{BWR57}}
\end{claim}

The \emph{ancestry gap} is the fewest required pairs that a sound one-parent
hierarchy leaves outside.  Retain as many required pairs as possible while
their class groups fit one tree:
\[
 \PaperNotationLink{PaperGap}{d_{\downarrow}}(\mathcal A)=
 \min_{\substack{\PaperNotationLink{PaperClassGroup}{A_f^{\rm nat}}\subseteq \PaperNotationLink{PaperClassGroup}{A_f}\\
                  \{\PaperNotationLink{PaperClassGroup}{A_f^{\rm nat}}:f\in \PaperNotationLink{PaperImplementations}{F}\}\ \mathrm{laminar}}}
       \sum_{f\in \PaperNotationLink{PaperImplementations}{F}}|\PaperNotationLink{PaperClassGroup}{A_f}\setminus \PaperNotationLink{PaperClassGroup}{A_f^{\rm nat}}|.
\]
A deleted $1$ leaves code or tooling to supply an implementation the class still
needs but no longer inherits.
A crossing describes required groups; the gap counts pairs that sound one-parent ancestry cannot derive.
The definition optimizes tables; Theorem~\ref{thm:delegation-lower-bound}
shows that every program design in its scope reduces to this optimization.
\leanmeta{\LHrng{NHF}{6}{9}, \LHrng{NHF}{13}{14}, \LHrng{NHF}{20}{21}}

\paragraph{Counting the hierarchy.}
The three coordinates count extra identity names, outside memberships, and
outside implementation pairs. Figure~\ref{fig:native-capacity-ladder}'s rungs leave:
\begin{center}
\PaperDenseTableSetup
\begin{tabular*}{\linewidth}{@{\extracolsep{\fill}}lcccl@{}}
\toprule
Native capacity & \shortstack{Extra\\identity names} & \shortstack{Outside\\memberships}
& \shortstack{Outside\\code pairs} & Computing the minimum \\
\midrule
Structural only & $n_{\rm id}$ & $|\PaperNotationLink{PaperMembership}{M^{\mathrm{req}}}|$ & $|\PaperNotationLink{PaperRequired}{R^{\mathrm{req}}}|$ & Graph coloring (\S3) \\
\midrule
One contract parent & $0$ & $\PaperNotationLink{PaperGap}{d_\downarrow}(\PaperNotationLink{PaperMembership}{M^{\mathrm{req}}})$ & $|\PaperNotationLink{PaperRequired}{R^{\mathrm{req}}}|$ & Deletion search (\S4.2) \\
\midrule
Multiple contracts & $0$ & $0$ & $|\PaperNotationLink{PaperRequired}{R^{\mathrm{req}}}|$ & Count required pairs \\
\midrule
One implementation parent & $0$ & $0$ & $\PaperNotationLink{PaperGap}{d_\downarrow}(\PaperNotationLink{PaperRequired}{R^{\mathrm{req}}})$ & Deletion search (\S4.2) \\
\midrule
Multiple implementation parents & $0$ & $0$ & $0$ & Direct edges (Prop.~\ref{prop:derivation-capacity}) \\
\bottomrule
\end{tabular*}
\end{center}
Deciding whether one parent can leave at most $\ell$ pairs outside is NP-complete;
a four-way search finds the minimum (\S4.2). The last column concerns finding an optimal representation;
runtime execution is priced separately.
Here $n_{\rm id}=\max\{\PaperNotationLink{PaperKeys}{\chi(G^{\rm req})}-1,0\}$; $\PaperNotationLink{PaperMembership}{M^{\mathrm{req}}}$ lists required memberships.
Across supported changes, structural $|\PaperNotationLink{PaperMembership}{M^{\mathrm{req}}}|$ requires a reachable unwanted match per contract
(Proposition~\ref{prop:structural-membership}). Real languages combine capacities off this chain (Table~\ref{tab:implementation-carriers});
Lemma~\ref{lem:capacity-optimum-order} orders capacities coordinatewise.
Contract and implementation graphs can differ; Theorem~\ref{thm:contract-membership-gap}
prices sharing one, and Theorem~\ref{thm:laminar-growth-reparenting} covers updates.

\Needspace{2\baselineskip}
\leanmetapending{\LHrng{SCT}{1}{4}, \LHrng{SCT}{20}{26}, \LHrng{SCT}{30}{32}, \LHrng{SCT}{52}{55}}
\begin{theorem}[The exact price of inheritance incompleteness]
\label{thm:semantic-completion-tax}
For every finite required provider--class relation, retaining distinct providers even
when their required sets are equal:
\begin{enumerate}
  \item adding missing pairs restores exact supply exactly when ancestry is
        sound.  Every such repair maintains a distinct
        answer for each required pair that ancestry misses;
  \item inheriting only contracts supplies no implementation pair, so its
        outside cost is $|\PaperNotationLink{PaperRequired}{R^{\mathrm{req}}}|$;
  \item for every sound one-parent hierarchy $\PaperAncestry$,
        $|\PaperNotationLink{PaperRepairSet}{\mathcal D(H,R^{\mathrm{req}})}|=\PaperNotationLink{PaperGap}{d_\downarrow}+\PaperNotationLink{PaperExcess}{e_H}$, where $\PaperNotationLink{PaperExcess}{e_H}\geq 0$ counts
        this hierarchy's excess outside supply and vanishes exactly at the optimum.
        Refactoring removes $\PaperNotationLink{PaperExcess}{e_H}$; sound one-parent supply always costs at least $\PaperNotationLink{PaperGap}{d_\downarrow}$.
  \item multi-parent ancestry with distinct providers supplies exactly every
        finite required relation and has outside cost zero.
        \leanmeta{\LH{BWR42}, \LH{SCT23}, \LHrng{SCT}{30}{34}}
\end{enumerate}
\end{theorem}

\leanmetapending{\LHrng{CAP}{2}{7}, \LHrng{CPL}{1}{4}}
\begin{theorem}[Contract membership obeys the same gap law]
\label{thm:contract-membership-gap}
Let $\PaperNotationLink{PaperMembership}{M^{\mathrm{req}}}\subseteq T\times \PaperNotationLink{PaperClasses}{C}$ declare nominal contract memberships.
One-parent graphs derive every membership exactly when member sets are tree-shaped.
Over all sound one-parent graphs and membership-preserving outside layers, the
exact outside minimum is $\PaperNotationLink{PaperGap}{d_\downarrow}(\PaperNotationLink{PaperMembership}{M^{\mathrm{req}}})$; multiple parents leave zero.
Additions cannot cancel unwanted inherited membership, hence soundness.
Independent contract and implementation graphs have joint minimum
$\PaperNotationLink{PaperGap}{d_\downarrow}(\PaperNotationLink{PaperMembership}{M^{\mathrm{req}}})+\PaperNotationLink{PaperGap}{d_\downarrow}(\PaperNotationLink{PaperRequired}{R^{\mathrm{req}}})$, counting disjoint answer types.

If one parent graph carries both relations, its exact minimum is
$\PaperNotationLink{PaperGap}{d_\downarrow}(\PaperNotationLink{PaperMembership}{M^{\mathrm{req}}}\uplus \PaperNotationLink{PaperRequired}{R^{\mathrm{req}}})\ge
\PaperNotationLink{PaperGap}{d_\downarrow}(\PaperNotationLink{PaperMembership}{M^{\mathrm{req}}})+\PaperNotationLink{PaperGap}{d_\downarrow}(\PaperNotationLink{PaperRequired}{R^{\mathrm{req}}})$.  For a strict witness, let $\PaperNotationLink{PaperMembership}{M^{\mathrm{req}}}$ have
the single row $\{W,\PaperFileReader\}$ and $\PaperNotationLink{PaperRequired}{R^{\mathrm{req}}}$ the single row $\{\PaperFileReader,B\}$:
separate gaps zero, shared gap one. A declaration serving contract class set $A$
and implementation class set $B$ leaves at least $|A\mathbin{\triangle}B|$ outside;
keeping $A\cap B$ is optimal. An implementation-only declaration separates the
graphs, as AnyIO's private file base does.
\end{theorem}

Contract and implementation incompleteness obey the same one-parent law.
Registries or generated tables still determine omitted memberships.
The hierarchy's four strict steps use Figure~\ref{fig:field-deprecation}'s identity
collision, AnyIO's crossing declared as memberships, any nonempty
$\PaperNotationLink{PaperRequired}{R^{\mathrm{req}}}$, and Figure~\ref{fig:workflow-crossing}'s implementation crossing.
\leanmeta{\LH{PD1}, \LH{CAP2}, \LH{NHF52}, \LH{OWM2}, \LHrng{SCT}{31}{32}}

\leanmetapending{\LHrng{SCT}{31}{32}}
A tree-shaped map has zero gap but its updates can force deletion, forwarding,
or removal (Theorem~\ref{thm:laminar-growth-reparenting});
Corollary~\ref{cor:factorization-economics} prices the recurring obligation.
\leanmeta{\LHrng{NHF}{8}{14}, \LH{BWR9}}

\subsection{Inheritance supplies candidates; MRO orders them}

Python's C3 method resolution order lists ancestors; member lookup uses
the first definition and cooperative \texttt{super()} continues through that
order~\cite{barrett1996c3}. A resolver cannot choose a missing implementation.
Supplement~\S\ref{supp-dual-axis-resolution} proves that coherent resolution
follows one strict precedence order.
Multiple parents supply every candidate; only genuine conflicts need ordering.
A one-parent refactoring must preserve lookup choices while supplying missing implementations.

Specified lookup winners, state sharing, and initialization order are representation-invariant,
outside the supply triples (Theorem~\ref{cor:coverage-resolution}; Supplement
Theorem~\ref{supp-state-answer-invariance}). With the same state requirements,
multiple parents still permit every one-parent and composition design
(Corollary~\ref{supp-state-capacity-containment}). A shared ancestor creates a diamond;
conflicts require competing definitions. C++ virtual bases and Python's instance namespace
decide sharing. Zero outside supply while meeting state requirements needs a
compatibility proof; Python's lookup/initialization coupling costs exactly one
outside answer in the two-class witness (Supplement~\S\ref{supp-state-coordination}).
\leanmeta{\LHrng{SCI}{1}{3}, \LH{NHF10}, \LH{NHF16}, \LH{NHF19}}
The specification fixes these answers for every representation; layout has a separate runtime price.

Sequential mixin application creates a chain and orders lookup
~\cite{bracha1990mixin,flatt1998classes}; repeated applications create replicas
(Theorem~\ref{thm:replica-completion-law}).

\subsection{Crossings force pairs outside ancestry}

The file reader needs both file-lifecycle and iteration code; the writer needs
only file code, and the buffered reader only iteration. With one parent per class,
at least one of these four required pairs needs forwarding or other code.
The best design leaves exactly one outside inheritance. The following formula
counts the minimum for any two implementations.

\leanmetapending{\LHrng{AVG}{1}{7}}
\begin{theorem}[Exact gaps for two implementations and a row--column grid]
\label{thm:two-role-average-gap}
For two required sets $A,B$, let
$N_{10}=|A\setminus B|$, $N_{01}=|B\setminus A|$, and
$N_{11}=|A\cap B|$.  Then
$\PaperNotationLink{PaperGap}{d_\downarrow}(A,B)=\min\{N_{10},N_{01},N_{11}\}$.
The three terms count the pairs dropped to fit one tree: drop $A$'s implementation
from $A$-only classes, drop $B$'s from $B$-only classes, or drop one implementation
from classes needing both. The retained sets then satisfy $A\subseteq B$,
$B\subseteq A$, or disjointness.
For the deterministic $h\times h$ row--column domain, give each of the
$n=h^2$ classes its row implementation and its column implementation. No implementation
inheritance leaves all $2n$ pairs outside, one implementation parent leaves
$\PaperNotationLink{PaperGap}{d_\downarrow}=h(h-1)=n-\sqrt n$, and multiple implementation parents leave
zero.
\end{theorem}

For independent classes, cell probabilities $q_{00},q_{10},q_{01},q_{11}$ mean
neither, only $A$, only $B$, or both. With $q_{\min}=\min\{q_{10},q_{01},q_{11}\}>0$,
zero-gap probability lies between $(1-q_{\min})^n$ and $3(1-q_{\min})^n$
(exact inclusion--exclusion proof: Supplement~\S\ref{supp-two-role-row-column-gaps}).
\footnote{\LHrng{AGP}{1}{7}}
Figure~\ref{fig:average-gap-growth} plots this exact law:
the crossing probability reaches 0.99 at $n=20$ with balanced regions; making the shared region 20 times rarer
moves that point to 367; a source with
$\Pr[\PaperNotationLink{PaperGap}{d_\downarrow}>0]=0.05$ at $n=256$ reaches
0.99 by $n=22{,}984$ (off-axis). Guaranteed avoidance must exclude a crossing region,
restricting requirements to tree shape.
\begin{figure}[htbp]
\centering
\PaperDenseTableSetup
\begin{minipage}[t]{0.49\linewidth}
\centering
\begin{tikzpicture}
\begin{axis}[
  width=0.98\linewidth,
  height=2.85cm,
  xmode=log,
  log basis x=2,
  xmin=3.5,
  xmax=1120,
  y coord trafo/.code={\pgfmathparse{ln(0.01+#1)-ln(0.01)}},
  y coord inv trafo/.code={\pgfmathparse{exp(#1+ln(0.01))-0.01}},
  ymin=0,
  ymax=260,
  xtick={4,16,64,256,1024},
  xticklabels={4,16,64,256,1024},
  ytick={0,.1,1,10,256},
  yticklabels={0,0.1,1,10,256},
  xlabel={consumer classes $n$},
  ylabel={expected gap},
  title={Required pairs left outside},
  title style={font=\scriptsize\bfseries},
  grid=major,
  tick label style={font=\scriptsize},
  label style={font=\scriptsize},
  legend entries={balanced,shared $\times0.05$,$\Pr_{256}=0.05$,multiple parents (left only)},
  legend to name=gap-source-legend,
  legend columns=4,
  legend style={font=\tiny,draw=none,fill=none,legend cell align=left,
    /tikz/every even column/.append style={column sep=3pt}},
]
\input{content/skewed_two_role_main_gap_plots_auto.tex}
\addplot+[teal!75!black,mark=*,thick] coordinates {
  (4,0) (8,0) (16,0) (32,0) (64,0) (128,0) (256,0) (512,0) (1024,0)
};
\end{axis}
\end{tikzpicture}
\end{minipage}\hfill
\begin{minipage}[t]{0.49\linewidth}
\centering
\begin{tikzpicture}
\begin{axis}[
  width=0.98\linewidth,
  height=2.85cm,
  xmode=log,
  log basis x=2,
  xmin=3.5,
  xmax=1120,
  ymin=0,
  ymax=1.03,
  xtick={4,16,64,256,1024},
  xticklabels={4,16,64,256,1024},
  ytick={0,.25,.5,.75,1},
  xlabel={consumer classes $n$},
  ylabel={crossing probability},
  title={Probability one parent misses a pair},
  title style={font=\scriptsize\bfseries},
  grid=major,
  tick label style={font=\scriptsize},
  label style={font=\scriptsize},
]
\input{content/skewed_two_role_main_probability_plots_auto.tex}
\end{axis}
\end{tikzpicture}
\end{minipage}

\ref{gap-source-legend}
\caption{Expected gap (left) and crossing probability (right) for three
distributions with independent class draws. Cells (neither, left, right, both) use
$q_{00}=.25$, $q_{10}=.5-q_{11}$, $q_{01}=.25$, and
$q_{11}=.25,.0125,\SkewedCalibratedBothMass$; the last gives
$\Pr[\PaperNotationLink{PaperGap}{d_\downarrow}>0]=.05$ at $n=256$. Left scale:
$\log(1+\text{gap}/.01)$, retaining zero. Multiple parents: zero outside.
Specified synthetic distributions determine these curves.}
\Description{Two side-by-side plots compare exact one-parent gap and crossing
probability under balanced requirements, a shared crossing region made twenty
times less likely, and a source chosen to cross with probability five percent
at 256 consumer classes.  A fourth line in the gap panel shows that
multiple-parent outside count remains zero. The gap axis uses a log-one-plus
scale to separate rare cases while retaining zero; tick labels show actual gaps.}
\label{fig:average-gap-growth}
\end{figure}

Tree shape decides whether one parent derives every pair in a finite requirement
table. Deleting at most $\ell$ pairs to reach it is NP-complete: Chen et al.'s
deletion-flip problem on complete $0/1$
tables~\cite[Theorem~2.2 and Corollary~3.1]{chen2006minimum}.
Their four-way search deletes one of a crossing's four pairs
~\cite[Theorem~5.1]{chen2006minimum}. Lean checks this step; the audit replays the search.
Supplement~\S\ref{supp-python-factorization-results} gives both finder runtimes
and distinguishes deletion-only from minimum-flip search.
\leanmeta{\LHrng{DEC}{2}{4}, \LH{NHF31}}

\subsection{The table minimum holds over every program design}

Java hierarchy work treats classes and interfaces separately
~\cite{moore1996guru,snelting2000concept,streckenbach2004kaba,balatsouras2013complementation};
FCA and phylogeny characterize tree fit~\cite{godin2005fca,gusfield1991efficient}.
Chen et al. minimize table deletions~\cite{chen2006minimum}; Theorem~\ref{thm:delegation-lower-bound}
lifts their minimum to programs with helpers and generated forwarding.
\leanmeta{\LH{NHF52}}
Felleisen studies local, structure-preserving translations~\cite{felleisen1991expressive};
we count facts left to code and repairs across updates. Assignment expressions illustrate the
boundary (Supplement~\S\ref{supp-supp:walrus-boundary}).

One-parent ancestry and one-chain mixins or traits
\cite{schaerli2003traits,ducasse2006traits,Scala3Traits} put each class's providers on a chain
in the global ancestor order $\preceq$. C3 orders Figure~\ref{fig:workflow-crossing}'s
\texttt{\_BaseFileStream} and \texttt{ByteReceiveStream}; they remain incomparable
in $\preceq$. Unlike selected type-constraint repairs~\cite{tip2003generalization},
the minimum covers every design with these ancestry constraints.

For a design $R$ with sound ancestry, let
$A^{\rm nat}_{f,R}=\{c:\PaperNotationLink{PaperProvider}{o_f}\preceq_R c\}$ be the classes reached through
ancestry and $G_f=\PaperNotationLink{PaperClassGroup}{A_f}\setminus A^{\rm nat}_{f,R}$ the classes it misses.

\leanmetapending{\LH{SCT24}, \LHrng{SCT}{27}{34}}
\begin{theorem}[The ancestry gap is the minimum outside supply]
\label{cor:minimized-delegation}
\label{thm:delegation-lower-bound}
Fix any finite required map. Consider every design preserving exact supply that keeps
implementation providers distinct and inherits only required implementations, with
any two providers inherited by one class comparable in the derivation order.
Allow arbitrary reparenting, helper nodes, delegation graphs, and forwarding
maps.  Providers stay distinct even when their required sets agree; the helper
set need not be finite.  Then
\[
 \{(f,c):c\in G_f\}\subseteq\operatorname{dom}\PaperNotationLink{PaperRoutes}{\varphi_R},
 \qquad
 \min_R|\operatorname{dom}\PaperNotationLink{PaperRoutes}{\varphi_R}|=\PaperNotationLink{PaperGap}{d_\downarrow}(\mathcal A).
\]
This attained minimum is zero iff the required sets are tree-shaped, positive
at every crossing, and strictly below contract-only cost when $\PaperNotationLink{PaperRequired}{R^{\mathrm{req}}}\ne\varnothing$.
\end{theorem}

\begin{proof}
Comparable providers have nested sets; incomparable ones share no class by the
one-chain hypothesis. Even infinite helpers induce a finite tree-shaped subtable;
missing pairs need forwarding. Inclusion represents minimizing sets, empty sets
stay off-path, and forwarding attains the bound
(Supplement~\S\ref{supp-minimum-one-parent-completion}).
Behavioral preservation is separate (Theorem~\ref{thm:behavioral-applicability-product}),
checked under AnyIO's local edit contract (Supplement~\S\ref{supp-anyio-refactoring-application}).
\end{proof}

Helpers added to fit one chain are slop. With unlimited helpers left uncharged,
the best sound design still leaves exactly $\PaperNotationLink{PaperGap}{d_\downarrow}$ outside: positive at crossings, versus zero
with multiple parents.

\Needspace{4\baselineskip}
\leanmetapending{\LHrng{SCT}{56}{59}, \LHrng{BWR}{58}{62}, \LHrng{SDN}{17}{21}}
\begin{theorem}[Copying cannot lower the ancestry gap]
\label{thm:replica-completion-law}
Same-identity declarations agreeing in meaning and required classes are replicas.
Choose one as original: every choice has the same required sets and gap; copies
supply pairs outside its ancestry.
Let $\PaperNotationLink{PaperAncestry}{H}\subseteq \PaperNotationLink{PaperRequired}{R^{\mathrm{req}}}$ be sound
one-parent ancestry and $E\subseteq \PaperNotationLink{PaperRequired}{R^{\mathrm{req}}}$ its replica-, registry-, forwarding-, or
generator-supplied pairs. Completeness implies
\[
  E\setminus \PaperNotationLink{PaperAncestry}{H}=\PaperNotationLink{PaperRequired}{R^{\mathrm{req}}}\setminus \PaperNotationLink{PaperAncestry}{H},
  \qquad |E\setminus \PaperNotationLink{PaperAncestry}{H}|\ge \PaperNotationLink{PaperGap}{d_\downarrow}.
\]
This includes replica-only supply. Independently changing copies mean distinct
implementations, changing $\PaperNotationLink{PaperRequired}{R^{\mathrm{req}}}$ (Corollary~\ref{cor:specification-or-payment}).
Independently edited copies can disagree.
\end{theorem}

\Needspace{4\baselineskip}

\leanmetapending{\LHrng{SEF}{1}{25}}
\begin{theorem}[Merge, share, or separate]
\label{thm:exact-semantic-factoring}
For classes $A,B$, let $m_A,m_B$ be their full state-to-answer tables over the
declared reachable states and $i_A,i_B$ their implementation answers. An answer-preserving
representation minimizing distinctions in full meanings and implementations merges them iff $m_A=m_B$;
if only $i_A=i_B$, it shares their
implementation but keeps their identities distinct; if $i_A\ne i_B$, it keeps
their implementations separate.  No equal structural profile, even the full
inheritance graph, can decode distinct answer histories. In the finite supply case,
let $i_A,i_B$ be the sets of required implementations. If both classes must use one set $X$,
the minimum number of answer changes is
\[
 \min_X\bigl(|i_A\mathbin{\triangle}X|+|i_B\mathbin{\triangle}X|\bigr)
 =|i_A\mathbin{\triangle}i_B|.
\]
It vanishes exactly when the answers agree; otherwise an answer-preserving replacement of
shared code changes at least one class assignment.
\end{theorem}

Identical code today can need different changes tomorrow. Sharing implementations
requires agreement through every supported change. For finite sets,
the attained minimum changes one answer per implementation needed by only one class.
Shared code need not merge identities.

Figure~\ref{fig:semantic-factoring-law} shows how to factor a declared shared combination;
Supplement~\S\ref{supp-semantic-factoring} proves all three cases and establishes when a finite migration cost is repaid.
\begin{figure}[!hb]
\centering
\PaperDenseTableSetup
\setlength{\fboxsep}{1.2pt}
\setlength{\fboxrule}{0.45pt}
\newcommand{\FactoringEdit}[1]{\fcolorbox{red!70!black}{red!4}{\strut\texttt{#1}}}
\begin{tabularx}{\linewidth}{@{}
  >{\hsize=.98\hsize\linewidth=\hsize}X
  >{\hsize=1.02\hsize\linewidth=\hsize}X@{}}
\multicolumn{2}{c}{%
  \texttt{class MixA: attr1 = 1}\qquad
  \texttt{class MixB: attr2 = 2}\qquad
  \texttt{class MixC: attr3 = 3}}\\[2pt]
\begin{tabular}[t]{@{}l@{}}
\textrm{\bfseries repeated parent combination} \\
\textit{before}\quad\texttt{class A(MixA, MixB): kind = "A"} \\
\quad\texttt{class B(MixA, MixB): kind = "B"} \\
\textit{after}\quad\texttt{class A(MixA, MixB}\FactoringEdit{, MixC}\texttt{): kind = "A"} \\
\quad\texttt{class B(MixA, MixB}\FactoringEdit{, MixC}\texttt{): kind = "B"} \\
\textbf{two changed headers}
\end{tabular}
&
\begin{tabular}[t]{@{}l@{}}
\textrm{\bfseries declared shared combination} \\
\textit{before}\quad\texttt{class Shared(MixA, MixB): pass} \\
\quad\texttt{class A(Shared): kind = "A"} \\
\quad\texttt{class B(Shared): kind = "B"} \\
\textit{after}\quad\texttt{class Shared(MixA, MixB}\FactoringEdit{, MixC}\texttt{): pass} \\
\quad\texttt{class A(Shared): kind = "A"} \\
\quad\texttt{class B(Shared): kind = "B"} \\
\textbf{one changed header}
\end{tabular}
\end{tabularx}
\caption{When classes must keep the same parent combination, declare it once.
Adding $m$ later common parents across $n$ classes then takes $m$ header edits
instead of $mn$. Identical base lists today do not establish that shared requirement.}
\Description{Two unboxed Python listings compare repeated multiple
inheritance with a factored shared parent class. On the left, classes A and B
both inherit MixA and MixB, which provide attr1 and attr2, so adding a later
common parent MixC changes both class headers. On the right, A and B inherit a
Shared class that inherits MixA and MixB. The after listing changes only the
Shared header to inherit MixC; the A and B headers remain unchanged. The
caption states that m common additions across n classes require m times n
repeated header edits and m factored header edits.}
\label{fig:semantic-factoring-law}
\end{figure}

\leanmetapending{\LHrng{SCT}{3}{7}, \LHrng{SCT}{20}{26}, \LHrng{BWR}{45}{49}, \LH{RBF3}, \LHrng{RLC}{7}{10}}
\begin{proposition}[Generator inputs distinguish outside answers]
\label{prop:tooling-completion-law}
Fix ancestry $\PaperAncestry$ and a generator. Producing $M\geq1$ different outside sets
requires at least $\lceil\log_2 M\rceil$ fixed-width input bits. All subsets of
$d$ outside pairs require $d$ bits regardless of source length; correlation may
lower the bound. A complete tool supplies the missing pairs of
Theorem~\ref{thm:replica-completion-law}; a complete checker must identify them
(Supplement~\S\ref{supp-independent-outside-bundles}).
\end{proposition}

\leanmetapending{\LH{BWR42}, \LH{SCT24}, \LH{SCT34}, \LH{SCT59}, \LHrng{OLC}{1}{5}, \LH{OLC9}}
\begin{corollary}[Passing runs and ancestry hide requirement shape; supply routes reveal it]
\label{cor:exact-one-parent-dichotomy}
For finite $\PaperNotationLink{PaperRequired}{R^{\mathrm{req}}}$, tree shape permits exact ancestry-only one-parent supply
with $\PaperNotationLink{PaperGap}{d_\downarrow}=0$. Otherwise every one-parent representation preserving required supply
with sound ancestry supplies at least $\PaperNotationLink{PaperGap}{d_\downarrow}>0$ pairs outside.
Both branches admit programs supplying every required pair
with empty ancestry and identical successful outputs; no rule using only
these observations distinguishes zero from positive gap.
Conversely, a design supplying exactly the required relation, with ancestry $\PaperAncestry$, removal set $\PaperRemovals$, and
outside routes $\PaperNotationLink{PaperRoutes}{\varphi}$ has $\PaperNotationLink{PaperRequired}{R^{\mathrm{req}}}=(\PaperNotationLink{PaperAncestry}{H}\setminus\PaperRemovals)\cup\operatorname{dom}\PaperNotationLink{PaperRoutes}{\varphi}$.
The full supply map reveals requirement shape and determines $\PaperNotationLink{PaperGap}{d_\downarrow}(\PaperNotationLink{PaperRequired}{R^{\mathrm{req}}})$;
views omitting part of it can hide that shape. Within
Theorem~\ref{thm:delegation-lower-bound}'s scope, $\PaperRemovals=\varnothing$ and
$|\operatorname{dom}\PaperNotationLink{PaperRoutes}{\varphi}\setminus \PaperNotationLink{PaperAncestry}{H}|\ge \PaperNotationLink{PaperGap}{d_\downarrow}(\PaperNotationLink{PaperRequired}{R^{\mathrm{req}}})$.
\end{corollary}

\subsection{Even relaxed supply keeps crossings costly}

\paragraph{Overrides preserve superclass promises.}
Overrides must preserve superclass contracts; typechecking alone is insufficient
(Section~\ref{sec:behavioral-contracts}). Removing unwanted implementation supply
cannot cancel a superclass promise. The preceding results count missing pairs;
here we also charge removal. Allowing unwanted pairs instead changes requirements
(Section~\ref{sec:motivation}).
\leanmeta{\LHrng{BWR}{33}{36}, \LHrng{BLK}{1}{8}}

Let $B_f$ list the classes inheriting $f$ before removal.
Forwarding supplies $\PaperNotationLink{PaperClassGroup}{A_f}\setminus B_f$; removing $B_f\setminus \PaperNotationLink{PaperClassGroup}{A_f}$
costs $\lambda\in\mathbb N$ per pair:
\[
 \PaperNotationLink{PaperRelaxedGap}{d_{\lambda}}(\mathcal A)=
 \min_{\{B_f:f\in \PaperNotationLink{PaperImplementations}{F}\}\ \mathrm{laminar}}
 \sum_{f\in \PaperNotationLink{PaperImplementations}{F}}\Bigl(
   |\PaperNotationLink{PaperClassGroup}{A_f}\setminus B_f|+\lambda|B_f\setminus \PaperNotationLink{PaperClassGroup}{A_f}|
 \Bigr).
\]

$d_1$ is Chen et al.'s minimum-flip distance to a tree-shaped family
~\cite{chen2006minimum}; $\lambda=0$ makes removal free.
One rule can remove several pairs.

\Needspace{2\baselineskip}
\leanmetapending{\LHrng{BLK}{1}{10}}
\begin{theorem}[Even relaxed supply keeps crossings costly]
\label{thm:paid-blocking-gap}
The relaxed finite one-parent supply minimum is $\PaperNotationLink{PaperRelaxedGap}{d_\lambda}(\mathcal A)$ when
forwarded required pairs cost one and removed unwanted pairs cost $\lambda$.
An optimum forwards $\PaperNotationLink{PaperClassGroup}{A_f}\setminus B_f$ and removes $B_f\setminus \PaperNotationLink{PaperClassGroup}{A_f}$.
Free removal gives $d_0(\mathcal A)=0$. Always
$\PaperNotationLink{PaperRelaxedGap}{d_\lambda}(\mathcal A)\le \PaperNotationLink{PaperGap}{d_\downarrow}(\mathcal A)$, with equality when
$\lambda\ge \PaperNotationLink{PaperGap}{d_\downarrow}(\mathcal A)$. If $\lambda>0$ and $\mathcal A$ is
nonlaminar, $\PaperNotationLink{PaperRelaxedGap}{d_\lambda}(\mathcal A)>0$
(proof: Supplement~\S\ref{supp-relaxed-supply-proof}).
\end{theorem}

Exact native supply has a provider-preserving multiple-parent presentation.
A registry, generator, or forwarding layer still determines every required
provider--class pair it supplies outside the original provider's ancestry.

\Needspace{2\baselineskip}
\leanmetapending{\LHrng{BWR}{9}{11}, \LHrng{NHF}{13}{14}, \LHrng{MGH}{9}{11}, \LHrng{UMS}{1}{6}}
\begin{corollary}[Complete supply characterizes correct maintenance]
\label{cor:strict-derivation-dominance}
On any finite crossing with distinct providers, multi-parent ancestry is at
$(0,0)$; every exact-supply one-parent design with sound ancestry is at $(0,m)$,
$m\ge \PaperNotationLink{PaperGap}{d_\downarrow}>0$. The coordinates are (violations, outside required answers).
Multiple parents strictly Pareto-dominate on these counts: equal violations and
zero outside pairs against a positive count. Fixed precedence conflicts do not change this order.
Among the declared supply rungs, with nominal identities fixed, a mechanism attains $\PaperMaintenanceIdeal$
on every finite declared history iff it has multiple contract and implementation parents.
The complete construction has zero outside supply and $\PaperNotationLink{PaperBurden}{B_{\mathrm{rep}}}=0$; lower rungs fail snapshot witnesses
(Supplement~\S\ref{supp-monotone-multiparent-attainment}).
\end{corollary}

Among behaviorally correct designs, multiple parents leave zero outside
pairs against exactly $\PaperNotationLink{PaperGap}{d_\downarrow}>0$ in the best sound one-parent design.
Price that work by the fact, by the batch, or with bulk discounts. Multiple
parents remain strictly cheaper on every crossing exactly when doing any
outside work costs more than doing none
(Supplement~Theorem~\ref{supp-thm:accounting-characterization}).
\leanmeta{\LH{OPT1}, \LHrng{BWR}{25}{27}, \LHrng{ACF}{1}{4}}

Even zero-gap updates can force an old parent-edge deletion, forwarding, or removal
(Theorem~\ref{thm:laminar-growth-reparenting}).
\leanmeta{\LHrng{NHF}{59}{62}} External routing creates choices that ancestry
no longer fixes, so required supply must be preserved in every reachable configuration.

Independent omission of either route in Figure~\ref{fig:storage-encodings}
gives four configurations, only one complete
(Theorem~\ref{thm:external-completeness-singleton}).  The caption states why
the monolith loses the required provider identities.

\paragraph{Required supply across configurations.}
Two representations \emph{preserve the same required supply} iff, at every reachable
state $t$, every reachable setup supplies every required implementation and no
forbidden one: $\PaperNotationLink{PaperSupplied}{R_D}=\PaperNotationLink{PaperRequired}{R_t^{\mathrm{req}}}$. One working setup does not establish this.
\leanmeta{\LHrng{BWR}{40}{41}}

\leanmetapending{\LHrng{BWR}{19}{21}, \LH{BWR24}, \LHrng{BWR}{29}{30}, \LHrng{BWR}{37}{38}, \LHrng{BWR}{45}{49}, \LH{SCT23}, \LHrng{SCT}{30}{34}}
\begin{theorem}[Reachable omissions violate required supply]
\label{thm:external-completeness-singleton}
Partition the $d$ required answers supplied separately into $b$ nonempty groups,
each enabled by one control.  A configuration is complete iff all $b$ controls
are enabled; a design preserves required supply iff every reachable configuration is complete.
Independent controls admit $2^b$ subsets, with $2^b-1$ incomplete.  The grouping
rule determines all $d$ answers.
Distinguishing the complete map from every single-pair omission exercises
every outside pair: $|\PaperNotationLink{PaperRequired}{R^{\mathrm{req}}}|$ without implementation inheritance,
$\PaperNotationLink{PaperGap}{d_\downarrow}$ in an optimal sound one-parent design, and zero with multiple
parents. Every crossing has a positive one-parent outside count.
\end{theorem}

Preserving supply is a prerequisite for correct maintenance; code, a checker, or
a generator must determine all $d$ missing pairs. Independent groups
need $b$ fixed-width bits (Supplement~\S\ref{supp-independent-outside-bundles}).
\leanmeta{\LHrng{BWR}{37}{39}, \LHrng{BWR}{45}{49}, \LH{BWR57}}

\subsection{Strict insertion forces repair at tree-shaped updates}

Adding functionality can make a class need code from a provider it did not inherit.
Even when requirements grow and stay tree-shaped, inserting that provider strictly
between existing ancestors forces extra changes under one parent.
The history below repeats this at every update while the gap stays zero.
Classes $a,b,c$ all need top; only $a$ needs left and only $b$ needs right.
Adding right to $a$'s requirements puts right strictly between left and top:
part~(0) uses $x=\mathrm{left}$, $g=\mathrm{right}$, $q=\mathrm{top}$.
One parent redirects left's parent edge from top to right; multiple parents keep it
and add right (Figure~\ref{fig:laminar-growth-transition}).

\begin{figure}[htbp]
\begin{minipage}[c]{0.64\linewidth}
\centering
\begingroup\PaperFixedDiagramSetup
\begin{tikzpicture}[
  x=1.05cm,
  every node/.style={font=\scriptsize},
  role/.style={draw,rounded corners=2pt,inner xsep=2pt,inner ysep=1.2pt,
    minimum width=10mm},
  old/.style={thick},
  removed/.style={red!70!black,dashed,thick},
  added/.style={green!45!black,very thick},
  change label/.style={font=\scriptsize,fill=white,inner sep=1.2pt}
]
  \node[font=\scriptsize\bfseries] at (0,1.38) {state $k$};
  \node[role] (t1) at (0,0.94) {top};
  \node[role] (l1) at (-0.72,0.20) {left};
  \node[role] (r1) at (0.72,0.20) {right};
  \draw[old] (t1)--(l1);
  \draw[old] (t1)--(r1);

  \node[font=\scriptsize\bfseries] at (2.72,1.38) {one parent at $k+1$};
  \node[role] (t2) at (2.72,0.94) {top};
  \node[role] (r2) at (2.72,0.39) {right};
  \node[role] (l2) at (2.72,-0.16) {left};
  \draw[old] (t2)--(r2);
  \draw[added,->,>=stealth] (r2.south)--(l2.north);
  \draw[removed] (t2.west) to[out=195,in=165]
    node[pos=.25,left=2pt,change label,text=red!70!black]{remove} (l2.west);

  \node[font=\scriptsize\bfseries] at (5.75,1.38) {multiple parents at $k+1$};
  \node[role] (t3) at (5.75,0.94) {top};
  \node[role] (l3) at (5.03,0.20) {left};
  \node[role] (r3) at (6.47,0.20) {right};
  \draw[old] (t3)--(l3);
  \draw[old] (t3)--(r3);
  \draw[added,->,>=stealth] (r3.south) to[out=235,in=305]
    node[midway,above=3pt,change label,text=green!45!black]{add} (l3.south);
\end{tikzpicture}
\endgroup
\end{minipage}\hfill
\begin{minipage}[c]{0.33\linewidth}
\caption{Same tree-shaped update.  Every ancestry-only tree changes at least
two direct-parent edges; multiple parents retain both and add one.}
\Description{Three diagrams show an initial top node with left and right
children, an updated one-parent chain that removes top--left, and an updated
multi-parent triangle that retains both old edges and adds right--left.}
\label{fig:laminar-growth-transition}
\end{minipage}
\end{figure}

Let $\PaperEdgeFootprint$ be the pairs whose old ancestry paths use edge $e$.
A deletion is charged iff $\PaperEdgeFootprint\cap \PaperNotationLink{PaperRequired}{R^{\mathrm{req}}_{t+1}}\ne\varnothing$.
One-parent paths are unique: a charged deletion destroys one for a still-required pair,
even if re-derived; addition-only updates retain old paths
(Supplement~\S\ref{supp-derivation-retention}).
$\PaperNotationLink{PaperBurden}{B_{\mathrm{rep}}(n)}$ sums charged deletions once and
missing or removed pairs at both update endpoints. An unchanged missing pair counts twice.
\leanmetapending{\LHrng{SIN}{1}{5}, \LHrng{IPE}{1}{28}, \LHrng{IPE}{33}{38}, \LHrng{SIP}{1}{7}, \LHrng{NRH}{1}{6}, \LHrng{NRH}{8}{9}, \LHrng{OPT}{3}{4}, \LHrng{BWR}{50}{56}, \LHrng{DRP}{3}{4}}
\begin{theorem}[Strict insertion forces one-parent repair]
\label{thm:laminar-growth-reparenting}
\textbf{(0) Strict insertion.} Let $\PaperNotationLink{PaperRequired}{R^{\mathrm{req}}_t}\subseteq \PaperNotationLink{PaperRequired}{R^{\mathrm{req}}_{t+1}}$ be both tree-shaped,
with fixed providers $x,g,q$ satisfying $\PaperNotationLink{PaperClassGroup}{A^t_x},\PaperNotationLink{PaperClassGroup}{A^{t+1}_x}\ne\varnothing$,
$\PaperNotationLink{PaperClassGroup}{A^t_x}\subsetneq \PaperNotationLink{PaperClassGroup}{A^t_q}$, $\PaperNotationLink{PaperClassGroup}{A^t_x}\nsubseteq \PaperNotationLink{PaperClassGroup}{A^t_g}$, and
$\PaperNotationLink{PaperClassGroup}{A^{t+1}_x}\subsetneq \PaperNotationLink{PaperClassGroup}{A^{t+1}_g}\subsetneq \PaperNotationLink{PaperClassGroup}{A^{t+1}_q}$.
Every one-parent update preserving required supply on a fixed set of nodes deletes an old parent
edge carrying a still-required pair or needs forwarding or removal before or
after the update, even with helpers.
\end{theorem}

\begin{proof}[Proof of (0)]
Absent endpoint forwarding or removal, $q$ is initially above $x$ and $g$ must
end between them. The initial $\PaperNotationLink{PaperClassGroup}{A^t_x}\nsubseteq \PaperNotationLink{PaperClassGroup}{A^t_g}$ excludes $g$ from that path.
Inserting it deletes an old edge carrying a still-required pair to $q$, even with helpers.
\end{proof}

\begin{theoremcontinuation}[Repeated history]
For every $n$, fix providers $f_0,\ldots,f_n$, top, and classes $c_0,\ldots,c_{n+1}$.
At state $k$, top is needed by all classes; $f_i$'s required set is
$\{c_0,\ldots,c_i\}$ for $i\le k$ and $\{c_i\}$ otherwise.
Every one-parent policy preserving required supply on a fixed finite node set, allowing forwarding,
removal, and arbitrary preallocated helpers (later helpers start isolated), has:
\begin{enumerate}
  \item \textbf{Static fit.} Every snapshot is tree-shaped and has ancestry
        gap zero;
  \item \textbf{Additive multi-parent update.} A direct exact-supply multi-parent
        graph retains every old edge and adds one per step, without helpers,
        forecast, forwarding, or removal.
        Python lists $f_i$ before top in $f_{i-1}$'s bases; the reverse fails C3;
  \item \textbf{Repair at every one-parent update.} Every step deletes an old parent
        edge or needs forwarding or removal at an endpoint;
  \item \textbf{Excess over the semantic floor.} Every update needs an operation
        (Proposition~\ref{prop:forced-semantic-unit}); multiple parents use one addition.
        Let $\delta$ count charged deletions and $\pi$ count updates with missing or
        removed pairs at either endpoint, once per update. One parent has
        $\delta+\pi\ge n$, hence $\PaperNotationLink{PaperBurden}{B_{\mathrm{rep}}(n)}\ge n$. Standing pairs need not change.
        The direct graph has $\delta=\pi=\PaperNotationLink{PaperBurden}{B_{\mathrm{rep}}(n)}=0$;
  \item \textbf{Quadratic repair when the parent map returns.} If final and initial
        parent maps agree, final missing plus initially removed unwanted pairs
        number at least $n(n+1)/2$, regardless of intermediate maps.
\end{enumerate}
\end{theoremcontinuation}

\begin{proof}[Proof of (1)--(5)]
Initially $c_i\to f_i\to\mathrm{top}$ ($i\le n$), $c_{n+1}\to\mathrm{top}$;
add $f_{i-1}\to f_i$ at update $i$. Increasing indices prove (1)--(2).
Apply (0) to $(f_{i-1},f_i,\mathrm{top})$ and sum for (3)--(4);
the top-only class makes insertion strict. Returning parent maps force initial
removal or final forwarding for all $\Pi=\{(f_i,c_j):0\le j<i\le n\}$,
$|\Pi|=n(n+1)/2$, proving (5) (paths: Supplement~\S\ref{supp-immediate-parent-evolution}).
\end{proof}

Zero gap in every version does not mean repair-free evolution.
Ancestry-only policies make at least $n$ deletions and $2n$ parent changes.
Multiple-parent additions use the current graph and $f_i$; (3) holds for every
prefix and every supply-preserving one-parent policy.

\leanmetapending{\LHrng{SCT}{38}{39}, \LHrng{IPE}{29}{32}, \LHrng{LRR}{1}{7}}
\begin{corollary}[An unchanged hierarchy leaves new requirements outside]
\label{cor:stable-exact-carrier-conservation}
Let a finite hierarchy supply exactly $\PaperNotationLink{PaperRequired}{R^{\mathrm{req}}_0}$ and stay unchanged when
requirements become $\PaperNotationLink{PaperRequired}{R^{\mathrm{req}}_T}$.  Its missing and unwanted pairs
correspond one-to-one with $\PaperNotationLink{PaperRequired}{R^{\mathrm{req}}_T}\mathbin{\triangle}\PaperNotationLink{PaperRequired}{R^{\mathrm{req}}_0}$.  In
this update sequence, if the initial one-parent hierarchy has no outside pair and
the final parent map equals the initial map, the final outside count is exactly
$n(n+1)/2$.

For fixed one-parent ancestry $\PaperNotationLink{PaperAncestry}{H}\subseteq \PaperNotationLink{PaperRequired}{R^{\mathrm{req}}_0}\subseteq \PaperNotationLink{PaperRequired}{R^{\mathrm{req}}_T}$, every added
pair increases outside supply by one. Excess $\PaperNotationLink{PaperExcess}{e_H}$ cannot decrease and rises
strictly when the gap grows by fewer pairs than the requirement
(algebra: Supplement~\S\ref{supp-immediate-parent-evolution}).
\end{corollary}

Composition obeys this bound; Proposition~\ref{prop:forced-semantic-unit} bounds the operations needed to supply these pairs.
\leanmeta{\LHrng{OPT}{3}{4}}

Parking top avoids deletions and route edits but leaves $n+2$ outside pairs at
every snapshot, missing the zero-outside optimum.
Rechecking those routes at every update is a separate protocol choice (Supplement~\S\ref{supp-fixed-top-routing};
Corollary~\ref{cor:factorization-economics}).

\leanmetapending{\LHrng{BWR}{22}{23}, \LH{BWR28}, \LHrng{BWR}{43}{44}, \LHrng{BWR}{50}{56}, \LHrng{GRI}{1}{4}, \LHrng{SCC}{1}{9}}
\begin{proposition}[Batching shares operations across facts]
\label{prop:forced-semantic-unit}
The gap counts required yes-or-no answers.
\begin{enumerate}
  \item Zero on the empty set and additivity over disjoint answer sets make
        $Q(S)$ the sum of individual charges. Charging one per answer uniquely
        gives $Q(S)=|S|$; positive charges make nonempty sets costly.
        Semantic edits count changed required answers, the incidence tables' Hamming distance:
        $|\PaperNotationLink{PaperRequired}{R^{\mathrm{req}}_0}\mathbin{\triangle}\PaperNotationLink{PaperRequired}{R^{\mathrm{req}}_1}|$.
  \item Adding unrelated required pairs can make
        $\PaperNotationLink{PaperGap}{d_\downarrow}/|\PaperNotationLink{PaperRequired}{R^{\mathrm{req}}}|$ arbitrarily small without reducing
        $\PaperNotationLink{PaperGap}{d_\downarrow}$.
  \item Fix an operation type changing at most $b>0$ required answers per operation.
        Sound one-parent supply needs at least $\lceil \PaperNotationLink{PaperGap}{d_\downarrow}/b\rceil$ operations;
        changing requirements needs $\lceil |\PaperNotationLink{PaperRequired}{R^{\mathrm{req}}_0}\mathbin{\triangle}\PaperNotationLink{PaperRequired}{R^{\mathrm{req}}_1}|/b\rceil$.
        A generator can preserve the answers with only the specification edited by a human.
        One reevaluation can cover the entire change; every faithful representation
        must execute a meaning-changing step (Supplement~Theorem~\ref{supp-thm:semantic-change-positive-cost}).
        Batching shares an operation's positive price across its answers
        (Corollary~\ref{cor:factorization-economics}).
        A human-edit bound needs finite $b$.
\end{enumerate}
\end{proposition}

\runinhead{Requirements are fixed before anyone writes code}
Fixing a bug to meet unchanged $\PaperNotationLink{PaperRequired}{R^{\mathrm{req}}}$ restores required facts or behavioral correctness.
The semantic-edit count is zero iff $\PaperNotationLink{PaperRequired}{R^{\mathrm{req}}_0}=\PaperNotationLink{PaperRequired}{R^{\mathrm{req}}_1}$; this does not erase standing
outside supply. We compare updates whose endpoints preserve required facts and
behavioral correctness. One specification
edit can trigger a generator determining every affected pair. Zero-error information theory
likewise fixes the required information before pricing its representation
~\cite{witsenhausen1976zero}.

New behavior is defined at its source. Slop restates facts native mechanisms would derive from that source.
Metaclasses, descriptors, \texttt{\_\_getattribute\_\_}, and generated ancestry can
reuse a rule iff equal inputs always require equal results. A different required result for
the same input forces a rule change (exact minimum: Supplement~\S\ref{supp-exact-rule-revisions}).
\leanmeta{\LHrng{RBF}{3}{6}}

Asking whether class $c$ gets implementation $f$ does not mean one can list everything $c$ gets
(Supplement~\S\ref{supp-finite-rule-generation}). Grouping or generating
routes does not lower the ancestry gap.
\leanmeta{\LHrng{BWR}{1}{2}, \LHrng{BWR}{8}{9}, \LH{BWR12}, \LHrng{SCT}{24}{25}, \LHrng{SDN}{13}{17}}

\leanmetapending{\LHrng{BLK}{1}{10}, \LHrng{NHF}{53}{58}, \LHrng{NHF}{64}{68}}
\begin{theorem}[Conflicts survive answer-preserving refactoring]
\label{cor:coverage-resolution}
Fix required pairs, implementation member names, and the required winner at every
lookup. Let $\operatorname{Conf}$ be consumer lookups with two distinct implementations
of the same name. Every answer-preserving representation, with multiple parents,
one parent, or no inheritance, has the same $\operatorname{Conf}$ and
$k=|\operatorname{Conf}|$ winner answers.
Multiple parents supply every required pair, giving count $k$.
Sound one-parent ancestry misses a set $G$ with $|G|\ge \PaperNotationLink{PaperGap}{d_\downarrow}(\mathcal A)$,
giving at least $k+\PaperNotationLink{PaperGap}{d_\downarrow}(\mathcal A)$. Unwanted ancestry with unit-cost removal
instead gives $k+d_1(\mathcal A)>k$ at a crossing. At $k=0$, multiple parents need
no precedence and leave zero answers outside; a positive sound one-parent gap still adds $\PaperNotationLink{PaperGap}{d_\downarrow}$.
\end{theorem}

Required sets can cross without same-name conflicts; $k$ and $\PaperNotationLink{PaperGap}{d_\downarrow}$ vary
independently. Qualifying provider names changes the member interface and contract.
Inheritance stores one inspectable lookup rule $\PaperLookup$; a shared resolver can
also store it, while separate overrides repeat decisions and can diverge.
Their maintenance cost is separate: SQLAlchemy's filtered component has no retained
lookup conflict and gap \SQLAlchemyComponentTax{} (Section~5).

Inherited code calls methods on the final object, so \texttt{self.m()} can reach
a subclass override: \emph{open recursion}~\cite{cook1994denotational}.
A wrapper calling on the wrapped object stays there.
\leanmetapending{\LHrng{SLF}{1}{7}}
\begin{proposition}[Self-calls move answers between routing and precedence]
\label{prop:self-call-ledger}
Fix a code provider exposing nonempty member set $E$, one consumer class, and a specified
choice $P(s)\in\{\mathsf{base},\mathsf{derived}\}$ for each call $s$ on itself.
Inheritance dispatches through the subclass, defaulting to $\mathsf{derived}$:
counts $(0,|P^{-1}(\mathsf{base})|)$.  A wrapper keeping these calls on the
wrapped object defaults to $\mathsf{base}$:
$(|E|,|P^{-1}(\mathsf{derived})|)$.  The coordinates count forwarded members and lookup choices differing from the
default; the constructions achieve both optima.  The subclass dominates exactly
when base lookup choices do not outnumber subclass lookup choices; otherwise the designs are incomparable.  One rule may batch several sites but must still determine every
lookup choice
(Proposition~\ref{prop:forced-semantic-unit}).
\end{proposition}

These calls need specified lookup choices.  Otherwise a base revision can
add unspecified call sites: the fragile-base-class problem
~\cite{bloch2018effective,kiczales1992classlibraries,
mikhajlov1998fragile,aldrich2004selective}.  Supplement~\S\ref{supp-dual-axis-resolution}
derives the counts and works through the counting-set and callback examples.

\subsection{Changing the implementation groups changes the specification}

\leanmetapending{\LH{SDN18}, \LHrng{NHF}{35}{41}}
\begin{corollary}[Changing required implementations changes the comparison]
\label{cor:specification-or-payment}
Preserving $\PaperNotationLink{PaperRequired}{R^{\mathrm{req}}}$ and its independently changing implementations leaves
$\PaperNotationLink{PaperGap}{d_\downarrow}$ unchanged. Changing either, including merging independent implementations,
changes the specification: name the required answer or independent change removed.
Changing the product may be valid; the original program's lower bound remains.
\end{corollary}

Family polymorphism deliberately lets related types vary together.  Choosing
that unit can change the implementation groups or $\PaperNotationLink{PaperRequired}{R^{\mathrm{req}}}$; it does not lower the gap for the
already chosen relation~\cite{ernst2001family}.

A helper can bundle outside pairs~\cite{theis2011bicliques}; Theorem~\ref{thm:delegation-lower-bound}
fixes their minimum.
\leanmeta{\LHrng{BWR}{3}{8}, \LHrng{BWR}{12}{18}, \LHrng{BWR}{31}{32}, \LHrng{SDN}{13}{17}, \LHrng{NHF}{29}{30}, \LHrng{NHF}{47}{48}}

\subsection{No algorithm certifies every permanently tree-shaped future}
\label{sec:open-world-laminarity}

\leanmetapending{\LHrng{UND}{31}{38}, \LH{UND41}}
\begin{theorem}[No algorithm certifies every permanently tree-shaped future]
\label{thm:open-world-laminarity}
In a fixed computable two-implementation, three-class requirement-stream language,
permanent tree shape is co-r.e.-complete: a finite state witnesses each failure,
but no procedure certifies exactly all safe generators. Multiple parents
derive the required pairs in every state of the same stream.
\end{theorem}

At state $t$, simulate $e(0)$ for $t$ steps: emit nonoverlapping one-class sets
before halting and the fixed crossing afterward. Every table is checkable;
all are tree-shaped exactly when $e(0)$ never halts. This explicit reduction
establishes the boundary; Rice's theorem supplies broader context~\cite{rice1953classes}.

A computable checker that never accepts an unsafe generator must leave some
permanently tree-shaped generator uncertified. A different generator producing
the same history may still be certified.
\leanmeta{\LH{UND39}}

Forwarding, registries, and adapters supply a crossing's missing pairs without
changing $\PaperNotationLink{PaperRequired}{R^{\mathrm{req}}}$ or restoring native capacity.

\paragraph{Finite tests do not certify arbitrary open-ended behavior.}
Passing tests can miss reachable cases; coverage is complete iff $|V\setminus T|=0$,
which a separate proof can establish.
One-test completeness in a fixed
computable trace family is co-r.e.-complete (proof: Supplement~\S\ref{supp-behavioral-substitution}).
\footnote{\LHrng{TST}{1}{10}}

\section{Measured Ancestry Gaps in Python}\label{sec:motivation}
Extraction fixes the counted pairs. Designs attaining checked lower bounds
establish objective, exact minima.
We count one pair when a class inherits a method body from a code provider. Abstract
methods and bodies replaced by overrides do not count. Several methods from one
provider still count as one pair. These recovered implementation requirements form $R_{\mathrm{obs}}$.

\paragraph{Requiredness witness: AnyIO omissions.}
AnyIO's whole-project study replaces omitted routes with failing method stubs.
After a \MutationBaselinePassed{}-test baseline, the pinned suite
reaches and rejects all \MutationDenominator{} single-route omissions,
witnessing requiredness. Each two-route omission runs both route-specific tests.
All \MutationPairDetected{}
omissions are detected, both routes reached in \MutationPairBothMarkers{}
cases (Supplement~\S\ref{supp-anyio-mutation-study}).

Theorem~\ref{thm:delegation-lower-bound} minimizes over every sound one-parent design.
Changing $R_{\mathrm{obs}}$ changes a named implementation--class pair
(Corollary~\ref{cor:specification-or-payment}).

\leanmetapending{\LH{DEC7}}
\begin{proposition}[Omitting required pairs can raise or lower the gap]
\label{prop:dropped-requirements-unsigned-gap}
For recovered relations $R\subseteq \PaperNotationLink{PaperRequired}{R^{\mathrm{req}}}$, neither
$\PaperNotationLink{PaperGap}{d_\downarrow}(R)\le \PaperNotationLink{PaperGap}{d_\downarrow}(\PaperNotationLink{PaperRequired}{R^{\mathrm{req}}})$ nor
$\PaperNotationLink{PaperGap}{d_\downarrow}(R)\ge \PaperNotationLink{PaperGap}{d_\downarrow}(\PaperNotationLink{PaperRequired}{R^{\mathrm{req}}})$ holds in general.
\end{proposition}
Supplement~\S\ref{supp-python-population-protocol} gives both minimal witnesses and their checked gaps.

\paragraph{Public-name contract.}
Each selected public method must still resolve to its original concrete provider,
including dunders. This public-name contract forces
\PublicApiPairs{} provider--class pairs with certified exact gap \PublicApiGap{} under sound ancestry
(Table~\ref{tab:audit-input-results};
Supplement~\S\ref{supp-python-public-api-contract}).
\leanmeta{\LHrng{APC}{1}{2}}

Calls, fields, registries, and generated code can hide crossings from ancestry.
Unlike inheritance-usage and fixed-hierarchy studies
~\cite{tempero2008inheritance,tempero2013programmers,ducournau2009empirical}, we optimize
every preserving one-parent hierarchy.
\leanmeta{\LH{BWR12}, \LH{SDN18}}

\paragraph{Audit inputs.}
The primary input groups method bodies by their most-derived declaring class; the separate-method
input lets methods change independently. SQLAlchemy's filtered input excludes
order-sensitive lookups; \emph{Closed-C3} keeps locally determined C3 choices.
Each comparison fixes different implementation requirements (Table~\ref{tab:audit-input-results}).

A unique concrete body among incomparable most-derived declarations owns the
lookup; competing bodies stay unattributed (Theorem~\ref{cor:coverage-resolution}).
External or unresolved bases, custom MRO,
runtime mutation, and descriptors are excluded here, though the model covers
descriptors (Supplement~\S\ref{supp-python-population-protocol}: exclusions and input hashes).

\begin{table}[H]
\centering
\PaperDenseTableSetup
\begin{tabular*}{\linewidth}{@{\extracolsep{\fill}}lrr@{}}
\toprule
Implementation requirements & Required pairs & Minimum outside pairs \\
\midrule
20 projects, methods grouped by class & \PythonPopulationProviderIncidences & \PythonProviderOptimumTotal \\
\midrule
20 projects, public-name contract & \PublicApiPairs & \PublicApiGap \\
\midrule
20 projects, methods separate & \PythonMethodIncidences & \PythonMethodSolverLowerTotal--\PythonMethodUpperTotal \\
\midrule
SQLAlchemy filtered component & \SQLAlchemyComponentProviderIncidences & \SQLAlchemyComponentTax \\
\midrule
SQLAlchemy Closed-C3 component & \SQLAlchemyComponentClosedCThreeProviderIncidences & \SQLAlchemyComponentClosedCThreeTax \\
\bottomrule
\end{tabular*}
\caption{Optima are certified except the separate-method row: 19 solver-optimal values plus
SQLAlchemy's \textsc{feasible} interval. Rows fix different requirements; components are project parts.}
\label{tab:audit-input-results}
\end{table}

Table~\ref{tab:python-factorization-population} totals preserve distinct providers
and replicas (Proposition~\ref{prop:role-normalization};
Theorems~\ref{cor:minimized-delegation}, \ref{thm:replica-completion-law}).
In recorded optima, \ProviderStatePairs{}/\ProviderStateTotal{} outside pairs
(\ProviderStatePercent{}\%) have state-bearing providers: they or resolved ancestors
declare nonempty \texttt{\_\_slots\_\_} or write instance attributes.
Stateless defaults cannot supply this state. Re-optimizing those providers' requirements
alone forces \StatefulGap{} outside pairs, certified exact on one sound stateful chain,
granting other providers defaults with zero outside supply, keeping each implementation and its state together in its provider
(Supplement~\S\ref{supp-provider-state-audit}).
\leanmeta{\LHrng{SRA}{1}{2}}

\paragraph{Project selection fixed before optimization.}
From the pinned \path{hugovk/top-pypi-packages} table's first 150
non-yanked universal pure-Python wheels (1~August~2026), \PythonPopulationSizeQualified{} have at least
50 classes; \PythonCrossBranchCensusProjects{} have cross-branch reuse.
Positive gaps occur in \PythonPopulationPositiveProviderTax{}/20 selected for at least
four cross-branch classes, and \PythonCrossBranchCensusPositive{}/\PythonCrossBranchCensusProjects{}
overall: at least \PythonCrossBranchCensusFrameLowerPercent{}\% of
\PythonPopulationSizeQualified{}, with \PythonCrossBranchCensusUnscored{} unscored.
Multiple-parent use need not cross: pip (10 MI classes) and aiobotocore have gap zero.
Unresolved external bases are unscored (exclusions: Supplement~\S\ref{supp-python-population-protocol}).
A second fixed rule gives exact gap 25 across 601 classes in nine pyapp-kit libraries
(Supplement~\S\ref{supp-pyappkit-portfolio}).

\Needspace{5\baselineskip}
\paragraph{Searching for forwarding through fields.}
Forwarding syntax alone does not identify shared code providers: in size-qualified
wheels without locally resolved multiple-parent classes, \texttt{self.field.m}
flags 23 crossing candidates, all rejected by tracing inherited members.
The detector finds all \PaperOneForwardingRecallDetectedPairs{} seeded AnyIO
pairs written this way, checking recall for this forwarding syntax.
Supplement~\S\ref{supp-external-route-control} gives selection, anchors, and blind spots.
Inherited routes also remove the sole candidate across three Apache Commons
libraries (Supplement~\S\ref{supp-java-forwarding-check}).

\begingroup
\begin{table}[htbp]
\centering
\PaperDenseTableSetup
\renewcommand{\arraystretch}{0.96}
\begin{tabular*}{\linewidth}{@{\extracolsep{\fill}}lrrrr@{\hspace{10pt}}lrrrr@{}}
\toprule
Project & cls. & MI & $|R|$ & $\PaperNotationLink{PaperGap}{d_\downarrow}$
& Project & cls. & MI & $|R|$ & $\PaperNotationLink{PaperGap}{d_\downarrow}$ \\
\midrule
\PythonPopulationHeadlineCompactPairedRows
\bottomrule
\end{tabular*}
\caption{Eight largest source-class gaps and two zero controls; all twenty:
Supplement~\S\ref{supp-python-factorization-results}. MI counts cross-branch concrete-method reuse.
Daggers mark Lean-checked components above 10; undaggered totals sum components at most 10.
Seven certificates cover six primary components (two SQLAlchemy) plus Closed-C3.}
\label{tab:python-factorization-population}
\end{table}
\endgroup

\textbf{Source-declared first-parent baseline.}
Global optimization reduces the first-parent outside count nearly twelvefold,
from \PythonFixedHierarchyTotal{} to \PythonProviderOptimumTotal{}.
Of the \PythonFixedHierarchyTotal{}, 91.6\% are removable excess; one-parent
incompleteness forces the remaining 8.4\%, the \PythonProviderOptimumTotal{}.

Figure~\ref{fig:sqlalchemy-crossing-scopes}'s classes independently combine SQL expression
semantics (\texttt{ColumnOperators}, \texttt{FromClause}) and executability
(\texttt{Executable}), producing overlap without containment.

\textbf{Primary component optimum.}
SQLAlchemy's largest component has 223 code providers, 4,831 required pairs, and gap
\SQLAlchemyComponentTax{} of the project's \PythonLongitudinalLastGap{}.
A relation-level hierarchy and its outside routes attain \SQLAlchemyComponentTax{};
executable integration is separate. A Lean-checked LRAT certificate rules out
one fewer outside pair under sound one-parent ancestry.
\textbf{Bundles for the recorded design.}
A bundle is a rectangle of missing pairs: implementations grouped with classes
needing all of them. The \emph{minimum biclique cover} of $\PaperNotationLink{PaperRequired}{R^{\mathrm{req}}}\setminus \PaperNotationLink{PaperAncestry}{H}$
for this design is \SQLAlchemyComponentBooleanRank{} bundles~\cite{theis2011bicliques};
other gap-optimal hierarchies can need different counts.
Lean-checked LRAT excludes \IndependentHolderExcludedBound{} bundles;
Supplement~\S\ref{supp-sqlalchemy-component} gives the maximal-rectangle construction.
Each bundle adds uncovered pairs, so independent enabling makes all
$2^{\SQLAlchemyComponentBooleanRank{}}$ subsets distinct, exactly one complete
(Theorem~\ref{thm:external-completeness-singleton}; Supplement~\S\ref{supp-sqlalchemy-component}).
\leanmeta{\LHrng{HCC}{1}{2}}

\textbf{Supply and member lookup choices.}
Excluding \SQLAlchemyComponentOmittedMroSensitiveLookups{} order-sensitive
lookups leaves 520 outside pairs with no retained conflict.

Closed-C3 keeps locally resolved lookup choices:
$(k,\PaperNotationLink{PaperGap}{d_\downarrow})=(\SQLAlchemyComponentClosedCThreeSelectorConflicts{},
\SQLAlchemyComponentClosedCThreeTax{})$ over
\SQLAlchemyComponentClosedCThreeProviderIncidences{} pairs. A design attains this bound;
Lean certifies it. One parent keeps every choice and adds the
outside pairs.
SQLAlchemy changed collection instrumentation to follow Python's MRO under
unpredictable subclass choices: precedence maintenance, separate from the 520
supply count~\cite{SQLAlchemyMROFix}.

\begin{figure}[htbp]
\begin{minipage}[c]{0.28\linewidth}
\centering
\begin{tikzpicture}[font=\scriptsize,every node/.style={inner sep=1pt}]
  \fill[blue!18,opacity=.75] (-0.34,0.15) circle (0.65);
  \fill[orange!24,opacity=.75] (0.34,0.15) circle (0.65);
  \fill[green!22!black,opacity=.28] (0,-0.39) circle (0.65);
  \draw[blue!65!black,thick] (-0.34,0.15) circle (0.65);
  \draw[orange!80!black,thick] (0.34,0.15) circle (0.65);
  \draw[green!45!black,thick] (0,-0.39) circle (0.65);
  \node[align=center,font=\scriptsize\bfseries,text=blue!60!black]
    at (-0.72,1.08) {\texttt{Column}\\\texttt{Operators}};
  \node[align=center,font=\scriptsize\bfseries,text=orange!75!black]
    at (0.72,1.08) {\texttt{Executable}};
  \node[align=center,font=\scriptsize\bfseries,text=green!45!black]
    at (0,-1.25) {\texttt{FromClause}};
  \node at (-0.67,0.20) {\SQLAlchemyComponentCrossingOnlyA};
  \node at (0.67,0.20) {\SQLAlchemyComponentCrossingOnlyB};
  \node at (0,-0.76) {\SQLAlchemyComponentCrossingOnlyC};
  \node at (0,0.49) {\SQLAlchemyComponentCrossingOnlyAB};
  \node at (-0.36,-0.30) {\SQLAlchemyComponentCrossingOnlyAC};
  \node at (0.36,-0.30) {\SQLAlchemyComponentCrossingOnlyBC};
  \node[font=\scriptsize\bfseries] at (0,-0.02) {\SQLAlchemyComponentCrossingAll};
\end{tikzpicture}
\end{minipage}\hfill
\begin{minipage}[c]{0.69\linewidth}
\caption{Three crossing SQLAlchemy groups share
\SQLAlchemyComponentCrossingAll{} classes; exclusive regions are
\SQLAlchemyComponentCrossingOnlyA{}, \SQLAlchemyComponentCrossingOnlyB{}, and
\SQLAlchemyComponentCrossingOnlyC{}, with pair-only regions zero. Regions count classes; lookup choices have their own count.
}
\Description{A three-circle diagram shows ColumnOperators, Executable, and
FromClause sharing 53 consumers while separately reaching 71, 61, and 17
consumers; all three pair-only regions are labelled zero.}
\label{fig:sqlalchemy-crossing-scopes}
\end{minipage}
\end{figure}

The source-class gap is $739/\text{\PythonPopulationProviderIncidences{}}=5.7\%$;
unrelated pairs change this fraction without changing the gap
(Proposition~\ref{prop:forced-semantic-unit}). Hidden routes must also be recovered
(Corollary~\ref{cor:exact-one-parent-dichotomy}).

\Needspace{4\baselineskip}
\paragraph{Missing inherited pairs across versions.}
As recovered requirements evolve across \PythonLongitudinalSnapshots{} pinned SQLAlchemy versions, pairs missing
from each class's first parent change in all \PythonLongitudinalTransitions{}
transitions:
\PythonLongitudinalCompletionEdits{} additions or deletions, including
\PythonLongitudinalStableCompletionChurn{} on still-required pairs.
The independently reconstructed edit count and initial gap both equal
\PythonLongitudinalFirstGap{}; no result relates them.
Supplement~\S\ref{supp-longitudinal-completion} gives count trajectories, versions, hashes, and replay.

\textbf{Strict-insertion trigger audit.}
Strict insertion forces repair (Theorem~\ref{thm:laminar-growth-reparenting}(0)).
The preregistered \StrictInsertionPipSourceClassSnapshots{} pip and
\StrictInsertionAiobotocoreSourceClassSnapshots{} aiobotocore versions are tree-shaped.
Their \StrictInsertionPipSourceClassTransitions{} and \StrictInsertionAiobotocoreSourceClassTransitions{}
transitions contain no trigger on persisting providers/classes, nor does
aiobotocore's separate-method input.
Supplement~\S\ref{supp-strict-insertion-audit} gives protocol and replay.

\textbf{Rewiring and relaxed supply.}
The random comparison changes which code providers each class needs. Each class
still needs the same number of providers, and each provider still serves the same
number of classes. Across \PythonNullSamples{} samples, verified lower bounds are
\PythonNullTotalMinimum--\PythonNullTotalMaximum{}: every sample's gap exceeds
the observed \PythonNullObservedTotal{}.

With unit-cost forwarding and removal, the source-class minimum is
\PythonPaidBlockingAggregate{} (solver lower; checked design upper),
against the sound minimum \PythonProviderOptimumTotal{}.
Each provider--class pair is declared required, forbidden, or tolerated:
tolerated pairs may be inherited or omitted without charge.
Tolerating omission of $t$ recovered pairs lowers the gap by at most $t$.
The reported lower bounds hold whichever unrequired classes are tolerated,
within the per-implementation allowance
(Supplement~\S\ref{supp-python-factorization-results}).
\footnote{\LH{TOL4}, \LH{TOL6}}
Supplement~\S\ref{supp-history-rewrite-screen} screens pinned MI-to-SI syntax histories.

\textbf{Changing what counts as one implementation.}
After merging providers needed by exactly the same classes, declared
first parents leave \EqualScopeFirstParentOutside{} outside pairs;
one-parent incompleteness forces the certified floor \EqualScopeGap{}
(\EqualScopeFirstParentForcedPercent{}\%).
Any other merge changes a named implementation--class pair
(Theorem~\ref{thm:exact-semantic-factoring}; Corollary~\ref{cor:specification-or-payment};
Supplement~\S\ref{supp-equal-scope-collapse}).
Source-class grouping favors one parent. Separate method identities change
$R^{\mathrm{req}}$ beyond Proposition~\ref{prop:role-normalization}'s fixed groups;
pip's gap becomes 43 (solver-optimal), versus provider gap 0.

\FloatBarrier

\section{Discussion and Language-Design Implications}\label{sec:language-consequences}
\subsection{Behavioral correctness assumes which implementations are needed}
\label{sec:behavioral-contracts}

The Liskov substitution principle (LSP) requires subclasses to preserve superclass
promises. Blocking an inherited promise violates LSP; forwarding that
preserves behavioral correctness does not.
\leanmeta{\LHrng{BAP}{20}{25}}

In the checked counterexample, two code providers initially serve $\{W,\PaperFileReader\}$. Change the second's required
classes to $\{\PaperFileReader,B\}$, leaving the first unchanged. All four required pairs
remain behaviorally valid, but the gap rises from zero to one
(Theorems~\ref{thm:two-role-average-gap}, \ref{thm:behavioral-applicability-product}). The gap changes because the required provider changes. Split providers
into roles rather than make unwanted promised methods raise; multiple parents supply the roles.

Permitting all behavior leaves required pairs fixed. Exact supply still requires
$\PaperNotationLink{PaperSupplied}{R_D}=\PaperNotationLink{PaperRequired}{R^{\mathrm{req}}}$.
\leanmeta{\LH{RCN3}, \LHrng{RCN}{10}{11}}

For implementation requirements alone, classes needing the same providers
can share a supply node; a class needing a subset of those providers can be the supply parent.
Grouping implementation supply does not decide contract membership or erase
distinct identities.\footnote{\LHrng{RCN}{1}{7}, \LH{SCT18}, \LH{BWR14}, \LH{BWR17}}
Storage does not change this grouping; identity, lifetime, and substitution
stay separate~\cite{cook1990inheritance}.

\leanmetapending{\LHrng{BAP}{1}{25}, \LHrng{INT}{8}{9}}
\begin{theorem}[Per-pair behavioral checks cannot choose the required pairs]
\label{thm:behavioral-applicability-product}
Exact supply, $\PaperNotationLink{PaperSupplied}{R_D}=\PaperNotationLink{PaperRequired}{R^{\mathrm{req}}}$, and behavioral correctness are separate prerequisites
for correct maintenance. Semantic substitution requires that
for each required pair, the specification must permit every consumer observation
from reachable provider states or histories~\cite{liskov1994behavioral}. Neither implies the other;
even complete per-pair checks cannot choose $\PaperNotationLink{PaperRequired}{R^{\mathrm{req}}}$.
Supplement~\S\ref{supp-behavioral-substitution} gives observation sets;
Liskov--Wing's proof rules remain external.
\end{theorem}

\subsection{Is-a and has-a are not opposites}

\begin{claim}{}{prop:has-is-collapse}
\textbf{Implementation supply.}
\(\PaperNotationLink{PaperProvider}{o_f}\preceq c\Longrightarrow f@c\):
inheriting an implementation causes a class to have it.

Having an implementation does not exclude containing component objects.
Treating ``is-a'' and ``has-a'' as opposites is therefore a category error.
Ancestry and composition can supply the same required pairs, natively or through routes.
\leanmeta{\LHrng{CND}{1}{7}}
\end{claim}

\textbf{Contract membership.} A contract admits an object separately from code or
component fields. Other composition accounts state limits
~\cite{gamma1994designpatterns,bloch2018effective,lowe2015composition,
pietschmann2026inheritance}. The cited tutorial replaces \texttt{Report}'s inheritance
from \texttt{DatabaseServer} and \texttt{Logger} with fields, arguing a report
``is neither''~\cite{arjancodes2026goodoop}.
For logger behavior without identity, $\PaperNotationLink{PaperMembership}{M^{\mathrm{req}}}$ drops membership but
$\PaperNotationLink{PaperRequired}{R^{\mathrm{req}}}$ keeps the pair (Theorem~\ref{thm:contract-membership-gap}); composition supplies it outside.
The report's output agrees, but losing required memberships changes the specification
(Section~2). Outside supply is $|\PaperNotationLink{PaperRequired}{R^{\mathrm{req}}}|$ pairs without ancestry,
$\PaperNotationLink{PaperGap}{d_\downarrow}$ under optimal sound one-parent ancestry.
Theorem~\ref{thm:exact-semantic-factoring} counts changes to required facts on merging;
Proposition~\ref{prop:self-call-ledger} counts self-call forwarding and lookup choices.

\texttt{isinstance} and \texttt{hasattr} branches store nominal and structural
membership facts in control flow. Same-looking extensions can break \texttt{hasattr}
(Proposition~\ref{prop:identity-capacity}); in Figure~\ref{fig:storage-encodings},
inheriting \texttt{MixA} derives \texttt{x.attr1}.

Forwarding, registries, and adapters supply missing implementations; independently edited
copies can disagree. Exclude every invalid reachable state.
\leanmeta{\LHrng{BWR}{58}{62}}

\subsection{Names and implementation inheritance solve independent problems}
\label{sec:naming-reach-independence}

Even with all code inherited, look-alike records needing different handling need
extra identity keys (Theorem~\ref{thm:future-conflict}):
$m$ pairwise confusable meanings need $m-1$ even at zero gap. Independently,
distinct identities do not remove crossings: the $h\times h$ row--column family
needs $h(h-1)$ outside pairs but no extra name.
\leanmeta{\LH{NHF15}, \LH{NHF28}, \LH{EVO17}, \LHrng{EVO}{23}{30}, \LH{EVO40}, \LHrng{AVG}{1}{7}}

Successful outputs and ancestry cannot identify the gap; ancestry, removals,
and outside routes recover $\PaperNotationLink{PaperRequired}{R^{\mathrm{req}}}$ (Corollary~\ref{cor:exact-one-parent-dichotomy}).
\leanmeta{\LH{BWR12}, \LHrng{BWR}{29}{30}, \LHrng{BWR}{37}{39}, \LH{BWR42}, \LHrng{SCT}{1}{9}, \LHrng{SCT}{33}{34}, \LHrng{OLC}{1}{5}, \LHrng{OLC}{7}{8}, \LHrng{NPS}{1}{9}}

\subsection{Language mechanisms differ in what they derive}

\begin{table}[!ht]
\centering
\PaperDenseTableSetup
\renewcommand{\arraystretch}{0.94}
\begin{tabularx}{\linewidth}{@{}>{\raggedright\arraybackslash}m{0.25\linewidth}XXXX@{}}
\toprule
Language mechanisms & Identity & Contract membership & Stateless supply & Stateful supply \\
\midrule
Python classes~\cite{PythonDataModel} & nominal & multiple nominal & multiple bases & multiple bases \\
\midrule
C++ classes~\cite{CppWorkingDraft} & nominal & multiple nominal & multiple bases & multiple bases \\
\midrule
Scala traits~\cite{ducasse2006traits,Scala3Traits,Scala3TraitParameters} & nominal & multiple nominal
  & multiple traits & multiple traits \\
\midrule
Java: classes, interfaces~\cite{JavaLanguageSpecificationInterfaces} & nominal & multiple nominal
  & multiple defaults & one class chain \\
\midrule
C\#: classes, interfaces~\cite{CSharpInterfaces} & nominal & multiple nominal
  & multiple defaults & one class chain \\
\midrule
Rust: traits, \texttt{Deref}~\cite{RustTraitsReference,RustImplementationsReference} & nominal
  & multiple explicit \texttt{impl}s & multiple defaults & one \texttt{Deref} chain~\cite{RustDerefReference} \\
\midrule
Go: types, embedding~\cite{GoSpecification} & named types & structural; maintained seals
  & multiple promotions & multiple embeddings \\
\midrule
TypeScript~\cite{TypeScriptDocs,TypeScriptSymbols} & structural; maintained brands
  & structural; erased & one chain + mixins & one chain + mixins \\
\midrule
Haskell type classes~\cite{Haskell2010Report} & nominal types & multiple instances
  & multiple defaults & separate data \\
\bottomrule
\end{tabularx}
\caption{Language mechanisms in the correct-maintenance partial order
(Lemma~\ref{lem:capacity-optimum-order}; Figure~\ref{fig:native-capacity-ladder}). Implementation paths supply state; contract edges admit classes;
brands and seals maintain identity evidence. Sharing and initialization are specified
answers (Section~4.1); layout is a runtime cost. State requirements preserve this order
(Corollary~\ref{supp-state-capacity-containment}). Workarounds supply missing facts
(Supplement~\S\ref{supp-language-encodings}).}
\label{tab:implementation-carriers}
\vspace{-1.2\baselineskip}
\end{table}

Table~\ref{tab:implementation-carriers} locates the required facts slop must restate:
stateful crossings in Java, C\#, and Rust; inherited state in Haskell;
identity evidence in Go seals and TypeScript brands, and TypeScript's crossing supply.
Every answer-preserving representation pays member lookup choices (Theorem~\ref{cor:coverage-resolution});
unwanted supply costs extra (Theorem~\ref{thm:paid-blocking-gap}).
Haskell type classes supply multiple contracts through superclass constraints and
multiple code providers through defaults, but no inherited state~\cite{Haskell2010Report}.
Objects combine with functional programming through typed lambda-calculus encodings
~\cite{pierce1994simple} and complement abstract data types~\cite{cook2009abstraction}.
Java admits multiple nominal contracts but one state-bearing class chain; Go embeds
multiple types but unwanted structural matches need identity evidence. Neither is
higher on all coordinates. Promotion obeys selector ambiguity
~\cite{GoSpecification} (Theorem~\ref{thm:contract-membership-gap}).

Java, Go, and CLR restrict supply; C++, Eiffel, and CLOS use the upper rung
~\cite{JavaLanguageEnvironment,GoFAQ,DotNetCLRDesign,CppWorkingDraft,
EiffelStandard,CommonLispMOP,stroustrup1989multiple}. Python, C++, and R are July 2026
TIOBE top-ten languages with state-bearing multiple parents~\cite{TIOBEJuly2026}.

\subsection{Incompleteness leaves required facts for code to supply}
\label{sec:discussion-rationales}

In SlopCodeBench, agents pass checkpoints while verbosity and erosion rise;
guidance lowers initial levels while degradation rates persist~\cite{orlanski2026slopcodebench}.
These heuristics parallel removable excess, forced outside supply, and strict-insertion
obligations. Passing runs cannot reveal outside supply
(Corollary~\ref{cor:exact-one-parent-dichotomy}).

\runinhead{Language-design rationales} If designs keep their costs, a ban cannot lower
the best cost, even across projects (Lemma~\ref{lem:capacity-optimum-order}).
A ban lowers actual costs only if people would otherwise choose a design costlier than the best available.
Removing the only built-in derivation leaves code to state the required fact. A ban justified by excluding
invalid states must name those it excludes automatically and code must otherwise reject.

Dropping requirements changes the specification (Corollary~\ref{cor:specification-or-payment});
enforced preconditions exclude conflicting future changes. No algorithm exactly certifies tree-shaped
open futures (Theorem~\ref{thm:open-world-laminarity}).
Dhall omits some total computable functions to derive termination without per-program proof
~\cite{DhallSafetyGuarantees,rogers1987computability}.
A multiple-parent ban removes crossing derivations; lookup choices remain required
(Theorem~\ref{cor:coverage-resolution}); strict insertions force repair
(Theorem~\ref{thm:laminar-growth-reparenting}).
The cited Java and Go rationales measure no comparison and derive no exclusion
~\cite{JavaLanguageEnvironment,GoFAQ}.

Adoption costs $\PaperNotationLink{PaperAdoption}{K_a}$ once and at most $\PaperNotationLink{PaperAdoptionRate}{r_a}$ per update; the compared workaround
procedure costs at least $q$ per obligation after batching shares the positive
operation cost (Corollary~\ref{cor:factorization-economics}).
Graham's Blub paradox offers an analogy for unfamiliar capacity mistaken for complexity
~\cite{graham2001beating}. Social proof alone establishes neither lower cost nor derived exclusions.

Python's syntax ban left binding to \texttt{for y in [expr]}:
``No new syntax needed''~\cite{pythonideas2018temporary}.
Assignment expressions arrived in Python~3.8~\cite{pep572};
3.9 still optimized the workaround~\cite{python39temporarybinding}.
Pinned CPython has \CPythonAssignmentExpressionCount{} non-test standard-library
assignment expressions~\cite{CPythonLib5c7445b}; the obligation remains
(Supplement~\S\ref{supp-supp:walrus-boundary}).

\leanmetapending{\LHrng{IPE}{23}{27}, \LH{NRH6}, \LHrng{OPT}{3}{16}, \LHrng{BWR}{37}{39}, \LHrng{BWR}{45}{49}, \LH{BWR63}}
\begin{corollary}[Finite adoption cost cannot erase recurring workaround cost]
\label{cor:factorization-economics}
On the program evolution in Theorem~\ref{thm:laminar-growth-reparenting}, every supply-preserving
one-parent policy has $\PaperNotationLink{PaperBurden}{B_{\mathrm{rep}}(n)}\ge n$.
Fix a physical procedure covering every counted obligation, including standing routes,
with each additional operation costing at least $q_0>0$ and covering at most $b>0$ obligations.
Its cost is at least $qB_{\mathrm{rep}}(n)$, with $q=q_0/b>0$ even after batching
(parked-top policy: Supplement~\S\ref{supp-fixed-top-routing}; pricing:
\S\ref{supp-bounded-realization-cost} and Theorem~\ref{supp-thm:semantic-change-positive-cost}).

Subtract common costs $\PaperNotationLink{PaperCommonCost}{K_c(n)}$. Feasible adoption $a$ costs at most $\PaperNotationLink{PaperAdoption}{K_a}+\PaperNotationLink{PaperAdoptionUpkeep}{E_a(n)}$,
with finite setup $\PaperNotationLink{PaperAdoption}{K_a}$ and recurring $\PaperNotationLink{PaperAdoptionUpkeep}{E_a(n)}\le r_an$. It is cheaper if
$\PaperNotationLink{PaperAdoptionRate}{r_a}<q$ and $n>\PaperNotationLink{PaperAdoption}{K_a}/(q-\PaperNotationLink{PaperAdoptionRate}{r_a})$; one parent is not optimal.
Proof: Supplement~\S\S\ref{supp-repair-retention-pricing}--\ref{supp-supp:acquisition-policy-optimality}.
\end{corollary}

\leanmetapending{\LH{OPT13}, \LH{OPT20}, \LHrng{CSD}{1}{8}, \LHrng{SIN}{1}{5}, \LHrng{IPE}{21}{28}}
\begin{corollary}[Conservative capacity additions]
\label{cor:conservative-capacity-addition}
Adding capacity while retaining old answer- and behavior-preserving designs,
histories, costs, and answers (identity, membership, supply, lookup) keeps every
old choice available; the cheapest cannot become more expensive
(Lemma~\ref{lem:capacity-optimum-order}).
Multiple parents strictly lower outside supply at crossings and repair at strict insertions
(Corollary~\ref{cor:strict-derivation-dominance}; Theorem~\ref{thm:laminar-growth-reparenting}(0)).
\end{corollary}
Python~2.3 changed lookups and rejected old hierarchies:
nonconservative~\cite{simionato2003mro}.

\section{Conclusion}
With nominal identities fixed, only multiple contract and implementation parents
attain the optimum as code and interface requirements evolve through any finite sequence
(Corollary~\ref{cor:strict-derivation-dominance}).
At crossings, both designs exhibit behavioral correctness, but multiple parents
leave zero outside implementation pairs; the best sound one-parent design leaves
exactly $\PaperNotationLink{PaperGap}{d_\downarrow}>0$, even with helpers or generators, and fails correct maintenance
(Theorem~\ref{thm:delegation-lower-bound}). Strict insertion breaks an old parent
edge still carrying a required pair, or needs forwarding or removal before or after the update.
Our construction repeats this while every version fits a tree; multiple parents add one
edge per update, retaining old derivations (Theorem~\ref{thm:laminar-growth-reparenting}).
Refactoring removes slop for excess pairs $\PaperNotationLink{PaperExcess}{e_H}$; native derivation removes it for forced pairs $\PaperNotationLink{PaperGap}{d_\downarrow}$
(Theorem~\ref{thm:semantic-completion-tax}); conservative additions never raise optima
(Corollary~\ref{cor:conservative-capacity-addition}).
For the \PythonAblationCases{} projects' recovered requirements:
\PythonPopulationProviderIncidences{} (no inheritance) $\to$
\PythonFixedHierarchyTotal{} (declared first parents) $\to$
\PythonProviderOptimumTotal{} (best sound one-parent, exact) $\to$ 0 outside implementation pairs (multiple parents;
Proposition~\ref{prop:derivation-capacity}).
Across histories, the stable-key minimum is the confusability graph's chromatic
number; contract memberships obey the gap law
(Theorems~\ref{thm:future-conflict}, \ref{thm:contract-membership-gap}).
Design choice is taught as preference: ``favor composition''~\cite{gamma1994designpatterns}.
Objective counts decide correct maintenance as requirements evolve. Price
outside work however you like: multiple parents remain strictly cheaper on
every crossing exactly when doing any outside work costs more than doing none
(Supplement~Theorem~\ref{supp-thm:accounting-characterization}).

\runinhead{Data-Availability Statement}~

The artifact supplies Lean sources, the claim-to-Lean handle mapping, fixed inputs,

LRAT certificates, verifiers, mutation logs, and regeneration scripts for
Artifact Evaluation.  \label{sec:mechanization}
Lean checks every numbered result and tagged claim, with the external inputs
listed below, through the claim-to-Lean
handle mapping (Supplement~\S\ref{supp-claim-handle-mapping}). Direct replay proves optima
for $\PaperNotationLink{PaperGap}{d_\downarrow}\le10$. For the \IndependentGapCertificateCount{} larger components,
replay checks a matching repair and Lean checks LRAT refutations excluding one
fewer deletion. The separate public-name contract adds
\PublicApiCertificates{} independently checked component certificates.
The equal-scope audit certifies its floor by replaying matching certificates and adding
\EqualScopeNewCertificates{} certificates for changed larger components.
The stateful-row audit adds \StatefulCertificates{} component certificates.
A further LRAT refutation certifies the
\IndependentHolderMinimum{}-bundle minimum.
From the recovered relation and bound, Lean generates every checked CNF,
including the 520-pair crossing-packing residual and bundle-cover formula.
CP-SAT searches for candidates; separate-method solver-optimal values rely on
its optimizer (Supplement Table~\ref{supp-tab:python-method-sensitivity}). Other declarations support proofs and
encoders; source extraction and Liskov--Wing's proof rules
(Theorem~\ref{thm:behavioral-applicability-product}) remain external.
Lean also checks the exact inclusion--exclusion law and its asymptotic bound;
the synthetic distributions are fixed by the stated cell probabilities and crossing rates;
their correspondence to program populations remains external.
\leanmeta{\LHrng{SPN}{1}{20}}

For the identity-history, open-world, and finite-test results (Sections~3
and~\ref{sec:open-world-laminarity}), Lean checks r.e. membership of the failure sets and many-one
reductions from halting and nonhalting. These establish membership and hardness;
``co-r.e.-complete'' uses the standard class definitions~\cite{rogers1987computability}.
Section~6.5's cited diagonalization fact remains external. The axiom report names
\texttt{Lean.ofReduceBool} and \texttt{Lean.trustCompiler} for native Boolean
certificate evaluation; the compiler is trusted to evaluate the checked Boolean computation correctly.

The \texttt{ancestry-gap} command and independent replay procedure are
documented in Supplement~\S\ref{supp-analyzer-new-input}.

\par\smallskip\noindent The artifact is available on Zenodo:
\url{\PaperZenodoURL}.

\par\smallskip\noindent\runinhead{AI Disclosure} Generative AI tools and agents
helped translate informal arguments into formal statements, draft prose, run
empirical studies, build research infrastructure, and conduct adversarial mock
reviews.  Lean~4 compiled the resulting proofs without unproved placeholders
(\texttt{sorry}).

\bibliographystyle{plain}
\bibliography{references}

\end{document}


\title{Supplementary Material\\[0.4ex]\PaperTitleAuto}

\PaperPublicationAuthor{article}

\date{\today}
\maketitle

\section*{Notation}
\label{notation}
Read this table in order to build the notation from classes and code providers to
the bounds; the plain-language column ties it to the main-text AnyIO example.
\global\PaperNotationTablePrintedtrue
\begingroup
\PaperNotationLinksfalse
\PaperTableSetup
\newlength{\PaperNotationSymbolWidth}
\newlength{\PaperNotationMeasureWidth}
\settowidth{\PaperNotationSymbolWidth}{Symbol}
\newcommand{\PaperNotationMeasure}[1]{%
  \settowidth{\PaperNotationMeasureWidth}{#1}%
  \ifdim\PaperNotationMeasureWidth>\PaperNotationSymbolWidth
    \setlength{\PaperNotationSymbolWidth}{\PaperNotationMeasureWidth}%
  \fi}
\PaperNotationMeasure{\PaperClasses}
\PaperNotationMeasure{\PaperImplementations}
\PaperNotationMeasure{\PaperProvider}
\PaperNotationMeasure{\PaperFileReader}
\PaperNotationMeasure{\PaperClassGroup}
\PaperNotationMeasure{\PaperRequired}
\PaperNotationMeasure{\PaperPublicRequired}
\PaperNotationMeasure{\PaperMembership}
\PaperNotationMeasure{\PaperAncestry}
\PaperNotationMeasure{\PaperRoutes}
\PaperNotationMeasure{\PaperRoutingStep}
\PaperNotationMeasure{\PaperRoutingGraph}
\PaperNotationMeasure{\PaperRoutingClosure}
\PaperNotationMeasure{\PaperRemovals}
\PaperNotationMeasure{\PaperSupplied}
\PaperNotationMeasure{\PaperSound}
\PaperNotationMeasure{\PaperTreeShaped}
\PaperNotationMeasure{\PaperGap}
\PaperNotationMeasure{\PaperExcess}
\PaperNotationMeasure{\PaperRepairSet}
\PaperNotationMeasure{\PaperRelaxedGap}
\PaperNotationMeasure{\PaperOutsideAtState}
\PaperNotationMeasure{\PaperBurden}
\PaperNotationMeasure{\PaperChargedDeletions}
\PaperNotationMeasure{\PaperOutsideUpdates}
\PaperNotationMeasure{\PaperGrowthPairs}
\PaperNotationMeasure{\PaperEdgeFootprint}
\PaperNotationMeasure{\PaperMaintenanceIdeal}
\PaperNotationMeasure{\PaperSafety}
\PaperNotationMeasure{\PaperInput}
\PaperNotationMeasure{\PaperAnswerSpace}
\PaperNotationMeasure{\PaperEncodingSpace}
\PaperNotationMeasure{\PaperAnswer}
\PaperNotationMeasure{\PaperConfusability}
\PaperNotationMeasure{\PaperKeys}
\PaperNotationMeasure{\PaperLookup}
\PaperNotationMeasure{\PaperLookupChoices}
\PaperNotationMeasure{\PaperSharingQuestions}
\PaperNotationMeasure{\PaperDistinctSharing}
\PaperNotationMeasure{\PaperInitializationQuestions}
\PaperNotationMeasure{\PaperRecurringCharge}
\PaperNotationMeasure{\PaperAdoption}
\PaperNotationMeasure{\PaperAdoptionUpkeep}
\PaperNotationMeasure{\PaperAdoptionRate}
\PaperNotationMeasure{\PaperCommonCost}
\PaperNotationMeasure{\PaperBatchLimit}
\setlength{\LTleft}{0pt}\setlength{\LTright}{0pt}
\begin{longtable}{@{}c*{2}{>{\centering\arraybackslash}p{\dimexpr(\linewidth-\PaperNotationSymbolWidth-4\tabcolsep)/2\relax}}@{}}
\toprule
Symbol & In words & Precise meaning \\
\midrule\endhead
\bottomrule\endlastfoot
\hypertarget{paper-notation:PaperClasses}{\PaperClasses} & The classes: $W$ (writer), $\mathrm{FR}$ (file reader), and $B$ (buffered reader) in the main-text AnyIO example. & Declared classes; for identity questions, persistent meanings. \\
\midrule
\hypertarget{paper-notation:PaperImplementations}{\PaperImplementations} & Independently changing shared code: file lifecycle and byte iteration in the main-text AnyIO example. & Independently changing implementations. \\
\midrule
\hypertarget{paper-notation:PaperProvider}{\PaperProvider} & The class declaring the shared code: \texttt{\_BaseFileStream} or \texttt{ByteReceiveStream}. & The distinct code-provider declaration containing implementation $f$. \\
\midrule
\hypertarget{paper-notation:PaperFileReader}{\PaperFileReader} & The file reader that needs both code providers; the other two classes need one each. & The file-reader class in the main-text AnyIO example; $W$ is the writer and $B$ the buffered reader. \\
\midrule
\hypertarget{paper-notation:PaperClassGroup}{\PaperClassGroup} & Who needs the code: file lifecycle $\{W,\mathrm{FR}\}$; iteration $\{\mathrm{FR},B\}$. & Classes requiring implementation $f$: from code provider to consumer classes. \\
\midrule
\hypertarget{paper-notation:PaperRequired}{\PaperRequired} & The complete list of implementation needs, written as (code, class) pairs: four here. & Required implementation--class pairs; $R_t^{\mathrm{req}}$ is the relation at state $t$. \\
\midrule
\hypertarget{paper-notation:PaperPublicRequired}{\PaperPublicRequired} & The separate audit contract keeping selected public lookups, including Python special methods. & Pairs witnessed by selected public-name lookups under the separate resolution-preservation contract, including dunders. \\
\midrule
\hypertarget{paper-notation:PaperMembership}{\PaperMembership} & Which classes must satisfy which interfaces; this is separate from receiving their code. & Required class--contract memberships, separate from implementation supply. \\
\midrule
\hypertarget{paper-notation:PaperAncestry}{\PaperAncestry} & The needs inheritance supplies: all four with multiple parents, at most three with sound one-parent ancestry here. & Pairs supplied by ancestry; sound ancestry has $H\subseteq R^{\mathrm{req}}$. \\
\midrule
\hypertarget{paper-notation:PaperRoutes}{\PaperRoutes} & Which classes receive required code through forwarding instead of inheritance: one pair in the best one-parent design here; generated routes count too. & Outside supply routes; $\operatorname{dom}\varphi$ lists the pairs they supply. \\
\midrule
\hypertarget{paper-notation:PaperRoutingStep}{\PaperRoutingStep} & One forwarding step from a class to a destination; ancestry follows separately declared parent edges. & One routing edge; ancestry is the separate order $\preceq$. \\
\midrule
\hypertarget{paper-notation:PaperRoutingGraph}{\PaperRoutingGraph} & The forwarding steps through which code or tooling supplies outside pairs. & Routing graph whose edges are $\to$; distinct from ancestry. \\
\midrule
\hypertarget{paper-notation:PaperRoutingClosure}{\PaperRoutingClosure} & Follow no routing step, one step, or several steps to the destination. & Reflexive--transitive routing closure: zero or more routing edges. \\
\midrule
\hypertarget{paper-notation:PaperRemovals}{\PaperRemovals} & Unwanted inherited pairs removed in the relaxed model; superclass promises remain binding. & Removed ancestry pairs in the relaxed-supply model (main-text \S4.4). \\
\midrule
\hypertarget{paper-notation:PaperSupplied}{\PaperSupplied} & Everything supplied by either route: exactly the four required pairs in a design supplying all and only those pairs. & Pairs design $D$ supplies; exact supply requires $R_D=R^{\mathrm{req}}$. \\
\midrule
\hypertarget{paper-notation:PaperSound}{\PaperSound} & Inheritance gives no class an implementation it does not need. & Sound ancestry satisfies $H\subseteq R^{\mathrm{req}}$. \\
\midrule
\hypertarget{paper-notation:PaperTreeShaped}{\PaperTreeShaped} & Any two need-groups nest or do not overlap. The main-text AnyIO example instead crosses: the groups share $\mathrm{FR}$ and neither contains the other. & For all providers $f,g$, $A_f\subseteq A_g$, $A_g\subseteq A_f$, or $A_f\cap A_g=\varnothing$. \\
\midrule
\hypertarget{paper-notation:PaperGap}{\PaperGap} & The fewest pairs code or tooling must supply outside sound one-parent inheritance: one here. & Fewest required pairs outside any sound one-parent ancestry preserving the providers. \\
\midrule
\hypertarget{paper-notation:PaperExcess}{\PaperExcess} & Outside pairs beyond the gap: zero in the best hierarchy here. Refactoring removes excess; every sound one-parent hierarchy still leaves the gap. & Removable excess above the ancestry gap for a fixed sound hierarchy $H$. \\
\midrule
\hypertarget{paper-notation:PaperRepairSet}{\PaperRepairSet} & What ancestry misses, together with what it supplies wrongly. & Missing required pairs together with unwanted supplied pairs. \\
\midrule
\hypertarget{paper-notation:PaperRelaxedGap}{\PaperRelaxedGap} & Allow unwanted inherited pairs too, charging $\lambda$ to remove each one; minimize that cost plus missing pairs. & Relaxed-supply minimum with removal charge $\lambda$. \\
\midrule
\hypertarget{paper-notation:PaperOutsideAtState}{\PaperOutsideAtState} & How many required type identities, interface memberships, and implementation--class pairs are left to code or tooling at this state. & Outside-answer counts for design history $D$ at state $t$; infima are taken separately for each coordinate. \\
\midrule
\hypertarget{paper-notation:PaperBurden}{\PaperBurden} & Broken links still carrying required code, plus pairs supplied or removed at updates. Unchanged pairs count again because required supply persists even when route code stays unchanged. The main-text AnyIO example has no updates. & Over $n$ updates: deleted parent edges carrying still-required pairs, plus missing or removed pairs at both endpoints, including unchanged pairs. \\
\midrule
\hypertarget{paper-notation:PaperChargedDeletions}{\PaperChargedDeletions} & Count old inheritance links broken while they still carry required code. & Charged parent-edge deletions in main-text Theorem~4.14: each cuts a still-required pair. \\
\midrule
\hypertarget{paper-notation:PaperOutsideUpdates}{\PaperOutsideUpdates} & Count updates with required pairs still supplied outside inheritance or unwanted pairs removed, even if no route changes. & Updates with missing or removed pairs at either endpoint, counted once per update; $\delta+\pi\ge n$ in main-text Theorem~4.14. \\
\midrule
\hypertarget{paper-notation:PaperGrowthPairs}{\PaperGrowthPairs} & All new implementation needs in the constructed every-step history. & Added pairs $\{(f_i,c_j):0\le j<i\le n\}$ in the staged growth history; $|\Pi|=n(n+1)/2$. \\
\midrule
\hypertarget{paper-notation:PaperEdgeFootprint}{\PaperEdgeFootprint} & Which pairs use an old inheritance edge. Its deletion is charged only if one remains required. & Pairs whose old ancestry paths use edge $e$; its deletion is charged only if at least one remains required afterward. \\
\midrule
\hypertarget{paper-notation:PaperMaintenanceIdeal}{\PaperMaintenanceIdeal} & The best score on each count. A semantically complete language supplies mechanisms that reach all these supply and repair scores together, even when requirements shrink or return. & Coordinatewise outside-answer and repair-obligation infima; complete supply jointly attains zero on every finite declared supply history (main-text Corollary 4.12). \\
\midrule
\hypertarget{paper-notation:PaperSafety}{\PaperSafety} & Every state the program can reach satisfies the specification. & Safety: every reachable state is valid. \\
\midrule
\hypertarget{paper-notation:PaperInput}{\PaperInput} & The cases a program must handle; for example, a client using an AnyIO stream. & Interface inputs or cases in the decoder equation (main-text \S2). \\
\midrule
\hypertarget{paper-notation:PaperAnswerSpace}{\PaperAnswerSpace} & The results a case may be required to produce, including a complete set of permitted outputs. & Required-answer space in the decoder equation (main-text \S2). \\
\midrule
\hypertarget{paper-notation:PaperEncodingSpace}{\PaperEncodingSpace} & The representations from which the decoder must recover each case's specified result. & Encoding space: $e:X\to\mathcal E$ and $r:\mathcal E\to\mathcal V_{\mathrm{ans}}$. \\
\midrule
\hypertarget{paper-notation:PaperAnswer}{\PaperAnswer} & The specified result for a case; a test checks its observed output against this requirement. & The answer required for case $x$; $e$ encodes it and $r$ decodes it. \\
\midrule
\hypertarget{paper-notation:PaperConfusability}{\PaperConfusability} & Join meanings that can look identical while needing different handling. This identity question is separate from the main-text AnyIO supply crossing. & Graph joining look-alike cases that require different answers. \\
\midrule
\hypertarget{paper-notation:PaperKeys}{\PaperKeys} & The fewest labels keeping joined meanings distinct; one shared label leaves this many minus one additional labels. & Fewest stable key values; one shared value leaves $\chi-1$ additional values. \\
\midrule
\hypertarget{paper-notation:PaperLookup}{\PaperLookup} & Which code provider wins when definitions share a name. Every design preserving the specified member lookups supplies these choices. & Required lookup map choosing among same-name contenders; $k$ counts these choices. \\
\midrule
\hypertarget{paper-notation:PaperLookupChoices}{\PaperLookupChoices} & How many same-name conflicts the specification resolves; every representation preserving the specified lookups supplies those decisions. & Number of required lookup choices in the fixed lookup map $P$. \\
\midrule
\hypertarget{paper-notation:PaperSharingQuestions}{\PaperSharingQuestions} & Which state accesses must see the same object or distinct objects; required in every representation. & Specified state-sharing questions; the state-coordination supplement defines their domain. \\
\midrule
\hypertarget{paper-notation:PaperDistinctSharing}{\PaperDistinctSharing} & How many distinct-state requirements depart from a native shared-state default; implementation pairs are counted separately in the required supply relation. & Specified distinct-state answers on common providers in the supplement's sharing-default proposition. \\
\midrule
\hypertarget{paper-notation:PaperInitializationQuestions}{\PaperInitializationQuestions} & Which state component must initialize before another; required in every representation. & Specified initialization-order questions; representations preserving the specified state answers resolve the same $|S|+|I|$ questions. \\
\midrule
\hypertarget{paper-notation:PaperRecurringCharge}{\PaperRecurringCharge} & Each repair obligation handled by this procedure costs at least this much, even when work is grouped. & Derived bound $q=q_0/b>0$: operations cost at least $q_0$ and cover at most $b$ obligations, with all standing-route endpoints covered (main-text Corollary 6.2). \\
\midrule
\hypertarget{paper-notation:PaperAdoption}{\PaperAdoption} & The one-time cost of adopting a mechanism or tool. & Finite adoption cost for plan $a$; $E_a(n)$ is its additional cost through $n$ updates. \\
\midrule
\hypertarget{paper-notation:PaperAdoptionUpkeep}{\PaperAdoptionUpkeep} & The additional cost after setup under adoption plan $a$, through $n$ updates. & Additional cost through $n$ updates under adoption plan $a$; the pricing corollary assumes $E_a(n)\le r_an$. \\
\midrule
\hypertarget{paper-notation:PaperAdoptionRate}{\PaperAdoptionRate} & The adopted mechanism's recurring cost per update; the threshold compares it with the workaround's lower bound. & Upper bound on adoption plan $a$'s additional cost per update: $E_a(n)\le r_an$. \\
\midrule
\hypertarget{paper-notation:PaperCommonCost}{\PaperCommonCost} & Costs both compared policies share; subtract them before comparing. & Costs common to the compared policies, subtracted before comparing adoption and recurring repair. \\
\midrule
\hypertarget{paper-notation:PaperBatchLimit}{\PaperBatchLimit} & How many answers one operation can change or entries one probe can read, in the stated model. & Maximum required answers changed or entries read per operation, in the stated operation model. \\
\end{longtable}
\endgroup

\section{What Each Result Counts}
\label{scope-ledger}

\paragraph{Each missing capacity forces an irreducible semantic obligation.}
\label{cor:capacity-removal-descent}
The main-text economics corollary prices strict-insertion repair; its arity example
shows equal outputs with unequal answer counts.
\leanmeta{\LHrng{OLC}{1}{5}, \LHrng{CSD}{1}{8}, \LH{CAP7}, \LH{SCT31}, \LHrng{INC}{1}{9}}

The identity results ask how many identity keys remain sufficient as visible records
change. Table~\ref{tab:result-scope-supp} records which meanings persist or arrive,
what the decoder can observe, and which future changes the allocator knows.
The second table separates implementation supply from lookup and self-calls.

\begin{table}[h]
\centering
\PaperTableSetup
\begin{tabularx}{\textwidth}{@{}>{\RaggedRight\arraybackslash}m{0.19\textwidth}
  >{\RaggedRight\arraybackslash}m{0.15\textwidth}
  >{\RaggedRight\arraybackslash}m{0.22\textwidth}Y@{}}
\toprule
Result & Meanings present & Visible evidence & Future changes and information \\
\midrule
Future-conflict price & persistent, arrivals allowed & arbitrary equivalence states & realized trace or declared state family; zero recourse \\
\midrule
Clone-upgrade recourse & structural then independent identity evidence & one observational clone & normalized tag change; additive and replacement calibrations \\
\midrule
Refinement price & persistent, arrivals allowed & arbitrary equivalences & causal allocation; refinements only \\
\midrule
Coarsening gap & persistent, arrivals allowed & arbitrary equivalences & online versus offline; monotone coarsening \\
\midrule
Matching recourse & fixed $n$ & one matching of pair merges & matching revealed after commitment; at most $r$ migrations \\
\midrule
Deprecation pair test & fixed population & Boolean presence/method sets & downward-closed coordinate deletions; offline robust price \\
\midrule
One-deletion stability & fixed population & Boolean presence/method sets & at most one safe coordinate deletion; tight multiplicity \\
\midrule
Field menu & fixed $\mathbb F_2^k$ population & Boolean linear fields & full menu followed by literal field deletion; offline robust price \\
\midrule
Two-state threshold & fixed population & two arbitrary partitions & both states known offline \\
\bottomrule
\end{tabularx}
\caption{Scope of the identity-capacity results.  Only the last four rows use
coordinate-set profiles; the earlier rows range over abstract observation
equivalences.}
\label{tab:result-scope-supp}
\end{table}

\begin{table}[h]
\centering
\PaperTableSetup
\begin{tabularx}{\textwidth}{@{}>{\RaggedRight\arraybackslash}m{0.20\textwidth}
  >{\RaggedRight\arraybackslash}m{0.18\textwidth}
  >{\RaggedRight\arraybackslash}m{0.23\textwidth}Y@{}}
\toprule
Result & Native mechanism & Workarounds & Assumptions \\
\midrule
Derivation capacity & contract-only, one-parent, or multi-parent ancestry
& none in the capacity theorem
& arbitrary required role--consumer relation; distinct provider endpoints \\
\midrule
Ancestry gap $\PaperNotationLink{PaperGap}{d_\downarrow}$ & raw native reach supplies only required pairs
& forwarding of every missed required pair
& unwanted supply forbidden; arbitrary reparenting and helpers allowed \\
\midrule
Paid-removal gap $\PaperNotationLink{PaperRelaxedGap}{d_\lambda}$ & raw native rows can over-reach
& forwarding plus one negative pair decision per unwanted native pair
& incidence cost; source declarations can compress several pair decisions \\
\midrule
Dynamic retraction & one fixed finite carrier across states
& forwarding and unwanted-supply removal decisions
& monotone scope growth; persistent providers, consumers, and preallocated helpers \\
\midrule
Coupled contract and implementation graphs & one shared one-parent graph
& workarounds supplying missed memberships and implementation pairs
& the two row families must be laminar jointly; separate-graph minima remain lower bounds \\
\midrule
Lookup choices & one fixed winner contract $P$
& winner-preservation costs remain separate from the semantic gap
& both preserve the specified winners; their update costs are priced separately \\
\midrule
Self-call choices & open-recursive subclass or closed wrapper
& exceptional winner decisions; wrapper also routes every exposed member
& requires a declared self-use site set and winner at each site \\
\bottomrule
\end{tabularx}
\caption{Scope of the derivation results.  Paid removal relaxes exact supply
without changing the required-only inheritance theorem or permitting promise-blocking.}
\label{tab:derivation-scope-supp}
\end{table}

\subsection{Cost units}

The main paper counts one maintained semantic answer at a time and keeps source
artifacts, operations, and developer time separate.  Table~\ref{tab:cost-units}
records the exact conversions and the hypotheses each conversion requires.

\begin{table}[t]
\centering
\small
\PaperTableSetup
\begin{tabularx}{\linewidth}{@{}>{\RaggedRight\arraybackslash}m{0.22\linewidth}YY@{}}
\toprule
Quantity & Unit & Assumption for a lower bound \\
\midrule
Permanent-key minimum (\S3) & names & declared future states \\
\midrule
Ancestry gap (\S4) & implementation-to-class pairs & exact one-parent representation \\
\midrule
Paid-removal gap (\S4) & forwarded pairs plus $\lambda$ times removed pairs & relaxed exact supply; cancel unwanted implementation supply while preserving superclass promises \\
\midrule
Hierarchy change (\S4) & deleted old direct-parent edges & monotonically growing requirements; exact consecutive states on one fixed finite node set \\
\midrule
External-map minimum (\S4) & states or bits & fixed ancestry view \\
\midrule
Repair lower bound (\S4) & operations & one operation covers at most $b>0$ pairs \\
\midrule
Total-cost model (\S4) & declared project cost units & each priced resource has a stated positive marginal cost \\
\bottomrule
\end{tabularx}
\caption{The paper keeps distinct quantities in distinct units.  Source lines,
human time, and comprehension enter only when a stated model measures them.}
\label{tab:cost-units}
\end{table}

\section{Required Pairs, Inheritance, and Workarounds}
\label{atomic-native-completion}

First fix which implementation each class needs. Then ask which of those pairs
inheritance derives and which code must supply. We define semantic consequence
by its usual requirement: every model satisfying the declarations must satisfy
the consequence. For the positive atomic declarations below, this proves that
the consequences are exactly the declared pairs. The later minimum then equals
both the fewest deletions needed to obtain tree shape and the fewest required
implementation--class pairs supplied by forwarding. Selecting among competing method bodies remains separate.

Fix code providers $\PaperNotationLink{PaperImplementations}{F}$, consumer classes $\PaperNotationLink{PaperClasses}{C}$, and an independently declared required relation
$\PaperNotationLink{PaperRequired}{R^{\mathrm{req}}}\subseteq \PaperNotationLink{PaperImplementations}{F}\times \PaperNotationLink{PaperClasses}{C}$.  An applicability model $M$ assigns a truth value to
every atomic proposition $f@c$.  It satisfies $\PaperNotationLink{PaperRequired}{R^{\mathrm{req}}}$ when every pair in $\PaperNotationLink{PaperRequired}{R^{\mathrm{req}}}$ is
true.  Define semantic consequence in the usual all-models form:
\[
  \PaperNotationLink{PaperRequired}{R^{\mathrm{req}}}\models f@c
  \quad\Longleftrightarrow\quad
  \text{$M\models R^{\mathrm{req}}$ implies $M\models f@c$ for every $M$.}
\]
The canonical model makes exactly the pairs in $\PaperNotationLink{PaperRequired}{R^{\mathrm{req}}}$ true.  Hence, for this
positive atomic theory,
\[
  \PaperNotationLink{PaperRequired}{R^{\mathrm{req}}}\models f@c \quad\Longleftrightarrow\quad (f,c)\in \PaperNotationLink{PaperRequired}{R^{\mathrm{req}}}.
\]
The canonical model proves this equivalence from the all-models definition.

Native ancestry supplies a relation $\PaperNotationLink{PaperAncestry}{H}\subseteq \PaperNotationLink{PaperImplementations}{F}\times \PaperNotationLink{PaperClasses}{C}$, as in the main text,
with scope $\PaperNotationLink{PaperAncestry}{H}(f)=\{c:(f,c)\in \PaperNotationLink{PaperAncestry}{H}\}$ for each provider. We abbreviate native reach by
\[
  \frac{c\in \PaperNotationLink{PaperAncestry}{H}(f)}{\PaperNotationLink{PaperAncestry}{H}\vdash f@c}.
\]
Thus the representation is sound when $\PaperNotationLink{PaperAncestry}{H}(f)\subseteq \PaperNotationLink{PaperRequired}{R^{\mathrm{req}}}(f)$ for every $f$, and complete
when every semantic consequence is derivable.  Define exact native supply by
\[
  \PaperNotationLink{PaperAncestry}{H}\vdash f@c\quad\Longleftrightarrow\quad \PaperNotationLink{PaperRequired}{R^{\mathrm{req}}}\models f@c
  \qquad(f\in \PaperNotationLink{PaperImplementations}{F},c\in \PaperNotationLink{PaperClasses}{C}).
\]
Exact native supply is therefore equivalent to soundness plus completeness.  Validity
is fixed before any hierarchy, resolver, or tie-breaking policy is chosen.

The native deficit and excess are
\[
 \Delta_H^+=\PaperNotationLink{PaperRequired}{R^{\mathrm{req}}}\setminus \PaperNotationLink{PaperAncestry}{H},
 \qquad
 \Delta_H^-=\PaperNotationLink{PaperAncestry}{H}\setminus \PaperNotationLink{PaperRequired}{R^{\mathrm{req}}},
 \qquad \PaperNotationLink{PaperRepairSet}{\mathcal D(H,R^{\mathrm{req}})}=\Delta_H^+\uplus\Delta_H^-.
\]
Completeness is $\Delta_H^+=\varnothing$; soundness is
$\Delta_H^-=\varnothing$. An add-only outside workaround preserving exact supply adds a relation $E$
such that $(\PaperNotationLink{PaperAncestry}{H}\vdash f@c)\lor E(f,c)$ agrees with semantic validity.
Every member of $\Delta_H^+$ must then map injectively to a distinct supplied
decision. The canonical repair of sound ancestry takes $E=\Delta_H^+$ and attains the
bound.

Copies supply the same implementations to the same classes, but their authoritative source matters.
For any grouping of declarations into roles, either all declarations in a role
mean the same thing or two of them witness a distinction. Lean checks both
cases. Keeping the distinction splits the role and changes the required relation.
When the declarations agree, choosing any one as that role's authority gives
the same required class sets and ancestry gap.

For any such authority choice, let $\PaperNotationLink{PaperAncestry}{H}$ contain only pairs reached from those authorities and let $E$
contain all pairs supplied outside that ancestry, including through other
declarations or forwarding. Exactness gives $E\setminus \PaperNotationLink{PaperAncestry}{H}=\PaperNotationLink{PaperRequired}{R^{\mathrm{req}}}\setminus \PaperNotationLink{PaperAncestry}{H}$.
In the replica-only case, $E$ is the replica-supplied relation.  Lean identifies
these outside decisions with the native deficit, identifies their indexed cardinality with
the selected ancestry gap, and proves that every such count is at least the
global laminar minimum.  A replica can cover many consumers syntactically;
the relation records the applicability answer it supplies at each consumer.
Independently writable replicas additionally expose precisely the disagreement
states excluded by deterministic exact expansion.
\leanmeta{\LHrng{SCT}{56}{59}, \LHrng{BWR}{58}{62}, \LHrng{SDN}{17}{21}}

\subsection{Exact semantic factoring}
\label{semantic-factoring}
Two classes may merge only if no requirement distinguishes them throughout their
supported evolution. If only their required implementations agree throughout that
evolution, share the code and retain their identities. If their required implementations
differ, keep separate implementations.

Lean checks these three cases using one table of answers across reachable
states. Its canonical quotient groups entities with identical complete answer
histories; every answer-preserving encoding may merge only within a group.
Splitting the table into meaning and implementation questions gives full
equality, implementation equality alone, or implementation inequality. Every minimal exact
implementation encoding must share equal implementation columns; distinct full
meanings still require distinct full-answer codes.  A two-entity witness
has the same implementation column under both a mergeable and a distinct full
specification, so no implementation-only rule decides which case holds.
More generally, Lean lets a structural profile contain any data, including a
complete inheritance graph, and proves that no decoder through that profile is
exact when it maps two meanings with different required answers to the same
profile value.
For finite answer sets $A,B$, the symmetric-difference triangle inequality
proves
\[
 |A\mathbin{\triangle}B|
 \le |A\mathbin{\triangle}X|+|B\mathbin{\triangle}X|,
\]
and $X=A$ attains equality.  Thus their symmetric difference is the exact
correction floor for a forced shared answer set: it counts present disagreements
even before any future edit is chosen.
Any current encoding that merges two distinct columns must change at least one
of their code assignments to become exact.  The changed semantic answer is
required by every representation; this recoding is the merger's additional
distance from an exact factoring.

Shared parent combinations obey the same rule.  For finite sets $J$ of later
parent additions and $\PaperNotationLink{PaperClasses}{C}$ of classes declared to receive all of them, direct
repetition stores the product $J\times \PaperNotationLink{PaperClasses}{C}$.  A shared factor stores $J$ once
and expands it across the already inherited class set $\PaperNotationLink{PaperClasses}{C}$.  Lean proves that
both representations derive exactly $J\times \PaperNotationLink{PaperClasses}{C}$, while their edit counts are
$|J||\PaperNotationLink{PaperClasses}{C}|$ and $|J|$.  The exact saving is $|J|(|\PaperNotationLink{PaperClasses}{C}|-1)$; it is positive exactly
when $J$ is nonempty and at least two classes share the factor.  For a finite
migration charge $K$, every later addition count $|J|>K$ makes
$K+|J|<|J||\PaperNotationLink{PaperClasses}{C}|$ when $|\PaperNotationLink{PaperClasses}{C}|\ge2$.  A separate two-state witness gives the same
parent lists initially but common evolution in one specification and divergent
evolution in the other.  Thus current structural equality cannot decide
whether the shared class is semantically exact.
\leanmeta{\LHrng{SEF}{1}{25}}

\subsection{Minimum one-parent workaround count}
\label{minimum-one-parent-completion}
Multiple parents can supply the whole required relation, $\PaperNotationLink{PaperAncestry}{H}=\PaperNotationLink{PaperRequired}{R^{\mathrm{req}}}$. The canonical provider-preserving
multi-parent graph realizes that relation, so every required relation has a
sound and complete native representation.  One-parent ancestry restricts the family
$\{\PaperNotationLink{PaperAncestry}{H}(f)\}$ to be tree-shaped. Consequently, a sound and complete one-parent
representation would force the required family $\{\PaperNotationLink{PaperRequired}{R^{\mathrm{req}}}(f)\}$ itself to be tree-shaped.

The main-text AnyIO relation gives the explicit counterexample to completeness:
file-lifecycle code is required by $W$ and $\PaperFileReader$, while iteration code is required by
$\PaperFileReader$ and $B$.  The two required scopes cross.  Therefore every sound
one-parent native representation omits some required provider--class pair;
which pair is omitted may depend on the chosen tree.  External delegation
can derive the missing pair behaviorally, but then the derivation is no
longer native.

For finite indexed families, minimizing $|\Delta_H^+|$ over all sound
one-parent syntaxes gives exactly $\PaperNotationLink{PaperGap}{d_\downarrow}$.  The index retains distinct
providers even when their required scopes are equal.  Minimizing the decision set
of every add-only repair preserving exact supply gives the same value, as does independently
minimizing explicit forwarding over every provider-preserving
one-parent-plus-delegation representation. That representation may choose its
nodes, helpers, provider and leaf placements, ancestry, delegation graph,
and partial forwarding map, and supplies exactly the required relation. Contract-only syntax has $\PaperNotationLink{PaperAncestry}{H}=\varnothing$ and leaves
$|\PaperNotationLink{PaperRequired}{R^{\mathrm{req}}}|$ answers outside; whenever $\PaperNotationLink{PaperRequired}{R^{\mathrm{req}}}$ is nonempty, retaining the incidences at one consumer
already makes the optimal one-parent outside-answer count strictly smaller.  Unrestricted
multi-parent syntax pays zero.  The one-parent optimum is zero iff
the indexed required scopes are tree-shaped, and positive on every crossing
family. At each state, $\Delta_H^+$ lists missing required pairs and
$\Delta_H^-$ lists unwanted supplied pairs. Their cardinalities count what this
hierarchy misses or supplies wrongly. The staged deletion-or-compensation theorem and its quadratic
endpoint bound therefore count these same pairs.
\leanmeta{\LHrng{NPS}{1}{9}, \LHrng{SCT}{1}{35}, \LHrng{BWR}{43}{49}}

\section{Proofs for the Finite Identity Policies}
\label{finite-identity-policies}

These proofs compare ways to retain identity as records change. The first two
paragraphs compare adding identity evidence after a collision with retaining
names from the start. The remaining proofs fix particular kinds of record
change and compute their identity-key or renaming costs.

\paragraph{Adding identity evidence after a collision.}
Count zero changes when the persistent tag function stays fixed and one when
it changes. This is the normalized identity-change unit. If a finite extension contains two distinct
meanings with the same structural profile, every adequate continuation of a
structural discipline has cost one: retaining the old tag function would
retain the collision.  One global nominal discipline is adequate before and
after with cost zero.  For the separate one-artifact calibration, let $s$ be
the structural declaration and $n$ the independent identity evidence, and use
semantic symmetric-difference distance.  The direct path
$\varnothing\to\{n\}$ costs one.  Retaining $s$ gives
$\varnothing\to\{s\}\to\{s,n\}$ at cost two; replacing it gives
$\varnothing\to\{s\}\to\{n\}$ at cost three.  Clone extension forces the
positive recourse. Keeping $s$ attains the two-addition path, so deleting it
is an extra operation charged only on the three-change replacement path.
\leanmeta{\LHrng{IMR}{1}{5}}

\paragraph{Retaining names cannot increase extra-key demand.}
Index every reachable state through every supported evolution of requirements. Let $p_t$ be the observed
profile and $a_t$ the required answer; one permanent extra key must support a
resolver recovering the specified answers at every state. Retaining $(p_t,\sigma)$ instead of $p_t$
preserves every adequate key assignment. An injective nominal name $\sigma$
needs no extra key values. If one reachable state merges profiles whose required
answers differ, every structure-only repair needs at least two extra-key values;
helpers computed from the profile cannot separate them. A checked two-state
history starts with distinct profiles and then merges them: nominal identity
works unchanged, while structural identity needs extra evidence or recourse.
This comparison covers every supported evolution of requirements, requiring no forecast of a
particular collision.
\leanmeta{\LH{HVD1}, \LH{HVD3}, \LH{HVD4}, \LH{HVD7}, \LH{HVD8}, \LH{HVD11}, \LH{HVD12}}
The existing structural Protocol observer factors through its required-member
view, ignoring nominal ancestry. This checks the observation property; PEP~544
and the ABC documentation specify Python's explicit and implicit admission rules.
\leanmeta{\LH{PRT2}}
\cite{PEP544,PythonABC}

\paragraph{Keys when later observations only distinguish more cases.}
Suppose later refinements only split groups of same-looking meanings. Every
earlier meaning that can conflict with a new meaning $v$ must already look like
$v$ when it arrives. That group, or fiber, only splits later.
The number of earlier neighbors (the future backdegree) is therefore at most $B-1$, so greedy gives the
$B$-tag upper bound.  A fiber of size $B$ is a clique and gives the matching
lower bound.

\paragraph{Renaming when nonoverlapping pairs become indistinguishable.}
With $m$ initial tag classes of sizes $s_i$, an adversary can match
$\sum_i\lfloor s_i/2\rfloor\ge(n-m)/2$ monochromatic pairs.  Each needs one
change, so $m\ge n-2r$, and a possible pair also needs two tags.  Conversely,
use $m=\max(2,n-2r)$ tags, put $n-m+1$ declarations in one class, and leave
the others singleton.  Any matching has at most $r$ monochromatic edges;
changing one endpoint of each restores adequacy.

\paragraph{Monotone-coarsening online--offline gap.}
For part (a), color each final coarsening class separately with $B$ tags.  A
fiber of size $B$ proves exactness.  For part (b), each fixed pair-merge trace
has a two-tag assignment and contains a fiber of size two, so its offline price
is exactly two.  Online, first present $n$ distinct profiles.  Reusing a tag
fixes a pair that an oblivious adversary can compute from the common prefix and
merge in one later revision.  For a randomized allocator, the finitely many
pair-collision events cover the sample space, so one fixed pair collides with
positive probability; merging it violates probability-one adequacy.  Global
names attain $n$.

\paragraph{Direct downward-closed deprecation graph.}
Any deletion $D$ that merges $x$ and $y$ must hide every observation on which
they differ.  Downward closure admits the canonical subset
$D_{xy}=M_x\mathbin{\triangle}M_y$, whose finer partition preserves the
multiplicity bound.  Conversely, hiding $D_{xy}$ merges the pair.

\paragraph{Tight one-deletion stability.}
Collapse each current fiber to a vertex of weight $w(u)$.  Single-deletion
conflicts join profiles whose cardinalities differ by one, so the quotient
conflict graph is bipartite by the parity of $|M_u|$, the number of
observations in the profile.  Fiber weight plays no role in that parity.  Give the even side its
first $w(u)$ colors and the odd side its last $w(u)$ colors from a palette of
size $A$.  Admissibility gives $w(u)+w(v)\le A$ on every edge, so the two
intervals are disjoint.  A current fiber of size $A$ gives the lower bound.

\paragraph{Two deletions can raise the price.}
Use Boolean presence profiles $x=\varnothing$, $y=\{a,b\}$, and
$z=\{b,c\}$.  Their symmetric differences give the explicit odd cycle
\[
 x\ \overset{\{a,b\}}{\longleftrightarrow}\ y\
 \overset{\{a,c\}}{\longleftrightarrow}\ z\
 \overset{\{b,c\}}{\longleftrightarrow}\ x.
\]
Each label is the pair of fields hidden to merge its endpoints.  Two spare
declarations share profile $\{d,e,f,g\}$, so the current price is two.  The
triangle requires three colors, and the checked three-color assignment handles
every safe deletion of at most two fields.

\paragraph{Two versus three known states.}
For two states, largest fibers give the clique lower bound.  In the bipartite
state-incidence multigraph $\PaperNotationLink{PaperAncestry}{H}$, a tag class is a matching, so permanent tags
are edge colors and $\Delta(\PaperNotationLink{PaperAncestry}{H})$ is the largest fiber.  K\H{o}nig's theorem
gives $\chi'(\PaperNotationLink{PaperAncestry}{H})=\Delta(\PaperNotationLink{PaperAncestry}{H})$.  For three states, use the three $k=2$ branches
retaining $x_1$, $x_2$, and $x_1\mathbin\oplus x_2$.  Each state consists of
two size-two fibers, while their conflict matchings cover $K_4$ and therefore
require four colors.

\section{Supply, Behavior, and Update Proofs}
The proofs below separate three questions: whether the representation preserves
required answers, whether executions satisfy behavioral contracts, and how much
outside supply or change remains. The pricing arguments then state the operations
and costs used to convert answer counts into work.

\paragraph{Relaxed-supply proof.}
\label{relaxed-supply-proof}
One-parent descendant sets are tree-shaped. For any feasible family $\{B_f\}$,
exact supply requires forwarding every pair in $\PaperNotationLink{PaperClassGroup}{A_f}\setminus B_f$ and removing
every pair in $B_f\setminus \PaperNotationLink{PaperClassGroup}{A_f}$. Doing exactly this attains its stated cost;
minimizing over feasible families gives $\PaperNotationLink{PaperRelaxedGap}{d_\lambda}$.

A sound optimum, with $B_f\subseteq \PaperNotationLink{PaperClassGroup}{A_f}$, is feasible and has no removal term,
so $\PaperNotationLink{PaperRelaxedGap}{d_\lambda}\le \PaperNotationLink{PaperGap}{d_\downarrow}$. If $\lambda\ge \PaperNotationLink{PaperGap}{d_\downarrow}$, any family with an
unwanted pair costs at least $\lambda\ge \PaperNotationLink{PaperGap}{d_\downarrow}$, while every sound family
costs at least $\PaperNotationLink{PaperGap}{d_\downarrow}$ by definition. Hence equality holds at that threshold.
For $\lambda=0$, taking every $B_f=\PaperNotationLink{PaperClasses}{C}$ gives a tree-shaped family of cost zero.

If $\mathcal A$ is nonlaminar, $B_f=\PaperNotationLink{PaperClassGroup}{A_f}$ for all $f$ is infeasible. Every feasible
family therefore differs in at least one pair, costing at least
$\min\{1,\lambda\}\ge1$ when $\lambda\in\mathbb N$ and $\lambda>0$.
Thus $\PaperNotationLink{PaperRelaxedGap}{d_\lambda}>0$. The equalities and removal-charge threshold are checked by
\LHrng{BLK}{1}{10}; the full declaration ledger appears in
Section~\ref{lean-handle-ledger}.

\paragraph{One answer-preservation condition across the models.}
As in main-text \S2, an intent on cases $\PaperInput$ and answers $\PaperAnswerSpace$ is a function $a:\PaperInput\to\PaperAnswerSpace$. A visible
representation $e:\PaperInput\to\PaperEncodingSpace$ with resolver $r:\PaperEncodingSpace\to\PaperAnswerSpace$ preserves the required answers when
$r(e(x))=\PaperNotationLink{PaperAnswer}{a(x)}$ for every $x$.  Equality of required answers induces an
equivalence on $\PaperInput$.  The visible equivalence refines it iff an answer-preserving
resolver exists on the reachable visible values.  Schema adequacy instantiates
this result with $e=(p,\sigma)$.  Implementation reachability takes
$X=\PaperNotationLink{PaperImplementations}{F}\times \PaperNotationLink{PaperClasses}{C}$, lets $a(f,c)$ assert $(f,c)\in \PaperNotationLink{PaperRequired}{R^{\mathrm{req}}}$, uses the represented
membership bit as $e(f,c)$, and fixes $r$ to the identity.  The last choice
preserves the positive/negative orientation and makes answer preservation equivalent to
$\PaperNotationLink{PaperSupplied}{R_D}=\PaperNotationLink{PaperRequired}{R^{\mathrm{req}}}$: each named pair keeps its truth value, whereas equal induced partitions also permit swapping truth values. No finiteness
or decidability instance is assumed by this shared layer.
\leanmeta{\LHrng{INT}{1}{7}}

\subsection{Behavioral substitution and exact applicability}
\label{behavioral-substitution}
For each required pair $q$, let $\alpha_q$ map reachable provider states or
histories into the observation space of the consumer contract, and let
$\operatorname{Valid}_q$ be the permitted observations.  Lean proves the
three equivalent forms
\[
 \alpha_q(\operatorname{Reach}_q)\subseteq\operatorname{Valid}_q,
 \qquad
 \alpha_q(\operatorname{Reach}_q)\setminus
   \operatorname{Valid}_q=\varnothing,
 \qquad
 \forall Q,\ Q|_{\operatorname{Valid}_q}\Rightarrow
   Q|_{\alpha_q(\operatorname{Reach}_q)}.
\]
The final form is the semantic property-preservation requirement: the reverse
implication to set inclusion uses membership in $\operatorname{Valid}_q$ as
$Q$.  Thus the converse assumes an assertion language containing that
predicate; in a weaker language, inclusion still implies preservation but the
reverse need not follow.  Taking the observation type to be traces incorporates
history constraints into $\operatorname{Valid}_q$ without adding object layout,
initialization, or state-aliasing semantics to the ancestry model.

Two checks are needed: inheritance must supply exactly the required pairs,
$\PaperNotationLink{PaperAncestry}{H}=\PaperNotationLink{PaperRequired}{R^{\mathrm{req}}}$, and each required pair $q\in \PaperNotationLink{PaperRequired}{R^{\mathrm{req}}}$ must satisfy the behavioral condition
above. Finite witnesses show that neither check implies the other. Passing both
is necessary for correct maintenance, but does not establish its minimum.
Combining them with the one-parent lower bound shows that a behaviorally valid
crossing still has gap at least $\PaperNotationLink{PaperGap}{d_\downarrow}>0$.

Any sound proof method may establish the behavioral condition; the resulting
statement does not depend on which certificate was accepted. Any two sound and semantically
complete methods certify exactly the same pairs.  Incomplete methods may
reject valid pairs but cannot change the boundary.  The mechanization
identifies that semantic boundary; it does not mechanize Liskov and Wing's
particular invariant, method, constraint, or extension rules.

Checking each pair's behavior cannot choose the required pairs. A bijection
that preserves every pair's behavioral condition also preserves the result of
all those checks. Lean constructs such a bijection between two four-pair
relations in which every pair's behavior is permitted. The first family is laminar and has
minimum outside-answer count zero; the second crosses and has strictly positive
minimum. Hence no criterion checking only each pair's behavior, including behavioral
substitution, can determine or discount the ancestry gap.
Concretely, use the main-text AnyIO classes $W$, $\PaperFileReader$, and $B$,
with $s$ naming file-lifecycle code and $p$ iteration code. The laminar relation is
\[
 L=\{(s,W),(s,\PaperFileReader),(p,W),(p,\PaperFileReader)\},
\]
and the crossing relation is
\[
 R'=\{(s,W),(s,\PaperFileReader),(p,\PaperFileReader),(p,B)\}.
\]
The checked bijection fixes the two $s$-pairs and maps $(p,W)$ to $(p,\PaperFileReader)$ and
$(p,\PaperFileReader)$ to $(p,B)$.  All four behavioral fibers are universally valid in both
relations, while the ancestry gap changes from zero to positive.
\leanmeta{\LHrng{BAP}{1}{19}}

\paragraph{Valid overrides keep promised calls available.}
The specialization \texttt{MethodContract} fixes accepted inputs $D$ and
permitted results $V(i)$.  An implementation is
$m:I\to\operatorname{Option}(O)$: \texttt{none} denotes an unavailable operation.
Permitted exceptional results belong to $O$ and appear as \texttt{some} results,
so the promised call remains available. Its behavioral boundary keeps every $i\in D$ reachable and
observes $(i,m(i))$.  Lean proves
\[
 \operatorname{Substitution}(m)
 \quad\Longleftrightarrow\quad
 \forall i\in D,\ \exists o\in V(i),\ m(i)=\operatorname{some}(o).
\]
Thus a replacement preserves substitution when it keeps all promised calls
available with permitted results.  If $m(i)=\texttt{none}$ for one $i\in D$,
substitution fails, regardless of all other calls.  A present method also
fails if its result is forbidden.  This is a contract promising outcomes on
its accepted inputs; other methods have their own declared contracts.
The model does not mechanize language typechecking or Liskov--Wing's proof
rules.  The explicit Boolean witness permits either result: replacing
\texttt{false} with \texttt{true} is valid, but disabling the operation is not.

Lean separately proves that if behavioral subtyping preserves promised
operations, a child excluding an operation promised by its proposed supertype
cannot be its behavioral subtype.  Such requirements call for changing the
subtype relation, for example by splitting a bundled parent into independent
roles and composing only the required roles.  The role-normalization and
multiple-parent construction already establish the implementation-supply
coordinate; each selected role still has its behavioral contract.
The relaxed model cancels unwanted implementation--class pairs while keeping
supertype promises binding. Valid replacement preserves the contract;
promise-blocking violates it.
Requiring behavioral admissibility can restrict those repairs and raise their
minimum cost, but cannot lower the exact-supply minimum.
Legacy formal names \texttt{BlockingRepair}, \texttt{finiteBlockPairs}, and
\texttt{BLK} identify this set-level removal model; promised operations remain
subject to the behavioral contract.
\leanmeta{\LHrng{BAP}{20}{25}}

\paragraph{Declared property instances.}
The common contract is parameterized by four inputs: cases $X$, required
answers $a$, visible evidence (including an observer projection when one is
part of the model), and the admitted policy states.  Table~\ref{tab:property-instances-supp}
records selected instantiations of this contract.  Lean packages these
four inputs and proves that policy-wide answer preservation is equivalent to an answer-preserving
resolver at every supported state.
\leanmeta{\LH{BPI1}}

\begin{table}[h]
\centering
\PaperTableSetup
\begin{tabularx}{\textwidth}{@{}>{\RaggedRight\arraybackslash}m{0.14\textwidth}
  >{\RaggedRight\arraybackslash}m{0.16\textwidth}
  >{\RaggedRight\arraybackslash}m{0.18\textwidth}
  >{\RaggedRight\arraybackslash}m{0.20\textwidth}Y@{}}
\toprule
Property face & Cases $X$ & Required answer $a$ & Visible evidence or projection & Admitted policy \\
\midrule
Integrity & executions or configurations & intended result & trusted boundary state & supported versions and configurations \\
\midrule
Projection-limited non-extraction & possible worlds or secrets & protected answer & observer projection $\pi\circ e$ & declared projection and supported worlds \\
\midrule
Safety & finite trace prefixes & whether the bad state is absent & monitor-visible prefix & admitted traces \\
\midrule
Availability & request--state pairs & required service result & exposed deployment state & supported failure states \\
\midrule
Liveness & declared fair traces & whether the obligation is eventually met & trace projection & supported fair traces \\
\midrule
Maintainability & version--requirement pairs & required meaning or reachability bit & retained key, ancestry, and external routes & declared evolution policy \\
\bottomrule
\end{tabularx}
\caption{Instances of the shared boundary contract.  The safety, availability,
and liveness rows characterize preservation of the declared answer;
they do not add latency, probability, scheduler, or fairness claims not present
in the row's inputs.  All rows instantiate the generic Lean theorem BPI1; the
projection row additionally uses PJD1--4 for its exact-recovery/non-extraction
duality.}
\label{tab:property-instances-supp}
\end{table}

For the projection-limited row, trusted recovery of every required answer is equivalent to the
absence of a conflict inside a full-evidence fiber, whereas every projected
decoder fails somewhere iff the projected view has such a conflict.
The projection is the adversary hypothesis; side channels require a different
projection and therefore a different theorem instance.
\leanmeta{\LHrng{PJD}{1}{4}}

\paragraph{Uncomputability of unrestricted structural safety.}
For every fixed $B\ge1$, a total computable transformation maps each program
index $e$ to a total observer $f_e$ such that $f_e$ has a fiber larger than $B$
exactly when program $e$ halts on input zero.  Consequently the safety set,
restricted to codes produced by this total-observer transformation, lies
outside the recursively enumerable sets.
\leanmeta{\LHrng{UND}{5}{22}}

For the computability classification, map program index $e$ to the total
observer
\[
 f_e(x)=
 \begin{cases}
 0,&\text{if }\PaperNotationLink{PaperRoutes}{\varphi_e}(0)\text{ halts within }x\text{ steps},\\
 x+1,&\text{otherwise}.
 \end{cases}
\]
If $\PaperNotationLink{PaperRoutes}{\varphi_e}(0)$ halts by step $s$, inputs $s,\ldots,s+B$ map to zero;
otherwise the observer is injective.  Halting therefore reduces to
$\mathsf{HasBarrier}_B$.  For every fixed $B\ge1$, nonhalting reduces through
the same total observers to $\mathsf{MultiplicityBound}_B$, and that safe
predicate lies outside the recursively enumerable sets.  Non-enumerability
persists after restricting the safe set to total observers.
\leanmeta{\LHrng{UND}{13}{22}}

A sharper fixed-table construction uses the carrier $\operatorname{Fin}(B+1)$.
At state $s$, bounded evaluation chooses either the identity table or the
constant-zero table.  Each snapshot is total and decidable, but safety of the
whole trace is not recursively enumerable for any fixed $B\ge1$.  For $B=1$
the carrier has only two meanings.
\leanmeta{\LHrng{UND}{23}{30}}

The finite-table strengthening moves the clock from an observer input to the
trace index.  At state $s$, \texttt{finiteClockTrace} returns identity until
bounded evaluation sees the source halt and returns constant zero afterward.
Thus each state is an executable table on $B+1$ meanings, while the trace is
safe iff the source does not halt, including when the carrier has only
two meanings.
\leanmeta{\LHrng{UND}{23}{30}}

\paragraph{Open-world one-parent adequacy.}
The inheritance-side transfer uses two implementation roles and three
consumers.  Before bounded evaluation observes program $e$ halt, the stream
emits two disjoint singleton scopes.  Afterward it emits the fixed main-text AnyIO
crossing $\{W,\PaperFileReader\}$ and $\{\PaperFileReader,B\}$.  Every snapshot is one of these two finite
tables and has a decidable laminarity test.  The whole stream remains laminar
exactly when $e(0)$ does not halt, so the safe-generator set is not recursively
enumerable.  Unrestricted multi-parent ancestry realizes both tables by the
same provider-preserving construction.
\leanmeta{\LHrng{UND}{31}{38}}

Consequently, any computable checker that never accepts an unsafe generator
accepts only a proper recursively enumerable subset of the safe generators:
some safe generator remains uncertified. A decidable project policy can restrict
which generators it admits, but must exclude at least one safe future. It has
changed which futures are admitted; the original uncertified generators remain beyond that checker's coverage.
\leanmeta{\LH{UND39}}

\paragraph{Finite tests and their trust boundary.}
For a finite reachable scope $V$ and tested states $T$, Lean defines
$d_{\rm test}=|V\setminus T|$ and proves that it vanishes iff
$V\subseteq T$.  More generally, a correctly specified suite establishes
agreement with its specified outcomes iff it passes and every reachable violation lies in $T$.  On the
natural-number state space every finite $T$ omits some state; flipping only
that state's Boolean outcome constructs an incorrect behavior that still
passes every stored expectation.

For the computability transfer, prepend one fixed tree-shaped snapshot to
\nolinkurl{finiteRequirementClockTrace} and test only that first state.  The
nonempty singleton suite always passes and is complete iff every snapshot is
tree-shaped, iff the source program does not halt.  Thus the set of generators
for which even this one-test suite is complete is not
recursively enumerable.  The suite's expected outcomes are themselves an
external relation: verifying them moves authority to a checker.  Verification terminates in a declared trusted base or leaves an assumption.
\leanmeta{\LHrng{TST}{1}{10}}

\paragraph{Matching-coarsening recourse.}
For $n\ge2$ entities, one arbitrary revealed matching coarsening, strict zero
error, and an integer budget $r\ge0$ of endpoint retaggings after that matching
is revealed, the exact permanent alphabet size is
$\max(2,n-2r)$.  At $r=0$ this is global uniqueness; once
$r\ge\lceil(n-2)/2\rceil$, two reusable tags suffice.
\leanmeta{\LHrng{MRC}{1}{5}}

\paragraph{Explicit field-menu translation.}
For every $k\ge2$, a field-monotone family on $N=2^k$ entities uses exactly
$k+\binom{k}{2}=\binom{k+1}{2}=O(\log^2N)$ binary fields, starts from a
discrete profile, and keeps multiplicity at most two in every supported branch,
while its union conflict graph is $K_N$.  Every robust permanent assignment
therefore uses $N$ distinct tags.
\leanmeta{\LHrng{FLD}{1}{8}}

\paragraph{Two and three known states.}
For a finite declaration set and at most two known observation states, the
minimum permanent tag alphabet equals the largest same-looking group in either
state.  There also exist three states on four declarations where each state
costs two tags and the union costs four.
\leanmeta{\LHrng{PST}{1}{4}}

\paragraph{Minimum replacement state.}
Fix a chosen single-parent repair and let $b_c(f)=1$ exactly when the repair
misses implementation $f$ at class $c$.  If a decoder preserving these omission answers receives $f$ and
a residual code $z(c)$ while $c$ stays hidden, then
$z(c)=z(c')$ implies $b_c=b_{c'}$.  The equivalence classes of equal complete
answer patterns form an exact code, so their number is the exact minimum state
count.
\leanmeta{\LHrng{BWR}{3}{8}, \LHrng{BWR}{14}{18}}

\subsection{Independent outside bundles}
\label{independent-outside-bundles}
Grouping forwarding routes shortens their description, but each group still supplies
required implementations to specified classes. Fix $d$ required implementation--class
pairs supplied outside inheritance, partitioned among $b$ nonempty bundles.
Independent control exposes all $2^b$ bundle sets, of which only the full set
supplies every required pair. Any mechanism that distinguishes those sets therefore has at
least $2^b$ states and needs at least $b$ fixed-width bits.  A checker,
generator, or coupling rule can exclude the incomplete sets only by determining
all $d$ pairs and becoming their authority.
\leanmeta{\LHrng{RLC}{3}{4}, \LH{RLC6}, \LH{BWR37}}

For a fixed sound native relation $\PaperNotationLink{PaperAncestry}{H}$, let $\mathcal S$ be the finite family of
outside sets $\PaperNotationLink{PaperRequired}{R_t^{\mathrm{req}}}\setminus \PaperNotationLink{PaperAncestry}{H}$ admitted by the declared histories.  If a fixed
generator decodes each $S\in\mathcal S$ from its input, equal inputs imply equal
outside sets.  Its input therefore has at least $|\mathcal S|$ states and any
fixed-width binary input needs at least $\lceil\log_2|\mathcal S|\rceil$ bits.
If $\mathcal S$ contains all subsets of $d$ outside pairs, the bound is $d$.
Correlated requirements can shrink $\mathcal S$; source length does not change
this counting bound.  Any varying rule or additional state used by the decoder
is part of the input being counted.
\leanmeta{\LHrng{RLC}{7}{10}}

\paragraph{Setup-inclusive charged work.}
\label{charged-work-floor}
Compare designs with the same declarations and previously installed resources.
Among states sharing this context, suppose $M\ge1$ require different outside
sets. The decoder receives that shared context plus every additional varying
input, including changes to its own rule. Call that additional input its transcript.
Suppose $s$ setup actions and $u$ update actions carry at most $w>0$ bits each;
the padded transcript has width $w(s+u)$, including any stopping choices.
Then
\[
 M\le 2^{w(s+u)},\qquad
 s+u\ge A:=\left\lceil\frac{\lceil\log_2 M\rceil}{w}\right\rceil.
\]
Equal context and transcript must give the same required set, so the transcript
is injective on these $M$ states. Counting transcripts gives the first inequality;
the binary logarithm and integer rounding give the second. For $d$ independently
variable outside answers not determined by the context, $M=2^d$ and
$A=\lceil d/w\rceil$. Setup and updates contribute to the same action total,
so moving an action into setup leaves the bound unchanged. Let shared work cost $C_0$ and total cost $\PaperNotationLink{PaperClasses}{C}$ satisfy
$\PaperNotationLink{PaperClasses}{C}\ge C_0+q_0(s+u)$ for a declared per-action price $q_0>0$. Then
$\PaperNotationLink{PaperClasses}{C}-C_0\ge q_0 A$, and $M\ge2$ gives the coarse floor $q_0$.
This is a human-labor floor when the charged actions are human input actions;
machine actions use their own price and unit.
\leanmeta{\LHrng{CWF}{1}{6}}

Installing a required tool has a separate coarse floor. If a required batching tool
is unavailable before installation and available after $s$ charged installation
actions, $s>0$: zero actions leave the initial state unchanged. If these human
actions satisfy $\PaperNotationLink{PaperClasses}{C}\ge C_0+q_0s$, then $\PaperNotationLink{PaperClasses}{C}-C_0\ge q_0$, even when every later update
is automatic. Charge an existing batcher's construction at its acquisition
boundary; do not charge it again per use. The availability premise and labor
price are explicit, rather than inferred from the ancestry gap.
The two floors need not add: one setup action can satisfy both.
\leanmeta{\LHrng{CWF}{8}{9}}

Input supplied by the common specification is counted once. If that context already
determines every outside set, an exact generator needs zero additional input
bits: take the context and required answer to be the same set, and the transcript
to have width zero. Lean checks this construction for arbitrary outside sets.
Zero additional input bits means the shared context already supplies the information.
Construction and execution still have their own costs, and the protocol identifies
which actions supply that context or install the tool. The human-input bound charges
those actions; the semantic-answer bound counts the required facts even when a
generator receives them from shared context.
\leanmeta{\LH{CWF7}}

\paragraph{Adding unrelated pairs leaves the gap fixed while shrinking its fraction.}
For any finite required table and any $n$, add one new role that serves $n$
fresh classes and no old class.  This adds exactly $n$ required pairs but
does not change the old crossing or its ancestry gap.  Hence
$\PaperNotationLink{PaperGap}{d_\downarrow}/|\PaperNotationLink{PaperRequired}{R^{\mathrm{req}}}|$ can approach zero while the number of pairs a
one-parent rewrite must move stays fixed.
\leanmeta{\LHrng{GRI}{1}{4}}

\paragraph{Proof detail for keys after name erasure.}
Main-text Section~6.3 applies the stable-key coloring criterion.
Fix a finite nonempty set $\PaperNotationLink{PaperClasses}{C}$ of lookup cases, their required answers $a(c)$,
and the retained views $p_t(c)$ over the declared reachable states.
Join $c$ and $c'$ when some state gives $p_t(c)=p_t(c')$ but
$a(c)\ne a(c')$.  A permanent key is adequate exactly when it properly colors
this conflict graph $G$.  The minimum raw palette has $\chi(G)$ values;
the additional-name count beyond one shared constant is
$\chi_{\rm add}=\chi(G)-1$, so $0\le\chi_{\rm add}\le |\PaperNotationLink{PaperClasses}{C}|-1$.
It is positive exactly when $G$ has an edge.  An injective retained view at
every state needs no additional key even when class names are erased.
Thus neither name erasure nor a positive ancestry gap alone forces an extra
name. The upper bound counts lookup cases because each key distinguishes their
required interpretations; displaced implementation pairs have a separate count.
The prices are independent: a complete-conflict identity family on $m$ meanings
has additional-name price $m-1$ for $m\ge2$ with an empty implementation relation, while the $h\times h$
row--column implementation family has gap $h(h-1)$ with an empty identity
domain.
\leanmeta{\LH{NHF15}, \LH{NHF28}, \LH{EVO17}, \LHrng{EVO}{23}{30}, \LH{EVO40}, \LHrng{AVG}{1}{7}}

\paragraph{Minimum reusable adapters.}
When an adapter is selected before the requested implementation is known and
supplies only displaced dependencies, adapter representations preserving those dependencies translate
to and from all-ones rectangle covers of the gap matrix.  The minimum active
adapter count is therefore its biclique-cover number, equivalently its Boolean
rank; every fooling family gives a checkable lower bound.
\leanmeta{\LHrng{NHF}{29}{30}, \LHrng{NHF}{47}{48}}

\paragraph{Choosing one authoritative declaration per role.}
The formalization separates the intended relation from its source encoding.
For specification $\PaperNotationLink{PaperRequired}{R^{\mathrm{req}}}$, a declaration set $D$ supplies exactly the required pairs when its encoded
relation $\PaperNotationLink{PaperSupplied}{R_D}$ equals $\PaperNotationLink{PaperRequired}{R^{\mathrm{req}}}$.  Replacing $D$ by a denotationally equivalent
$D'$ therefore preserves exact supply, and a declaration whose denotation is
already covered can be erased.  These statements allow factoring without
identifying declarations with semantics.
\leanmeta{\LHrng{SDN}{1}{4}}

Fix a finite declaration-independent set of atomic roles, where one role
contains the supplied meaning and its required consumer scope.  Assume each
normalized declaration denotes exactly one whole role, every required role has
a candidate, and two candidates share a role exactly when their denotations
agree.  Exact supply is equivalent to coverage of every role; irredundancy chooses
exactly one candidate per role.
The irredundant choices supplying every role are therefore the bases of the partition matroid
whose parts are denotation fibers.  Every basis has one authority per role, and
basis exchange substitutes an authority only inside its denotation fiber.
\leanmeta{\LHrng{SDN}{5}{14}}

Because consumer scope belongs to the role denotation, every such basis
induces the same required scope family.  Hence any scope-dependent quantity,
including the ancestry gap, has one basis-invariant value.  There is no further
basis optimization: changing the required scope changes $\PaperNotationLink{PaperRequired}{R^{\mathrm{req}}}$, while exchanging
an authority preserves it.
\leanmeta{\LHrng{SDN}{15}{17}}

\paragraph{Role-normalization direction.}
Splitting independently changeable roles across distinct provider endpoints gives
the one-parent rewrite more freedom; it does not create its lower bound.  For
roles with one required scope, replacing their retained rows by a largest row
preserves nesting or disjointness and cannot increase deleted pairs.  Once a
candidate forest is fixed, choosing the largest descendant set contained in
each role's required scope minimizes that role's gap.  These two reductions
prove the numbered normalization proposition in the main text.  Forcing several
roles to remain on one source node adds a co-location constraint and therefore
cannot improve the optimum.
\leanmeta{\LHrng{NHF}{35}{41}}

\paragraph{Exact supply forbids over-coverage.}
The native-scope subset condition is load-bearing.  If unwanted native pairs
are permitted and the objective charges only omissions, assign every fact the
universal consumer set.  The resulting family is laminar and misses no
required pair, so every input relation has gap zero.
\leanmeta{\LHrng{BWR}{33}{34}}
The main model
instead requires the derived relation to equal the required relation at each
state.  Preinstalling a future path is therefore inexact whenever that path
creates a pair absent from the current requirement; this is exactly the
obstruction used by the immediate-parent construction below.

The closure formulation supplies a separate general extension.  Let
$\operatorname{cl}(S)$ contain the primitive responsibilities semantically
derivable from $S$.  A finite extensive, monotone, idempotent closure satisfying
exchange is a matroid closure; its irredundant complete sets are the bases and
share one rank.
\leanmeta{\LHrng{L}{10}{19}}
The normalized atomic-role model derives exchange
from its denotation partition.  More general semantic closures fall under the
same matroid result when they satisfy exchange; domains without exchange retain
the preceding denotational supply-preservation results without the basis theorem.

\subsection{Immediate-parent evolution}
\label{immediate-parent-evolution}
The snapshot theorem tells us which classes inherit which implementations.
It does not tell us whether an update can preserve the edges deriving them.
Here the code-provider nodes and consumer-class leaves stay fixed as requirements evolve.
Each node has at most one direct parent; following zero or more parent edges
defines ancestry. The fixed finite set of nodes may include arbitrary helpers
before the first state. An edge-monotone update keeps every old parent edge.

The path lemma says that if every edge of an old parent chain survives, every
node lying between its endpoints after the update already lay on that chain.
The local hypotheses therefore force repair at every strict insertion; the history below repeats that event.
Write $\PaperNotationLink{PaperClassGroup}{A_f^t}$ for the classes needing code provider $f$ at time $t$.
Suppose $\PaperNotationLink{PaperRequired}{R_t^{\mathrm{req}}}\subseteq R_{t+1}^{\mathrm{req}}$, both snapshots are tree-shaped, $A_x^{t+1}\ne\varnothing$,
$\varnothing\ne A_x^t\subsetneq A_q^t$, $A_x^t\nsubseteq \PaperNotationLink{PaperClassGroup}{A_g^t}$, and
$A_x^{t+1}\subsetneq \PaperNotationLink{PaperClassGroup}{A_g^{t+1}}\subsetneq A_q^{t+1}$.
Every one-parent update preserving required supply deletes an old parent edge or leaves a missing
or unwanted pair before or after the update, even with preallocated helpers.
If both endpoints have exact ancestry-only supply, strict nonempty scope inclusion
orders the providers: $q$ is above $x$ initially, and $g$ lies strictly between
them afterward. A class in $A_x^t\setminus \PaperNotationLink{PaperClassGroup}{A_g^t}$ proves that $g$ was off the old
path. The preserved-path lemma forces deletion of an edge on the old $x$--$q$
path. A class in $A_x^t$ reaches $q$ through that edge, and growth keeps this pair
required, so the deletion is charged under $\PaperNotationLink{PaperBurden}{B_{\mathrm{rep}}}$. With inexact
ancestry, a missing or unwanted endpoint pair already witnesses repair.
The Lean result does not need a separate laminarity hypothesis: exact one-parent
endpoints entail it.
\leanmeta{\LHrng{SIN}{1}{5}, \LHrng{NRH}{1}{5}}
Now take scopes

\[
\begin{array}{c|ccc}
 & a & b & c\\ \hline
\mathsf{top} & 1&1&1\\
\mathsf{left} & 1&0&0\\
\mathsf{right}_{0} & 0&1&0\\
\mathsf{right}_{1} & 1&1&0
\end{array}
\]

where only the right scope changes.  Both snapshots are laminar.  In the
second snapshot, right lies strictly between top and left.  It did not lie on
that chain in the first snapshot, so an edge-monotone one-parent update is
impossible.  A concrete tree preserving exact supply retracts top--left and adds right--left.
Exact supply forbids preallocating the final chain: putting left below right in
the first state would make right reach $a$.  The exact-supply multi-parent
construction retains every old edge and adds only right--left.

For the length-$n$ family, keep one permanent top role.  At state $k$, roles
through $k$ receive nested consumer prefixes, while later roles receive
disjoint singleton consumers.  Each state is laminar and each transition only
adds required pairs.  Every ancestry-only one-parent policy preserving required supply retracts at
least one old parent edge at each transition, so the accumulated retraction
count is at least $n$.  More generally, a one-parent policy preserving required supply with
forwarding and removal either retracts an edge or has a missing required pair
or an unwanted inherited pair at one endpoint.  If its final raw parent
function equals its initial one, even after arbitrary intermediate changes,
the final missing pairs together with the initial removed unwanted pairs contain
at least $n(n+1)/2$ pairs.  An exact-supply
multi-parent construction retains the initial leaf-to-provider and provider-to-top
edges.  At transition $i$, provider $f_{i-1}$, whose consumers are already
$\{c_0,\ldots,c_{i-1}\}$, also inherits $f_i$.  Induction gives precisely the
required prefixes.  Provider indices increase along every provider-to-provider edge,
so every snapshot is acyclic.  The update uses no helpers or forecast: it
retains the old graph and adds this one current edge.  It performs exactly $n$
additions and no retractions.
\leanmeta{\LHrng{IPE}{1}{28}, \LHrng{IPE}{33}{38}, \LHrng{SIP}{1}{7}}

An unchanged hierarchy has a simpler exact accounting, independent of the
one-parent condition.
If a finite native carrier is exact for $\PaperNotationLink{PaperRequired}{R^{\mathrm{req}}_0}$ and remains unchanged, its
later deficits and excesses are in bijection with
$\PaperNotationLink{PaperRequired}{R^{\mathrm{req}}_T}\mathbin{\triangle}\PaperNotationLink{PaperRequired}{R^{\mathrm{req}}_0}$: additions become deficits and removals become
excesses.  Hence their total count equals the number of changed semantic
answers.  In the staged family, if the initial one-parent carrier is gap-free
and the final parent function returns to the initial one, every final gap is a
unique strict lower-triangular pair and the final outside count is exactly
$n(n+1)/2$.
\leanmeta{\LHrng{SCT}{38}{39}, \LHrng{IPE}{29}{32}}

More generally, fix one-parent ancestry $\PaperNotationLink{PaperAncestry}{H}\subseteq \PaperNotationLink{PaperRequired}{R^{\mathrm{req}}_0}\subseteq \PaperNotationLink{PaperRequired}{R^{\mathrm{req}}_T}$.
Write $C_H(t)=|\PaperNotationLink{PaperRequired}{R^{\mathrm{req}}_t}\setminus \PaperNotationLink{PaperAncestry}{H}|$ and
$\PaperNotationLink{PaperExcess}{e_H}(t)=C_H(t)-\PaperNotationLink{PaperGap}{d_\downarrow}(\PaperNotationLink{PaperRequired}{R^{\mathrm{req}}_t})$. Then
\[
 C_H(T)=C_H(0)+|\PaperNotationLink{PaperRequired}{R^{\mathrm{req}}_T}\setminus \PaperNotationLink{PaperRequired}{R^{\mathrm{req}}_0}|,
 \qquad \PaperNotationLink{PaperExcess}{e_H}(T)\ge \PaperNotationLink{PaperExcess}{e_H}(0).
\]
The excess rises strictly when the gap grows by fewer pairs than the requirement;
if the gap is unchanged, every added pair increases $\PaperNotationLink{PaperExcess}{e_H}$ by one.
\leanmeta{\LHrng{LRR}{1}{7}}

An unchanged hierarchy also has a fixed coverage numerator. If it is exact for a
nonempty $\PaperNotationLink{PaperRequired}{R^{\mathrm{req}}_0}$, remains unchanged, and later requirements only add answers,
then its currently valid native answers are in bijection with $\PaperNotationLink{PaperRequired}{R^{\mathrm{req}}_0}$.
Consequently the numerator of its coverage fraction remains $|\PaperNotationLink{PaperRequired}{R^{\mathrm{req}}_0}|$, and
the fraction $|\PaperNotationLink{PaperRequired}{R^{\mathrm{req}}_0}|/|\PaperNotationLink{PaperRequired}{R^{\mathrm{req}}_t}|$ decreases strictly whenever the finite required
relation grows.  Independently, any view of answer capacity at most $M$ has
coverage at most $M/|\PaperNotationLink{PaperRequired}{R^{\mathrm{req}}_t}|$; for $M>0$ that ceiling strictly decreases as the
required-answer count increases.  The first statement concerns the actual
coverage of one fixed exact view; the second is an upper bound for arbitrary
views of bounded capacity.
\leanmeta{\LHrng{SCT}{40}{45}}

The required relation grows throughout this family.  Exact supply at the old
state means that every implementation-to-class pair derived through a
retracted edge was required, and monotonicity keeps every such pair required
after the transition.  Retraction therefore records a representation-induced
rederivation and verification obligation because the semantic requirement
persists. An atomic reparenting primitive still contains that retraction.

\paragraph{Missing and unwanted pairs split into forced gap and removable excess.}
For required relation $\PaperNotationLink{PaperRequired}{R^{\mathrm{req}}}$ and native relation $\PaperNotationLink{PaperAncestry}{H}$, define
\[
 \PaperNotationLink{PaperRepairSet}{\mathcal D(H,R^{\mathrm{req}})}=(\PaperNotationLink{PaperRequired}{R^{\mathrm{req}}}\setminus \PaperNotationLink{PaperAncestry}{H})\uplus(\PaperNotationLink{PaperAncestry}{H}\setminus \PaperNotationLink{PaperRequired}{R^{\mathrm{req}}}).
\]
Lean proves that exact native supply is equivalent both to $\PaperNotationLink{PaperAncestry}{H}=\PaperNotationLink{PaperRequired}{R^{\mathrm{req}}}$ and to
$\PaperNotationLink{PaperRepairSet}{\mathcal D(H,R^{\mathrm{req}})}=\varnothing$.  On finite relations its cardinality is the
sum of the deficit and excess cardinalities and is positive exactly when exact native
supply fails. There is therefore a missing required pair, an unwanted supplied
pair, or both. An add-only repair preserving exact supply exists precisely
when the excess is empty; it injectively owns the entire deficit.  Any excess
instead needs a negative repair or a changed specification.  Normalized
disjoint additivity makes cardinality the unique answer count, so this debt
object is defined before and independently of the one-parent optimization;
minimizing its sound one-parent instance yields $\PaperNotationLink{PaperGap}{d_\downarrow}$.  More strongly,
for every sound one-parent selection $\PaperNotationLink{PaperAncestry}{H}$, Lean constructs an equivalence
between $\PaperNotationLink{PaperRepairSet}{\mathcal D(H,R^{\mathrm{req}})}$ and its ancestry gap.  If $m=\PaperNotationLink{PaperGap}{d_\downarrow}$ and
$\PaperNotationLink{PaperExcess}{e_H}:=|\PaperNotationLink{PaperRepairSet}{\mathcal D(H,R^{\mathrm{req}})}|-m$, then
$|\PaperNotationLink{PaperRepairSet}{\mathcal D(H,R^{\mathrm{req}})}|=m+\PaperNotationLink{PaperExcess}{e_H}$, and $\PaperNotationLink{PaperExcess}{e_H}=0$ iff $\PaperNotationLink{PaperAncestry}{H}$ attains the minimum.
Thus the actual outside count splits into the forced ancestry gap and the
removable excess of this particular hierarchy.
\leanmeta{\LHrng{SCT}{3}{8}, \LHrng{SCT}{46}{55}, \LH{BWR52}}

\paragraph{Successful outputs and ancestry cannot reveal the gap.}
One program has tree-shaped requirements and another has a crossing. Both
supply every required answer through routes and expose the same empty ancestry.
A rule reading only that ancestry must give both the same classification,
so it cannot classify both correctly. Lean checks the general principle:
identical observations cannot distinguish worlds where the required property differs.
The witnesses in \S\ref{behavioral-substitution} instantiate it for tree shape
and crossing-triggered policies. A classifier that may abstain must abstain on
their shared observation to avoid a wrong answer. Identical successful outputs
add no distinguishing evidence.
\leanmeta{\LH{BWR42}, \LHrng{OLC}{1}{5}, \LHrng{OLC}{7}{8}}

Conversely, exact supply recovers the required relation from ancestry $\PaperNotationLink{PaperAncestry}{H}$,
removals $\PaperRemovals$, and outside routes $\phi$:
$\PaperNotationLink{PaperRequired}{R^{\mathrm{req}}}=(\PaperNotationLink{PaperAncestry}{H}\setminus\PaperRemovals)\cup\operatorname{dom}\phi$.
Thus the complete supply map determines requirement shape and its ancestry
gap; the ambiguity concerns observations that omit part of that map.
\leanmeta{\LH{OLC9}}

\paragraph{Capacity-removal descent.}
Fix the requirements and the five native-capacity rungs.  Write their outside
counts as
\[
 \max\{\PaperNotationLink{PaperKeys}{\chi(G^{\rm req})}-1,0\}+|\PaperNotationLink{PaperMembership}{M^{\mathrm{req}}}|+|\PaperNotationLink{PaperRequired}{R^{\mathrm{req}}}|,\quad
 \PaperNotationLink{PaperGap}{d_\downarrow}(\PaperNotationLink{PaperMembership}{M^{\mathrm{req}}})+|\PaperNotationLink{PaperRequired}{R^{\mathrm{req}}}|,\quad |\PaperNotationLink{PaperRequired}{R^{\mathrm{req}}}|,\quad
 \PaperNotationLink{PaperGap}{d_\downarrow}(\PaperNotationLink{PaperRequired}{R^{\mathrm{req}}}),\quad 0.
\]
Lean proves that removing the next capacity never lowers this count and raises
it strictly whenever the removed capacity is used.  A policy that fires on a
positive outside count therefore preserves its own trigger and reaches the
structural-only endpoint within four removals.  If practitioner skill enters
only through a per-occurrence price $q(\sigma)$, lowering that price lowers the
charge but changes neither the semantic trajectory nor its endpoint.
\leanmeta{\LHrng{CSD}{1}{8}, \LH{CAP7}, \LHrng{EVO}{17}{18}, \LH{EVO40}}

\paragraph{Ignoring workaround costs reverses the choice.}
Let \emph{acquire} pay any finite training, migration, or implementation cost
$K\in\mathbb N_{>0}$, and let \emph{decline} retain the incomplete mechanism's workaround stream. A
declared ledger that charges $K$ to acquire while assigning zero benefit to
avoided workarounds strictly ranks decline cheaper, independently of the
per-occurrence repair price $q$.  Lean also gives two exact
programs with identical successful native observation and hidden workaround
indicators zero and one; consequently no function of that observation
determines whether the minimum outside count is zero or positive.  For every
repair price $q\in\mathbb N_{>0}$, the staged-history lower bound supplies the explicit finite
threshold $N=\lfloor K/q\rfloor+1$, beyond which $K<qB_{\mathrm{rep}}(n)$ for every
one-parent policy preserving required supply. The correct-maintenance
ledger then strictly ranks acquisition cheaper while the declared ledger still
ranks decline cheaper. That threshold never increases as positive $q$ increases: lower
per-occurrence cost delays but cannot eliminate the finite reversal.  This
proves an incentive-order reversal without an
assumption about rational choice, practitioner cognition, or actual conduct.
The main text's primary records document cost-side omission rationales; its
pinned census documents installed availability rather than behavior.
\leanmeta{\LHrng{INC}{1}{8}, \LHrng{OLC}{1}{5}, \LH{SCT31}, \LHrng{OPT}{3}{12}}

Choosing implementation authorities and choosing inheritance edges are different
problems. The partition matroid records the first: choose one declaration per
role before choosing a hierarchy. Tree edges give the second, a graphic-matroid
choice: adding the third edge of the top--left--right triangle requires removing
an old edge. The immediate-parent theorem keeps the providers and classes
fixed while proving that this carrier exchange cannot be avoided by helper
nodes.  The staged family repeats that graphic exchange at every transition
while the authority basis stays fixed and every snapshot remains laminar.

\paragraph{Parked-top policy: standing supply without fresh repairs.}
\label{fixed-top-routing}
Start with $c_i\to f_i$, leave $c_{n+1}$ parentless, and route top to all
$n+2$ classes once. Update $i$ adds $f_{i-1}\to f_i$. At state $k$, each
$c_j$ with $j\le k$ reaches $f_j,\ldots,f_k$; later $c_j$ with $j\le n$ reach
only $f_j$. These are exactly the required non-top pairs. No edge is deleted
and no route changes. The same $n+2$ missing top pairs give
$\PaperNotationLink{PaperBurden}{B_{\mathrm{rep}}(n)}=2n(n+2)$ and positive outside supply at every state, so the policy
misses the maintenance ideal throughout: the required outside pairs persist even with unchanged route code.
More sharply, $\delta+\pi\ge n$, where $\delta$ counts charged deletions and $\pi$ counts
updates with endpoint missing or removed pairs, once per update.
\leanmeta{\LHrng{NRH}{8}{9}}
An ancestry-only policy has $\pi=0$, hence $\delta\ge n$; the parked policy has
$\delta=0$, $\pi=n$. The pricing corollary counts additional work covering
each endpoint occurrence, including standing routes. One invariant validates
unchanged routes; fresh configurations require fresh checks, as the protocol below proves.

\subsection{Admission checks for fresh routing configurations}
\label{sec:completion-admission-audit}
Fix the required relation and $m$ outside routing entries that can each be omitted independently.
Each deployment supplies a fresh configuration; the complete configuration
and each single omission remain possible regardless of earlier audit results.
The configuration stays fixed during admission and use.  The checker must
accept the complete configuration and reject every single omission.  Its only
evidence about enabled entries comes from probes, each asking whether all
entries in a chosen batch are enabled.  Batches contain at most $b>0$ entries;
the next batch may depend on previous answers.

On the complete input, any unread entry could instead be missing without
changing the observed answers or the accepted execution path.  Therefore
every entry must be read, and at least $\lceil m/b\rceil$ probes are necessary.
For $N$ independently supplied deployments the same argument applies to each
fresh input, giving at least $Nm/b$ probes in total.  This counts executed
queries: each fresh input can independently omit an entry, so earlier checks cannot certify it. Fix the
maintainer's positive per-probe price $\ell$ in a declared cost unit. The cumulative charge is
at least $\ell Nm/b$.  Thus this protocol supplies rate $q=\ell m/b>0$ when
$m>0$.  Acquisition cost and continuing disadvantages can be compared against
this rate using the finite-acquisition inequality below.
Fresh deployments force fresh probes and their recurring charge.
The staged count $\PaperNotationLink{PaperBurden}{B_{\mathrm{rep}}}$ prices a protocol covering its endpoint occurrences
(\S\ref{bounded-realization-cost}).\leanmeta{\LHrng{CAA}{1}{5}}

The artifact's \texttt{completion\_admission\_audit.py} runs this protocol on
the entry identities in the archived AnyIO omission report.  It reads fresh
Boolean configurations. Reusing AnyIO's entry identities fixes the input; the admission checker itself executes the probes.
For 21 archived entry identities, capacities 1, 4, 64 give respectively 21, 6, 1 probes on each accepted deployment, attaining the bound; 8 fresh deployments take 168, 48, 8 probes. Every single omission is rejected for each capacity.

Tests also exhaust all configurations
of zero through eight entries.  The report binds the historical input by
digest and retains every queried index.  These exact operation counts hold before prices; every positive per-probe
price gives the corresponding lower charge bound.

The comparison credits the same guarantees on both sides.  A checked native
construction or a proved routing generator can restrict deployments to complete
configurations; then this additional omission audit needs zero probes.
Building and checking that guarantee remains part of acquisition and upkeep.
Common specification and behavioral checks remain in both policies' costs.
With unrestricted probe capacity, one aggregate query also suffices, although
its implementation and cost still need accounting.  These cases are checked
alongside the lower bound.  Fresh independent omission possibilities, rather
than the word ``composition,'' create this recurring validation obligation.
\leanmeta{\LHrng{CAA}{6}{7}}

\subsection{Pricing repair and retention}
\label{repair-retention-pricing}
A ban leaves requirements in force while removing their native derivations. Workarounds supply outside answers,
with labor and replacement costs~\cite{boehm1984software,boehm2000roadmap,sullivan2001modularity}
shaped by debt, libraries, installed code, and experience~\cite{cunningham1992wycash,meyerovich2013adoption}.
Graham's adoption barrier and Ousterhout's \emph{tactical tornado} price learning
and deferred costs~\cite{graham2001beating,ousterhout2021philosophy}; neither erases workarounds.
\paragraph{Semantic deletions do not create repair obligations.}
\label{semantic-deletion-normalization}
For an old edge $e=(u,v)$, its footprint $\PaperEdgeFootprint$ consists of pairs
$(f,c)$ with an old path from $c$ to $u$, the edge $u\to v$, and an old path
from $v$ to $\PaperNotationLink{PaperProvider}{o_f}$. Charge its deletion exactly when
$\PaperEdgeFootprint\cap \PaperNotationLink{PaperRequired}{R^{\mathrm{req}}_{t+1}}\ne\varnothing$.
The test uses the old footprint: a replacement path restores supply while the
still-required answer loses its old derivation. Withdrawing only no-longer-required
supply is a semantic edit; removing an unused edge also creates no repair charge.
$\PaperNotationLink{PaperBurden}{B_{\mathrm{rep}}}$ adds charged deletions to missing and unwanted endpoint pairs.

\paragraph{Charged deletions interrupt retained derivations.}
\label{derivation-retention}
Keep the same nodes across the update, including any preallocated helpers.
Let $\operatorname{Der}_t(f,c)$ contain the finite ancestry edge paths from $c$
to provider $\PaperNotationLink{PaperProvider}{o_f}$ at state $t$. The pair retains a derivation iff
$\operatorname{Der}_t(f,c)\cap\operatorname{Der}_{t+1}(f,c)\ne\varnothing$.
These edge lists retain the answer's derivation: two paths can reach the same provider through different edges.

\leanmetapending{\LHrng{DRP}{1}{5}}
\begin{lemma}[Charged deletion is loss of a still-required derivation]
\label{charged-deletion-derivation-retention}
On a fixed acyclic one-parent carrier, including no self-parent edges,
$|\operatorname{Der}_t(f,c)|\le1$. An update has a charged deletion iff some
$(f,c)\in \PaperNotationLink{PaperAncestry}{H_t}\cap \PaperNotationLink{PaperRequired}{R^{\mathrm{req}}_{t+1}}$ retains no old derivation.
Any addition-only update retains every old path; in particular, the direct
multi-parent construction does so on every growing history.
\end{lemma}
\begin{proof}
The parent function fixes each next edge. Two paths with the same endpoints
therefore coincide: induction compares their first edges, and acyclicity
excludes a nonempty path competing with an empty one.
A charged edge carries a still-required pair. That pair's unique old path
uses the deleted edge and cannot survive. Conversely, take a still-required
pair with an old path but no retained derivation. Some edge of its old path
was deleted; otherwise the same edge list would remain a path. Its deletion
is charged because that pair lies in its old footprint. An addition-only
update preserves each edge of every old path.
\end{proof}
This prices derivation replacement even when the required answer and provider
identity stay fixed. The deletion count remains an edge count: several deleted
edges can interrupt one pair's path. One header edit may replace many paths;
the answer, operation, and labor counts remain separate (main-text
Proposition~4.16). The lemma equates the per-update witnesses; several deleted edges on one path explain why their counts can differ.

For $R_0^{\mathrm{req}}=\{(f,c)\}$ and $R_1^{\mathrm{req}}=\varnothing$, raw edge-deletion accounting
gives direct ancestry $(O_0,B_{\mathrm{raw}})=(0,1)$. Forwarding also pays one
endpoint obligation for its old missing pair, giving $(1,1)$: the old outside pair still contributes.
This example does not refute joint attainment of the raw ideal. Normalization
instead separates a requested withdrawal from repair: deleting the direct edge
is uncharged and direct ancestry attains $(0,0)$.
\leanmeta{\LHrng{MGH}{9}{11}, \LH{NRH7}}

\paragraph{Joint attainment on arbitrary finite supply histories.}
\label{monotone-multiparent-attainment}
Fix finite provider and consumer sets $\PaperNotationLink{PaperImplementations}{F},\PaperNotationLink{PaperClasses}{C}$, using separate nodes even when one
source declaration has both roles. At each state give providers no parents and
include $c\to \PaperNotationLink{PaperProvider}{o_f}$ exactly when $(f,c)\in \PaperNotationLink{PaperRequired}{R_t^{\mathrm{req}}}$. Paths have at most one edge,
so the graph is acyclic and supplies exactly the required relation. Each deleted
edge carried only its now-unrequired pair. There are no charged deletions,
missing pairs, or unwanted pairs, even when requirements shrink and later return.
Thus one design history jointly attains zero outside supply and $\PaperNotationLink{PaperBurden}{B_{\mathrm{rep}}}=0$ on every
finite prefix. The same construction applies independently to contract edges;
separate graphs attain both supply coordinates together.
The flat graph satisfies main-text Proposition~4.2's base-order condition.
Lean checks these abstract graphs and zero counts; Python's class construction remains external.
\leanmeta{\LHrng{MGH}{1}{3}, \LHrng{MGH}{9}{11}, \LHrng{UMS}{5}{6}}

For the supply capacities (none, one parent, multiple parents), define universal
attainment as attaining this joint ideal on every finite declared history.
Multiple parents attain it by the direct construction. No implementation parents
fail on one required pair; one parent fails on the two-provider, three-class
crossing. The negative proofs allow arbitrary helper nodes. Instantiating the
same statement for contracts and implementations proves the iff in main-text
Corollary~4.12. Identity histories retain their separate key minimum (main-text
Theorem~3.2); this characterization concerns the supply rungs.
\leanmeta{\LHrng{UMS}{1}{6}}

\paragraph{Exactly which accountings preserve the strict separations.}
\label{price-invariant-separations}
Multiple parents leave no outside work on a crossing. Every sound one-parent
design leaves some. Changing the prices preserves every crossing separation
exactly when any nonempty batch costs more than doing nothing. Prices may
include bulk discounts and may fall when more facts are supplied.

\leanmetapending{\LHrng{ACF}{1}{4}}
\begin{theorem}[Every crossing survives exactly when nonempty outside work costs positively]
\label{thm:accounting-characterization}
Let $U$ be an infinite universe of fact labels and let
$Q:\{S\subseteq U:S\text{ finite}\}\to\mathbb R$, with
$Q(\varnothing)=0$. Give distinct required implementation--class pairs
distinct labels. The following are equivalent:
\begin{enumerate}
\item For every finite crossing relation, the zero-outside multiple-parent
design costs strictly less than every correct sound one-parent design.
\item $Q(S)>0$ for every finite nonempty $S\subseteq U$.
\end{enumerate}
For each finite declared relation, condition (2) also gives a uniform positive
lower bound on all its nonempty outside sets, hence a strictly positive
one-parent infimum. No additivity or monotonicity is required.
\end{theorem}
\begin{proof}
For (2)$\Rightarrow$(1), a sound one-parent design of a crossing leaves a
nonempty outside set (main-text Theorem~4.6). Its price is positive, against
$Q(\varnothing)=0$. The declaration has only finitely many nonempty subsets;
the minimum of their positive prices is positive.

For the converse, fix any finite nonempty $S$ and choose $a\in S$ and three
distinct fresh labels $b,c,d\notin S$. Give provider $f$ the class set
$\{x,y\}$ and provider $g$ the class set $\{y,z\}$; label their four pairs
$(f,x),(f,y),(g,y),(g,z)$ by $a,b,c,d$, respectively. For each other label in
$S$, add one provider required by one fresh class. Retain just the three
crossing pairs labelled $b,c,d$ in ancestry and forward exactly $S$.
The retained scopes are $\{y\}$, $\{y,z\}$, and empty sets, so they are
tree-shaped. Native supply is sound and forwarding completes every required
pair. Its outside set is exactly $S$, so (1) forces $Q(S)>0$.
The Lean construction also checks that all required pairs have distinct labels.
\end{proof}

Thus an accounting can erase a strict crossing separation only by assigning
zero or negative price to some nonempty outside set. This characterizes the
price of the same required facts; labels identify those facts regardless of
their storage or source syntax. Bundling changes how they are encoded.
Every other zero-versus-nonempty witness obeys the same sufficient condition.
The earlier monotone-price result below is a special case: its assumptions
imply positive prices for all nonempty sets.

\leanmetapending{\LHrng{PIS}{1}{6}}
\begin{theorem}[Strict separations survive non-additive prices]
\label{thm:price-invariant-separations}
Fix a finite nonempty universe $A$ of required facts and a common price
$Q:2^A\to\mathbb R$ for their outside supply. Suppose
$Q(\varnothing)=0$, $Q(\{a\})>0$ for every $a\in A$, and
$S\subseteq T$ implies $Q(S)\le Q(T)$.
There is $\varepsilon>0$ such that every nonempty outside set costs at least
$\varepsilon$. Consequently, if an upper-rung design leaves no facts outside
and every lower-rung design leaves some, the upper design costs zero and the
infimum over all lower designs is at least $\varepsilon>0$.
In particular, every sound one-parent design for a crossing costs positively
under every such price; exact multiple-parent supply costs zero.
\end{theorem}
\begin{proof}
Take $\varepsilon=\min_{a\in A}Q(\{a\})>0$.
Every nonempty $S$ contains an $a$, so
$Q(S)\ge Q(\{a\})\ge\varepsilon$. The zero-outside design costs
$Q(\varnothing)=0$. For the implementation case, suppose a sound one-parent
selection leaves nothing outside. Every retained provider scope then equals
its required scope. Retained scopes are tree-shaped, contradicting the crossing.
The multiple-parent construction supplies every pair natively.
\end{proof}
Each strict step in the main-text hierarchy has this zero-versus-positive
witness. The theorem allows unequal fact prices and bulk discounts, including
a fixed positive charge for an entire nonempty batch. Such a charge need not
add across facts. Unit charges recover the exact counts; changing prices
preserves these strict separations. Assigning zero cost to a required fact
removes its contribution to the price; it does not remove the required fact.
Changing the intended behavior is a
specification change (main-text Corollary~4.19).

\paragraph{Every faithful representation must realize a semantic change.}
\label{semantic-change-positive-cost}
A representation may be program text, a generated graph, or a compressed
description together with the inputs and state that determine its behavior.
Let $r$ decode it to the declared meaning. A faithful change of representation
$e$ satisfies $r'(e(p))=r(p)$; it preserves whether meanings differ.

\leanmetapending{\LHrng{SCC}{1}{9}}
\begin{theorem}[A semantic change requires positively priced execution]
\label{thm:semantic-change-positive-cost}
Let a finite operation sequence take representation $p$ to $p'$ with
$r(p)\ne r(p')$. It contains a step that changes the decoded meaning.
If each executed operation has positive price, their total price is positive,
regardless of how many facts one operation changes. For the finite sequence
of $u>0$ operations, its smallest operation price $q_0$ is positive and its
total cost is at least $q_0u$. These conclusions hold for every decoder and
every faithful re-encoding of the same declared meanings.
\end{theorem}
\begin{proof}
An empty sequence leaves the representation unchanged. If every step preserved
the decoded meaning, induction would give $r(p)=r(p')$. Thus some executed step
changes it. A finite nonempty list of positive prices has a positive minimum
$q_0$, and summing the individual bounds gives $\PaperNotationLink{PaperClasses}{C}\ge q_0u>0$.
Faithfulness gives $r'(e(p))\ne r'(e(p'))$ exactly when $r(p)\ne r(p')$.
\end{proof}
This derives positive total execution cost without assigning a price to each
changed fact. A single operation can change them all and still has its positive
price. For a uniform per-fact rate across executions, the machine and protocol
supply the capacity and price bounds used next. Additional maintenance cost
subtracts operations shared with the compared design.

\paragraph{Compression and bounded-capacity realization.}
\label{bounded-realization-cost}
Compression can shorten the description of outside answers without changing
how many required answers the program supplies. The input bound in
\S\ref{independent-outside-bundles} counts distinguishable outside sets. Executing a decoded set
requires operations even when its description is short.
The update index can select the outside set in the main text's prescribed growing
history; independently variable outside answers are a separate premise.
For a finite execution or materialization trace, let $m$ answer occurrences
be covered by $u$ operations, each covering at most $b>0$ occurrences.
An occurrence is counted once, even if several operations cover it.
The coverage bound gives $u\ge\lceil m/b\rceil$.
If those operations cost at least $q_0>0$ each, above a common baseline $C_0$, then
\[
 \PaperNotationLink{PaperClasses}{C}-C_0\ge q_0u\ge \frac{q_0}{b}m,
 \qquad q:=\frac{q_0}{b}>0.
\]
Lean proves the positive rate and cost bound from the finite coverage, capacity,
and charging premises. The operation-cover bound permits each operation to
realize up to $b$ answers, so batching is already included.
Batching divides the positive operation price across at most $b$ occurrences;
the resulting price $q_0/b$ is still positive. Description compression and
zero additional input bits do not cancel this execution charge.
\leanmeta{\LH{BWR22}, \LH{BWR63}}

To set $m=\PaperNotationLink{PaperBurden}{B_{\mathrm{rep}}(n)}$, the protocol must cover every occurrence that $\PaperNotationLink{PaperBurden}{B_{\mathrm{rep}}}$ counts,
including a pair at both endpoints when it remains missing. Retaining one route
need not execute it twice. Requiredness over reachable executions likewise does
not require every pair to be exercised in one execution. The AnyIO suite reaches
and rejects all \MutationDenominator{} single-route omissions
(\S\ref{anyio-mutation-study}); these executions witness requiredness.
The machine and protocol determine the price of realizing those required facts.
Native implementations also execute required behavior. A gross execution cost
is incremental only when the charged operations are additional to those of the
comparator; shared work belongs in $C_0$.

The fixed machine and executed protocol supply the operation capacity and price.
Landauer's bound prices logically irreversible information loss; applying it
requires identifying such an operation~\cite{landauer1961irreversibility,bennett2003notes}.
Retention bounds depend on the memory model, duration, reliability, and refresh
protocol; Chiuchi\`u et al. show that preservation cost in their model can be
arbitrarily reduced at fixed duration and error probability by sufficiently
frequent refresh~\cite{chiuchiu2018remembering}.
These physical results remain external to Lean. A fixed machine and protocol can supply $b$ and
$q_0$ in declared energy, time, or monetary units. A labor or monetary comparison
prices its own additional operations; energy uses its own unit and comparator.

Across the staged $n$-transition history, $\PaperNotationLink{PaperBurden}{B_{\mathrm{rep}}(n)}$ counts charged parent-edge deletions
plus missing required pairs and removed unwanted pairs at both update endpoints.
$\PaperNotationLink{PaperBurden}{B_{\mathrm{rep}}(n)}\ge n$. Lean checks this bound with the normalized charge.
\leanmeta{\LHrng{NRH}{3}{6}}
If incremental charges after batching are at least $qB_{\mathrm{rep}}(n)$
for $q>0$, a policy that keeps one parent has cost at least
$\PaperNotationLink{PaperCommonCost}{K_c(n)}+qB_{\mathrm{rep}}(n)\ge \PaperNotationLink{PaperCommonCost}{K_c(n)}+qn$, where $\PaperNotationLink{PaperCommonCost}{K_c(n)}$ contains costs common to the compared
policies. Keeping the mechanism incomplete creates this additional term.
Common project costs stay in $\PaperNotationLink{PaperCommonCost}{K_c(n)}$; expertise changes occurrence prices while
the history fixes $\PaperNotationLink{PaperBurden}{B_{\mathrm{rep}}}$'s semantic count.

The acquisition comparison separates setup from continuing cost. A feasible
plan records one finite initial cost $\PaperNotationLink{PaperAdoption}{K_a}$ for training,
supervised adoption, implementation, and migration, plus continuing excess
$\PaperNotationLink{PaperAdoptionUpkeep}{E_a(n)}\le r_an$. If $\PaperNotationLink{PaperAdoptionRate}{r_a}<q$, it is cheaper than permanently keeping one parent after
$\lfloor \PaperNotationLink{PaperAdoption}{K_a}/(q-\PaperNotationLink{PaperAdoptionRate}{r_a})\rfloor+1$.  Familiarity, installed base, turnover,
retraining, ecosystem integration, and unavailable tooling can raise $\PaperNotationLink{PaperAdoption}{K_a}$ or
$\PaperNotationLink{PaperAdoptionRate}{r_a}$; they describe path dependence produced by earlier availability and
adoption. Complete capacity retains the option to use the same familiar design,
so these contextual prices leave the irreducible supply gap intact. Applying the theorem to
each feasible policy gives the quantified claim in the main text; an unavailable
path is absent from the policy family rather than assigned infinite cost.
The operation bound
additionally assumes an allowed operation vocabulary, nonnegative
per-operation cost $p$, and capacity $b>0$.

\subsection{Learning belongs in the policy comparison}
\label{supp:acquisition-policy-optimality}
Dynamic optimization compares policies using their present and future
costs~\cite{bertsekas2015planning}.  Current skills describe the starting
state; feasible training and migration can change it.  Let
$\mathcal P_{\rm stay}\ne\varnothing$ contain policies preserving the required answers and behavior that retain
one parent, and let $\mathcal P_{\rm all}$ also admit feasible acquisition
policies.  Both sets preserve the same requirements and start from the same
state.  Use one cumulative cost function $C_n$ at horizon $n$, including
acquisition and all continuing costs.  Set inclusion gives
\[
 \inf_{p\in\mathcal P_{\rm all}}C_n(p)
 \le \inf_{p\in\mathcal P_{\rm stay}}C_n(p).
\]
The generic order result requires the smaller set to be nonempty and the
larger cost set to be bounded below.  Nonnegative costs satisfy the latter.
Allowing acquisition retains every staying policy, so it cannot worsen the
best available cost.\leanmeta{\LHrng{OPT}{13}{14}}

\paragraph{Optimum versus realized choice.}
Fix the requirements and a cost $\PaperNotationLink{PaperClasses}{C}$ per design. Let $\mathcal D'\subseteq
\mathcal D$, with chosen designs $d\in\mathcal D$ and $d'\in\mathcal D'$.
If $d$ minimizes $\PaperNotationLink{PaperClasses}{C}$ on $\mathcal D$, then $\PaperNotationLink{PaperClasses}{C}(d)\le \PaperNotationLink{PaperClasses}{C}(d')$, since
$d'$ remains feasible in $\mathcal D$. Thus $\PaperNotationLink{PaperClasses}{C}(d')<\PaperNotationLink{PaperClasses}{C}(d)$ implies that
$d$ was suboptimal. This applies unchanged when a design is a joint choice
across projects and $\PaperNotationLink{PaperClasses}{C}$ is a well-defined aggregate cost. A burden reduction
under these fixed requirements and costs therefore needs evidence about
design selection; it cannot follow from a lower restricted optimum.
\leanmeta{\LHrng{OPT}{17}{19}}

For the strict result, use the acquisition model's nonnegative integer cost
units.  Suppose each staying policy $s$ implements the staged history and its
execution or validation protocol incurs
$C_n(s)\ge \PaperNotationLink{PaperCommonCost}{K_c(n)}+qB_{\mathrm{rep},s}(n)$ after batching, with the common baseline $\PaperNotationLink{PaperCommonCost}{K_c(n)}$
and $q>0$.  Suppose a feasible acquisition policy $a$ has
$C_n(a)\le \PaperNotationLink{PaperCommonCost}{K_c(n)}+\PaperNotationLink{PaperAdoption}{K_a}+\PaperNotationLink{PaperAdoptionUpkeep}{E_a(n)}$, where $\PaperNotationLink{PaperAdoptionUpkeep}{E_a(n)}\le r_an$ and $\PaperNotationLink{PaperAdoptionRate}{r_a}<q$.
The staged theorem gives $B_{\mathrm{rep},s}(n)\ge n$ for every $s$.
For every $n\ge\lfloor \PaperNotationLink{PaperAdoption}{K_a}/(q-\PaperNotationLink{PaperAdoptionRate}{r_a})\rfloor+1$,
\[
 \inf_{p\in\mathcal P_{\rm all}}C_n(p)
 \le C_n(a)\le \PaperNotationLink{PaperCommonCost}{K_c(n)}+\PaperNotationLink{PaperAdoption}{K_a}+\PaperNotationLink{PaperAdoptionUpkeep}{E_a(n)}
 <\PaperNotationLink{PaperCommonCost}{K_c(n)}+qn
 \le\inf_{s\in\mathcal P_{\rm stay}}C_n(s).
\]
Thus the best cost strictly improves, and no staying policy minimizes cost
over the enlarged set.  The proof uses a uniform lower bound over all staying
policies; it does not assume that the larger set has a minimizing policy.
The comparison can therefore exclude permanently keeping one parent without finding
the globally best acquisition policy.\leanmeta{\LHrng{OPT}{15}{16}}

\subsection{Tools must supply every missing implementation--class pair}\label{supp:tooling-completion}
Let $\PaperNotationLink{PaperAncestry}{H}\subseteq \PaperNotationLink{PaperRequired}{R^{\mathrm{req}}}$ be the judgments supplied by a sound native mechanism,
and let a tool supply $T\subseteq \PaperNotationLink{PaperRequired}{R^{\mathrm{req}}}$ such that $\PaperNotationLink{PaperAncestry}{H}\cup T=\PaperNotationLink{PaperRequired}{R^{\mathrm{req}}}$.  Exact supply
forces
\[
  T\setminus \PaperNotationLink{PaperAncestry}{H}=\PaperNotationLink{PaperRequired}{R^{\mathrm{req}}}\setminus \PaperNotationLink{PaperAncestry}{H}.
\]
For each required pair the native mechanism misses, the tool must determine how that
class receives that implementation. A
registry, adapter, generator, macro system, code-generation pipeline, or
language server can compress its syntax, but its rule still determines every
missing pair. For one-parent ancestry, minimizing over all native trees and
all such tools gives exactly $\PaperNotationLink{PaperGap}{d_\downarrow}$; multiple-parent ancestry can make
the deficit empty.  A checker is the dual case: to enforce exact supply it must
reject every incomplete reachable map, so its acceptance rule owns the same
pairs. These statements are independent of the tool's implementation
language or presentation and are mechanized by the native-deficit
injection and bundled-topology theorems.
\leanmeta{\LHrng{SCT}{3}{7}, \LHrng{SCT}{20}{26}, \LHrng{BWR}{45}{49}}

\paragraph{Assignment expressions separate local rewrites from whole-program repairs.}
\label{supp:walrus-boundary}
In February 2018, Stephan Houben answered a request for comprehension bindings
with \texttt{for y in [expr]}: ``No new syntax needed''
~\cite{pythonideas2018temporary}. PEP~572 documents recomputation and control
flow restructured into statements~\cite{pep572}. Learning assignment expressions adds
adoption cost ($\PaperNotationLink{PaperAdoption}{K_a}$, Section~\ref{supp:acquisition-policy-optimality}).
Banning them leaves the required behavior to be implemented another way. Van Rossum accepted the PEP, then withdrew
from decision-making, citing the fight~\cite{vanrossum2018transfer}.
Python~3.9 later optimized the workaround to assignment speed, without exporting
its variable to the enclosing scope~\cite{python39temporarybinding}.
Optimizing a workaround does not eliminate the obligation it supplies.

At the pinned CPython revision~\cite{CPythonLib5c7445b}, the census finds
\CPythonAssignmentExpressionCount{} assignment expressions in
\CPythonAssignmentExpressionFiles{} non-test files. The artifact's
\path{language_capacity_census/count_cpython_assignment_expressions.py}
counts AST \texttt{NamedExpr} nodes in Python files under \texttt{Lib/}, excluding
directories named \texttt{test}, \texttt{tests}, and \texttt{idle\_test}; it retains
the implementation of \texttt{unittest}. Its locked revision, source hashes,
included and excluded file lists, and source locations make the count replayable.
Every included file must parse; unsupported grammar aborts the scan so every
reported file contributes checked syntax nodes. The recorded replay used a parser built from that same
revision. The count locates installed use; optimality follows from comparing
preserving implementations under the chosen cost.

Python's \texttt{:=} supplies a binding inside an expression and, inside a
comprehension, can export it to the containing function
~\cite{pep572}. Fix one Python version's semantics and remove only assignment
expressions to compare the two syntax sets. Comparing Python 3.7 directly with
3.8 would also change dictionary-comprehension evaluation order. Under one
cost function and specification, adding syntax preserves every old realization,
so its attained minimum cannot increase. Strict improvement needs a witness
on that same cost coordinate. \leanmeta{\LH{WAL5}}

\emph{Reuse alone does not force duplication.}
A local binding can be encoded as
\texttt{(lambda y: body)(f(x))}; Felleisen's call-by-value \emph{let}
translation already permits this~\cite{felleisen1991expressive}.
A comprehension can instead use \texttt{for y in [f(x)]}.
Python 3.9 optimizes that idiom to assignment speed while retaining its
non-escaping scope~\cite{python39temporarybinding}.
Both have one authoritative occurrence of \texttt{f(x)}. One lambda can bind
several values; an existing iteration binder may already supply the required
reuse. Even \texttt{0 < f(x) < 10} uses one evaluation in two comparisons.
Consequently, neither a duplicated occurrence per consumer nor an added
construct per reused value is a universal lower bound.
Counts of syntax constructs are separate from counts of changed required facts
and from the ancestry gap. The lexical model checks both binder encodings.
\leanmeta{\LHrng{WAL}{3}{4}}

\emph{Lazy branch selection also admits a shared workaround.}
PEP 572's \texttt{copy.py} example and Dropbox observations contrast a flat
\texttt{if}/\texttt{elif} chain with eager evaluation or nested statements
~\cite{pep572}. A loop over the candidate functions can preserve first-success
order, bind each tested value once, and stop immediately, using one shared
loop for any number of tests. Where the enclosing function already returns
the selected result, assignment statements followed by \texttt{if}/\texttt{return}
also keep the chain flat. Neither needs one helper per branch. Re-indentation
changes source layout while preserving each required answer, so its cost belongs to the encoding.
A chain-preserving bound must declare its representation rule and structural
cost. The binding footprint supplies the local obstruction below.

\emph{An escaping binding has a genuine local obstruction.}
In the first-order lexical core, primitive operators preserve enclosing
bindings. Lambda parameters occupy a private function store, and
comprehension targets occupy a private iteration store. Reflection, dynamic
compilation, and callbacks writing captured locals are separate capabilities.
Induction over expression syntax proves that every assignment-free expression
preserves the enclosing store, including arbitrarily nested binders.
\leanmeta{\LH{WAL1}}
Start with \texttt{x = 0} and hold the surrounding program fixed:
evaluate the expression hole, then execute \texttt{while not x: pass}.
The source hole \texttt{(x := 1)} makes the context terminate. Every
assignment-free replacement in this core leaves \texttt{x} zero, so the same
context diverges. Thus no replacement of the expression alone preserves
termination. This is a local non-eliminability result with Felleisen's
termination observation in the stated core: restructuring the surrounding program
changes the permitted rewrite and can supply the binding elsewhere.
Lean checks the invariant, the changed-binding witness, and the separation
for every replacement expression. \leanmeta{\LH{WAL2}, \LHrng{WAL}{6}{7}}

PEP 572's short-circuit witness makes the same scope distinction concrete:
\texttt{any((comment := line).startswith('\#') for line in lines)} leaves
the last examined line in the containing scope. On success it is the witness;
on failure it is the last tested line, if any. A loop with \texttt{break},
or an initialized mutable cell populated inside the generator, preserves this
behavior by changing the enclosing representation. Ordinary lambda parameters
and iteration targets alone do not export that binding. Full Python also has
preexisting \texttt{nonlocal} writers, and Python 3.13 exposes write-through
frame locals~\cite{pep667}; the non-interference invariant does not cover those
operations. The primitive contract must list them because they can change
the enclosing store and therefore the non-interference proof's hypotheses.

The artifact's assignment-expression regression suite checks single evaluation,
lazy order, successful and unsuccessful searches, empty inputs, batched
binders, and these counterexamples. It executes on the installed Python;
grammar parsing checks accepted syntax, while interpreter execution checks its behavior.
\texttt{Paper1PL/AssignmentExpressions.lean} proves the invariant for the declared lexical calculus;
the Python regressions exercise its concrete examples. This example separates preservation of an expression's
lexical effects from whole-program computability, while retaining the paper's
distinction between semantic-answer counts and representation-specific work.

Figure~\ref{fig:storage-encodings} compares native derivation with explicit routes.
\begin{figure}[htbp]
\centering
\PaperDenseTableSetup
\providecommand{\PaperCodeContinuation}[1]{}
\begin{tabularx}{\linewidth}{@{}
  >{\hsize=.50\hsize\linewidth=\hsize}X
  >{\hsize=1.20\hsize\linewidth=\hsize}X
  >{\hsize=1.40\hsize\linewidth=\hsize}X
  >{\hsize=.90\hsize\linewidth=\hsize}X@{}}
\begin{tabular}[t]{@{}l@{}}
\textrm{\bfseries monolithic} \\
class MonoA: \\
\quad attr1 = 1 \\
\quad attr2 = 2
\end{tabular}
&
\begin{tabular}[t]{@{}l@{}}
\textrm{\bfseries multi-parent} \\
class MixA: attr1 = 1 \\
class MixB: attr2 = 2 \\
class NativeA(\PaperCodeContinuation{\quad}MixA, MixB):\PaperCodeContinuation{\quad} pass
\end{tabular}
&
\begin{tabular}[t]{@{}l@{}}
\textrm{\bfseries fields + forwarding} \\
class RoutedA: \\
\quad def \_\_init\_\_(s): \\
\quad\quad s.\_a=MixA() \\
\quad\quad s.\_b=MixB() \\
\quad attr1=property(\PaperCodeContinuation{\quad\quad}lambda s:s.\_a.attr1) \\
\quad attr2=property(\PaperCodeContinuation{\quad\quad}lambda s:s.\_b.attr2)
\end{tabular}
&
\begin{tabular}[t]{@{}l@{}}
\textrm{\bfseries fields + assignment} \\
class AssignedA: \\
\quad def \_\_init\_\_(s): \\
\quad\quad s.\_a=MixA() \\
\quad\quad s.\_b=MixB() \\
\quad\quad s.attr1=s.\_a.attr1 \\
\quad\quad s.attr2=s.\_b.attr2
\end{tabular}
\end{tabularx}
\caption{Same shown outputs, different facts for code to maintain as providers evolve.
Inheritance derives both pairs; fields need routes. The monolith
loses \texttt{MixA} and \texttt{MixB}'s required independent identities.}
\Description{Four short Python encodings place the same two snapshot
attributes in one class, two inherited mixins, two contained provider objects
with property forwarding, or two contained provider objects with constructor
assignments.}
\label{fig:storage-encodings}
\end{figure}

\paragraph{Invalid-state necessity.}
For state $t$ and design $D$, behavioral safety requires
$\PaperNotationLink{PaperSafety}{\operatorname{Reach}_t(D)\subseteq\operatorname{Valid}_t}$.
When the boundary is finite, define
$I_t(D)=|\operatorname{Reach}_t(D)\setminus\operatorname{Valid}_t|$;
then $I_t(D)=0$ is equivalent to behavioral safety.
The safety condition needs no finiteness assumption. For
any positive real denominator $m$, $I_t(D)/m=0$ iff $I_t(D)=0$, so
normalization adds no information at the target.  It can only dilute a
positive count by adding unrelated valid behavior, exactly the denominator
failure isolated by the main paper's gap-ratio corollary.  Correct maintenance
requires this safety condition at every change, preservation of required answers, and attainment of the outside-answer and repair minima. Security against a
declared invalid behavior requires the same necessary condition: that behavior
cannot remain reachable.  The Hamming and physical-edit results then price the
change from an invalidated old representation to a new behaviorally safe one.
\leanmeta{\LHrng{BWR}{38}{39}, \LH{BWR57}, \LHrng{BWR}{54}{56}}

More formally, let $\mathcal E(D)$ be any allowed family of physical encodings
of a required external relation $D$, and let $a(e)\ge0$ price the work of
selecting, coordinating, and policing encoding $e$.  If each displaced answer
has positive marginal cost at least $q$, every one-parent design preserving required supply obeys
\[
  q|D|+a(e)\ \ge\ q|D|\ \ge\ qd_\downarrow.
\]
Choosing forwarding rather than a registry, adapter, or generator can change
$a(e)$ and the presentation of $D$; it cannot turn the forced relation or the
coordination around it into a benefit of restricting ancestry. Every encoding
supplies the same missing pairs and adds its own nonnegative coordination cost.
\leanmeta{\LHrng{OPT}{3}{5}}

For an evolving hierarchy, let $\rho_R$ count retracted old direct-parent
edges.  If this coordinate is priced separately, the total ledger adds its
declared marginal cost.  The staged immediate-parent family has $d_t=0$ at
every state and multi-parent point $(\rho_M,D_M)=(0,\varnothing)$.  An
ancestry-only one-parent policy has $\rho_S\ge n$; a policy preserving required supply and freezing
its initially sound parent function has $|D_n|\ge n(n+1)/2$; every mixed policy satisfies the per-transition
retraction-or-external-pair dichotomy.  Summing over $n$ transitions gives at
least $n$ charged occurrences.  At positive marginal cost $q$, a fixed
one-time saving $K$ can offset them only while $K\ge qn$; some finite horizon
violates that inequality.
\subsection{Time-weighted acquisition costs}
\label{time-weighted-acquisition}
More generally, let $K_0<\infty$ be a one-time prevention cost, let
$k_t\ge0$ be recurring intervention-side cost, and let $\delta_t\ge0$ weight
the $t$th avoided charge $q_t>0$.  Established intertemporal accounting chooses
the intervention exactly when
\[
  K_0 + \sum_{t=1}^{T}\delta_t k_t
  < \sum_{t=1}^{T}\delta_t q_t.
\]
The paper's cumulative semantic-work ledger is the undiscounted case
$\delta_t=1$.  A discount schedule can change monetary present value but not
the required edit count. Migration and initial training enter $K_0$~\cite{becker1962human};
turnover, hiring premiums, review, and tool upkeep enter $k_t$. Prevention removes
the compared repair stream; any remaining charges stay in the ledger.
\leanmeta{\LH{IPE23}, \LHrng{OPT}{3}{6}}

Within the atomic-role model, same-role authority exchange preserves the scope
family.  Changing a scope changes $\PaperNotationLink{PaperRequired}{R^{\mathrm{req}}}$; omitting a required pair or admitting
an unwanted one breaks exact supply.  Separate copies become distinct authorities.
A policy that admits separate edits admits divergence, while a common generator
or propagation checker carries the authority relation externally.  The final
case stores the dependency relation outside ancestry, where the exact cost
theorem applies.

\subsection{Exact two-role and row--column gaps}
\label{two-role-row-column-gaps}
For two required class sets $A,B$, partition the consumer classes into the left-only,
right-only, shared, and outside regions, of sizes $N_{10},N_{01},N_{11},N_{00}$.
Two retained rows are laminar only if at least one of the first three regions
is emptied.  Deleting the left-only region makes the left row a subset of the
right; deleting the right-only region gives the reverse containment; deleting
the shared region leaves no overlap. Therefore
\[
 \PaperNotationLink{PaperGap}{d_\downarrow}(A,B)=\min\{N_{10},N_{01},N_{11}\}.
\]
Lean checks both the lower bound over every repair and all three attaining
constructions.
\leanmeta{\LHrng{AVG}{1}{3}}

Lean also checks the exact zero-gap probability and its asymptotic bound. For arbitrary
cell probabilities $q_{00},q_{10},q_{01},q_{11}$, zero gap means that at least one
of cells $10,01,11$ is absent. Inclusion--exclusion therefore gives
\[
 \Pr[\PaperNotationLink{PaperGap}{d_\downarrow}=0]=\sum_r(1-q_r)^n
 -\sum_{r<s}(1-q_r-q_s)^n+q_{00}^n,
 \qquad r,s\in\{10,01,11\}.
\]
For fixed $q_{\min}=\min\{q_{10},q_{01},q_{11}\}>0$, absence of a least-probable
cell gives the lower bound $(1-q_{\min})^n$; the union bound gives three times
that quantity as an upper bound. Thus the probability is
$\Theta((1-q_{\min})^n)$.
\leanmeta{\LHrng{AGP}{1}{7}}

Suppose next that membership in each of the two scopes is independent across
providers and across $n$ consumers, with probability $p\in(0,1)$.  Put
$q=p(1-p)$, $r=p^2$, and $s=(1-p)^2$.  Then
$(N_{10},N_{01},N_{11},N_{00})$ is multinomial with cell probabilities
$(q,q,r,s)$, so the finite expected gap is
\[
 \mathbb E[\PaperNotationLink{PaperGap}{d_\downarrow}]
 =\!\sum_{\substack{a,b,c\ge0\\a+b+c\le n}}
 \min\{a,b,c\}
 \frac{n!}{a!b!c!(n-a-b-c)!}
 q^{a+b}r^c s^{n-a-b-c}.
\]
Inclusion--exclusion gives the exact crossing probability
\[
\begin{split}
 \Pr[\PaperNotationLink{PaperGap}{d_\downarrow}>0]={}&1-2(1-q)^n-(1-p^2)^n
 +(p^2+(1-p)^2)^n\\
 &+2(1-p)^n-(1-p)^{2n}.
\end{split}
\]
Every subtracted base is strictly below one, hence this probability tends to
one.  The strong law applied to the three region counts gives
$\PaperNotationLink{PaperGap}{d_\downarrow}/n\to p\min\{p,1-p\}$ almost surely; bounded convergence gives
the same limit in expectation.  The relation itself has expected size $2np$.
At $p=1/2$, write $D_i=N_i-n/4$ for the three relevant cells.  Since
$\operatorname{Var}(D_i)=3n/16$, Cauchy--Schwarz and
$\min_iD_i\ge-\sum_i|D_i|$ give
\[
 \mathbb E[\PaperNotationLink{PaperGap}{d_\downarrow}]\ge n/4-(3\sqrt3/4)\sqrt n.
\]
For $X=N_{10}-N_{01}$, $\mathbb E[X^2]=n/2$ and
$\mathbb E[X^4]=(3n^2-n)/4$.  H\"older's inequality gives
$\mathbb E|X|\ge\sqrt{n/6}$, while
$\min(N_{10},N_{01})=(N_{10}+N_{01}-|X|)/2$; hence
\[
 \mathbb E[\PaperNotationLink{PaperGap}{d_\downarrow}]\le n/4-\sqrt n/(2\sqrt6).
\]
Therefore the one-parent cost is $n/4-\Theta(\sqrt n)$.  The multinomial
central limit theorem sharpens this to
$n/4-(3/(4\sqrt\pi))\sqrt n+o(\sqrt n)$.
Thus the three implementation-reachability costs are
$n$, $n/4-\Theta(\sqrt n)$, and zero for no implementation inheritance, one parent, and
multiple parents respectively. The declared source distribution determines
these rates; a codebase population would need its own measured distribution.

\paragraph{Arbitrary skew and changing population size.}
The balanced law is one point in an exact four-cell family.  Let
$(q_{00},q_{10},q_{01},q_{11})$ be any declared distribution over neither role,
left only, right only, and both roles.  The same pointwise theorem gives
\[
 \PaperNotationLink{PaperGap}{d_\downarrow}=\min\{N_{10},N_{01},N_{11}\},
 \qquad
 \frac{\PaperNotationLink{PaperGap}{d_\downarrow}}{n}\longrightarrow
 \min\{q_{10},q_{01},q_{11}\}
\]
almost surely and in expectation.  Inclusion--exclusion gives
\[
\begin{split}
 \Pr[\PaperNotationLink{PaperGap}{d_\downarrow}>0]={}&1-(1-q_{10})^n-(1-q_{01})^n-(1-q_{11})^n\\
 &+(q_{00}+q_{10})^n+(q_{00}+q_{01})^n
 +(q_{00}+q_{11})^n-q_{00}^n.
\end{split}
\]
Positive-support skew changes only the waiting time and the linear rate; it
removes neither crossing risk nor asymptotic certainty.  The gap is identically zero only under the additional closed-world
premise that one of $q_{10},q_{01},q_{11}$ is exactly zero forever.  That is a
tree-shaped support limitation: it excludes a crossing cell altogether. No finite history
of non-occurrence proves zero probability, and the main paper's open-world
theorem rules out an algorithm recognizing exactly the safe arbitrary future
generators.\leanmeta{\LHrng{UND}{31}{38}} With
positive mass in all three crossing cells, the crossing probability tends to
one.
Define
\[
 n_\alpha=\min\{n:\Pr[\PaperNotationLink{PaperGap}{d_\downarrow}>0]\ge\alpha\}.
\]
Table~\ref{tab:skewed-crossing-thresholds} gives the exact population at which
each skewed source reaches 50, 90, and 99 percent crossing probability;
Table~\ref{tab:skewed-gap-severity} gives exact finite severity.  The checked
script evaluates the multinomial sum directly rather than sampling it.  Finite
expectation and limiting rate are computed separately; in the skewed rows the
rare crossing cell is almost always the minimum, so their rounded values at
$n=256$ nearly equal $256$ times the limiting rate.  The
final source removes only $q_{11}$, transfers its mass to $q_{10}$, and
solves for a 5 percent crossing probability at $n=256$. This reference population
matches the \SQLAlchemyComponentAffectedConsumers{} classes touched by
SQLAlchemy's certified optimum. That match fixes population size: the component contains
\SQLAlchemyComponentProviderFacts{} providers and
\SQLAlchemyComponentProviderIncidences{} incidences, whereas the synthetic law
isolates two roles.

This source's 99\% threshold can also be reconstructed without rounded cell
probabilities. Keeping the rare-cell term gives
$(1-q_{11})^{256}\approx0.95$, hence
\[
 n_{.99}\approx\left\lceil
 \frac{256\ln(0.01)}{\ln(0.95)}\right\rceil
 =\lceil22{,}983.97055\ldots\rceil=22{,}984.
\]
The full inclusion--exclusion calculation verifies the same integer:
the crossing probability is below $0.99$ at $22{,}983$ and above it at $22{,}984$.

The audited SQLAlchemy and OpenAI wheels parse
\PythonSQLAlchemyParsedClasses{} and \PythonOpenAIParsedClasses{} classes, so
256 is not a program-size ceiling.  Total parsed classes are not, however, the
consumer population of one fixed two-role law.  The twenty projects have
different role counts and assignments, so the audit does not infer synthetic
$n$ or $\PaperNotationLink{PaperGap}{d_\downarrow}$ from total class count.

\begin{table}[t]
\centering
\PaperDenseTableSetup
\begin{tabular}{@{}lrrrrrrrr@{}}
\toprule
Source & $q_{00}$ & $q_{10}$ & $q_{01}$ & $q_{11}$ & $n_{.5}$ & $n_{.9}$ & $n_{.99}$ & $\lim d_\downarrow/n$ \\
\midrule
\textcolor{blue!75!black}{balanced} & 0.250 & 0.250 & 0.250 & 0.250 & 6 & 12 & 20 & 0.250 \\
\midrule
\textcolor{orange!90!black}{moderate skew} & 0.250 & 0.500 & 0.150 & 0.100 & 10 & 24 & 45 & 0.100 \\
\midrule
\textcolor{magenta!75!black}{95\% cell suppression} & 0.250 & 0.487 & 0.250 & 0.013 & 56 & 184 & 367 & 0.013 \\
\midrule
\textcolor{violet!80!black}{extreme skew} & 0.250 & 0.740 & 0.009 & 0.001 & 695 & 2302 & 4603 & 0.001 \\
\midrule
\textcolor{green!50!black}{5\% at $n=256$} & 0.250 & 0.500 & 0.250 & .0002003444 & 3460 & 11492 & 22984 & .0002003444 \\
\bottomrule
\end{tabular}
\caption{Exact waiting thresholds under skew.  The four cells are neither, left only, right only, and both.  $n_{\alpha}$ is the first population size with $\Pr[d_\downarrow>0]\ge\alpha$.}
\label{tab:skewed-crossing-thresholds}
\end{table}

\begin{figure}[t]
\centering
\begin{tikzpicture}
\begin{axis}[
  width=0.82\textwidth,
  height=5.2cm,
  xmode=log,
  log basis x=2,
  xmin=1, xmax=280,
  ymin=0, ymax=1.02,
  xtick={1,2,4,8,16,32,64,128,256},
  xlabel={current consumers $n$},
  ylabel={$\Pr[d_\downarrow>0]$},
  grid=major,
  tick label style={font=\small},
  label style={font=\small},
]
\addplot+[blue!75!black,mark=*,thick] coordinates {(1,0.0000000000) (2,0.0000000000) (4,0.2343750000) (8,0.7113647461) (16,0.9699779889) (32,0.9996986434) (64,0.9999999697) (128,1.0000000000) (256,1.0000000000) (512,1.0000000000) (1024,1.0000000000)};
\addplot+[orange!90!black,mark=*,thick] coordinates {(1,0.0000000000) (2,0.0000000000) (4,0.1125000000) (8,0.3941142187) (16,0.7504547118) (32,0.9602503902) (64,0.9987905960) (128,0.9999986090) (256,1.0000000000) (512,1.0000000000) (1024,1.0000000000)};
\addplot+[magenta!75!black,mark=*,thick] coordinates {(1,0.0000000000) (2,0.0000000000) (4,0.0228515625) (8,0.0822916778) (16,0.1799299894) (32,0.3313259360) (64,0.5529308715) (128,0.8001292003) (256,0.9600516634) (512,0.9984041304) (1024,0.9999974532)};
\addplot+[violet!80!black,mark=*,thick] coordinates {(1,0.0000000000) (2,0.0000000000) (4,0.0000999000) (8,0.0004889796) (16,0.0020148820) (32,0.0077045918) (64,0.0269430440) (128,0.0820942773) (256,0.2034497306) (512,0.3969158178) (1024,0.6409670682)};
\addplot+[green!50!black,mark=*,thick] coordinates {(1,0.0000000000) (2,0.0000000000) (4,0.0003754952) (8,0.0013754484) (16,0.0031578486) (32,0.0063902958) (64,0.0127414549) (128,0.0253205655) (256,0.0500000000) (512,0.0975000000) (1024,0.1854937500)};
\node[anchor=south east,font=\scriptsize,text=green!50!black] at (axis cs:256,0.05) {$q_{11}=0.020\%$};
\end{axis}
\end{tikzpicture}
\caption{At the audited component size $n=256$, the green source reaches only 5\% because one crossing cell is suppressed by 99.92\% (1,248$\times$ below balanced).  With positive support in all three crossing cells, a crossing eventually appears with probability approaching one; skew changes its arrival time.}
\Description{Five exact crossing-probability curves through 256 consumers. Balanced combinations cross fastest. The green source reaches five percent at 256 with one crossing combination about 1,248 times rarer than balanced. Every curve rises because every crossing combination remains possible.}
\label{fig:skewed-crossing-probability}
\end{figure}

\begin{table}[t]
\centering
\PaperDenseTableSetup
\begin{tabular}{@{}lrrrrrr@{}}
\toprule
Source & $\mathbb E d_{64}$ & $\Pr[d_{64}\ge4]$ & $\mathbb E d_{256}$ & $\Pr[d_{256}\ge4]$ & $\Pr[d_{256}\ge16]$ & $\mathbb E|R^{\mathrm{req}}_{256}|$ \\
\midrule
\textcolor{blue!75!black}{balanced} & 12.668 & 0.9999 & 57.279 & $> .9999$ & $> .9999$ & 256.0 \\
\midrule
\textcolor{orange!90!black}{moderate skew} & 5.933 & 0.8850 & 25.421 & $> .9999$ & 0.9869 & 217.6 \\
\midrule
\textcolor{magenta!75!black}{95\% cell suppression} & 0.800 & 0.0086 & 3.200 & 0.3978 & $< .0001$ & 195.2 \\
\midrule
\textcolor{violet!80!black}{extreme skew} & 0.027 & $< .0001$ & 0.223 & $< .0001$ & $< .0001$ & 192.3 \\
\midrule
\textcolor{green!50!black}{5\% at $n=256$} & 0.013 & $< .0001$ & 0.051 & $< .0001$ & $< .0001$ & 192.1 \\
\bottomrule
\end{tabular}
\caption{Exact gap severity under the same skewed sources.  Skew changes the waiting time and slope while each displaced pair remains one outside semantic answer.}
\label{tab:skewed-gap-severity}
\end{table}

The same law covers shrinking populations.  Suppose additions draw a combination from $q$,
deletions choose one extant consumer uniformly, and the add/delete choice does
not inspect its combination.  Conditional on the current population $N_t=n$,
exchangeability gives the same multinomial law at every time $t$, regardless
of the preceding additions and deletions.  This follows by induction: an
independent addition appends one $q$-draw, while uniform deletion from an
i.i.d. sample leaves an i.i.d. sample.  Figure~\ref{fig:skewed-crossing-probability}
therefore reads in either direction as $N_t$ rises or falls.

Elapsed time alone cannot determine exposure when population can shrink: a
process can hold $N_t$ small by deleting.  The exact path quantities are the
current $N_t$ and the accumulated semantic variation
\[
 V_T=\sum_{t=1}^{T}|\PaperNotationLink{PaperRequired}{R_t^{\mathrm{req}}}\mathbin{\triangle}R_{t-1}^{\mathrm{req}}|.
\]
Every required pair inserted or deleted changes $V_T$ by one semantic-answer
unit.  Under the exchangeable process above,
\[
 \mathbb E[V_T]=T(q_{10}+q_{01}+2q_{11}).
\]
For arrival-only growth, $N_t=t$ and the $n_\alpha$ values are also exact
first-crossing quantiles.  Deletions can postpone the crossing only by changing
$\PaperNotationLink{PaperRequired}{R^{\mathrm{req}}}$; $V_T$ counts those deletions and does not erase earlier variation.

The row--column family removes the distributional premise.  On consumers
$[h]\times[h]$, give each cell its row role and column role.  Let $a$ retained
row scopes and $b$ retained column scopes have cardinality at least two.
Every other scope omits at least $h-1$ incidences.  A large row and large
column cannot both retain their common cell: their intersection contains only
that cell, so laminarity would otherwise force a set of size at least two into
a singleton intersection.  The $ab$ large row--column pairs therefore force
at least $ab$ further omissions.  Consequently every one-parent repair has
\[
 \PaperNotationLink{PaperGap}{d_\downarrow}\ge
 (h-a)(h-1)+(h-b)(h-1)+ab\ge h(h-1).
\]
Keeping every row and one diagonal cell from each column attains the bound.
The exact costs are therefore $2h^2$, $h(h-1)$, and zero across the same three
mechanisms.  Lean checks the cross-intersection lemma, the double-counting
bound, and the attaining construction.
\leanmeta{\LHrng{AVG}{4}{7}}

\paragraph{Fixed-description factoring statement.}
In a fixed description model, successful factoring produces a shorter program preserving the model's specified answers
after paying for the abstraction, so its expanded form exceeds that
model's minimum~\cite{LiVitanyi2008}.  The statement depends on the chosen
description language; a helper, generator, table, interpreter, or inherited
implementation can hold the shared factor.
\leanmeta{\LH{BWR13}, \LHrng{CRS}{3}{4}}

\paragraph{Exact expansion preserves the valid domain.}
An effective exact factorization supplies a factored authority, a deterministic
expansion function, soundness of every generated explicit state, and exact
reconstruction of every valid explicit state.  Lean proves that the expansion
range is exactly the validity set.  Relative to unrestricted manual expansion,
the additional write-time representation states are therefore precisely the
invalid ones.  Here a representation state may be a source tree, generated
file set, registry, or configuration.  This claim concerns coherence before
execution; operational runtime validity is a separate predicate.  For an
arbitrary indexed family of duplicated answers, Lean also constructs the
one-answer factorization and proves that its generated range is exactly the
states in which every copy agrees.  Unrestricted manual copies add precisely
the states containing a disagreement.
\leanmeta{\LHrng{BWR}{58}{62}}

\section{Linear Field-Menu Bounds}
\label{linear-field-menu-bounds}

The field bounds count the binary observations needed to expose every future
identity conflict while each supported deletion merges at most two meanings.
The calculations below give
matching asymptotic bounds within the binary-linear model. They are paper
proofs; the next section gives an explicit mechanized construction.
Let the entities be $\mathbb F_2^k$, and sample $r$
nonzero linear functionals.  A direction $v\ne0$ is safe when the sampled
functionals in $v^\perp$ span that hyperplane.  Failure means missing
$\PaperNotationLink{PaperAncestry}{H}\setminus \PaperNotationLink{PaperAncestry}{H}'$ for a codimension-one subspace $\PaperNotationLink{PaperAncestry}{H}'$ of a hyperplane $\PaperNotationLink{PaperAncestry}{H}$.
One uniform nonzero draw hits that set with probability greater than $1/4$,
and there are fewer than $2^{2k}$ pairs $(\PaperNotationLink{PaperAncestry}{H}',\PaperNotationLink{PaperAncestry}{H})$.  Hence
\[
  \Pr[\text{some branch fails}]<2^{2k}(3/4)^r<1
  \qquad\text{when}\qquad
  r>\frac{2k}{\log_2(4/3)}\approx4.82k.
\]
Deduplicating the draws yields an existence bound of at most
$\lceil5k\rceil$ fields.  This paper contributes the deprecation-menu
interpretation of that finite probabilistic construction.

Within the binary-linear class, every retained set must contain at least
$k-1$ functionals in each hyperplane $v^\perp$, while each nonzero functional
belongs to exactly $2^{k-1}-1$ such hyperplanes.  Double counting gives
\[
  m(2^{k-1}-1)\ge(2^k-1)(k-1),
\]
so $m>2(k-1)=2\log_2N-2$.  These asymptotic calculations remain at manuscript
level.  The numbered coordinate-and-parity construction below is explicit.
\leanmeta{\LHrng{FLD}{1}{8}}

\section{Explicit Coordinate-and-Parity Menu}
\label{explicit-coordinate-parity-menu}

The construction makes every pair of meanings look alike in some supported
deletion, while no individual deletion merges more than two. It uses entities
$x\in\mathbb F_2^k$, the $k$ coordinate functionals $x_i$, and the
$\binom{k}{2}$ pair parities $x_i+x_j$.  The complete translation is:

\begin{center}
\begin{tabularx}{\linewidth}{@{}XX@{}}
\toprule
Evolving-observation object & Finite-geometry object \\
\midrule
entity & vector $x\in\mathbb F_2^k$ \\
\midrule
optional binary field & nonzero linear functional $u$ \\
\midrule
deprecation direction & vector $v\ne0$ \\
\midrule
merged pairs & matching $x\leftrightarrow x+v$ \\
\midrule
fields retained by that branch & master-menu points in $v^\perp$ \\
\midrule
multiplicity at most two & retained points span $v^\perp$ \\
\midrule
complete future conflict & every hyperplane is spanned \\
\midrule
minimum binary-linear field count & size of a strong blocking set \\
\bottomrule
\end{tabularx}
\end{center}

For each nonzero direction $v$, retain coordinate $i$ when $v_i=0$ and
every parity $ij$ when $v_i=v_j=1$; all other fields are deprecated once and
never return.  The retained tests lie in and span $v^\perp$.  To see spanning,
choose a pivot $p$ with $v_p=1$.  Equal branch profiles agree on
every zero-direction coordinate and on every parity $x_i+x_p$ for which
$v_i=1$.  Once $x_p$ is fixed, these observations determine every coordinate,
so each fiber has size at most two.  Conversely, $x$ and $y$ have the same
branch profile when $v=x-y\ne0$: they agree on every zero position, and
flipping both endpoints preserves every retained parity.  Thus branch $x-y$
merges every distinct pair $x,y$, making the union conflict graph complete.
The coordinate fields make the initial profile discrete, and their count is
$k+\binom{k}{2}$.

For $k=3$, the construction uses six fields on eight entities.  Each of the
seven nonzero directions pairs $x$ with $x+v$, while every branch retains
fibers of size two.  The optimized resource is the number of binary
observations: the menu still has $N-1$ direction-indexed branches and uses
correlated linear tests.  Menu-description complexity, branch count, and the
naturalness of these observations in an ordinary schema remain separate.

If every branch instead retains one arbitrary finite-valued field and
$N\ge3$, each multiplicity-two conflict relation is a matching.  Covering
$K_N$ needs at least $N-1$ matchings for even $N$ and $N$ for odd $N$; a
one-factorization attains those bounds.  The exact singleton-menu field count
is therefore $N-1$ for even $N$ and $N$ for odd $N$.  Allowing a branch to
retain several correlated fields makes a sublinear menu possible.  The
explicit coordinate-and-parity construction uses $\Theta(\log^2 N)$ fields;
the preceding probabilistic construction and double-counting lower bound give
the optimal $\Theta(\log N)$ order in the binary-linear model.

\section{Known Schema Families and Lattice Closure}
\label{known-schema-families}

Knowing every future schema is different from knowing only how schemas may
change. The key bounds depend on which of these facts the allocator receives.
Suppose a finite catalog of entities and every admissible schema state are
known before allocation.  The robust price is the chromatic number of the
union conflict graph.  If the schema family is known but entities arrive in
an unknown order, and the allocator can compute each arriving entity's future
conflicts with earlier entities, the future-backdegree theorem gives the
greedy bound for that order.  The optimal guarantee over admissible orders is
the corresponding online chromatic number.  If even the future incidence of
unseen entities is unknown, the adaptive lower bound applies.  Thus ``known
schema'' must distinguish a labeled future universe from a transition grammar.

For one state, a reusable tag class is independent in the partition matroid
whose parts are profile fibers.  For two states, common independent sets are
the matchings of a bipartite multigraph: its two sides are the two collections
of fibers, and each entity is an edge between its two fibers.  K\H{o}nig line
coloring therefore makes the exact offline price the maximum fiber size.  For
a fully declared labeled universe, precoloring also attains that price
causally.  With stream-revealed endpoints, greedy uses at most $2B-1$ colors;
the classical online edge-coloring lower bound makes this palette tight in the
low-degree regime, even for bipartite instances
~\cite{barNoy1992online}.  This
allows the intersection of the two partition matroids to violate matroid axioms;
the matchings of a three-edge path already have maximal members of sizes one
and two.  With three states the incidence object is instead a tripartite
3-uniform hypergraph.  The $k=2$ coordinate-and-parity menu gives exactly three
states with snapshot price two and union price four.

Neither chromatic parameter is controlled by transition depth alone.  A
constant-depth menu containing the discrete state and one pair-merge state for
each pair has complete future-conflict graph, although every allowed state has
multiplicity at most two.

A literal lattice of partitions behaves differently.  Order it from fine to
coarse and suppose every lattice state is admissible and satisfies the
multiplicity bound.  Its top element is a supported coarsest schema, every
possible conflict lies within one of its fibers, and $B$ tags suffice.  The
constant-depth counterexample permits many incomparable pair merges but
excludes their join because that join violates the multiplicity promise.  Its
promise-respecting state family therefore fails join closure.  This separates
schema depth from supported coarsening closure.

\section{Lookup Order, State Sharing, and Initialization}
\label{dual-axis-resolution}

\paragraph{Established inheritance semantics.}
Cook and Palsberg model a class by a generator $G$ from self to a method record,
and a subclass by a wrapper $W$ taking self and super~\cite{cook1994denotational}:
\[
  (W\mathbin{\triangleright}G)(s)=W(s)(G(s))\oplus G(s).
\]
Here $\oplus$ is left-biased record merge; the object is the least fixed point
of the combined generator. Their equivalence proof connects this model to
method lookup. Passing the same self to both generators explains the self-call
default below; our result counts departures from that default.
For supply, a single parent gives each code provider a descendant subtree, so the
class groups nest or do not overlap. The ancestry gap prices required pairs that no
such tree derives (main-text Theorem~4.6).

\paragraph{Supply and winner selection count different answers.}
Fix a required coverage relation, a conflict set of size $k$, and its winner
policy $\PaperLookup$.  Every carrier preserving the specified lookup answers realizes the same $\PaperLookupChoices$ winner answers.
Multi-parent derivation adds no outside coverage answer, so its combined
coverage--resolution cost is $k$ under any resolver preserving those answers.  Every
one-parent repair preserving the required coverage and lookup answers retains those $k$ answers and adds at least $\PaperNotationLink{PaperGap}{d_\downarrow}$
outside coverage answers, for cost at least $k+\PaperNotationLink{PaperGap}{d_\downarrow}$.  At $k=0$, no
precedence mechanism is needed: multi-parent coverage costs zero while every
positive-gap one-parent rewrite still pays $\PaperNotationLink{PaperGap}{d_\downarrow}$.  Candidate supply
does not appear in the resolver theorem below; precedence is an independent
mechanism, and MRO is its ancestry-derived instance.
\leanmeta{\LHrng{NHF}{53}{58}, \LHrng{NHF}{64}{68}, \LH{NHF10}, \LH{NHF19}}

\paragraph{One graph can couple otherwise separate relations.}
When one parent graph carries contract membership and implementation reach,
the union of their native row families must be laminar.  Restricting any joint
repair to either relation remains admissible, and the omitted-pair counts add;
hence the joint optimum is at least the sum of the independently chosen
optima.  For a strict witness, let the contract relation have the single row
$\{W,\mathrm{FR}\}$ and implementation reach the single row $\{\mathrm{FR},B\}$.  Each relation
alone has gap zero, while their shared graph has gap one.  If one
declaration itself carries required sets $A$ and $B$ for the two roles, one
sound native descendant set lies in $A\cap B$.  Retaining that intersection
attains the exact outside floor $|A\mathbin{\triangle}B|$.
\leanmeta{\LHrng{CPL}{1}{4}}

\paragraph{Self-use reverses the native winner default.}
For a fixed declared self-call site set, open recursion defaults each site to
the derived implementation; a closed wrapper defaults it to the base
implementation.  A repair preserving the specified self-call winners must materialize every site where the
required winner differs from that default, and the canonical repair
materializes exactly those sites.  The subclass therefore has ledger
$(0,|P^{-1}(\mathsf{base})|)$, while a wrapper around a provider exposing $E$
has $(|E|,|P^{-1}(\mathsf{derived})|)$.  If $E$ is nonempty, the wrapper never
dominates on both coordinates.  The subclass dominates exactly when its
base-winner count is no larger; otherwise the ledgers are incomparable.
\leanmeta{\LHrng{SLF}{1}{7}}

The comparison requires a declared self-use contract.  In Bloch's counting-set
example, a base-winner \texttt{addAll}$\to$\texttt{add} site costs the subclass
one exceptional winner answer; the wrapper avoids that answer but routes every
exposed member, so neither ledger dominates.  If \texttt{add} is instead a
declared extension callback, the derived winner is native to the subclass.
When self-use is undeclared, a base revision may add sites whose required
winners were never stated; that is the fragile-base-class boundary
~\cite{bloch2018effective,kiczales1992classlibraries,
mikhajlov1998fragile,aldrich2004selective}.

\paragraph{Coherent resolution is one precedence order.}
Let a resolver select one member of every finite nonempty provider set and
preserve its choice whenever the selected provider survives removal of rejected
candidates.  Define $a\prec b$ when the resolver selects $a$ from $\{a,b\}$.
Pairwise choice gives irreflexivity and totality.  Choice preservation gives
transitivity and places the selected member of any finite set before every
rejected member.  The resolver therefore selects the first member of one strict
total order.  If every ancestor--descendant pair selects the descendant, this
order extends reverse ancestry.
\leanmeta{\LH{NHF10}, \LH{NHF19}}

Configuration lookup can order both providers and context scopes. Searching
providers first need not select the same declaration as searching scopes first.
The criterion below says exactly when the loop orders agree.
Fix one semantic key $k$, let
$D_k\subseteq T\times S$ be the nonempty set of sites that define it, and order
provider types $T$ and scopes $S$ from most to least specific.  Write
$t^\star$ and $s^\star$ for the coordinate minima.  Provider-first and
scope-first lexicographic resolution select the same provenance site exactly
when $(t^\star,s^\star)\in D_k$.  If their choices agree, each loop order
forces one coordinate of the common site to its minimum; conversely the joint
minimum is first under both orders.
\leanmeta{\LH{NHF16}}

Changing other key columns leaves $D_k$ fixed.  With both orders and all
existing declaration sites fixed, inserting a provider that declares no keys
also leaves provider-first resolution unchanged.
\leanmeta{\LHrng{NHF}{17}{18}}
A source-language wrapper that
recomputes C3 can change the provider order and lies outside the latter claim.
For the crossed pair $\{(t_0,s_1),(t_1,s_0)\}$, the two loop orders disagree;
the absent joint minimum is exactly why axis priority becomes observable
semantic policy.

\subsection{State sharing and cooperative initialization}
\label{state-coordination}

Two routes may be required to see the same state object or distinct objects.
Constructors may be required to initialize one component before another.
Every representation must preserve the required state sharing and initialization
order; its native rules determine which requirements need additional code. Fix a class $c$.
A state access $(f,p)$ means code provider $f$'s
code accessing provider $p$'s state in an instance of $c$. A partial map
$\sigma_c^*$ assigns \emph{same} or \emph{distinct} to specified unordered
pairs of accesses. A finite relation $\prec_c^*$ specifies which state
components must initialize before which others. Preserving these sharing and ordering requirements requires the
realized sharing to agree on $\operatorname{dom}\sigma_c^*$ and the realized
schedule to contain $\prec_c^*$. Unspecified pairs are not counted.
Write $k_\sigma=\sum_c|\operatorname{dom}\sigma_c^*|$ and
$k_\prec=\sum_c|\prec_c^*|$.

\begin{theorem}[Specified state answers survive answer-preserving refactoring]
\label{state-answer-invariance}
Every representation preserving the specified state answers supplies the same $k_\sigma+k_\prec$ specified
state answers, with multiple parents, one parent, or no inheritance.
When the specified lookup answers are also preserved, the combined count is $k+k_\sigma+k_\prec$.
\leanmeta{\LHrng{SCI}{1}{3}, \LH{SCI11}, \LHrng{NHF}{53}{58}}
\end{theorem}
\begin{proof}
Answer preservation fixes each specified answer; the specification fixes the questions.
Sharing queries and initialization queries are separate types in the checked
model, so no query belongs to both and their counts add.
\end{proof}
The specification fixes these required questions; source edits and input bits
depend on how their answers are encoded. The native and outside parts can differ.
A state requirement can constrain feasible supply designs, so joint attainment
of the unconstrained supply minimum needs a compatibility argument.

\leanmetapending{\LHrng{SCI}{4}{5}}
\begin{lemma}[Cooperative initialization follows and reverses the MRO]
\label{cooperative-init-order}
Suppose each participating initializer terminates normally and executes
\texttt{pre; super().\_\_init\_\_(); post}, calling the next participating
initializer exactly once. The pre segments run in MRO order; the post segments
run in reverse order. For components $s,t$ in distinct providers, each placed
once in its provider's pre or post segment, $s$ initializes before $t$ iff:
\begin{enumerate}
\item $s$ is pre and $t$ is post;
\item both are pre and $s$'s provider precedes $t$'s;
\item both are post and $t$'s provider precedes $s$'s.
\end{enumerate}
\end{lemma}
\begin{proof}
Induction on the participating MRO suffix gives the trace
$\operatorname{pre}(L)\mathbin{+\!+}\operatorname{post}(\operatorname{reverse}L)$.
A provider at rank $r$ in a chain of length $n$ has pre rank $r$ and post rank
$2n-1-r$. Comparing these ranks gives the three cases.
\end{proof}
Lean checks this list model and rank comparison. Python's correspondence is
external: \texttt{super()} starts searching after the calling class in the
instance's MRO~\cite{PythonCooperativeSuper,barrett1996c3}. Initializers that
skip, repeat, or conditionally select calls need a different execution model.
The replay script \texttt{proofs/checks/state\_coordination\_replay.py} checks
all four placements and both provider orders in Python.

\begin{proposition}[A shared-state default prices distinct-state exceptions]
\label{sharing-default-exceptions}
Restrict to specified queries whose two accesses name one common provider.
Suppose accesses through ancestry share that provider's state. If a query
requires distinct state, at least one access must depart from that default.
Thus at least
$\PaperNotationLink{PaperDistinctSharing}{k_{\sigma,\mathrm{distinct}}}=\sum_c|\{q\in\operatorname{dom}\sigma_c^*: \sigma_c^*(q)=\mathsf{distinct}\}|$
sharing answers are determined outside that default. A field-only composition
baseline with no native sharing rule for these queries determines all
$k_\sigma$ answers through construction and reference passing.
\leanmeta{\LHrng{SCI}{6}{8}}
\end{proposition}
\begin{proof}
Two native accesses would give \emph{same}, contradicting the required
\emph{distinct}. The distinct-query set is therefore contained in the
outside-default query set. If no query has a native answer, that set is the
whole specified domain.
\end{proof}
Python's ordinary instance namespace does not create a separate subobject for
each diamond route. C++ shares a common \emph{virtual} base subobject; repeated
nonvirtual bases can instead be distinct~\cite{PythonDataModel,CppWorkingDraft}.
The lower bound follows from the shared default; nonvirtual C++ arrangements
can derive distinct state directly. Sharing queries and implementation pairs
have separate counts; attaining $\PaperNotationLink{PaperDistinctSharing}{k_{\sigma,\mathrm{distinct}}}$
jointly with a supply optimum requires a separate compatibility argument.
Language-native delegation or state-sharing mechanisms are not the field-only
baseline.

\begin{proposition}[Lookup and initialization can require one outside answer]
\label{state-lookup-coupling}
Let distinct providers $L,R$ both define method $m$ and own a state component.
Two classes $c_1,c_2$ both require $L$'s method. The first requires $L$'s state
before $R$'s; the second requires the reverse. Fix each provider's placement
as pre or post under Lemma~\ref{cooperative-init-order}. Over both base orders
and all four shared placements, the minimum number of lookup or initialization
answers supplied outside the defaults is exactly one.
\leanmeta{\LHrng{SCI}{9}{10}}
\end{proposition}
\begin{proof}
Without an outside lookup answer, first-definition lookup puts $L$ before $R$
in both classes. Every shared placement then gives both the same initialization
order, so at least one required answer needs repair. Lean checks every placement,
base order, and subset of the four possible lookup/initialization repairs.
For attainment put both components pre, use $c_1(L,R)$ and $c_2(R,L)$, and
declare \texttt{m = L.m} in $c_2$. This one lookup answer restores the required
winner while the two MROs derive opposite initialization orders.
\end{proof}
The Python replay checks this attaining design. Exhausting this finite witness
establishes its exact minimum; computing the minimum for other inputs needs their own search. Composition can
state the orders and lookup winners separately. C++ likewise separates its
base-construction order from first-definition lookup: direct bases initialize
in declaration order, virtual bases in traversal order, and genuinely ambiguous
member lookup requires explicit resolution~\cite{CppInitializationOrder,CppWorkingDraft}.

\begin{corollary}[State requirements preserve capacity containment]
\label{state-capacity-containment}
If adding multiple parents preserves every old design's supply, sharing,
initialization, and lookup answers, every feasible one-parent or composition
design and its outside-answer ledger remains feasible. Adding capacity cannot
worsen the optimum with the state specification fixed. Existing strict
crossing witnesses survive by taking the state specification empty; inheriting
one parentless provider into one class saves one supply answer against
no-inheritance supply without adding a state answer.
\leanmeta{\LH{OPT20}, \LHrng{OPT}{17}{19}, \LHrng{CSD}{1}{8}, \LH{SCT23}, \LHrng{SCT}{30}{34}}
\end{corollary}
\begin{proof}
The conservative embedding retains old feasible designs and their complete
answer ledgers. The main-text capacity-monotonicity lemma gives monotonicity.
The existing crossing and nonempty-supply constructions give strict witnesses.
\end{proof}
State requirements therefore add answers while capacity containment retains
every old feasible design. A ban's justification still requires the derived
exclusion tested in the main-text language-design discussion.

\section{Additional Audit Detail}
\label{audit-detail}

\begin{table}[htbp]
\centering
\small
\PaperTableSetup
\begin{tabularx}{\linewidth}{@{}m{0.20\linewidth}XX@{}}
\toprule
Measurement & Fixed input & Result and boundary \\
\midrule
Python provider gap
& source-closed classes with unique concrete providers
& preserves the observed provider-bundle map; excludes external bases,
descriptors, ambiguous providers, custom MROs, and runtime mutation \\
\midrule
Python sensitivities
& same extracted map, with method roles, pair removal, or tolerated over-reach
& pair costs do not equal source declarations; method-level and removal
sensitivities retain their stated solver intervals \\
\midrule
SQLAlchemy Closed-C3
& fully local \SQLAlchemyComponentClosedCThreeProviderFacts{}-provider,
\SQLAlchemyComponentClosedCThreeConsumers{}-consumer subcomponent with
\SQLAlchemyComponentClosedCThreeProviderIncidences{} pairs
& its \SQLAlchemyComponentClosedCThreeSelectorConflicts{} conflict sites and
\SQLAlchemyComponentClosedCThreeTax{}-pair exact gap are independently
certified; bases outside the parsed wheel remain excluded \\
\midrule
SQLAlchemy longitudinal workaround map
& six pinned releases under the source-declared first-parent policy
& measures recurrence under that policy; the source nodes can change, while the growth theorem fixes its nodes \\
\midrule
History screen
& direct-base syntax in 54 commits
& does not recover semantic scopes, forwarding, registries, generated code,
or continuity across renames \\
\bottomrule
\end{tabularx}
\caption{Fixed inputs, evidence, and boundaries of the inheritance audits.}
\label{tab:audit-scope-ledger}
\end{table}

\subsection{SQLAlchemy's largest independently certified component}
\label{sqlalchemy-component}

This audit establishes two different minima: missing implementation--class
pairs over every sound one-parent hierarchy, and reusable forwarding bundles
for one recorded optimal hierarchy. The second does not minimize jointly over
all optimal hierarchies.

The component contains \SQLAlchemyComponentProviderFacts{} implementation
providers and \SQLAlchemyComponentProviderIncidences{} required dependencies.  Its
exact optimum for the extracted provider relation is
\SQLAlchemyComponentTax{}, spread across
\SQLAlchemyComponentAffectedProviders{} providers and
\SQLAlchemyComponentAffectedConsumers{} classes.  The optimized gap has
\SQLAlchemyComponentAnswerPatterns{} distinct answer patterns; the complete
SQLAlchemy repair has \SQLAlchemyComponentProjectAnswerPatterns{}.
For the pinned repair $\PaperNotationLink{PaperAncestry}{H}$, its gap matrix has exact Boolean rank, equivalently
minimum biclique cover $\operatorname{bc}(\PaperNotationLink{PaperRequired}{R^{\mathrm{req}}}\setminus \PaperNotationLink{PaperAncestry}{H})$,
\SQLAlchemyComponentBooleanRank{}~\cite{theis2011bicliques}.  The report enumerates all
\SQLAlchemyComponentBooleanRankCandidates{} maximal all-ones rectangles, solves
their exact set-cover problem, stores the selected cover, and verifies that the
objective and solver bound both equal \SQLAlchemyComponentBooleanRank{}.
Each holder is selected before the requested implementation is known and
supplies an entire rectangle of missing pairs. Minimizing those holders groups
the 520 missing implementation--class pairs without reducing their count.

Every holder in a minimum cover supplies a pair no other holder supplies;
otherwise it could be removed.  These private pairs distinguish all subsets
of enabled holders, even when the holders overlap elsewhere.  Thus a minimum
cover of size $b$ gives $2^b$ distinct workaround maps, and only the full subset
covers the gap.  Lean checks the private-pair argument, injectivity, count, and
unique complete subset.  Applied to this fixed repair,
$b=\IndependentHolderMinimum{}$.
\leanmeta{\LHrng{HCC}{1}{4}}
All these maps are reachable only if the program permits every holder subset.
The cover certificate establishes the minimum cover. Reaching every holder
subset additionally requires independent controls; a generator can exclude incomplete configurations. The
\IndependentHolderMinimum{}-holder
minimum fixes $\PaperNotationLink{PaperAncestry}{H}$ and does not optimize holder count over alternative hierarchies.

Three large scopes provide a compact crossing witness.
\texttt{ColumnOperators}, \texttt{Executable}, and \texttt{FromClause} share
\SQLAlchemyComponentCrossingAll{} consumers while separately reaching
\SQLAlchemyComponentCrossingOnlyA{},
\SQLAlchemyComponentCrossingOnlyB{}, and
\SQLAlchemyComponentCrossingOnlyC{} consumers.  Their pair-only regions are
all zero.  Hence every pair intersects, and each provider also reaches a class the
other does not; no pair of scopes nests.

The protocol selected the first five displaced dependencies in lexicographic
provider--consumer order.  The table lists every concrete method attributed to
the provider at that consumer.  A direct forwarding replacement must name these
targets or invoke a mechanism that derives the same answer.

\begingroup
\PaperTableSetup
\begin{longtable}{@{}>{\RaggedRight\arraybackslash}m{0.29\textwidth}
  >{\RaggedRight\arraybackslash}m{0.29\textwidth}
  >{\RaggedRight\arraybackslash}m{0.34\textwidth}@{}}
\caption{Five source-anchored displaced dependencies from the pinned
SQLAlchemy repair.}\\
\toprule
Implementation provider & Receiving class & Inherited methods \\
\midrule
\endhead
\SQLAlchemyExampleRows
\bottomrule
\end{longtable}
\endgroup

Each row shows what a direct encoding inside the optimized one-parent repair
must supply: every method in the rightmost cell to the receiving class in the
middle cell, with the leftmost class remaining its authoritative implementation
source.  An adapter or generated shim can supply the same bindings, and then
that mechanism owns their continued presence.  The rows come from the bundled
repair; SQLAlchemy itself is unchanged.

Each row is regenerated from the wheel source and the pinned repair.
The report stores both declaration paths and line anchors; its independent
verifier reparses the pinned wheel and rejects any changed pair, method set, or
source digest.  Under the explicitly independent configuration policy, the
\SQLAlchemyComponentConfigurationExponent{} displaced dependencies admit
$2^{\SQLAlchemyComponentConfigurationExponent}$ possible subsets.  This count
describes the replacement boundary's worst-case state space.  The independently
checked feasible repair and the \SQLAlchemyComponentAnswerPatterns{} answer
patterns remain the measured quantities.  An independently generated LRAT
refutation, checked by Lean, proves that no smaller repair exists.

\subsection{Workaround changes under the declared first-parent policy}
\label{longitudinal-completion}

This audit follows one concrete policy: keep each class's source-declared first
parent and supply its missing pairs outside ancestry. It reconstructs that hierarchy
at \PythonLongitudinalSnapshots{} pinned SQLAlchemy releases, from
\PythonLongitudinalFirstRelease{} through
\PythonLongitudinalLastRelease{}.  All
\PythonLongitudinalTransitions{} adjacent transitions edit the workaround map.
Across transitions it contains \PythonLongitudinalCompletionEdits{} outside-pair additions
or deletions, including \PythonLongitudinalStableCompletionChurn{} on pairs
required on both sides of the release boundary.  The 511 edit total and the
equal first-release gap below are independently reconstructed quantities.
Required pairs grow from
\PythonLongitudinalFirstRequired{} to \PythonLongitudinalLastRequired{},
outside pairs from \PythonLongitudinalFirstCompletion{} to
\PythonLongitudinalLastCompletion{}.  Separately, the exact snapshot gap grows from
\PythonLongitudinalFirstGap{} to \PythonLongitudinalLastGap{}.  Removable
arrangement excess therefore changes from
\PythonLongitudinalFirstExcess{} to \PythonLongitudinalLastExcess{}.

The lock records every wheel digest, recovered relation digest, optimum
witness, and adjacent set difference.  Its verifier reparses all six wheels,
reconstructs every map, checks each witness, and rejects any changed count or
digest.  This measures recurrence under one source-declared policy.  It does
not infer why any edit occurred, and its source carrier changes, whereas the
main paper's fixed-carrier corollary holds that carrier fixed.

\subsection{Preregistered strict-insertion history audit}
\label{strict-insertion-audit}

The protocol fixes pip and aiobotocore, the two projects with zero gap when methods
are grouped by declaring class, before extraction. It also fixes aiobotocore's
separate-method input. For each project it selects every non-prerelease,
non-yanked release with a universal pure-Python wheel uploaded in the
365-day window ending at the population's pinned wheel, and no version later
than that endpoint. When several wheels qualify, the filename's lexical order
selects one. This gives \StrictInsertionPipSourceClassSnapshots{} pip releases
ending at 26.2.1 and \StrictInsertionAiobotocoreSourceClassSnapshots{}
aiobotocore releases ending at 3.9.0.

The registration was committed before extraction at \texttt{47cdbf76}.
\nolinkurl{strict_insertion_protocol.lock.json} records the UTC date, window,
wheel URLs and SHA-256 hashes, analyzer fingerprint, and Python interpreter
binary hash. Its SHA-256 is
\nolinkurl{c37dd3975b5886822f2b5337905dc961aa71263b66cb3b99efa3599e93d45787}.
No additional projects or components were admitted after seeing results.

Extraction reuses the population audit's MRO-invariant projection. Provider
rows and classes match by fully qualified name, without rename heuristics.
Only provider rows present in both versions enter a transition's search;
their class groups are restricted to declarations present at both versions.
Both full recovered snapshots must first be tree-shaped. The search records
every triple satisfying the mixed-update conditions in
\S\ref{immediate-parent-evolution},
including a nonempty new lower group; growth is not required.
It flags any triple touching a class whose extraction eligibility changes.
The relation contains resolved declarations only; unresolved program requirements
must be recovered before their gap can be classified.

All registered snapshots are tree-shaped. The
\StrictInsertionPipSourceClassTransitions{} pip transitions and
\StrictInsertionAiobotocoreSourceClassTransitions{} aiobotocore transitions
contain zero triples, under both registered aiobotocore inputs.
The null result therefore supplies no strict-insertion witness in these
histories. It does not classify other histories or rule out repair caused by
other updates.

\nolinkurl{python_strict_insertion/report.json} retains every recovered
snapshot, shared carrier, unmatched declaration, eligibility change, and
triple list. Its derived \nolinkurl{audit.md} gives the per-transition table
and wheel hashes. An independent verifier reparses the raw hashed wheels,
recomputes tree shape and exhaustive triples without importing the producer's
search, and agrees on all snapshots and transitions. With no measured
triples, there are no measured witness declarations to add to the claim map;
the general mixed-update result is checked by \LH{SIN4} and \LH{SIN5}.
\leanmeta{\LHrng{SIN}{4}{5}}

\subsection{Pinned pyapp-kit portfolio}
\label{pyappkit-portfolio}

The portfolio rule selects all nine libraries named on the pyapp-kit landing
page at commit \texttt{cdd24545f564}.  Each repository is pinned before the
class graph is parsed.  The MRO-invariant projection resolves local bases and
retains only a unique concrete implementation provider.  Multi-parent classes
with an unresolved direct base remain in the syntax count and in the external
boundary column; they cannot contribute implementation dependencies to the
recovered local implementation--class relation.

\begin{table}[h]
\centering
\small
\PaperTableSetup
\begin{tabular}{@{}lrrrrrr@{}}
\toprule
Project & classes & syntax MI & external MI & impl.-MI & dependencies & gap \\
\midrule
\PyAppKitRows
\bottomrule
\end{tabular}
\caption{All libraries on the pinned pyapp-kit landing page.  ``External MI''
counts multi-parent classes with at least one unresolved direct base.  The
exact gap prices the source-local provider map and therefore excludes those
external branches.}
\label{tab:pyappkit-portfolio}
\end{table}

Across \PyAppKitPythonFiles{} Python files and \PyAppKitClasses{} classes, the
screen finds \PyAppKitSyntacticMI{} syntactic multi-parent declarations,
including \PyAppKitExternalBoundaryMI{} at an external boundary.  The
source-local relation has \PyAppKitImplementationMI{} implementation-MI
classes, \PyAppKitProviderDependencies{} dependencies, and exact gap
\PyAppKitExactGap{}.  An independent rerun checks every repository commit,
rebuilds every class graph, and byte-compares the report.

\subsection{History screen for MI-to-SI rewrites}
\label{history-rewrite-screen}

The history study searches for a directly visible change: a class drops from
multiple declared bases to at most one. Its protocol selects the five largest positive provider-bundle gaps
from the locked Python population, then follows the mainline history to the
pinned release.  A candidate keeps the same qualified class name and path while
its direct-base list falls from at least two entries to at most one.  The scan
records the commit, subject, line anchors, and both base lists.

\begin{table}[h]
\centering
\small
\PaperTableSetup
\begin{tabular}{@{}lrrrr@{}}
\toprule
Project & candidates & commits & parse failures & fallback \\
\midrule
\HistoryRows
\bottomrule
\end{tabular}
\caption{Locked mainline syntax screen for the five largest positive
provider-bundle gaps.  The scan finds 13 relocated nominal factors in total;
the table reports candidate, commit, parser-failure, and fallback counts by
project.  Historical files outside the parser model remain counted as parse
failures.}
\label{tab:history-rewrite-screen}
\end{table}

The scan finds \HistoryBaseReductions{} candidates in
\HistoryBaseReductionCommits{} commits, including
\HistoryRelocatedNominalFactors{} relocated nominal factors, and records
\HistoryParseFailures{} parse failures.  The candidates are evidence about the
observable class syntax. Hidden forwarding, registry, and generated-code
rewrites are not measured by this base-list screen. The report is pinned to the
protocol digest and every repository commit; the independent structural check
recomputes the project totals, candidate counts, and source-path constraints.

A separate SQLAlchemy fix changed collection instrumentation to follow
Python's MRO after unpredictable subclass choices~\cite{SQLAlchemyMROFix}.
The case separates the two resources used in the main analysis: ancestry
supplies implementations, while MRO orders overlapping providers.

\subsection{An end-to-end AnyIO refactoring decision}
\label{anyio-refactoring-application}

This worked application starts with public clients, fixes which code provider
changes must reach them, and uses the analyzer to choose a preserving refactoring.
It uses AnyIO~4.14.2 at commit
\texttt{c384f99687c64c59ed8a}\allowbreak\texttt{11c3a0f11a2d57daff71};
the controlled source copy supplies refactoring evidence while the population
still counts AnyIO once. The artifact file
\texttt{practical\_impact/}\allowbreak\texttt{ANYIO\_REFACTORING.md} supplies commands and source
anchors.  The runner checks the commit, unchanged tracked sources and tests,
and the source actually imported by Python.

\paragraph{Clients and requirements.}
\texttt{FileReadStream} reads bytes, seeks, and reports its file position;
\texttt{FileWriteStream} writes bytes.  Their common
\texttt{\_BaseFileStream} owns initialization, closure, and typed file metadata.
\texttt{BufferedByteReceiveStream} can wrap an in-memory receiver and has its
own buffer and lifecycle code.  It shares concrete async-iteration methods
with the file reader through \texttt{ByteReceiveStream}.
The unchanged upstream suite exercises these APIs;
the added application client writes a temporary file, reads and iterates it,
checks its metadata and closure, and iterates a separate memory stream.

\begin{center}
\small
\begin{tabular}{@{}lccc@{}}
\toprule
Required provider & File reader & File writer & Buffered reader \\
\midrule
File initialization, closure, metadata & yes & yes & no \\
\midrule
Byte-receive iteration & yes & no & yes \\
\bottomrule
\end{tabular}
\end{center}

The two exclusions have operational meanings.  The write-only stream must not
gain the receive/iteration API.  The memory-backed buffered reader must not
acquire an independent file handle or execute the file provider's lifecycle.
It may still forward metadata from a wrapped file; that does not make it the
provider of a separate file.  Effective method lookup, including inherited
methods and overrides, confirms every cell.  The extraction compares actual
source-member identities, so a selected base counts only through the implementation it supplies.

The maintenance contract retains the three public class names and their
distinct meanings, public receive/send membership, factories, instance APIs,
exceptions, typed metadata, and each selected implementation's source and
receiver.  File-lifecycle and byte-iteration bodies may change independently
while retaining their member names and descriptor kinds.  Replacement happens
before client construction.  Private file-base identity tests, existing-instance
migration, custom MROs, and cooperative-\texttt{super} extensions lie outside this
application; the transformer rejects methods with \texttt{\_\_class\_\_}
closures.  New member names require regeneration and validation.

\paragraph{What the analyzer changes.}
All three crossing regions have size one.  The analyzer returns
$|\PaperNotationLink{PaperRequired}{R^{\mathrm{req}}}|=\AnyIOApplicationPairs{}$ and
$\PaperNotationLink{PaperGap}{d_\downarrow}=\AnyIOApplicationGap{}$, checked by a feasible repair and a
matching conflict-packing certificate.  The two-role formula gives the same
minimum.  Its implementation-only objective permits several optimal repairs;
the application's public membership obligation selects one that retains
\texttt{ByteReceiveStream} natively.  Removing the private file provider from
the reader leaves three native pairs and one outside pair.  The artifact
checks this chosen repair against the original relation, independently of
which optimum the solver first returns.

The following shows the changed declaration.  The original factory,
\texttt{receive}, \texttt{seek}, and \texttt{tell} method objects remain unchanged;
the executable transformer derives the forwarding members from the provider's
own declarations.
\begin{lstlisting}[language=Python,basicstyle=\ttfamily\small]
# Before: FileReadStream(_BaseFileStream, ByteReceiveStream)
class FileReadStream(ByteReceiveStream):
    __init__ = live_member(_BaseFileStream, "__init__")
    aclose = live_member(_BaseFileStream, "aclose")
    extra_attributes = live_member(_BaseFileStream, "extra_attributes")
    # Original factory, receive, seek, and tell retained unchanged.
\end{lstlisting}
\texttt{live\_member} reads the current provider member at lookup time and
binds the original stream as its receiver.  Its property variant retains
read-only data-descriptor behavior.  No provider body or file state is copied.
Both providers remain independently owned, and updates to existing members
need no additional route edit.  A captured-method assignment is tested
separately: it retains the old function after replacement, while live forwarding
uses the new one.  The generator supplies the one missing provider--reader
pair.  This realizes the selected relation with one-parent reach;
other AnyIO classes and providers remain unchanged.

\paragraph{Executable checks.}
Separate processes run the full unchanged upstream suite on the native and
forwarded reader.  Each passes \AnyIOApplicationUpstreamPassed{} cases, skips
the same \AnyIOApplicationUpstreamSkipped{}, and has the same
\AnyIOApplicationUpstreamExpectedFailures{} expected failures.  The runner
compares every individual outcome.
The \AnyIOApplicationContractPassed{} added tests check the exact relation,
attaining repair, method sources and receivers, public membership, property
write protection, real file/memory clients, and independent provider updates.
All four combinations of file and iteration body replacement are exercised
under both representations, with identical affected client sets.

The omission check executes all eight subsets of the three generated bindings.
Only the full subset passes the same read/metadata/close client; the other seven
fail construction, abstract-method, or metadata checks.  These three member
bindings implement one provider--client pair.  The report keeps those units
separate and saves source hashes, test logs, JUnit outcomes, and an independently
verifiable gap certificate.  The finite relation minimum is exact. Source checks and executions support
the stated Python behavior; Lean certifies the relation minimum.

\subsection{Displaced-dependency mutation study}
\label{anyio-mutation-study}

The mutation study uses the smallest archived wheel by byte size whose
provider-bundle gap lies between 10 and 25, a nontrivial window capping
two-route omission cases at $\binom{25}{2}=300$: AnyIO 4.14.2, with gap 21. The
denominator contains every single provider-consumer omission in the independently
certified repair.  For each
omission, the runner shadows every uniquely attributed method with a
\texttt{MutationError} stub and runs the complete pinned test suite.  A marker
distinguishes a suite that reached the mutation from infrastructure failure
before exposure.  Before the retained campaign, one pilot exceeded seven
baseline-suite durations; the amendment discarded that pilot campaign and
reran every mutant with a 60-second per-test and 600-second whole-suite cutoff.
Project selection, denominator, mutation operator, and completed pilot outcomes
were unchanged.

\begin{table}[h]
\centering
\small
\PaperTableSetup
\begin{tabular}{@{}lr@{}}
\toprule
Outcome & count \\
\midrule
Killed & \MutationKilled \\
\midrule
Timed out after exposure & \MutationTimedOutAfterExposure \\
\midrule
Survived & \MutationSurvived \\
\midrule
Unloadable & \MutationUnloadingFailures \\
\midrule
Infrastructure failure & \MutationInfrastructureFailures \\
\bottomrule
\end{tabular}
\caption{AnyIO mutation outcomes for all \MutationDenominator{} generated
single omissions.  The baseline passed \MutationBaselinePassed{} tests, and
the pinned suite detected all \MutationDetected{} omissions.}
\label{tab:anyio-mutation}
\end{table}

All \MutationDetected{} mutations were detected, and no mutation survived.
Omitting any one measured displaced route causes a detected test failure
in the pinned test configuration; this does not estimate
the cost of maintaining those routes.

The joint-omission sensitivity selects one marker-confirmed test for each of
the \MutationPairRoutes{} routes, runs both selected tests for every unordered
pair, and independently verifies the raw logs.  It detects all
\MutationPairDetected{} of \MutationPairDenominator{} pairs.  Both omission
markers are reached in \MutationPairBothMarkers{} pairs; at least one is reached
in all \MutationPairAtLeastOneMarker{}.  Thus no tested pair masks both routes.  Triple omissions, the full suite under each pair, omissions during a
refactoring, and future program states require separate evidence.  The artifact
stores the baseline, mutant, witness, and pair logs; typed frozen records own
their schemas, shared base classes own common execution and verification, and
subclasses supply only semantic hooks.  Independent verifiers check every
digest, configuration, outcome, marker, denominator, and summary.

The packaged \texttt{pytest\_counts\_replay.json} ledgers reparse the retained
logs without changing the historical reports.  Four single-omission logs gain
one previously uncounted error each; detection totals are unchanged, and no
new executions are recorded.  The pairwise ledger needs no count corrections.

\subsection{Python population selection and extraction protocol}
\label{python-population-protocol}
\paragraph{Omitted requirements can raise or lower the gap.}
With two implementations and three classes, Theorem~4.5 gives
$\PaperNotationLink{PaperGap}{d_\downarrow}(A,B)=\min\{N_{10},N_{01},N_{11}\}$.
For $A=\{1,2\}$ and $B=\{1,2,3\}$, the region counts are $(0,1,2)$,
so the gap is zero. Dropping $(B,1)$ leaves $B=\{2,3\}$:
all three counts become one, so the gap rises to one.
Conversely, start with $A=\{1,2\}$ and $B=\{2,3\}$, with gap one.
Dropping $(A,1)$ leaves $A=\{2\}\subseteq B$, with counts $(0,1,1)$
and gap zero. A crossing needs all three regions nonempty, so two implementations
and three classes are minimal. Both directions are machine-checked, up to
renaming classes and exchanging the providers.\leanmeta{\LH{DEC7}}

The Python factorization study screens a locked population before measuring
the repair outcome.  The locked ranking is
\texttt{hugovk/top-pypi-packages} commit \texttt{e935455}: of its first 150
non-yanked universal pure-Python wheels, 65 have at least 50 locally parsed
classes, 31 have a cross-branch class, and 20 have at least four.  The source
lock contains the ranking snapshot, wheel URL, release, and SHA-256 digest for
every candidate.  Each selected project has at
least 50 locally parsed classes and at least four consumer classes that inherit
surviving concrete methods from two or more direct-base branches.  Marker and
abstract-only bases, shared diamond ancestors, unresolved bases, competing
concrete providers, and direct overrides fall outside the selection rule.
When incomparable most-derived declarations contain exactly one concrete body,
that body is attributed: coverage is forced even though no precedence answer
is available or needed.
For source-class grouping, overriding one method does not remove the provider
if another survives: if \texttt{C} overrides \texttt{P.f} but inherits
\texttt{P.g}, the method relation keeps $(\texttt{P.g},\texttt{C})$ and drops
$(\texttt{P.f},\texttt{C})$; the grouped relation retains $(\texttt{P},\texttt{C})$.
The primary threshold of four selects 20 projects.  Repeating the complete
analysis for the retained threshold-three screen selects 23, of which 20 have
positive gap; it externalizes 747 of 14,249 pairs (5.24\%), or 200 of
8,436 (2.37\%) after excluding SQLAlchemy.  The artifact exports and verifies
all 23 result bundles.

After the primary audit, a census extension scores every size-qualified project
with at least one screened cross-branch class, without changing the primary
lock or the 20-project ladder. Of these \PythonCrossBranchCensusProjects{}
projects, \PythonCrossBranchCensusPositive{} have positive recovered gaps.
The \PythonCrossBranchCensusAdditionalProjects{} newly scored projects below
use the same pinned wheels and extraction rule; their optima are checked by
independent direct replay; all their component gaps are at most 10.
The remaining \PythonCrossBranchCensusUnscored{}
size-qualified projects remain unscored: their unmeasured gaps remain unknown.
\nolinkurl{cross_branch_census.json} binds each certificate's bytes to the
original lock; counts and this table are generated from those certificates.

\begin{table}[htbp]
\centering
\PaperTableSetup
\begin{tabular}{@{}lrrr@{}}
\toprule
Project & Cross-branch classes & Required pairs & Gap \\
\midrule
\PythonCrossBranchCensusAdditionalRows
\bottomrule
\end{tabular}
\caption{Census extension to the eleven projects below the primary selection
threshold. All are scored, including zero-gap projects.}
\end{table}

\subsubsection{Public-name preservation contract}
\label{python-public-api-contract}
This audit asks what the library must retain under a separate, reproducible contract:
preserve the original concrete provider resolution of every selected public
member on every statically exposed public class. It neither infers documented
intent nor changes the primary twenty-project selection.

A public member has no leading underscore or is a dunder \texttt{\_\_x\_\_};
this includes \texttt{\_\_enter\_\_}, \texttt{\_\_iter\_\_},
\texttt{\_\_eq\_\_}, and \texttt{\_\_init\_\_}. Single-underscore names
\texttt{\_x} and mangled \texttt{\_\_x} without trailing double underscores
are private. A public class is exposed in a module path with no
underscore-prefixed segment, or publicly re-exported there. A literal
\texttt{\_\_all\_\_} governs a module's exports; otherwise non-underscore
class and imported-symbol names are candidates. Explicit re-exports resolve
their named bindings; star re-exports follow the target module's export list.
Aliases retain declaration identity and do not multiply pairs.
Dynamic export lists and conflicting static targets are omitted rather than
guessed. Table~\ref{tab:public-api-contract} records dynamic-list exclusions;
the source extractor's other exclusions remain in force.

Filter the existing MRO-invariant concrete-method lookups by these rules.
Define $\PaperPublicRequired$ to contain $(f,c)$ exactly when a selected public
lookup on $c$ resolves to provider $f$. Each pair retains a witness member,
public class alias, and original provider in the archived source-derived
report. A design preserving these specified resolutions must supply every such
pair: apply lookup preservation to the witness member, then lookup-to-supply
implication. This argument allows forwarding equally with inheritance.
Lean checks the implication from the declared preservation contract. Source
extraction supplies its public-member witnesses, which can then be exercised by clients.\leanmeta{\LH{APC1}, \LH{APC2}}

The artifact binds this relation to the original pinned wheels and to
\path{plans/public-api-contract-audit.md}. It independently replays each
component with gap at most ten; \PublicApiCertificates{} further components
have feasible repairs and Lean-checked LRAT refutations excluding one fewer
deletion. This contract has \PublicApiPairs{} pairs and exact gap
\PublicApiGap{}. Its gap is neither a lower nor an upper bound on the
primary 739: deleting pairs can move the gap in either direction.
A zero means the selected public lookups fit ancestry; other API requirements
define their own relation and gap.

\begin{table}[htbp]
\centering
\PaperTableSetup
\begin{tabular}{@{}lrrr@{}}
\toprule
Project & Public-name pairs & Exact gap & Dynamic export modules omitted \\
\midrule
\PublicApiRows
\bottomrule
\end{tabular}
\caption{Separate public-name resolution contract on the original twenty
projects. Dunders count as public; provider declarations may be private.
Zero-pair rows are retained.}
\label{tab:public-api-contract}
\end{table}

\subsubsection{Collapsing equal provider scopes}
\label{equal-scope-collapse}
Merging every equal-scope group gives the certified equal-scope floor
\EqualScopeGap{} across the
locked twenty projects. Define $f\equiv g$ exactly when $\PaperNotationLink{PaperClassGroup}{A_f}=\PaperNotationLink{PaperClassGroup}{A_g}$, and keep one
provider for each equivalence class. The quotient reduces
\EqualScopeProvidersBefore{} providers to \EqualScopeProvidersAfter{}, with
\EqualScopePairs{} required pairs; \EqualScopePositive{} projects still cross
(Table~\ref{tab:equal-scope-collapse}).

This test removes every distinction between code providers that serve precisely the
same classes. Such a merge preserves each class's current code bundle; replacing
unequal scopes with one shared scope changes at least $|\PaperNotationLink{PaperClassGroup}{A_f}\mathbin{\triangle}\PaperNotationLink{PaperClassGroup}{A_g}|$ answers
(main-text Theorem~4.8; Corollary~4.19). The test allows providers with identical
class sets to share one identity, and recomputes the minimum after merging them.
The surviving gap therefore does not require keeping equal-scope providers
distinct.

All twenty projects are retained. The artifact reconstructs scopes from the
locked wheels and archives every equivalence class and original provider name.
Every quotient component has a checked feasible repair. Exhaustive replay
certifies components with gap at most ten; larger components use Lean-checked
LRAT refutations excluding one fewer deletion. An existing certificate is
reused only after its input family and source digest match the reconstructed
component, and its proof is replayed. Changed larger components have
\EqualScopeNewCertificates{} new certificates. Counts and rows below are
generated from this checked report.

\begin{table}[htbp]
\centering
\PaperTableSetup
\begin{tabular}{@{}lrrrr@{}}
\toprule
Project & Original providers & Equal-scope groups & Quotient pairs & Exact gap \\
\midrule
\EqualScopeRows
\bottomrule
\end{tabular}
\caption{Equal-scope relaxation on the locked twenty-project cohort. Only
identical class sets merge; every quotient optimum is independently certified.}
\label{tab:equal-scope-collapse}
\end{table}

\paragraph{Provider state in archived optimal repairs.}
\label{provider-state-audit}
Refining the state rule to exclude empty slots and initializer-only declarations
and include inherited state changes the earlier count from 457 (61.8\%) to
\ProviderStatePairs{} (\ProviderStatePercent{}\%).
Source-class grouping yields \PythonPopulationProviderFacts{} providers and
\PythonPopulationProviderIncidences{} pairs from
\PythonPopulationInheritedMethodIncidences{} inherited method--class pairs.
Of the primary repairs' \ProviderStateTotal{} outside pairs,
\ProviderStatePairs{} (\ProviderStatePercent{}\%) have providers carrying instance
state: the provider or a locally resolved ancestor declares nonempty
\texttt{\_\_slots\_\_} or writes an instance attribute.
The detector counts stores on an instance method's receiver and
\texttt{setattr(receiver, ...)}; it excludes reads, class/static methods,
nested function scopes, and annotation-only declarations. Empty slots and an
initializer that only calls \texttt{super()} do not qualify by themselves;
inherited state does. Ancestry follows the extraction's locally resolved graph;
unresolved external bases supply no inferred state. It classifies the archived
optimal repair choices; it does not minimize or maximize state-bearing pairs
over all optima.
\path{python_factorization_audit/provider_state_audit.py} verifies each pinned
wheel and source digest against the archived case, resolves every externalized
provider, retains its state-source declarations, and derives both the report
and manuscript counts from them.

\leanmetapending{\LHrng{SRA}{1}{2}}
\begin{proposition}[Restricting provider rows bounds the stateful gap]
\label{stateful-row-projection-bound}
Let $F_{\rm state}$ be the providers carrying instance state under the rule above.
Restrict any sound laminar retention to those rows. It stays laminar, so its
missing pairs bound $\PaperNotationLink{PaperGap}{d_\downarrow}(\PaperNotationLink{PaperRequired}{R^{\mathrm{req}}}|_{F_{\rm state}})$ from above.
In the archived primary repairs this gives
$\PaperNotationLink{PaperGap}{d_\downarrow}(\PaperNotationLink{PaperRequired}{R^{\mathrm{req}}}|_{F_{\rm state}})\le\StatefulUpper$.
\end{proposition}
\begin{proof}
Every pair of retained rows already nests or is disjoint. Removing other rows
preserves that property; its missing cells are exactly the original missing
cells whose provider is retained. The minimum cannot exceed this feasible cost.
\end{proof}

Reoptimizing the full state-bearing subrelation gives the independently certified
aggregate minimum \StatefulGap{}. The \StatefulUpper{} state-bearing pairs in
the original full optimum give a feasible upper bound; reoptimization includes all qualifying rows, including
those with no original outside pair. State detection uses the same declaration
scan and locally resolved ancestry as the annotation above; a positive instance
write resolves statefulness even if a slots expression is not statically evaluated.
At fixed provider/state carriers and sound ancestry, this is a lower bound for Table~\ref{tab:supp-language-carriers}'s
one-stateful-chain placement: optimistically granting every other provider free
default-method supply cannot remove the stateful rows' laminar obstruction.
Defaults that cannot express an additional provider can only increase the cost
~\cite{JavaLanguageSpecificationInterfaces}.
Every translation preserving these provider/state carriers and sound one-chain
supply must meet this bound. Provider grouping also retains methods that never access state.

\begin{table}[t]
\centering\small
\begin{tabular}{lrrrr}
\toprule
Project & Stateful providers & Pairs & Projected upper & Exact gap \\
\midrule
\StatefulRowsTable
\bottomrule
\end{tabular}
\caption{State-bearing provider rows, with fixed identities, scopes, and sound ancestry. The upper
column restricts the primary repair; the exact column reoptimizes this subrelation.
Components of gap at most 10 use exhaustive replay; larger ones use Lean-checked LRAT.}
\label{tab:stateful-provider-gap}
\end{table}

\subsubsection{Recovering unique concrete bodies among abstract declarations}
\label{python-extraction-strengthening}
The earlier frozen extraction rule dropped a lookup whenever its most-derived declaration
set was incomparable.  The strengthened rule first removes abstract
declarations from that set.  It attributes the lookup when exactly one concrete
declaration remains and still declines to guess when two concrete declarations
compete.  Thus the rule changes candidate supply without inventing a winner.

The artifact reruns both rules from the same locked wheels, loading the earlier frozen
rule from its retained source bytes and comparing normalized implementation--class
relations by SHA-256 digest.  It recovers
\PythonExtractionStrengtheningRecovered{} method lookups and reduces recorded
ambiguities from \PythonExtractionStrengtheningAmbiguousBefore{} to
\PythonExtractionStrengtheningAmbiguousAfter{}.  Nineteen of twenty relations
are identical to that earlier frozen-rule baseline.  In opentelemetry-api, six recovered lookups add three
implementation--class pairs; an exact rerun leaves its gap at one.  Consequently all
reported optima and every LRAT-backed relation remain unchanged, while the
primary aggregate relation grows by three pairs.  The complete 20-project
comparison is archived in
\path{generated/python_extraction_strengthening/results.json}; Table~\ref{tab:python-extraction-strengthening}
shows the seven named end-to-end checks.

\begin{table}[h]
\centering
\small
\PaperTableSetup
\begin{tabular}{@{}lrrrr@{}}
\toprule
Project & recovered lookups & ambiguities & current $|R|$ & current digest \\
\midrule
\PythonExtractionStrengtheningRows
\bottomrule
\end{tabular}
\caption{Frozen-rule versus strengthened-rule extraction.  ``Ambiguities'' is
before--after.  Equal relation digests are recorded in the complete JSON; the
displayed digest is the strengthened relation's SHA-256 prefix.}
\label{tab:python-extraction-strengthening}
\end{table}

Figure~\ref{fig:abstract-concrete-boundary} gives the boundary case in which
recovery can change the relation.  The concrete provider sets are
$\{\texttt{Value},\texttt{Pointer}\}$ for \texttt{owner} and
$\{\texttt{Value},\texttt{Interface}\}$ for \texttt{attach}, so their exact
gap is one.  Attributing \texttt{AttachMixin.attach} through \texttt{Value}
does not require choosing it over \texttt{Receiver.owner}: the members differ.

\begin{figure}[h]
\centering
\begin{minipage}{0.72\textwidth}
\begin{lstlisting}[language=Python,basicstyle=\ttfamily\small,frame=single]
class Receiver:
    def owner(self): ...
class AttachMixin:
    def attach(self): ...
class Value(AttachMixin, Receiver): pass
class Pointer(Receiver):
    def attach(self): ...
class Interface(AttachMixin):
    def owner(self): ...
\end{lstlisting}
\end{minipage}
\caption{A five-class abstract/concrete boundary witness.  Coverage is forced
when one incomparable candidate supplies the queried member; genuinely
competing concrete declarations remain unattributed.}
\label{fig:abstract-concrete-boundary}
\end{figure}

\subsubsection{Run the analyzer on another input}
\label{analyzer-new-input}
From the artifact root, \texttt{python3 -m pip install ./experiments} installs
\path{ancestry-gap}.  It accepts Python source without importing its code,
or a user-declared finite implementation--class relation:
\begin{verbatim}
ancestry-gap python /path/to/package --out gap.json
ancestry-gap relation requirements.json --out declared-gap.json
ancestry-gap verify gap.json
\end{verbatim}
The report contains the relation and its size $|R|$, source hashes and
declaration locations, omissions, a feasible outside repair, and crossing
witnesses.  Matching independently checked lower and upper bounds establish
an exact gap; otherwise the report retains their interval.  CP-SAT status is
separate from independent verification.  The verifier rechecks feasibility
and disjoint crossing witnesses and, where needed, reruns the four-way search
under a declared node limit.  Reaching that limit reports incomplete
verification. New inputs use feasibility, packing, and bounded exhaustive search;
larger archived instances add Lean-checked LRAT certificates. The installed examples have $|R|=4$ and
$\PaperNotationLink{PaperGap}{d_\downarrow}=1$; the tool guide defines the input format and scope.

\subsubsection{External-route syntax control}
\label{external-route-control}
The control reuses the complete population lock and selects every parsed wheel
with at least 50 locally parsed classes and zero locally resolved multiple-parent
classes, before scanning routes.  This gives \ExternalRouteControlEligible{}
projects.  The scanner records a pair $(\texttt{field.member},c)$ when a method
or property named \texttt{member}, declared directly on class $c$, accesses or
calls \texttt{self.field.member} in its body.  It compares roles only when they
share the same field name.  An independent verifier reparses all eligible
wheels, checks wheel and source digests, reconstructs every role set, and checks
the selected crossing regions.  Table~\ref{tab:external-route-controls} lists
one lexicographically first route-spelling crossing per project with such a crossing; every listed consumer
declaration has at most one syntactic base.

This grammar groups matching route spellings. Resolving provider identity
also requires inherited forwarding and receiver domains. The pinned follow-up checks all 23 selected
route-pair candidates.  Pydantic's selected crossings disappear when inherited
members are included; httpcore's matching spellings cross distinct receiver
domains and optional adapters; BigQuery's selected spellings resolve to
distinct job-specific getter providers.  No selected candidate therefore
establishes a shared-provider crossing. The control shows that matching route
spellings can name distinct providers or hide inherited routes.

\paragraph{Seeded forwarding recall.}
The same scanner finds \PaperOneForwardingRecallDetectedPairs{} of
\PaperOneForwardingRecallPairs{} locked AnyIO provider--consumer pairs rewritten
as direct \texttt{self.field.member} declarations, matching all
\PaperOneForwardingRecallDetectedMembers{} member declarations. Pair recall
requires every listed member to match. The experiment retains the original
ancestry, fixing a positive control for the scanner's declared syntax.
Optimal repairs and naturally occurring forwarding require their own resolved inputs. Joint witness execution binds
\PaperOneForwardingRecallTracedPairs{} pairs; three abstract-class wrappers are
shadowed by concrete wrappers. Both original and seeded witness suites pass,
the full upstream suite is outside this syntax control. The artifact records
per-declaration scanner matches, selected-provider binding traces, witnesses,
source hashes, and the JSON from which manuscript counts are derived.

\begin{table}[h]
\centering
\small
\PaperTableSetup
\setlength{\tabcolsep}{3pt}
\begin{tabular}{@{}lllrrrr@{}}
\toprule
Project & left member & right member & left only & both & right only & syntax gap \\
\midrule
\ExternalRouteControlRows
\bottomrule
\end{tabular}
\caption{Route-spelling crossings among
\ExternalRouteControlPositive{} of \ExternalRouteControlEligible{} size-qualified
zero-local-MI wheels. Each row's two member spellings share the field name
recorded in the verified JSON. The final column applies the two-role formula
$\min\{N_{10},N_{11},N_{01}\}=1$ to this syntactic relation.
Provider resolution rejects every candidate: matching spellings either use distinct
providers or omit inherited routes, which resolves the apparent crossings.}
\label{tab:external-route-controls}
\end{table}

\subsubsection{Java forwarding check}
\label{java-forwarding-check}
A fixed three-library check scans Apache Commons Collections~4.5.0,
IO~2.20.0, and Lang~3.18.0 source releases.  Its direct-declaration grammar
records same-named calls through a resolved instance field with unchanged
arguments, preserving the field's declaring identity and written method
signature.  Collections has one crossing pair in that declared relation;
the other two libraries have none.  The candidate concerns \texttt{entrySet}
and \texttt{put} in \texttt{AbstractInputCheckedMapDecorator},
\texttt{FixedSizeMap}, and \texttt{PredicatedMap}.  However,
\texttt{PredicatedMap} inherits the first route.  An exploratory sensitivity
including inherited routes through the nearest local method declaration
removes the candidate and detects no crossing in the three libraries.
The source hashes, complete route reports, exclusions, and replay commands
retain both analyses.  This check supplies no positive Java semantic witness:
direct declarations omit inherited capabilities, and the expanded grammar
still omits accessor-mediated and unresolved routes.  Complete requirement
shape remains unmeasured.

\subsection{Python factorization results and certificates}
\label{python-factorization-results}
The main MRO-invariant projection keeps only a unique most-derived concrete
provider.  A separate sensitivity applies standard C3 when every explicit
transitive base resolves within the parsed source.  It measures
\PythonPopulationClosedCThreeTax{} external pairs among
\PythonPopulationClosedCThreeIncidences{} incidences, compared with
\PythonPopulationClosedInvariantTax{} among
\PythonPopulationClosedInvariantIncidences{} under the invariant projection on
the identical consumer carrier.
For the largest SQLAlchemy component, the invariant projection retains zero
multi-provider selector sites but omits
\SQLAlchemyComponentOmittedMroSensitiveLookups{} MRO-sensitive lookups among
its consumers.  Its fully local Closed-C3 subcomponent contains
\SQLAlchemyComponentClosedCThreeProviderFacts{} providers,
\SQLAlchemyComponentClosedCThreeConsumers{} consumers, and
\SQLAlchemyComponentClosedCThreeProviderIncidences{} pairs.  C3 resolves
\SQLAlchemyComponentClosedCThreeSelectorConflicts{} conflict sites there, and
its exact one-parent gap is \SQLAlchemyComponentClosedCThreeTax{}; hence the
matched coordinates are $(71,139)$.  Because 139 exceeds the FPT replay ceiling
of 10, exactness here is proved by an independent LRAT refutation checked in
Lean together with the feasible 139-pair retained-laminar witness.  The
remaining sites depend on bases
outside the parsed wheel boundary.  Across the project the invariant
projection omits \SQLAlchemyComponentProjectOmittedMroSensitiveLookups{} such
lookups.  Thus $(0,520)$ and $(71,139)$ are separate matched measurements on
the conservative and locally C3-closed relations, respectively.
Their gap fractions are $520/4{,}831=10.8\%$ and $139/1{,}173=11.8\%$;
the corresponding conflict counts are zero and 71.

Each counterfactual retains the measured provider--consumer relation and
allows arbitrary reparenting, synthetic factor nodes, and optimal forwarding.
An independent checker reparses the source and verifies each wheel digest and
retained-laminar upper witness.  For every component with $k_i\le10$, the
four-way FPT replay proves the matching lower bound.  For the other
\IndependentGapCertificateCount{} components, a decision CNF asks whether one
fewer deletion suffices, CaDiCaL
emits an LRAT refutation, and Lean validates that proof against the CNF.
Lean generates every checked gap CNF from the recovered finite relation and
bound; completeness is proved (a laminar retention within budget implies
satisfiability).  Source extraction remains external.  Exhaustive small-instance
tests agree with brute-force gap on the families stated in the artifact.
For the holder bound, Lean checks an intersection-closure certificate, derives
all \SQLAlchemyComponentBooleanRankCandidates{} maximal rectangles, generates
the \IndependentHolderExcludedBound{}-holder decision CNF, and checks
its LRAT refutation.
CP-SAT's \texttt{OPTIMAL} status cross-checks the search; the generated CNF,
repair-to-satisfiability proof, and checked refutation establish the lower bound independently.

For the Closed-\PaperNotationLink{PaperClasses}{C}3 component, a separate 122-crossing packing makes the
Lean-native refutation compact; Lean checks it against the same recovered
family as the archived Python/PySAT cross-check. For the 520-pair component,
a separately generated packing contains
\FlagshipObstructionCount{} incidence-disjoint crossings.  Every laminar repair
deletes at least one incidence from each crossing, leaving only
\FlagshipResidualBudget{} further deletions available under the strict
\FlagshipStrictBudget{} budget.  Lean checks the packing, generates the
packing-factored residual CNF, proves that every repair within the strict
budget satisfies it, and checks its LRAT refutation.  Python supplies only the
recovered relation and packing candidate; neither is accepted without the
corresponding Lean-side checks.  The JSON also contains rectangle covers,
fooling families, and component-level search statistics.

\begin{table}[h]
\centering
\PaperTableSetup
\setlength{\tabcolsep}{3.5pt}
\begin{tabular}{@{}lrrrrrl@{}}
\toprule
Project, component & Providers & Classes & Pairs & Objective & Bound & Status \\
\midrule
AnyIO, 0           & 18  & 57  & 196  & 21  & 21  & \texttt{OPTIMAL} \\
\midrule
google-auth, 0      & 16  & 42  & 148  & 11  & 11  & \texttt{OPTIMAL} \\
\midrule
SQLAlchemy, 3       & 223 & 728 & 4831 & 520 & 520 & \texttt{OPTIMAL} \\
\midrule
SQLAlchemy, \PaperNotationLink{PaperClasses}{C}3 max  & 92  & 183 & 1173 & 139 & 139 & \texttt{OPTIMAL} \\
\midrule
SQLAlchemy, 51      & 17  & 96  & 224  & 11  & 11  & \texttt{OPTIMAL} \\
\midrule
redis, 2            & 51  & 68  & 545  & 87  & 87  & \texttt{OPTIMAL} \\
\midrule
azure-identity, 1   & 14  & 39  & 68   & 13  & 13  & \texttt{OPTIMAL} \\
\bottomrule
\end{tabular}
\caption{Archived OR-Tools CP-SAT 9.15.6755 search results for the seven
components with independent Lean-checked LRAT lower bounds.  Primary component
indices are zero-based;
``\PaperNotationLink{PaperClasses}{C}3 max'' is the maximum-gap fully local Closed-\PaperNotationLink{PaperClasses}{C}3 subcomponent.  The
independent checker reparses each retained-laminar witness.  The archived
solver status is not a premise of exactness.}
\label{tab:python-cpsat-trust}
\end{table}

The random comparison swaps provider--class pairs in each project's recovered
requirements. It preserves each provider's class count and each class's provider count,
as well as $|F|$, $|C|$, and $|R|$.  Consecutive samples are separated by 20
accepted switches per incidence and use the fixed seed recorded in the report.
For each sample, disjoint local crossing witnesses give a verified lower bound;
the null run does not invoke CP-SAT and therefore does not label those bounds
as exact optima.  Across \PythonNullSamples{} samples, the observed aggregate
gap \PythonNullObservedTotal{} is below every aggregate null lower bound
(\PythonNullTotalMinimum--\PythonNullTotalMaximum). Samples follow the fixed swapping
procedure on recovered requirements; uniform sampling over all such graphs is not established.

\paragraph{Declared tolerance: a third cell value.}
A pair can be required, forbidden, or tolerated. A tolerated pair may be
inherited or omitted without charge; the gap is minimized over this declared
three-valued relation. Table~\ref{tab:python-specification-sensitivity} measures
overreach tolerance: the optimizer chooses up to $k$ extra classes per
implementation. Every specific declaration within that budget restricts this
choice, so its solver lower bounds apply to every declaration at the stated
budget. These bounds rely on the optimizer; the primary exact optima additionally
have independently checked replay or refutation certificates.\leanmeta{\LH{TOL5}, \LH{TOL6}}

Omission tolerance has no measured row. Its loss is nevertheless bounded:
let $T\subseteq R^{\mathrm{req}}$ mark recovered pairs that may be omitted, while
remaining legal to inherit. Write
\[
 d_T=\min_{\substack{\PaperNotationLink{PaperAncestry}{H}\subseteq \PaperNotationLink{PaperRequired}{R^{\mathrm{req}}}\\H\ \mathrm{tree\text{-}shaped}}}
       |(\PaperNotationLink{PaperRequired}{R^{\mathrm{req}}}\setminus \PaperNotationLink{PaperAncestry}{H})\setminus T|.
\]
Deleting $T$ from the legal supply relation is a different operation and can
raise the gap (main-text Proposition~5.1).

\leanmetapending{\LHrng{TOL}{1}{4}}
\begin{lemma}[Omission tolerance saves at most one answer per tolerated pair]
\label{omission-tolerance-gap}
For finite $R^{\mathrm{req}}$ and $T\subseteq R^{\mathrm{req}}$,
$d_T\le d_\downarrow(R^{\mathrm{req}})\le d_T+|T|$.
\end{lemma}
\begin{proof}
An optimal sound hierarchy for the original relation remains admissible, and
tolerance only removes charges, proving the first inequality. Conversely,
take a minimizing hierarchy $H$ for $d_T$. Its original missing pairs are the
charged misses and at most $|T|$ tolerated misses, so
$d_\downarrow(R^{\mathrm{req}})\le|R^{\mathrm{req}}\setminus H|\le d_T+|T|$.
Lean also checks attainment of the finite minimum and the inclusion of every
declared overreach choice in its budgeted search space.
\end{proof}
Thus reducing the certified original gap \PythonProviderOptimumTotal{} by $m$ through omission tolerance
requires declaring at least $m$ recovered pairs dispensable.\leanmeta{\LH{TOL7}}

\begin{table}[h]
\centering
\small
\PaperTableSetup
\begin{tabular}{@{}lrr@{}}
\toprule
Sensitivity & all projects & SQLAlchemy \\
\midrule
No specification slack ($t=0$) & \PythonOverreachAggregate{0} & \PythonOverreachSQLA{0} \\
\midrule
At most one extra consumer per role & \PythonOverreachAggregate{1} & \PythonOverreachSQLA{1} \\
\midrule
At most two extra consumers per role & \PythonOverreachAggregate{2} & \PythonOverreachSQLA{2} \\
\midrule
At most five extra consumers per role & \PythonOverreachAggregate{5} & \PythonOverreachSQLA{5} \\
\midrule
Unit-cost forwarding plus removal & \PythonPaidBlockingAggregate{} & \PythonPaidBlockingSQLA{} \\
\bottomrule
\end{tabular}
\caption{Specification and removal sensitivities.  Each interval gives a
solver lower bound and a checked supply-repair upper bound.  For unit removal cost,
the aggregate upper repair contains \PythonPaidBlockingForwarded{} forwarded
pairs and \PythonPaidBlockingBlocked{} removed pairs.}
\label{tab:python-specification-sensitivity}
\end{table}

\begin{table}[h]
\centering
\scriptsize
\PaperTableSetup
\setlength{\tabcolsep}{3pt}
\begin{tabular}{@{}lrrrrrrrrr@{}}
\toprule
Project & $|C|$ & impl.-MI & $|F|$ & $|R|$ & 1st parent & gap & gap/$|R|$ & max gap & $|C/{\equiv_b}|$ \\
\midrule
\PythonPopulationRows
\bottomrule
\end{tabular}
\caption{All 20 within-project no-MI counterfactuals for the MRO-invariant
provider relation.  The fact
granularity is one nominal provider bundle;
the generated first-parent column sums to \PythonFixedHierarchyTotal{};
the gap $d_\downarrow(\mathcal A)$ is the exact number of provider--consumer pairs that
must leave ancestry; the final column counts distinct answer columns on the
full consumer carrier, including the all-zero column.}
\end{table}

Nineteen separate-method relations cross; only aiobotocore stays zero in both
provider and method inputs. Solver-dependent method bounds
\PythonMethodSolverLowerTotal--\PythonMethodUpperTotal{} span
\PythonMethodSolverLowerPercent--\PythonMethodUpperPercent\% of
\PythonMethodIncidences{} method--class pairs.

\begingroup
\PaperTableSetup
\begin{longtable}{@{}lrrrr@{}}
\caption{Method-slot sensitivity by project.  The method gap is the pinned
solver optimum when the reported status is \texttt{OPTIMAL}; otherwise the
entry is the checked lower--upper interval.  Solver optimality remains inside
the declared trust boundary.}\label{tab:python-method-sensitivity}\\
\toprule
Project & provider gap & method pairs & method gap & status \\
\midrule
\endfirsthead
\toprule
Project & provider gap & method pairs & method gap & status \\
\midrule
\endhead
\bottomrule
\endfoot
\PythonMethodRows
\end{longtable}
\endgroup

Source extraction parses packages without importing or executing them.  The parser resolves
module-local declarations, aliases, imports, relative imports, and re-exports.
Two selected projects have $k=0$.  Syntactic MI can therefore coexist with a
tree-shaped provider relation.  A positive independently checked crossing
lower bound certifies nonlaminarity of the measured MRO-invariant relation;
FPT replay or Lean-checked LRAT refutations prove the provider-level optimum
magnitudes.
Undocumented maintainer intent, source lines, and runtime operations remain
outside these results.

\paragraph{Weighted and fixed-forest factorization objectives.}
Let $\mathcal A^\circ$ contain the distinct applicability scopes and let $m(S)$
count independently factorable roles with scope $S$.  Equal-scope roles
share an optimal retained scope, so the indexed objective is
\[
 \min_{\substack{S^{\rm nat}\subseteq S\ (S\in\mathcal A^\circ)\\
                 \{S^{\rm nat}:S\in\mathcal A^\circ\}\ \mathrm{laminar}}}
      \sum_{S\in\mathcal A^\circ}m(S)|S\setminus S^{\rm nat}|.
\]
This is the weighting used by the provider-bundle and independently factorable
method-slot audits. Facts needed by nonoverlapping groups of classes form independent
components, so $d_\downarrow=\sum_i k_i$ and the artifact's componentwise FPT
running time is
$\sum_iO(4^{k_i}|F_i|^2|C_i|)$.  If a candidate forest $T$ is fixed, each fact
instead chooses its largest fitting descendant set independently:
\[
 d^T_\downarrow(\mathcal A)=
 \sum_f\left(|\PaperNotationLink{PaperClassGroup}{A_f}|-\max_{\operatorname{desc}_T(v)\subseteq \PaperNotationLink{PaperClassGroup}{A_f}}
 |\operatorname{desc}_T(v)|\right).
\]
Once descendant sets are known, this takes $O(|F||T|)$.  Enumerating
relation-level forwarding stubs is then mechanical.  Their behavioral
integration, review, and validation remain outside this objective.  Choosing
the tree, including reparenting and factor nodes, is the global search below.
\leanmeta{\LHrng{NHF}{34}{41}}

For a finite required table, checking pairs of sets decides whether one parent
suffices.  Chen et al.'s deletion-flip problem uses the same complete $0/1$
character matrix and permits only removing present pairs.  Their M-free graph
condition says exactly that the required sets are tree-shaped; their
Theorem~2.2 gives this equivalence and Corollary~3.1 proves deletion-flip
NP-complete, already on complete inputs~\cite{chen2006minimum}.  This directly
classifies our deletion-only decision problem, without transferring hardness
from their separate two-direction minimum-flip problem.
A local crossing has four present pairs and two absent pairs.  Their
Theorem~5.1 branches on the four deletions for deletion-flip, or on all six
changes for minimum-flip.  With Gusfield's crossing finder, their deletion
algorithm runs in $O(4^{d_\downarrow}|F||C|)$.  The artifact instead checks
each role pair directly and scans its classes, giving
$O(4^{d_\downarrow}|F|^2|C|)$.  Lean proves that every deletion-only repair
must take one of the four branches.  Exhaustive search returns a minimum
tree-shaped result.

When every required set has size two and no two are identical, the full sets
retained by a repair form a matching: two distinct full sets sharing a class
cross.  Each remaining set can retain one class without creating a crossing.
Hence $d_\downarrow=|F|-\nu$, where $\nu$ is the maximum matching size of the
graph whose edges are the required sets.  Duplicate sets give edge weights
equal to their multiplicities and replace $\nu$ by the maximum matching weight.
These size-two inputs are polynomially solvable; the argument here is a
manuscript proof.

\section{Protocol Buffer Permanent-Key Audit}
\label{protobuf-audit}
Protocol Buffers puts an integer field number on the wire. Old data can still
contain that number after the field is deleted. Reserving it prevents a new
field from reusing it and changing the interpretation~\cite{ProtobufProto3}.
This audit follows deletions, reservations, and later reuse with a different
field descriptor. The required distinction outlives the visible field.

\paragraph{Worked decision: can field number 7 be reused?}
The audit unit is a fully qualified message name plus field number.  A Protocol
Buffer wire tag also includes wire type, but the no-reuse rule protects the
number across those encodings.  The persistent slot therefore uses message
name and field number because the no-reuse rule persists across wire-type changes. The audit must distinguish fields with different
names, types, cardinalities, or oneof groups.
The Temporal history contains the three
meanings in Table~\ref{tab:supp-protobuf-key-seven}.

\begin{table}[h]
\centering
\small
\PaperTableSetup
\begin{tabularx}{\linewidth}{@{}XX
  >{\Centering\arraybackslash}m{0.08\linewidth}
  >{\Centering\arraybackslash}m{0.14\linewidth}@{}}
\toprule
Historical meaning & message & field & certificate color \\
\midrule
\path{visibility_archival_uri} & \path{NamespaceConfig} & 7 & $\alpha$ \\
\midrule
\path{custom_search_attribute_aliases} & \path{NamespaceConfig} & 7 & $\beta$ \\
\midrule
\path{next_retry_delay} & \path{RespondActivityTaskFailedRequest} & 7 & $\alpha$ \\
\bottomrule
\end{tabularx}
\caption{An end-to-end reuse certificate from the pinned Protocol Buffer
history.  The schema-slot key is (message, field number).}
\label{tab:supp-protobuf-key-seven}
\end{table}

The first two meanings share \texttt{(NamespaceConfig, 7)}, so the pair test
inserts one conflict edge and requires different permanent keys.  The third
meaning uses integer 7 in another message, so its slot differs and it can reuse
color $\alpha$.  The shared slot and two-coloring certify an exact price of two
keys.  Protocol Buffers makes the equivalent commitment by reserving the
deleted number; the graph does not reserve every occurrence of integer 7 in
every message.

\begin{sloppypar}
The artifact compiles \ProtoSnapshots{} mainline snapshots from Temporal's
public API history~\cite{TemporalAPIRepository}.  Across \ProtoFieldSlots{}
message--number slots, it finds \ProtoDeletions{} deletions.  Only
\ProtoReservedImmediate{} reserve the number immediately;
\ProtoNeverReserved{} remain unreserved, and \ProtoReuses{} later return with
different field descriptors.  Those reuses form \ProtoReuseConflictEdges{} disjoint
conflict edges and require exactly \ProtoReuseConflictPrice{} permanent keys.
Reservation prevents those conflicts when the schema is edited.
\end{sloppypar}

\paragraph{Boundary.}
The audit uses qualified descriptors on \ProtoSnapshots{} mainline snapshots.
It establishes deletion, reservation, and descriptor-distinct reuse.  Runtime
impact would additionally require traffic, retained bytes, reader versions,
and semantic continuity across renames.

\section{TypeScript Published-Brand History Audit}
\label{typescript-brand-audit}

This audit asks how many stable keys keep published meanings distinct after
their brand markers are erased from the visible declarations. A conflict needs
two meanings that coexist and look alike in a sampled release; the number of
brand declarations alone does not determine the key minimum.

The population lock records the first \TSBrandPopulationPackages{} packages
returned by npm's public search endpoint for \path{keywords:typescript}, with
popularity weight one and quality and maintenance weights zero
~\cite{NpmRegistryAPI}.  For each package, the lock selects at most sixteen
stable releases at equal rank intervals, including the first and last release,
and records publication times, tarball URLs, and registry integrity hashes.
The lock contains \TSBrandPinnedSnapshots{} snapshots;
\TSBrandPopulationSnapshots{} were available and passed their recorded hash.
The unavailable snapshot is \path{@graphql-codegen/cli@1.21.0}; the audit does
not substitute another release.

For each snapshot, extraction starts from the declared TypeScript entry points
and follows relative declaration imports without importing or executing the
package.  The finite relation contains named public declarations with a
computed property keyed by a locally resolvable \texttt{unique symbol}.  The
visible profile is exact normalized declaration syntax after deleting those
marker properties.  Two persistent meanings are adjacent when some sampled
release contains both with the same visible profile.  Thus the measured price
is the history graph's chromatic number: only conflicting declarations need different keys.

Within each package, the persistent key is the normalized module path,
declaration kind (\texttt{type} or \texttt{interface}), and declaration name.
Path normalization removes the package prefix, a leading
\texttt{dist/esm/}, \texttt{dist/cjs/}, \texttt{build/esm/},
\texttt{build/cjs/}, \texttt{lib/esm/}, or \texttt{lib/cjs/}, and the
source or declaration suffix.  Reusing this key retains the same history
vertex; a rename or move that changes it creates a new vertex.  Marker names
are recorded separately.  This convention fixes the sampled identity relation;
the audit does not infer semantic continuity across renamed declarations.
Marker resolution uses local declarations and unambiguous names in the
traversed declaration files.  Conflicting records with the same key in one
snapshot are excluded.

Sampling cannot inflate that price.  Precomposing a declared history with a
sampling map only deletes possible witnesses: its conflict graph is a subgraph
of the full-history graph, and every key assignment adequate on the full
history remains adequate on the sample.  The reported values are therefore
exact for the pinned sampled relation and lower bounds for the packages' full
published histories.
\leanmeta{\LHrng{EVO}{38}{39}}

Runtime evidence records whether the shipped program retains the marker beyond
its declarations. A meaning has verified runtime
backing only when shipped non-declaration TypeScript or JavaScript both
allocates the named symbol and uses it as a computed runtime property.  Of the
\TSBrandPersistentMeanings{} extracted meanings, \TSBrandRuntimeMeanings{} meet
that test and \TSBrandNoRuntimeEvidenceMeanings{} have no such evidence in the shipped
artifact. The scan records shipped marker evidence; a runtime failure concerns
what a reachable execution actually does and needs its own behavioral evidence.

Table~\ref{tab:typescript-brand-history} lists every package with an extracted
meaning.  Each of its \TSBrandExactGraphs{} history graphs has a proper coloring
and a clique of the same size.  Their raw package-local alphabet sizes sum to
\TSBrandChromaticPriceSum{} and have maximum \TSBrandChromaticPriceMax{}.
Because every package owns an independent alphabet and one constant value is
free, the normalized additive price is
$\sum_p(\chi_p-1)=\TSBrandNormalizedNamePrice{}$. Each package needs its own
alphabet; a disjoint-union coloring instead reuses colors across packages.
The subtraction counts one free constant per package, while the
\TSBrandPersistentMeanings{} meanings supply the graph's vertices.
Across the audit, \TSBrandHistoryOnlyEdges{} conflict edges in
\TSBrandHistoryOnlyPackages{} packages occur only before the latest sampled
release; \TSBrandHistoryPriceIncreasePackages{} raw alphabet sizes exceed their
latest-snapshot prices.  Positive additional-name prices occur in
\TSBrandPositiveAdditionalPricePackages{} packages; conflicts remain in the
latest release of \TSBrandLatestConflictPackages{}, and historical states
account for \TSBrandHistoryAdditionalNamePrice{} of the
\TSBrandNormalizedNamePrice{} additional names.

\begingroup
\small
\PaperTableSetup
\begin{longtable}{@{}>{\RaggedRight\arraybackslash}m{0.25\textwidth}
  >{\Centering\arraybackslash}m{0.11\textwidth}
  >{\Centering\arraybackslash}m{0.08\textwidth}
  >{\Centering\arraybackslash}m{0.08\textwidth}
  >{\Centering\arraybackslash}m{0.10\textwidth}
  >{\Centering\arraybackslash}m{0.06\textwidth}
  >{\Centering\arraybackslash}m{0.12\textwidth}@{}}
\caption{Pinned TypeScript brand-history receipt.  Prices are exact for the
sampled package-local graphs.  A latest price of zero means that the latest
sampled release exposes no meaning in the extracted relation; it does not erase
earlier conflicts from the history contract.}
\label{tab:typescript-brand-history}\\
\toprule
Package & \shortstack{sampled/\\pinned} & meanings & runtime &
\shortstack{no runtime\\evidence} & edges & \shortstack{$\chi$ history/\\latest} \\
\midrule
\endfirsthead
\toprule
Package & \shortstack{sampled/\\pinned} & meanings & runtime &
\shortstack{no runtime\\evidence} & edges & \shortstack{$\chi$ history/\\latest} \\
\midrule
\endhead
\TSBrandAuditRows
\bottomrule
\end{longtable}
\endgroup

The size-three certificate occurs in \path{@sinclair/typebox}.  Release
\texttt{0.24.22} places \texttt{TOptional}, \texttt{TReadonly}, and
\texttt{TReadonlyOptional} in the same erased profile; the three pair witnesses
form a clique, and the stored three-coloring matches it.  In \path{type-fest},
\texttt{MultiTagContainer} and \texttt{TagContainer} conflict in release
\texttt{4.3.3}, although the latest sampled release has price one.  These are
history witnesses: their archived earlier releases contain the conflicting declarations.

The extracted relation excludes string and enum brands, private-member nominal
origins, unexpanded generic instantiations, nonlocal or ambiguous symbol aliases,
and unsampled releases.  Those exclusions can remove meanings and graph edges;
they cannot make the displayed certificates invalid for the retained relation.
The artifact stores every release/profile edge witness and checks the population
lock and available tarball hashes, re-extracts every retained snapshot, rebuilds
the relations, and verifies the colorings and cliques.

\section{Go Direct-Method History and Deletion Counterfactual}
\label{go-direct-method-history}

Go keeps named types distinct while interface satisfaction follows method
sets~\cite{GoSpec}.  Frozen API ledgers expose Go~1.0--1.26 without rebuilding
old toolchains~\cite{GoAPILedgers}.  On the cumulative nonempty direct-method
population of 395 value forms and 759 pointer forms, the checked historical
union graphs require 41 permanent key values (40 extra names) for values and
76 (75 extra names) for pointers. Each audit uses one pooled decoder namespace:
package-qualified type names identify required answers, but observations contain
only direct-method signatures. Interface values can cross package boundaries,
so this decoder uses a global alphabet, unlike the TypeScript package-local
decoders~\cite{GoSpec}. Thus each extra-name count subtracts one shared
key because this decoder pools interface values across packages. Retaining package paths in the observation would define
a different graph. These
history-wide key counts equal the worst-release values. The ledgers do not reconstruct method
promotion, so the history audit is limited to direct methods.

Identity and implementation supply have separate required relations.
Embedding natively derives the field and promoted-method
incidences it carries; ambiguous or unembedded incidences still require
forwarding or another external authority.  Structural interface satisfaction
separately supplies no nominal provider.  The two prices accumulate on the
relations for which each capacity is absent.

The Go~1.26.4 snapshot audit treats complete method sets as profiles.  The empty
method set is a genuine profile; the nonempty columns isolate more informative
collisions.

\begin{table}[h]
\centering
\small
\PaperTableSetup
\begin{tabular}{@{}lrrrrrr@{}}
\toprule
Form & entities & profiles & $A_\pi$ & bits & nonempty $A_\pi$ & bits \\
\midrule
named value & 2461 & 447 & 1751 & 11 & 59 & 6 \\
\midrule
named pointer & 2218 & 994 & 782 & 10 & 75 & 7 \\
\bottomrule
\end{tabular}
\caption{Method-set fibers in the audited Go~1.26.4 standard library.
``Bits'' is $\lceil\log_2 A_\pi\rceil$ when the method set remains visible.}
\end{table}

The deletion counterfactual uses terminal direct-method profiles, requiring
41 keys for values and 68 for pointers. The complete-method profiles above
include additional inherited methods and therefore define a different observation.
Hiding a method removes that observation
from every fixed profile.  Allowing at most $k$ hidden methods therefore gives
the future-conflict graph under the stated downward-closed policy and
multiplicity cap.  One hidden method preserves both minima; two raise them to
\GoAdversarialValueTwo{} and \GoAdversarialPointerTwo{}.

\begin{center}
\begin{tabular}{@{}lrrrrrr@{}}
\toprule
Form & current & $k\le1$ & $k\le2$ & $k\le5$ & $k\le10$ & any $k$ \\
\midrule
named value
  & \GoAdversarialValueZero
  & \GoAdversarialValueOne
  & \GoAdversarialValueTwo
  & \GoAdversarialValueFive
  & \GoAdversarialValueTen
  & \GoAdversarialValueAll \\
\midrule
named pointer
  & \GoAdversarialPointerZero
  & \GoAdversarialPointerOne
  & \GoAdversarialPointerTwo
  & \GoAdversarialPointerFive
  & \GoAdversarialPointerTen
  & \GoAdversarialPointerAll \\
\bottomrule
\end{tabular}
\end{center}

The values are reproduced from the generated artifact macros.  Each entry has
a weighted clique and entity-level coloring of equal size.  At $k=2$, the
value certificate contains \GoAdversarialValueTwoCliqueClasses{} profile
classes and the pointer certificate contains
\GoAdversarialPointerTwoCliqueClasses{}; every selected profile exposes one
method.  Distinct selected profiles therefore differ in exactly two
coordinates, and the edge verifier checks the multiplicity condition for each
canonical two-method deletion.

\section{What the Proofs and Executable Checks Establish}
\label{formalization-boundary}

\paragraph{Answer counts do not depend on a source-length convention.}
Take $m$ disjoint copies of the three-consumer crossing witness.
Additivity gives $d_\downarrow=m$, yet one indexed rule computes the residual
predicate for every copy; only its parameter grows.  A description language
can also install that predicate as a primitive of constant or zero charged
length.  Program-length bounds must therefore fix a language and size measure.
The answer-column quotient avoids that arbitrary choice: it is determined
solely by the visible input and required answer.
\leanmeta{\LH{BWR8}}

\paragraph{What single-omission checks must exercise.}
Let $G$ be a finite set of displaced dependencies.  Suppose each dependency
can be present or absent independently.  The grammar then contains exactly
$2^{|G|}$ omission configurations, and only the complete set contains every
required dependency.  For each $g\in G$, the single-omission mutant is
$G\setminus\{g\}$.  Any exercised set that detects every such mutant equals
$G$.  Thus complete single-omission coverage exercises all $|G|$ dependencies,
while one test may exercise several dependencies.
\leanmeta{\LHrng{BWR}{19}{21}}

Agreement in one complete routing setup is weaker than preserving answers
through every supported change. The latter quantifies over every reachable state and every
configuration reachable on either side; it therefore implies pointwise
agreement.  For residual routes it holds iff every incomplete
topology is unreachable.
\leanmeta{\LHrng{BWR}{37}{41}}

Let $E$ index semantic maintenance operations, and let
$\operatorname{touches}(e)\subseteq G$ contain the dependencies established or revised
by operation $e$.  If every operation touches at most $b>0$ dependencies and their
union covers $G$, then
\[
 |E|\ge \left\lceil\frac{|G|}{b}\right\rceil.
\]
Combining this finite-cover bound with the ancestry-gap optimum gives at least
$\lceil d_\downarrow/b\rceil$ operations for every rewrite preserving required supply.  If each
operation changes one dependency, $b=1$.  An unrestricted edit may have
$b=|G|$ and satisfy the bound with one operation.  This operation bound says
nothing about source length.
\leanmeta{\LHrng{BWR}{22}{23}, \LH{BWR8}}

\subsection{Rules can supply future pairs without new manual wiring}
\label{sec:rule-based-factoring}
A class-creation rule can generate bases or attributes.  A lookup rule can
compute an attribute on demand~\cite{PythonDataModel}.
An unchanged rule can handle later classes and updates from their declared
inputs.  Count edits to that rule, changes to the generated graph, and changed
required answers separately.  The staged-history theorem fixes classes and
code providers: its $n$ multi-parent additions remain $n$ graph changes even when
one generator update produces them all. Generation shares the human edit
across those changes while the graph ledger records each edge. Rule
evaluation, validation, and runtime costs enter their own comparison.

The finite-relation ladder already ends with a mechanism that realizes every
required relation.  Rules add a second question: how often must the source
change while those relations grow?  Give generated ancestry and programmable
lookup the same rule inputs.  For a finite provider set, a terminating rule that
decides an arriving class's required providers can generate its direct provider
edges.  The construction preserves distinct provider identities even when their
current requirements agree.

\leanmetapending{\LHrng{RBF}{1}{2}}
\begin{proposition}[One rule can generate exact ancestry at every input]
\label{prop:rule-generated-ancestry}
Let $g(s,x)$ give the required provider--class relation from source $s$ and input
$x$.  One fixed compiler applies the provider-preserving direct construction to
$g(s,x)$, deriving exactly that relation at every input.  If the compiled
program retains $s$, compilation preserves source-change events along every
finite history.  Finite provider and class sets give finite generated graphs.
\end{proposition}
\begin{proof}
Use disjoint provider and class nodes, with a provider above a class exactly when
$g(s,x)$ requires the pair.  Retaining $s$ makes the compilation
injective in the source, so two consecutive compiled programs differ in source
exactly when their sources differ.  Their generated graphs can change as $x$
changes while $s$ stays fixed.
\end{proof}

The Lean result checks the relation construction and source-change equality.
For a decidable finite table, evaluating its entries materializes the graph.
Python MRO, initialization, evaluation time, and refresh costs require their
own checks.  Programmable lookup adds no sixth finite-relation rung: both
mechanisms already derive every finite relation.  For an unbounded family,
compare a lookup rule with an ancestry generator that can create its future
classes, rather than with a graph forbidden to grow.

\subsection{When finite lookup code can generate the complete provider list}
\label{finite-rule-generation}
A program can look up one code provider without first finding every provider
the class needs.  Complete generation has a stronger stopping requirement:
it must return the whole list, including an empty list when none is required.
Fix the same class input on both sides.  Number possible providers by natural
numbers and let $L(x,p)$ be a total computable Boolean rule saying whether class
input $x$ requires provider $p$.  Computability here means execution by a finite
program that terminates on every admitted input; no running-time or source-size
bound is imposed.  A generator is exact when it returns a finite list $G(x)$
such that $p\in G(x)$ iff $L(x,p)=1$.

\leanmetapending{\LHrng{FDC}{1}{3}}
\begin{proposition}[Exact condition for complete provider-list generation]
\label{prop:finite-rule-generation}
A total computable exact generator exists iff a total computable bound $b(x)$
exists with
\[
  L(x,p)=1\ \Longrightarrow\ p<b(x).
\]
A fixed finite universe of possible providers supplies such a bound, uniformly
over all future class inputs.  Conversely, membership in any computably
generated complete list is a total computable lookup rule.
\end{proposition}
\begin{proof}
Given $b(x)$, test $L(x,p)$ for each $p<b(x)$ and return the successful indices.
Every test terminates and the bound covers every required provider.  Given an
exact generator, take one plus its largest returned index, or zero for an empty
list.  This is a computable bound.  A fixed finite provider universe has a
constant bound.  For the converse, generate the finite list and test membership.
The same input $x$ drives each construction.
\end{proof}

The bound concerns which provider indices may occur.  Knowing only how many
providers a class needs does not suffice, even when that number is at most one.

\leanmetapending{\LHrng{FDC}{4}{8}}
\begin{proposition}[One possible answer per class still need not be listable]
\label{prop:singleton-provider-separation}
There is a total computable lookup rule whose required provider set is empty
or a singleton for every input, but no total computable exact generator for
those sets.
\end{proposition}
\begin{proof}
Let the class input be a program $e$.  Bounded universal evaluation decides
whether $e(0)$ has succeeded within fuel bound $p$.  Define $L(e,p)=1$ when
evaluation first succeeds between bounds $p$ and $p+1$.  Each query runs two
bounded evaluations and terminates.  Success persists as fuel increases, so
there is at most one such $p$.  A provider exists exactly when $e(0)$ halts.
An exact terminating list generator would decide halting by testing whether
its output is empty, contradicting the halting theorem~\cite{turing1936computable}.
The Lean proof imports that undecidability theorem and checks the bounded
lookup, uniqueness, halting equivalence, and contradiction.
\end{proof}

Thus finite lookup code can describe an unbounded family of finite required
sets that no terminating complete-list generator computes uniformly.  Each
individual set still has a finite ancestry representation.  The obstruction
is producing the complete representation from the class input before use.
Allowing ongoing enumeration or delegating unanswered queries to executable
lookup changes that obligation.  The result compares total algorithms for
the same relation; it supplies neither a source-length advantage nor a new
rung in the finite-relation ladder.  It also leaves the Python runtime checks
below distinct from the computability proof.

Python descriptors can themselves be inherited.  Their lookup-time computation
does not imply an independently written route in each class.
\texttt{\_\_getattribute\_\_} customizes ordinary instance attribute access;
implicit special-method lookup can bypass it~\cite{PythonDataModel}.
Here $H$ records ancestry, so computed routes absent from $H$ remain in
$R^{\mathrm{req}}\setminus H$ even when one rule supplies them all.  A comparison that
includes programmable lookup in its measured mechanism must recompute $H$
and the outside count.  In either comparison, a rule needs inputs that
distinguish every required answer; it cannot anticipate an independently
changed requirement from unchanged inputs.

\paragraph{Python correspondence.}
The artifact's \path{nominal_factorization/rule_based_factoring.py}
uses a separate executable example with two workflow capabilities: selecting
items and choosing their destination. One family rule supplies both.
Each arriving class has two declared Boolean requirements.  The rule chooses
the corresponding providers; each provider owns one shared descriptor value.
Table~\ref{tab:rule-based-python} compares four executed routes through that
same rule.  The metaclass is also an ancestry generator, so the routes overlap
in their mechanisms.

\begin{table}[htbp]
\centering\PaperTableSetup
\begin{tabularx}{\linewidth}{@{}>{\RaggedRight\arraybackslash}p{0.19\linewidth}YY@{}}
\toprule
Route & How it supplies the required attributes & What changes for a new class \\
\midrule
Metaclass & \texttt{FamilyMeta} computes provider bases during class creation. & The generated base list; the family rule stays fixed. \\
\midrule
Generated ancestry & A factory passes the same computed bases to \texttt{type}. & The same base list, without a metaclass hook. \\
\midrule
Bound descriptors & A factory binds the selected shared descriptors in otherwise unrelated classes. & Generated attribute bindings; the descriptor providers stay fixed. \\
\midrule
Programmable lookup & \texttt{\_\_getattribute\_\_} uses the class's declared requirements to read the selected provider. & A new class and its request; no provider bases or per-attribute bindings. \\
\bottomrule
\end{tabularx}
\caption{One family rule, four Python routes. All four preserve separate class
identities and use the same descriptor providers. Class creation and its required
input still occur; an unchanged rule removes repeated wiring edits.}
\label{tab:rule-based-python}
\end{table}

The tests cover all four requirement combinations, absent attributes,
independent provider updates, and 256 new class identities per route without
a rule change.  In the first two routes Python inherits the descriptors;
the third binds those same objects directly.  Consumer writes are rejected,
keeping each value at its provider.  A separate test shows that a callable
returned for an explicit \texttt{obj.\_\_len\_\_()} lookup need not make
\texttt{len(obj)} work.  These executions check ordinary read-only attributes
with nonconflicting names and immutable class requirements.  Changing an
existing class's requirements requires refreshing a creation-time result;
on-demand lookup instead reevaluates whichever inputs its rule reads.
The executions accompany the abstract Lean results; they do not mechanize
the complete Python data model.

Two further executed variants read changing instance requirements.  One
inherits conditional descriptors; the other uses \texttt{\_\_getattribute\_\_}.
Both switch attribute availability without changing class identity or bases.
The descriptor code is inherited throughout; its predicate decides whether
the requested value is available.  Code reach and value availability are
different queries.  A creation-time ancestry example needs regeneration to
follow the changed requirement, and recreating its class changes identity.
Proposition~\ref{prop:rule-generated-ancestry} preserves abstract provider names;
it supplies no live-object migration procedure.

For chosen $R^{\mathrm{req}}$ and a grouping of independently changing implementations,
refactoring cannot change $d_\downarrow$: separate repairs and one indexed rule
for $m$ disjoint crossings both leave $m$ answers outside ancestry.
\leanmeta{\LHrng{BWR}{1}{2}, \LHrng{BWR}{8}{9}, \LH{BWR12}, \LHrng{SCT}{24}{25}, \LHrng{SDN}{13}{17}}

\subsection{Exactly when the rule must change}
\label{exact-rule-revisions}
Suppose one input appears twice but needs different answers.  Keeping the
same rule throughout that interval is impossible.  The converse also holds
on a finite declared history: put a rule replacement inside every such interval,
and each remaining segment admits one lookup function.

\leanmetapending{\LHrng{RBF}{3}{6}}
\begin{proposition}[Exact rule-revision count on a finite history]
\label{prop:exact-rule-revisions}
Fix inputs $x_0,\ldots,x_n$ and required answers $a_0,\ldots,a_n$ in a nonempty
answer space.  Allow arbitrary lookup functions $r_t$ with
$r_t(x_t)=a_t$.  Count a revision at $t$ when $r_{t+1}\ne r_t$.
For each $i<j$ with $x_i=x_j$ and $a_i\ne a_j$, form the interval of possible
revision positions $\{i,\ldots,j-1\}$.  The minimum revision count equals the
fewest positions meeting every such interval.  In particular, one unchanged
rule suffices exactly when equal inputs always require equal answers.  If
every consecutive pair has equal inputs and different answers, the minimum
is exactly $n$.
\end{proposition}
\begin{proof}
Without a revision between $i$ and $j$, their rules are equal, contradicting
the two different answers to the same input.  Conversely, cut at any set of
positions meeting every conflicting interval.  Within each segment, equal
inputs have equal answers; map each observed input to its required answer and
give unobserved inputs an arbitrary default.  Use that function until the next
cut.  This proves both the allowed-position criterion and equality of the
minimum counts.  In the last case every update position is required; storing
the current answer in a constant function attains $n$.
\end{proof}

The minimum permits a finite lookup table built with the whole declared
history available.  Distinct inputs can therefore give zero revisions even
when that initial table is large.  The result bounds changes to the function;
it does not promise a short program or predict unseen requirements.  A real
fixed family rule, as in the Python examples, supplies the stronger evidence
that later answers follow from already declared information.  Adding the
update number to the input alone does not supply that information for free.

In the operation-cover model above, a fewest-edit repair cannot contain a
dispensable edit. Every edit is therefore
the sole source of one dependency.  Remove that edit and the dependency is uncovered.  The
one-edit omission plans are pairwise distinct, so a fewest-edit repair with
$m$ edits has exactly $m$ distinct one-edit omission failures.
\leanmeta{\LH{BWR24}}

For a fixed residual predicate $b$, $C/{\equiv_b}$ is the least decoder state;
different optimal repairs can induce different $b$.  Applying the construction
to the complete relation gives repair-independent profiles $a_c\subseteq F$.
Put $[c]\sqsubseteq[c']$ when $a_c\subseteq a_{c'}$; every answer-preserving
hidden-consumer encoding preserves enough decoded answers to recover this
order.  Moreover, $\{f\}\subseteq a_c$ exactly when $(f,c)\in R$, recovering
the canonical incidence matrix.
\leanmeta{\LHrng{BWR}{14}{18}}

Several class hierarchies realize this order. Ancestry supplying exactly the required pairs can chain
providers of equal or nested scopes and add harmless internal comparabilities.
It natively \emph{factors} $R$ through reachability.  The typed direct
multi-parent construction realizes every matrix; a one-parent
factorization exists iff the scopes are laminar.  The factored relation and
optimal residue $d_\downarrow$ are canonical.  Concrete hierarchies can refine
that information object.  An infinite relation has a finite implementation
only through a finite presentation or generator, which then owns the answers
it produces.  Open-ended systems instead form unbounded sequences of finite
states.

\subsection{Checked kernels}
\label{checked-kernels}
The modules divide the proof work into identity, implementation supply, and
shared infrastructure. Their prefixes identify those responsibilities.
\texttt{Paper1IT} contains identity-over-time kernels,
\texttt{Paper1PL} contains the OOPSLA-facing language and factorization
kernels, and \texttt{AbstractClassSystem} supplies shared finite observation
and computability infrastructure.
\texttt{EvolvingObservation} proves the future-conflict equivalence, including
the arrival-aware sentinel totalization onto a persistent carrier, greedy
backdegree allocation, refinement preservation, terminal-coarsening control,
and the adaptive pair merge used by the coarsening lower bound.
\texttt{HistoryViewDominance} proves retained-name dominance across every
indexed history, strict collision bounds, helper non-recovery, and normalized
identity recourse, using the existing boundary-intent answer-preservation contract.
\texttt{FieldMenus} proves the coordinate-and-parity field count, initial
injectivity, exact direction-pair branch fibers, and complete union conflict
graph.  The probabilistic $O(k)$ argument remains manuscript-level.

\texttt{OneDeletionStability} proves the exact tight-contract result using a
constructive parity-interval coloring of weighted current profile classes.
\texttt{DelegationCompression} separates fact-aware forwarding from
fact-oblivious holders, translates the latter exactly to and from all-ones
rectangle covers, and injects every fooling family into the active holder
population.
\texttt{BehavioralRewrite} proves the stronger quotient statement used by
the main text's replacement-state result: when a decoder receives a fact and one residual code while the
nominal consumer stays hidden, equality of complete answer columns is the unique coarsest
answer-preserving identification.  It also proves that every single-parent rewrite preserving required supply
has ancestry gap at least $d_\downarrow$ even when its residual answers are
computed intensionally, and gives the formal counterexample to a
language-independent source-length bound.  It also checks the canonical
behavioral answer order, singleton incidence test, and fixed-description-model minimality claim
for the declared model.
\leanmeta{\LHrng{BWR}{1}{18}}

\texttt{RepresentationCanonicity} groups states that permit exactly the same
valid actions; this is the optimizer quotient. It proves that an encoding
preserving all valid-action answers must distinguish any two different groups.
For arbitrary binary answer columns, every
answer-preserving projected carrier likewise refines the canonical column quotient;
minimal answer-preserving carriers induce the same partition, support the same
fiber-saturated answers, and have equivalent represented-state quotients.
\texttt{CarrierNonDichotomy} then separates the required possession relation
from its carrier.  In an exact native representation, required implementation applicability
(``should have'') is equivalent to implementation ancestry (native ``has'').
In an exact routed representation, the same membership is equivalent to native
ancestry or an explicit route.  Exact native and routed carriers therefore
agree pair by pair, and an ancestry witness and the corresponding possession
hold simultaneously. More directly, ancestry derives supplied applicability by
the native branch of the supply mechanism, without using exactness. On every
inherited pair, declaring ancestry and possession mutually exclusive therefore
contradicts that derivation. This is the category error in ``is-a versus has-a'';
nominal membership and behavioral substitution are different relations.
\leanmeta{\LHrng{CND}{1}{7}}
\texttt{SemanticCompletionTax} realizes a complete implementation requirement
set as one valid action and proves that its optimizer quotient is exactly
equality of required-provider columns.
\leanmeta{\LHrng{RCN}{1}{11}, \LH{SCT18}}

\texttt{SemanticCompletionTax} also supplies the elementary positive-atomic
consequence relation, proves exact native supply equivalent to soundness plus
completeness, and
injects every native deficit into any outside workaround preserving exact supply.  Its finite
specialization preserves duplicate provider scopes and equates the least
auxiliary decision count, the indexed laminar deletion minimum, and the
independently defined universal delegation minimum.  It checks the strict
contract-only-to-one-parent improvement on every nonempty relation, the
provider-preserving multi-parent/zero endpoint, and translates
the dynamic signed repair bounds into native deficit and excess.
\leanmeta{\LHrng{SCT}{1}{32}}

\texttt{NominalFactorization} instantiates the observation, identity, ancestry,
precedence, and population axes.  It checks structural collapse, the
baseline-relative comparison, independence of nominal node identity from
ancestry, injectivity of answer-preserving providers on distinct applicability sets, both
directions of the ancestry-only/laminarity theorem, canonical multi-parent
derivation, nonempty external repair for every nonlaminar family, and
provider-sensitive coloring of repair records.  It
derives a strict total provider order from the stated choice axioms and, under
the separate ancestry-compatibility premise, proves that the order extends
reverse ancestry.  Applicability is represented extensionally as a set of
distinct consumer sets.  Normalizing equal scopes and restoring their
multiplicities as weights preserves the exact objective; the executable audit
expands the weighted witness back to typed provider and method facts.
\leanmeta{\LHrng{NHF}{35}{40}}

\texttt{ProductResolution} checks the commutation criterion for provider-first
and scope-first lookup, key-column noninterference, and invariance under a
declaration-empty nominal wrapper.  It retains the selected provider--scope
site with provider and scope orders fixed.  A wrapper that recomputes \PaperNotationLink{PaperClasses}{C}3 uses
a different contract.  \texttt{IdentityDiscipline} checks the fiber-union criterion, and
\texttt{BarrierComplexity} checks finite collision search, the total-observer
clock construction, and the many-one reductions.  The
arithmetical-hierarchy names and Python's complete \PaperNotationLink{PaperClasses}{C}3 algorithm are outside the
Lean boundary.

\texttt{PartitionStateThreshold} checks that one-state tag classes are
independent in the induced partition matroid, that two-state adequacy is common
independence and hence the matching condition in the bipartite incidence
multigraph, and that the three-state four-declaration witness has exact
snapshot price two and union lower bound four.  The general equality
$\chi'(H)=\Delta(H)$ for bipartite multigraphs is cited as external K\H{o}nig background.
Lean proves the reduction through the bipartite
matching condition but does not import that equality, so no numbered result's
handle depends on it.

\begin{center}
\small
\PaperTableSetup
\begin{tabular}{@{}lll@{}}
\toprule
Priced resource & Upper witness & Lower witness \\
\midrule
Permanent tags & proper coloring & clique \\
\midrule
Delegation decisions & laminar retained family & disjoint local conflicts \\
\midrule
Fact-oblivious holders & rectangle cover & fooling family \\
\bottomrule
\end{tabular}
\end{center}

\subsection{Executable audits}
\label{executable-audits}
The Go programs reconstruct snapshot and cumulative direct-method profiles,
retain departed endpoints, build the union graphs, and export proper
colorings.  The adversarial companion evaluates seven deprecation budgets and
exports profile classes, admissible edges, weighted cliques, and entity
colorings; a standard-library-only verifier rebuilds every edge and checks the
matching certificates.  In the Python study, retained-laminar families are
directly checked upper witnesses, bounded FPT replay supplies independent exact
lower bounds through $k_i\le10$.  Lean-checked LRAT refutations supply exact
lower bounds for the \IndependentGapCertificateCount{} larger primary or
Closed-\PaperNotationLink{PaperClasses}{C}3 components and the \IndependentHolderMinimum{}-holder cover.
The Protocol Buffer miner asks \texttt{protoc} for
descriptors at every measured mainline commit and separates deletion,
reservation, reuse, redefinition, and message removal.  The Python audit
resolves the 20 pinned public class graphs selected by the locked population
screen, computes the minimized single-parent gap component by component, and
emits independently checked
laminar-repair upper witnesses together with local-conflict, rectangle-cover,
and fooling-family certificates.  The TypeScript audit parses declaration
closures without executing packages, retains release/profile witnesses for
every history edge, and checks matching coloring and clique certificates while
keeping runtime-backed and declaration-only evidence separate.  All four
studies pin their source artifacts by repository revision or registry hash and
export machine-readable records, generated \LaTeX{} counts, and focused tests.

\section{Representative Language Encodings of the Factorization Gap}
\label{language-encodings}

Language mechanisms answer several different questions: whether an object
exposes a member, which declaration it denotes, where an implementation came
from, which declarations satisfy a contract, and which definition wins.
The same-looking-group criterion applies separately to each question below.

\begin{center}
\small
\PaperTableSetup
\begin{tabularx}{\linewidth}{@{}>{\raggedright\arraybackslash}m{0.15\linewidth}XXX@{}}
\toprule
Capability & Example query & Identity-dependent? & Required information \\
\midrule
Conformance & Does $x$ expose method $m$? & No, if constant on profile fibers & structural profile \\
\midrule
Identity & Is $x$ the declaration selected earlier? & Yes for structural clones & nominal handle \\
\midrule
Provenance & Which declaration supplied $m$? & Yes & nominal provider plus ancestry \\
\midrule
Enumeration & Which declarations implement $T$? & Yes in an open population & nominally scoped registry \\
\midrule
Resolution & Which definition wins? & Yes when clones have different providers & qualification, rejection, merge, or precedence \\
\bottomrule
\end{tabularx}
\end{center}

\begingroup
\setlength{\emergencystretch}{3em}
For implementation supply, ask whether the native mechanism records which
code provider supplies each class. Ancestry retains that provenance through its paths.
Surface syntax does not add a rung.  A trait, module, or effect system that
retains several provider edges is constrained multi-parent ancestry; a checked
\texttt{impl} table, registry, embedded-field remainder, generator, or forwarding
layer that stores individual incidences outside that graph is an outside
workaround.  Checking can establish exact supply, but it does not erase
the answers it owns: one-parent ancestry still leaves at least $d_\downarrow$,
and no implementation ancestry leaves $|R^{\mathrm{req}}|$.
At a crossing, preserving required supply and behavioral correctness still leaves a positive outside count
under sound one-parent ancestry, so it does not attain correct maintenance.
Formally, soundness plus an
empty native deficit forces the carrier's native scope to equal $R^{\mathrm{req}}$.
The main paper's complete-supply characterization gives a provider-preserving
multi-parent presentation of the same relation. Different graphs can realize
that relation; no unique graph topology is claimed. A carrier
that computes incidences outside native provenance is therefore a separate
workaround authority: it still owns the required incidences absent from native provenance.
\leanmeta{\LH{SCT37}, \LH{RCN2}, \LH{SCT18}, \LH{NHF50}}
The class and trait entries follow their language references
~\cite{CppWorkingDraft,EiffelStandard,CommonLispMOP,
PythonDataModel,barrett1996c3,Scala3Traits,Scala3TraitParameters,
Scala3TypeClasses,RSetClass2026}.  The interface and protocol entries follow their own
specifications~\cite{JavaLanguageSpecificationInterfaces,CSharpInterfaces,
KotlinInterfaces,RustTraitsReference,SwiftProtocols,
SwiftMultipleInheritance,GoSpecification,TypeScriptDocs,TypeScriptSymbols,Haskell2010Report}.
\par
\endgroup

\begingroup
\PaperTableSetup
\setlength{\tabcolsep}{3pt}
\begin{longtable}{@{}>{\raggedright\arraybackslash}m{0.11\linewidth}>{\raggedright\arraybackslash}m{0.18\linewidth}>{\raggedright\arraybackslash}m{0.18\linewidth}>{\raggedright\arraybackslash}m{0.16\linewidth}>{\raggedright\arraybackslash}m{0.22\linewidth}@{}}
\caption{Representative native carriers for implementation applicability.
These carriers locate the coordinatewise correct-maintenance capacities in
the main-text mechanism table; precedence and separate
answers remain part of each placement.}\label{tab:supp-language-carriers}\\
\toprule
Mechanism & Native applicability carrier & Code and state supplied & Precedence or constraint & Separate relation \\
\midrule
\endfirsthead
\toprule
Mechanism & Native applicability carrier & Code and state supplied & Precedence or constraint & Separate relation \\
\midrule
\endhead
\PaperNotationLink{PaperClasses}{C}++ classes & Multiple nominal bases & Virtual bases share one base subobject; nonvirtual bases retain distinct subobjects & Final-overrider rules, ambiguity rejection, and explicit qualification & The graph is native; shared-state identity is chosen per base edge. \\
\midrule
Eiffel classes & Multiple nominal parents & Features and state from several parents & \texttt{rename}, \texttt{redefine}, and \texttt{select} adapt conflicts explicitly & Retained parent clauses carry derivation; adaptation records only genuine clashes. \\
\midrule
CLOS classes & Multiple direct superclasses & Slots and generic methods are organized separately & Class precedence plus method selection and combination & The class graph retains membership; generic-function methods retain provider applicability. \\
\midrule
Python classes & Multiple nominal base edges & Methods and descriptors; cooperative initializers may establish state on the shared instance & \PaperNotationLink{PaperClasses}{C}3 linearization and cooperative \texttt{super()} & Facts derived from declared bases need no second applicability carrier; population enumeration remains separate. \\
\midrule
R S4 classes & Multiple \texttt{contains} superclasses & The object contains inherited slots; generic methods apply indirectly through superclass information & S4 method selection and compatible-slot constraints & Class metadata carries multiple superclass and state relations; generic-function tables carry method bodies. \\
\midrule
Scala 3 traits & Multiple nominal traits & Concrete members, values, and parameterized traits & Trait linearization & Trait-owned facts remain in the parent graph; type-class \texttt{given} instances instead store a separate incidence. \\
\midrule
Java interfaces & Multiple nominal interfaces & Default method bodies; no per-instance interface state & Most-specific override, rejection, or qualified \texttt{super} & State-bearing facts and population indices require another provider or record. \\
\midrule
\PaperNotationLink{PaperClasses}{C}\# interfaces & Multiple nominal interfaces & Default member bodies; no instance fields & Most-specific implementation or compile-time error & State and noninterface applicability remain outside interface ancestry. \\
\midrule
Kotlin interfaces & Multiple nominal interfaces & Method bodies and computed properties; no backing state & Required override and qualified \texttt{super} & Stored state and omitted applicability decisions remain external. \\
\midrule
Rust traits & Native \texttt{impl Trait for Type} incidence; one \texttt{Deref} target & Default methods; autoderef supplies target fields and methods along one chain & Coherence, orphan rules, visibility, and receiver constraints & Trait admission and target supply differ: dereferencing does not transfer the target's trait memberships. \\
\midrule
Haskell 2010 type classes & Multiple superclass constraints and explicit class--type instances & Default methods from several classes; no instance state & Superclass instances must exist; at most one matching instance & Instance declarations own admission; superclass entailment derives contract consequences. Stored state remains separate. \\
\midrule
Swift protocols & Multiple protocol conformances & Protocol-extension defaults; class state stays in the class hierarchy & Requirement dispatch and specialized constrained extensions & Conformance declarations and extensions store the incidence separately from single class ancestry. \\
\midrule
Go & Native embedded-field edges; structural interface satisfaction & Promoted methods from embedded values & Ambiguous selectors remain unpromoted & Embedding natively carries field and method composition; interface conformance supplies no nominal provider identity. \\
\midrule
TypeScript & One implementation superclass; multiple structural interfaces & Interface declarations are erased; mixins generate implementation & Explicit class or mixin order and user resolution & Brands, mixin inputs, registries, or adapters retain identity and applicability outside structural compatibility. \\
\bottomrule
\end{longtable}
\endgroup

\paragraph{Source sections for every main-table cell.}
Stateful supply follows instance fields; static storage supplies shared state with a different ownership relation. Each placement below
concerns the named mechanism; precedence, visibility, and admission constraints
remain in force.
\begin{description}
\item[Python.] Identity and nominal membership follow class declarations and bases;
code and state follow class and instance attribute lookup
\cite[Custom classes; Class instances]{PythonDataModel}.
\item[\PaperNotationLink{PaperClasses}{C}++.] Named class declarations establish identity; base classes establish
membership and supply both functions and data members
\cite[\texttt{[class.mi]}, \texttt{[class.virtual]}]{CppWorkingDraft}.
\item[Scala.] Classes and traits establish nominal identity and membership;
concrete trait methods, values, and parameters establish multiple code and state
providers\cite[Classes; Traits]{Scala3Traits}\cite{Scala3TraitParameters}.
\item[Java.] Class declarations and implemented interfaces establish nominal
identity and membership. Default methods supply code from multiple interfaces;
instance fields follow one superclass chain
\cite[\S\S8.1, 8.1.4--8.1.5, 8.3]{JavaLanguageSpecificationClasses}
\cite[\S\S9.1.3, 9.3--9.4]{JavaLanguageSpecificationInterfaces}.
\item[\PaperNotationLink{PaperClasses}{C}\#.] Class declarations establish identity and one state-bearing base;
implemented interfaces establish multiple nominal memberships and default bodies
\cite[\S\S15.2, 15.2.4, 15.5]{CSharpClasses}
\cite[\S\S19.2, 19.4]{CSharpInterfaces}.
\item[Rust.] Struct and enum declarations establish nominal types; explicit
\texttt{impl}s establish multiple trait memberships and default bodies
\cite{RustTypesReference,RustTraitsReference,RustImplementationsReference}.
The reference's \texttt{Deref} rules, field autoderef, and method candidate-receiver
search establish one target chain for field and method supply
\cite{RustDerefReference,RustFieldReference,RustMethodReference}.
Mutable access needs \texttt{DerefMut}; visibility and receiver constraints still
apply. One target declaration supplies the target's member set, so individual fields need no forwarding declarations.
\item[Go.] ``Type identity'' establishes named types; ``Interface types'' and
``Uniqueness of identifiers'' establish structural admission and package-scoped
unexported method keys. ``Struct types'' and ``Selectors'' establish multiple
embedded-field and promoted-method supply\cite{GoSpecification}.
Embedding can propagate an unexported-method key: a seal is not an absolute
prohibition on implementers outside the declaring package.
\item[TypeScript.] ``Type Compatibility'' and ``Symbols'' establish structural
identity and retained brands; interfaces are structural and erased
\cite{TypeScriptDocs,TypeScriptSymbols}. ``How Does A Mixin Work?'' composes
class-expression factories along one superclass chain. ``Alternative Pattern''
copies prototype descriptors and requires separate type declarations
\cite{TypeScriptMixins}. Descriptor copying creates replicas; class-expression factories instead retain their superclass edges.
Provider-preserving crossings and replica disagreement remain distinct questions.
\item[Haskell 2010.] Sections 4.2.1 and 4.2.3 establish nominal data types;
Sections 4.3.1--4.3.2 establish superclass constraints, explicit instances, and
default methods, without data inheritance\cite{Haskell2010Report}.
\end{description}

\paragraph{Supplying versus using a native carrier.}
For Haskell, use type constructors as consumers and type-class constraints as
contracts. An instance declaration establishes a class--type membership answer;
the superclass DAG then entails its superclass constraints. Each required
superclass instance must already exist: entailment uses that declaration.
Default class methods supply code when an instance omits the corresponding
binding. Admission, superclass entailment, and implementation supply therefore
have distinct authoritative declarations, just as membership and supply do in
the main model~\cite[\S\S4.3.1--4.3.2]{Haskell2010Report}.

Implementing a sound native carrier is itself engineering work.  Retrofitting
one into a runtime whose verifier, layout, dispatch, reflection, serialization,
ABI, and interlanguage contracts assume a technical limitation can be harder than
supplying it initially.  The CLR design notes make this distinction concrete:
runtime support is what makes inheritance type-safe and interoperable across
languages, while field-bearing multiple inheritance was not standardized
because languages give it different semantics~\cite{DotNetCLRDesign}.
This engineering cost belongs to implementing the carrier; once supplied,
the carrier derives the same required relation regardless of that setup price.

The compensation ledger can aggregate this boundary.  Let $K_L(T)$ collect
the platform-specific cost of implementing, standardizing, and maintaining the
carrier through horizon $T$, and let project $j$ otherwise pay cumulative
workaround burden $q_jB_j(n_j)$.  Ignoring terms shared by both choices,
supplying the carrier is cheaper once
\[
  K_L(T) < \sum_j q_j B_j(n_j).
\]
The inequality does not assert that a platform has reached that threshold.  It
states where the costs belong: $K_L(T)$ prices the mechanism's implementation;
$B_j$ prices the downstream answers left to each program while it is absent.
Neither term changes its derivation capacity after it is supplied.

\paragraph{Documented industrial artifacts.}
The primary sources below record shipped mechanisms and API behavior.  The
last column classifies each documented fact under the paper's definitions; it
does not infer why the platform adopted it or estimate prevalence.  The
tooling law fixes the accounting: a registry, adapter, generator,
or forwarding workaround preserving required supply with one-parent ancestry owns at least the
$d_\downarrow$ pairs missing from ancestry.  Native multi-parent ancestry can
leave no such deficit, so that workaround layer need not exist.  Retaining
overreach instead needs unwanted-supply removal or admits invalid states.

\begin{table}[htbp]
\centering
\PaperTableSetup
\setlength{\tabcolsep}{3pt}
\begin{tabularx}{\linewidth}{@{}>{\RaggedRight\arraybackslash}m{0.20\linewidth}
  >{\RaggedRight\arraybackslash}m{0.43\linewidth}Y@{}}
\toprule
Artifact & Primary-source fact & Classification in this model \\
\midrule
Java~8 and \PaperNotationLink{PaperClasses}{C}\#~8 interface defaults
& Both platforms later added method bodies to nominal interfaces; their
  specifications resolve or reject competing defaults
  ~\cite{JavaDefaultMethods,CSharpVersionHistory,CSharpInterfaces}.
& Multiple nominal contracts became native implementation providers.  The
  provider edges move method answers into a checked carrier; instance state
  remains elsewhere. \\
\midrule
Java \texttt{AbstractCollection}
& Oracle calls it a ``skeletal implementation'' that minimizes the effort to
  implement \texttt{Collection}; its default \texttt{add} throws
  \texttt{UnsupportedOperationException}
  ~\cite{JavaAbstractCollection25}.
& A companion one-parent class supplies implementation that the interface does
  not. The optional-operation exception fulfills the documented contract,
  so its implementation keeps the superclass promise. \\
\midrule
Java \texttt{Collection}
& The interface documents constructor conventions it cannot enforce and marks
  operations optional; unsupported implementations throw an unchecked
  exception~\cite{JavaCollection25}.
& Contract membership alone does not derive the constructor or operation
  applicability relation.  Each implementation or permitted exceptional result
  still needs a provider; disabling a promised operation would violate the contract. \\
\midrule
Java \texttt{Properties}
& Inherited \texttt{Hashtable.put} admits non-String keys or values even though
  property entries are specified as strings; later \texttt{store},
  \texttt{save}, \texttt{propertyNames}, or \texttt{list} calls can fail
  ~\cite{JavaProperties25}.
& The parent exposes behavior outside the narrower valid relation.  The
  documented failing states arise because inherited operations admit invalid values. \\
\midrule
Java \texttt{Timestamp}
& Oracle states that the \texttt{Date} edge denotes implementation rather than
  type inheritance and documents asymmetric equality between the two classes
  ~\cite{JavaTimestamp25}.
& Implementation reuse and substitutability are different relations.  One
  parent edge carries the first while exposing an invalid instance of the
  second. \\
\midrule
Eclipse adapter registry
& \texttt{IAdapterManager} maintains a registry of factories; plug-ins can
  register adaptable-class--adapter pairs in code or through an XML extension
  point~\cite{EclipseAdapterManager,EclipseAdapterExtension}.
& The registry is an outside workaround authority.  It owns the applicability
  pairs that class ancestry does not derive. \\
\bottomrule
\end{tabularx}
\caption{Primary-source instances of the carrier, replacement, overreach,
relation-splitting, and external-authority cases.  The classification is a
consequence of the fixed-relation model: shipped behavior determines the case,
and the platform's reasons for choosing it are a separate historical question.}
\label{tab:supp-documented-completion-artifacts}
\end{table}

These mechanisms show that a diamond is a topology.  Shared state and provider
conflicts require separate policies.  \PaperNotationLink{PaperClasses}{C}++, Eiffel, CLOS, Python, and Scala
carry broad implementation applicability in ancestry.  Java, \PaperNotationLink{PaperClasses}{C}\#, Kotlin,
Rust, and Swift carry useful behavior incidence through interfaces, traits,
conformance, and default implementations while keeping other resources in
separate carriers.
Java and \PaperNotationLink{PaperClasses}{C}\# added these subsets later: Java~8 introduced default methods for
interface evolution and binary compatibility, and \PaperNotationLink{PaperClasses}{C}\#~8 introduced default
interface members~\cite{JavaDefaultMethods,CSharpVersionHistory}.
Kotlin interfaces supplied method bodies from 1.0~\cite{Kotlin10Release,KotlinDefaultMethodsHistory}.
PHP~5.4 traits also supply properties, adding state to their method supply
~\cite{PHP54Features,PHPTraits}. Java and \PaperNotationLink{PaperClasses}{C}\# move
method bodies into multiple nominal providers while leaving instance state
elsewhere.  They add behavior carriers without adding full state-bearing
multiple class ancestry.  Go and TypeScript make the separation especially
visible.  For any declared mapping from a concrete construct to
provider--consumer judgments, the canonical-quotient proposition determines from
$R^{\mathrm{req}}$ which consumers need different providers and must remain distinguishable.
The table does not assert that conformance,
behavior, and state use identical role mappings across languages.  Memory layout,
compilation, dispatch latency, comprehensibility, and testing remain separate
coordinates.

\paragraph{Pinned availability snapshot.}
The artifact records the July 2026 TIOBE top ten and the classification boundary
in \texttt{experiments/language\_capacity\_census}.  Python, C++, and R expose
the state-bearing multi-parent coordinate; seven entries do not, including C
and SQL, which have no class-inheritance mechanism on this coordinate
~\cite{TIOBEJuly2026,RSetClass2026,ECMAScriptPrototype2026,
VisualBasicSingleInheritance}. This census locates available native capacity:
projects can keep a relation in ancestry only where that capacity exists.
Usage and adoption ask which available designs projects actually choose.

\section*{Related-Work Synthesis}
Opdyke and Johnson extract abstract superclasses by checked refactorings;
Godin and Mili, Moore, Dicky et al., Snelting and Tip, Tip and Sweeney, and
KABA construct, insert into, specialize, or refactor particular hierarchies
~\cite{opdyke1993superclasses,godin1993galois,moore1996guru,
dicky1996insertion,snelting2000concept,tip2000specialization,
streckenbach2004kaba}.  Their objectives include maximal factoring, minimal
member distribution, object size, and client-relative behavior preservation.
Balatsouras and Smaragdakis instead complete a partial subtype graph, give
separate single- and multiple-inheritance algorithms, decompose Java's class
and interface constraints, and prove minimum strata for their single-inheritance
construction~\cite{balatsouras2013complementation}.  Tip et al. and
constraint-based work construct selected behavior-preserving repairs
~\cite{tip2003generalization,steimann2018constraint}; Kegel and Steimann replace
inheritance with delegation~\cite{kegel2008refactoring}.

Those objectives do not minimize the required implementation--class pairs left
outside every sound program repair with one ancestor chain per class. Chen et al. optimize a fixed
table~\cite{chen2006minimum}; main-paper Theorem~4.6 identifies that table
optimum with the program-level floor despite arbitrary delegation, forwarding,
reparenting, and infinitely many helpers. Main-text Theorem~4.14 shows why fitting
a tree at each version does not avoid repair as requirements grow.
Trait work identifies qualitative reuse and conflict problems in
single, multiple, and mixin inheritance and proposes a compositional mechanism
~\cite{schaerli2003traits,ducasse2006traits}; we count the required pairs a
mechanism leaves outside.

Empirical inheritance studies count uses and conflict patterns
~\cite{tempero2008inheritance,tempero2013programmers,ducournau2009empirical};
design-pattern, mixin, family-polymorphism, and C3 work organizes composition or
orders candidates~\cite{gamma1994designpatterns,bracha1990mixin,
ancona1998mixin,ernst2001family,barrett1996c3}.
Behavioral subtyping checks the behavior of selected pairs; it does not choose $\PaperNotationLink{PaperRequired}{R^{\mathrm{req}}}$
~\cite{cardelli1988semantics,liskov1994behavioral}; Witsenhausen gives identity's
coloring law, and main-text Section~3 adds history~\cite{witsenhausen1976zero}.
\vspace{-0.25\baselineskip}

\section{Full Lean Handle Ledger}
\label{lean-handle-ledger}

This supplement provides the complete Lean handle ledger cited by the manuscript.
Unused numeric IDs are intentionally omitted.  IDs are stable and sparse.
It includes every handle identifier, declaration name, and source module path.

\IfFileExists{content/lean_handle_ids_auto.tex}{%
\begingroup
\scriptsize
\setlength{\tabcolsep}{4pt}
\renewcommand{\arraystretch}{1.06}
\ifdefined\PaperSupplementTableSetup\PaperSupplementTableSetup\fi
\setlength{\LTpre}{2pt}
\setlength{\LTpost}{2pt}
\setlength{\emergencystretch}{3em}
\sloppy
\urlstyle{tt}
\makeatletter
\newif\ifleanhandlelist
\if@twocolumn\leanhandlelisttrue\else\leanhandlelistfalse\fi
\@ifundefined{LeanHandleIndexListLayout}{}{\leanhandlelisttrue}
\ifleanhandlelist
\begin{list}{}{\leftmargin=0pt\itemindent=0pt\itemsep=2pt\parsep=0pt\topsep=4pt}
\item \textbf{\nolinkurl{ACF1}}\hypertarget{lh:ACF1}{}\enspace{\ttfamily Ssot.\allowbreak{}AccountingCharacterization.\allowbreak{}positive\_\allowbreak{}sets\_\allowbreak{}preserve\_\allowbreak{}crossings} {\tiny\ttfamily Paper1PL/\allowbreak AccountingCharacterization.lean}
\item \textbf{\nolinkurl{ACF2}}\hypertarget{lh:ACF2}{}\enspace{\ttfamily Ssot.\allowbreak{}AccountingCharacterization.\allowbreak{}finite\_\allowbreak{}declaration\_\allowbreak{}positive\_\allowbreak{}floor} {\tiny\ttfamily Paper1PL/\allowbreak AccountingCharacterization.lean}
\item \textbf{\nolinkurl{ACF3}}\hypertarget{lh:ACF3}{}\enspace{\ttfamily Ssot.\allowbreak{}AccountingCharacterization.\allowbreak{}every\_\allowbreak{}nonempty\_\allowbreak{}set\_\allowbreak{}is\_\allowbreak{}crossing\_\allowbreak{}residual} {\tiny\ttfamily Paper1PL/\allowbreak AccountingCharacterization.lean}
\item \textbf{\nolinkurl{ACF4}}\hypertarget{lh:ACF4}{}\enspace{\ttfamily Ssot.\allowbreak{}AccountingCharacterization.\allowbreak{}preserves\_\allowbreak{}all\_\allowbreak{}crossings\_\allowbreak{}iff\_\allowbreak{}positive} {\tiny\ttfamily Paper1PL/\allowbreak AccountingCharacterization.lean}
\item \textbf{\nolinkurl{AGP1}}\hypertarget{lh:AGP1}{}\enspace{\ttfamily Ssot.\allowbreak{}AncestryGapProbability.\allowbreak{}zeroGap\_\allowbreak{}iff\_\allowbreak{}regionMinimum\_\allowbreak{}zero} {\tiny\ttfamily Paper1PL/\allowbreak AncestryGapProbability.lean}
\item \textbf{\nolinkurl{AGP2}}\hypertarget{lh:AGP2}{}\enspace{\ttfamily Ssot.\allowbreak{}AncestryGapProbability.\allowbreak{}mass\_\allowbreak{}total} {\tiny\ttfamily Paper1PL/\allowbreak AncestryGapProbability.lean}
\item \textbf{\nolinkurl{AGP3}}\hypertarget{lh:AGP3}{}\enspace{\ttfamily Ssot.\allowbreak{}AncestryGapProbability.\allowbreak{}exact\_\allowbreak{}zero\_\allowbreak{}gap\_\allowbreak{}probability} {\tiny\ttfamily Paper1PL/\allowbreak AncestryGapProbability.lean}
\item \textbf{\nolinkurl{AGP4}}\hypertarget{lh:AGP4}{}\enspace{\ttfamily Ssot.\allowbreak{}AncestryGapProbability.\allowbreak{}zero\_\allowbreak{}gap\_\allowbreak{}bounds} {\tiny\ttfamily Paper1PL/\allowbreak AncestryGapProbability.lean}
\item \textbf{\nolinkurl{AGP5}}\hypertarget{lh:AGP5}{}\enspace{\ttfamily Ssot.\allowbreak{}AncestryGapProbability.\allowbreak{}zero\_\allowbreak{}gap\_\allowbreak{}isTheta} {\tiny\ttfamily Paper1PL/\allowbreak AncestryGapProbability.lean}
\item \textbf{\nolinkurl{AGP6}}\hypertarget{lh:AGP6}{}\enspace{\ttfamily Ssot.\allowbreak{}AncestryGapProbability.\allowbreak{}zero\_\allowbreak{}gap\_\allowbreak{}tendsto\_\allowbreak{}zero} {\tiny\ttfamily Paper1PL/\allowbreak AncestryGapProbability.lean}
\item \textbf{\nolinkurl{AGP7}}\hypertarget{lh:AGP7}{}\enspace{\ttfamily Ssot.\allowbreak{}AncestryGapProbability.\allowbreak{}mass\_\allowbreak{}complement} {\tiny\ttfamily Paper1PL/\allowbreak AncestryGapProbability.lean}
\item \textbf{\nolinkurl{APC1}}\hypertarget{lh:APC1}{}\enspace{\ttfamily Ssot.\allowbreak{}PublicApiContract.\allowbreak{}public\_\allowbreak{}pair\_\allowbreak{}has\_\allowbreak{}client\_\allowbreak{}witness} {\tiny\ttfamily Paper1PL/\allowbreak PublicApiContract.lean}
\item \textbf{\nolinkurl{APC2}}\hypertarget{lh:APC2}{}\enspace{\ttfamily Ssot.\allowbreak{}PublicApiContract.\allowbreak{}public\_\allowbreak{}resolution\_\allowbreak{}preservation\_\allowbreak{}forces\_\allowbreak{}supply} {\tiny\ttfamily Paper1PL/\allowbreak PublicApiContract.lean}
\item \textbf{\nolinkurl{ARI1}}\hypertarget{lh:ARI1}{}\enspace{\ttfamily Ssot.\allowbreak{}NativeCapacityLadder.\allowbreak{}fourBindings\_\allowbreak{}threeNativeSlots\_\allowbreak{}leave\_\allowbreak{}oneOutside} {\tiny\ttfamily Paper1PL/\allowbreak NativeCapacityLadder.lean}
\item \textbf{\nolinkurl{AVG1}}\hypertarget{lh:AVG1}{}\enspace{\ttfamily Ssot.\allowbreak{}AverageCaseAncestryGap.\allowbreak{}twoOwner\_\allowbreak{}crossing\_\allowbreak{}iff\_\allowbreak{}regions\_\allowbreak{}nonempty} {\tiny\ttfamily Paper1PL/\allowbreak AverageCaseAncestryGap.lean}
\item \textbf{\nolinkurl{AVG2}}\hypertarget{lh:AVG2}{}\enspace{\ttfamily Ssot.\allowbreak{}AverageCaseAncestryGap.\allowbreak{}twoOwnerRegionMinimum\_\allowbreak{}le\_\allowbreak{}gapCost} {\tiny\ttfamily Paper1PL/\allowbreak AverageCaseAncestryGap.lean}
\item \textbf{\nolinkurl{AVG3}}\hypertarget{lh:AVG3}{}\enspace{\ttfamily Ssot.\allowbreak{}AverageCaseAncestryGap.\allowbreak{}twoOwnerGap\_\allowbreak{}exact} {\tiny\ttfamily Paper1PL/\allowbreak AverageCaseAncestryGap.lean}
\item \textbf{\nolinkurl{AVG4}}\hypertarget{lh:AVG4}{}\enspace{\ttfamily Ssot.\allowbreak{}AverageCaseAncestryGap.\allowbreak{}RowColumnRepair.\allowbreak{}no\_\allowbreak{}big\_\allowbreak{}double} {\tiny\ttfamily Paper1PL/\allowbreak AverageCaseAncestryGap.lean}
\item \textbf{\nolinkurl{AVG5}}\hypertarget{lh:AVG5}{}\enspace{\ttfamily Ssot.\allowbreak{}AverageCaseAncestryGap.\allowbreak{}rowColumnGap\_\allowbreak{}lowerBound} {\tiny\ttfamily Paper1PL/\allowbreak AverageCaseAncestryGap.lean}
\item \textbf{\nolinkurl{AVG6}}\hypertarget{lh:AVG6}{}\enspace{\ttfamily Ssot.\allowbreak{}AverageCaseAncestryGap.\allowbreak{}canonicalRowColumnRepair\_\allowbreak{}gapCost} {\tiny\ttfamily Paper1PL/\allowbreak AverageCaseAncestryGap.lean}
\item \textbf{\nolinkurl{AVG7}}\hypertarget{lh:AVG7}{}\enspace{\ttfamily Ssot.\allowbreak{}AverageCaseAncestryGap.\allowbreak{}rowColumnGap\_\allowbreak{}exact} {\tiny\ttfamily Paper1PL/\allowbreak AverageCaseAncestryGap.lean}
\item \textbf{\nolinkurl{BAP1}}\hypertarget{lh:BAP1}{}\enspace{\ttfamily Ssot.\allowbreak{}BehavioralApplicability.\allowbreak{}BehavioralBoundary.\allowbreak{}behavioralSubstitution\_\allowbreak{}iff\_\allowbreak{}reachableInvalid\_\allowbreak{}eq\_\allowbreak{}empty} {\tiny\ttfamily Paper1PL/\allowbreak BehavioralApplicability.lean}
\item \textbf{\nolinkurl{BAP2}}\hypertarget{lh:BAP2}{}\enspace{\ttfamily Ssot.\allowbreak{}BehavioralApplicability.\allowbreak{}BehavioralBoundary.\allowbreak{}behavioralSubstitution\_\allowbreak{}iff\_\allowbreak{}forall\_\allowbreak{}reachable\_\allowbreak{}valid} {\tiny\ttfamily Paper1PL/\allowbreak BehavioralApplicability.lean}
\item \textbf{\nolinkurl{BAP3}}\hypertarget{lh:BAP3}{}\enspace{\ttfamily Ssot.\allowbreak{}BehavioralApplicability.\allowbreak{}noProductViolation\_\allowbreak{}iff\_\allowbreak{}productCorrect} {\tiny\ttfamily Paper1PL/\allowbreak BehavioralApplicability.lean}
\item \textbf{\nolinkurl{BAP4}}\hypertarget{lh:BAP4}{}\enspace{\ttfamily Ssot.\allowbreak{}BehavioralApplicability.\allowbreak{}behavioralCorrect\_\allowbreak{}does\_\allowbreak{}not\_\allowbreak{}imply\_\allowbreak{}exactApplicability} {\tiny\ttfamily Paper1PL/\allowbreak BehavioralApplicability.lean}
\item \textbf{\nolinkurl{BAP5}}\hypertarget{lh:BAP5}{}\enspace{\ttfamily Ssot.\allowbreak{}BehavioralApplicability.\allowbreak{}exactApplicability\_\allowbreak{}does\_\allowbreak{}not\_\allowbreak{}imply\_\allowbreak{}behavioralCorrect} {\tiny\ttfamily Paper1PL/\allowbreak BehavioralApplicability.lean}
\item \textbf{\nolinkurl{BAP6}}\hypertarget{lh:BAP6}{}\enspace{\ttfamily Ssot.\allowbreak{}BehavioralApplicability.\allowbreak{}minimumGap\_\allowbreak{}le\_\allowbreak{}behaviorallyValidExactSingleParentSystem\_\allowbreak{}gap} {\tiny\ttfamily Paper1PL/\allowbreak BehavioralApplicability.lean}
\item \textbf{\nolinkurl{BAP7}}\hypertarget{lh:BAP7}{}\enspace{\ttfamily Ssot.\allowbreak{}BehavioralApplicability.\allowbreak{}behaviorallyValid\_\allowbreak{}crossing\_\allowbreak{}still\_\allowbreak{}has\_\allowbreak{}positive\_\allowbreak{}ancestryGap} {\tiny\ttfamily Paper1PL/\allowbreak BehavioralApplicability.lean}
\item \textbf{\nolinkurl{BAP8}}\hypertarget{lh:BAP8}{}\enspace{\ttfamily Ssot.\allowbreak{}BehavioralApplicability.\allowbreak{}BehavioralBoundary.\allowbreak{}subset\_\allowbreak{}iff\_\allowbreak{}preserves\_\allowbreak{}every\_\allowbreak{}property} {\tiny\ttfamily Paper1PL/\allowbreak BehavioralApplicability.lean}
\item \textbf{\nolinkurl{BAP9}}\hypertarget{lh:BAP9}{}\enspace{\ttfamily Ssot.\allowbreak{}BehavioralApplicability.\allowbreak{}BehavioralBoundary.\allowbreak{}behavioralSubstitution\_\allowbreak{}iff\_\allowbreak{}preserves\_\allowbreak{}every\_\allowbreak{}valid\_\allowbreak{}property} {\tiny\ttfamily Paper1PL/\allowbreak BehavioralApplicability.lean}
\item \textbf{\nolinkurl{BAP10}}\hypertarget{lh:BAP10}{}\enspace{\ttfamily Ssot.\allowbreak{}BehavioralApplicability.\allowbreak{}noProductViolation\_\allowbreak{}iff\_\allowbreak{}exactApplicability\_\allowbreak{}and\_\allowbreak{}propertyPreservation} {\tiny\ttfamily Paper1PL/\allowbreak BehavioralApplicability.lean}
\item \textbf{\nolinkurl{BAP11}}\hypertarget{lh:BAP11}{}\enspace{\ttfamily Ssot.\allowbreak{}BehavioralApplicability.\allowbreak{}noProductViolation\_\allowbreak{}of\_\allowbreak{}exactApplicability\_\allowbreak{}and\_\allowbreak{}certificates} {\tiny\ttfamily Paper1PL/\allowbreak BehavioralApplicability.lean}
\item \textbf{\nolinkurl{BAP12}}\hypertarget{lh:BAP12}{}\enspace{\ttfamily Ssot.\allowbreak{}BehavioralApplicability.\allowbreak{}SoundBehavioralProofMethod.\allowbreak{}certifies\_\allowbreak{}iff\_\allowbreak{}behavioralSubstitution} {\tiny\ttfamily Paper1PL/\allowbreak BehavioralApplicability.lean}
\item \textbf{\nolinkurl{BAP13}}\hypertarget{lh:BAP13}{}\enspace{\ttfamily Ssot.\allowbreak{}BehavioralApplicability.\allowbreak{}SoundBehavioralProofMethod.\allowbreak{}complete\_\allowbreak{}methods\_\allowbreak{}certify\_\allowbreak{}iff} {\tiny\ttfamily Paper1PL/\allowbreak BehavioralApplicability.lean}
\item \textbf{\nolinkurl{BAP14}}\hypertarget{lh:BAP14}{}\enspace{\ttfamily Ssot.\allowbreak{}BehavioralApplicability.\allowbreak{}fiberLocalCorrect\_\allowbreak{}iff\_\allowbreak{}of\_\allowbreak{}equiv} {\tiny\ttfamily Paper1PL/\allowbreak BehavioralApplicability.lean}
\item \textbf{\nolinkurl{BAP15}}\hypertarget{lh:BAP15}{}\enspace{\ttfamily Ssot.\allowbreak{}BehavioralApplicability.\allowbreak{}hidden\_\allowbreak{}laminar\_\allowbreak{}crossing\_\allowbreak{}requiredIncidence\_\allowbreak{}card\_\allowbreak{}eq} {\tiny\ttfamily Paper1PL/\allowbreak BehavioralApplicability.lean}
\item \textbf{\nolinkurl{BAP16}}\hypertarget{lh:BAP16}{}\enspace{\ttfamily Ssot.\allowbreak{}BehavioralApplicability.\allowbreak{}hiddenRequiredPairEquiv\_\allowbreak{}preserves\_\allowbreak{}required} {\tiny\ttfamily Paper1PL/\allowbreak BehavioralApplicability.lean}
\item \textbf{\nolinkurl{BAP17}}\hypertarget{lh:BAP17}{}\enspace{\ttfamily Ssot.\allowbreak{}BehavioralApplicability.\allowbreak{}hiddenRequiredPairEquiv\_\allowbreak{}preserves\_\allowbreak{}behavioralSubstitution} {\tiny\ttfamily Paper1PL/\allowbreak BehavioralApplicability.lean}
\item \textbf{\nolinkurl{BAP18}}\hypertarget{lh:BAP18}{}\enspace{\ttfamily Ssot.\allowbreak{}BehavioralApplicability.\allowbreak{}fiberLocal\_\allowbreak{}behavior\_\allowbreak{}cannot\_\allowbreak{}determine\_\allowbreak{}ancestryGap} {\tiny\ttfamily Paper1PL/\allowbreak BehavioralApplicability.lean}
\item \textbf{\nolinkurl{BAP19}}\hypertarget{lh:BAP19}{}\enspace{\ttfamily Ssot.\allowbreak{}BehavioralApplicability.\allowbreak{}fiberLocal\_\allowbreak{}behavior\_\allowbreak{}same\_\allowbreak{}result\_\allowbreak{}different\_\allowbreak{}ancestryGap} {\tiny\ttfamily Paper1PL/\allowbreak BehavioralApplicability.lean}
\item \textbf{\nolinkurl{BAP20}}\hypertarget{lh:BAP20}{}\enspace{\ttfamily Ssot.\allowbreak{}BehavioralApplicability.\allowbreak{}MethodContract.\allowbreak{}substitution\_\allowbreak{}iff\_\allowbreak{}preserves\_\allowbreak{}promised\_\allowbreak{}calls} {\tiny\ttfamily Paper1PL/\allowbreak BehavioralApplicability.lean}
\item \textbf{\nolinkurl{BAP21}}\hypertarget{lh:BAP21}{}\enspace{\ttfamily Ssot.\allowbreak{}BehavioralApplicability.\allowbreak{}MethodContract.\allowbreak{}contract\_\allowbreak{}preserving\_\allowbreak{}override} {\tiny\ttfamily Paper1PL/\allowbreak BehavioralApplicability.lean}
\item \textbf{\nolinkurl{BAP22}}\hypertarget{lh:BAP22}{}\enspace{\ttfamily Ssot.\allowbreak{}BehavioralApplicability.\allowbreak{}MethodContract.\allowbreak{}disabling\_\allowbreak{}promised\_\allowbreak{}call\_\allowbreak{}violates\_\allowbreak{}substitution} {\tiny\ttfamily Paper1PL/\allowbreak BehavioralApplicability.lean}
\item \textbf{\nolinkurl{BAP23}}\hypertarget{lh:BAP23}{}\enspace{\ttfamily Ssot.\allowbreak{}BehavioralApplicability.\allowbreak{}MethodContract.\allowbreak{}forbidden\_\allowbreak{}result\_\allowbreak{}violates\_\allowbreak{}substitution} {\tiny\ttfamily Paper1PL/\allowbreak BehavioralApplicability.lean}
\item \textbf{\nolinkurl{BAP24}}\hypertarget{lh:BAP24}{}\enspace{\ttfamily Ssot.\allowbreak{}BehavioralApplicability.\allowbreak{}MethodContract.\allowbreak{}changing\_\allowbreak{}result\_\allowbreak{}allowed\_\allowbreak{}disabling\_\allowbreak{}not} {\tiny\ttfamily Paper1PL/\allowbreak BehavioralApplicability.lean}
\item \textbf{\nolinkurl{BAP25}}\hypertarget{lh:BAP25}{}\enspace{\ttfamily Ssot.\allowbreak{}BehavioralApplicability.\allowbreak{}excluded\_\allowbreak{}parent\_\allowbreak{}promise\_\allowbreak{}forbids\_\allowbreak{}subtyping} {\tiny\ttfamily Paper1PL/\allowbreak BehavioralApplicability.lean}
\item \textbf{\nolinkurl{BLK1}}\hypertarget{lh:BLK1}{}\enspace{\ttfamily Ssot.\allowbreak{}BlockingRepair.\allowbreak{}canonicalRepair\_\allowbreak{}decisionCost} {\tiny\ttfamily Paper1PL/\allowbreak BlockingRepair.lean}
\item \textbf{\nolinkurl{BLK2}}\hypertarget{lh:BLK2}{}\enspace{\ttfamily Ssot.\allowbreak{}BlockingRepair.\allowbreak{}forwardPairs\_\allowbreak{}subset} {\tiny\ttfamily Paper1PL/\allowbreak BlockingRepair.lean}
\item \textbf{\nolinkurl{BLK3}}\hypertarget{lh:BLK3}{}\enspace{\ttfamily Ssot.\allowbreak{}BlockingRepair.\allowbreak{}blockPairs\_\allowbreak{}subset} {\tiny\ttfamily Paper1PL/\allowbreak BlockingRepair.lean}
\item \textbf{\nolinkurl{BLK4}}\hypertarget{lh:BLK4}{}\enspace{\ttfamily Ssot.\allowbreak{}BlockingRepair.\allowbreak{}paidBlockingCost\_\allowbreak{}le\_\allowbreak{}decisionCost} {\tiny\ttfamily Paper1PL/\allowbreak BlockingRepair.lean}
\item \textbf{\nolinkurl{BLK5}}\hypertarget{lh:BLK5}{}\enspace{\ttfamily Ssot.\allowbreak{}BlockingRepair.\allowbreak{}bruteForcePaidBlockingGap\_\allowbreak{}spec} {\tiny\ttfamily Paper1PL/\allowbreak BlockingRepair.lean}
\item \textbf{\nolinkurl{BLK6}}\hypertarget{lh:BLK6}{}\enspace{\ttfamily Ssot.\allowbreak{}BlockingRepair.\allowbreak{}paidBlockingGap\_\allowbreak{}le\_\allowbreak{}blockerFreeGap} {\tiny\ttfamily Paper1PL/\allowbreak BlockingRepair.lean}
\item \textbf{\nolinkurl{BLK7}}\hypertarget{lh:BLK7}{}\enspace{\ttfamily Ssot.\allowbreak{}BlockingRepair.\allowbreak{}zeroWeight\_\allowbreak{}paidBlockingGap\_\allowbreak{}eq\_\allowbreak{}zero} {\tiny\ttfamily Paper1PL/\allowbreak BlockingRepair.lean}
\item \textbf{\nolinkurl{BLK8}}\hypertarget{lh:BLK8}{}\enspace{\ttfamily Ssot.\allowbreak{}BlockingRepair.\allowbreak{}positiveWeight\_\allowbreak{}paidBlockingGap\_\allowbreak{}pos\_\allowbreak{}of\_\allowbreak{}not\_\allowbreak{}laminar} {\tiny\ttfamily Paper1PL/\allowbreak BlockingRepair.lean}
\item \textbf{\nolinkurl{BLK9}}\hypertarget{lh:BLK9}{}\enspace{\ttfamily Ssot.\allowbreak{}BlockingRepair.\allowbreak{}FiniteBlockingRepair.\allowbreak{}decisionCost\_\allowbreak{}pos\_\allowbreak{}of\_\allowbreak{}not\_\allowbreak{}laminar} {\tiny\ttfamily Paper1PL/\allowbreak BlockingRepair.lean}
\item \textbf{\nolinkurl{BLK10}}\hypertarget{lh:BLK10}{}\enspace{\ttfamily Ssot.\allowbreak{}BlockingRepair.\allowbreak{}paidBlockingGap\_\allowbreak{}eq\_\allowbreak{}blockerFreeGap\_\allowbreak{}of\_\allowbreak{}gap\_\allowbreak{}le\_\allowbreak{}weight} {\tiny\ttfamily Paper1PL/\allowbreak BlockingRepair.lean}
\item \textbf{\nolinkurl{BPI1}}\hypertarget{lh:BPI1}{}\enspace{\ttfamily Ssot.\allowbreak{}BoundaryPropertyInstances.\allowbreak{}DeclaredBoundary.\allowbreak{}zeroError\_\allowbreak{}iff\_\allowbreak{}hasReachableResolvers} {\tiny\ttfamily Paper1IT/\allowbreak BoundaryPropertyInstances.lean}
\item \textbf{\nolinkurl{BWR1}}\hypertarget{lh:BWR1}{}\enspace{\ttfamily Ssot.\allowbreak{}BehavioralRewrite.\allowbreak{}ExactIndexedSingleParentBehavioralRewrite.\allowbreak{}residual\_\allowbreak{}of\_\allowbreak{}gap} {\tiny\ttfamily Paper1PL/\allowbreak BehavioralRewrite.lean}
\item \textbf{\nolinkurl{BWR2}}\hypertarget{lh:BWR2}{}\enspace{\ttfamily Ssot.\allowbreak{}BehavioralRewrite.\allowbreak{}minimumIndexedLaminarGapCost\_\allowbreak{}le\_\allowbreak{}every\_\allowbreak{}exactSingleParentRewrite\_\allowbreak{}ancestryGap} {\tiny\ttfamily Paper1PL/\allowbreak BehavioralRewrite.lean}
\item \textbf{\nolinkurl{BWR3}}\hypertarget{lh:BWR3}{}\enspace{\ttfamily Ssot.\allowbreak{}BehavioralRewrite.\allowbreak{}ExactBoundaryEncoding.\allowbreak{}equal\_\allowbreak{}joint\_\allowbreak{}state\_\allowbreak{}implies\_\allowbreak{}equal\_\allowbreak{}required} {\tiny\ttfamily Paper1PL/\allowbreak BehavioralRewrite.lean}
\item \textbf{\nolinkurl{BWR4}}\hypertarget{lh:BWR4}{}\enspace{\ttfamily Ssot.\allowbreak{}BehavioralRewrite.\allowbreak{}ExactBoundaryEncoding.\allowbreak{}requiredView\_\allowbreak{}factorsThrough\_\allowbreak{}jointState} {\tiny\ttfamily Paper1PL/\allowbreak BehavioralRewrite.lean}
\item \textbf{\nolinkurl{BWR5}}\hypertarget{lh:BWR5}{}\enspace{\ttfamily Ssot.\allowbreak{}BehavioralRewrite.\allowbreak{}ExactDecisionEncoding.\allowbreak{}column\_\allowbreak{}factorsThrough\_\allowbreak{}code} {\tiny\ttfamily Paper1PL/\allowbreak BehavioralRewrite.lean}
\item \textbf{\nolinkurl{BWR6}}\hypertarget{lh:BWR6}{}\enspace{\ttfamily Ssot.\allowbreak{}BehavioralRewrite.\allowbreak{}ExactDecisionEncoding.\allowbreak{}code\_\allowbreak{}eq\_\allowbreak{}implies\_\allowbreak{}quotient\_\allowbreak{}eq} {\tiny\ttfamily Paper1PL/\allowbreak BehavioralRewrite.lean}
\item \textbf{\nolinkurl{BWR7}}\hypertarget{lh:BWR7}{}\enspace{\ttfamily Ssot.\allowbreak{}BehavioralRewrite.\allowbreak{}canonicalDecisionQuotient\_\allowbreak{}card\_\allowbreak{}le\_\allowbreak{}code\_\allowbreak{}card} {\tiny\ttfamily Paper1PL/\allowbreak BehavioralRewrite.lean}
\item \textbf{\nolinkurl{BWR8}}\hypertarget{lh:BWR8}{}\enspace{\ttfamily Ssot.\allowbreak{}BehavioralRewrite.\allowbreak{}exists\_\allowbreak{}zero\_\allowbreak{}length\_\allowbreak{}exact\_\allowbreak{}description} {\tiny\ttfamily Paper1PL/\allowbreak BehavioralRewrite.lean}
\item \textbf{\nolinkurl{BWR9}}\hypertarget{lh:BWR9}{}\enspace{\ttfamily Ssot.\allowbreak{}BehavioralRewrite.\allowbreak{}strictDerivationalLoss\_\allowbreak{}of\_\allowbreak{}nonlaminar} {\tiny\ttfamily Paper1PL/\allowbreak BehavioralRewrite.lean}
\item \textbf{\nolinkurl{BWR10}}\hypertarget{lh:BWR10}{}\enspace{\ttfamily Ssot.\allowbreak{}BehavioralRewrite.\allowbreak{}multiParent\_\allowbreak{}strictlyDominates\_\allowbreak{}exactSingleParent\_\allowbreak{}on\_\allowbreak{}chargedAxes} {\tiny\ttfamily Paper1PL/\allowbreak BehavioralRewrite.lean}
\item \textbf{\nolinkurl{BWR11}}\hypertarget{lh:BWR11}{}\enspace{\ttfamily Ssot.\allowbreak{}BehavioralRewrite.\allowbreak{}dominanceRespectingPreference\_\allowbreak{}prefers\_\allowbreak{}multiParent} {\tiny\ttfamily Paper1PL/\allowbreak BehavioralRewrite.lean}
\item \textbf{\nolinkurl{BWR12}}\hypertarget{lh:BWR12}{}\enspace{\ttfamily Ssot.\allowbreak{}BehavioralRewrite.\allowbreak{}same\_\allowbreak{}native\_\allowbreak{}ancestry\_\allowbreak{}supports\_\allowbreak{}distinct\_\allowbreak{}required\_\allowbreak{}relations} {\tiny\ttfamily Paper1PL/\allowbreak BehavioralRewrite.lean}
\item \textbf{\nolinkurl{BWR13}}\hypertarget{lh:BWR13}{}\enspace{\ttfamily Ssot.\allowbreak{}BehavioralRewrite.\allowbreak{}shorter\_\allowbreak{}correct\_\allowbreak{}factorization\_\allowbreak{}excludes\_\allowbreak{}shortest} {\tiny\ttfamily Paper1PL/\allowbreak BehavioralRewrite.lean}
\item \textbf{\nolinkurl{BWR14}}\hypertarget{lh:BWR14}{}\enspace{\ttfamily Ssot.\allowbreak{}BehavioralRewrite.\allowbreak{}behavioralImplication\_\allowbreak{}iff\_\allowbreak{}profile\_\allowbreak{}subset} {\tiny\ttfamily Paper1PL/\allowbreak BehavioralRewrite.lean}
\item \textbf{\nolinkurl{BWR15}}\hypertarget{lh:BWR15}{}\enspace{\ttfamily Ssot.\allowbreak{}BehavioralRewrite.\allowbreak{}mutualBehavioralImplication\_\allowbreak{}iff\_\allowbreak{}sameDecisionColumn} {\tiny\ttfamily Paper1PL/\allowbreak BehavioralRewrite.lean}
\item \textbf{\nolinkurl{BWR16}}\hypertarget{lh:BWR16}{}\enspace{\ttfamily Ssot.\allowbreak{}BehavioralRewrite.\allowbreak{}ExactDecisionEncoding.\allowbreak{}reachableCodeImplication\_\allowbreak{}iff\_\allowbreak{}behavioralImplication} {\tiny\ttfamily Paper1PL/\allowbreak BehavioralRewrite.lean}
\item \textbf{\nolinkurl{BWR17}}\hypertarget{lh:BWR17}{}\enspace{\ttfamily Ssot.\allowbreak{}BehavioralRewrite.\allowbreak{}singletonOwner\_\allowbreak{}subset\_\allowbreak{}profile\_\allowbreak{}iff\_\allowbreak{}required} {\tiny\ttfamily Paper1PL/\allowbreak BehavioralRewrite.lean}
\item \textbf{\nolinkurl{BWR18}}\hypertarget{lh:BWR18}{}\enspace{\ttfamily Ssot.\allowbreak{}BehavioralRewrite.\allowbreak{}distinctSingletonOwners\_\allowbreak{}incomparable} {\tiny\ttfamily Paper1PL/\allowbreak BehavioralRewrite.lean}
\item \textbf{\nolinkurl{BWR19}}\hypertarget{lh:BWR19}{}\enspace{\ttfamily Ssot.\allowbreak{}BehavioralRewrite.\allowbreak{}independentResidualTopology\_\allowbreak{}card} {\tiny\ttfamily Paper1PL/\allowbreak BehavioralRewrite.lean}
\item \textbf{\nolinkurl{BWR20}}\hypertarget{lh:BWR20}{}\enspace{\ttfamily Ssot.\allowbreak{}BehavioralRewrite.\allowbreak{}detectsEverySingleOmission\_\allowbreak{}iff\_\allowbreak{}fullCoverage} {\tiny\ttfamily Paper1PL/\allowbreak BehavioralRewrite.lean}
\item \textbf{\nolinkurl{BWR21}}\hypertarget{lh:BWR21}{}\enspace{\ttfamily Ssot.\allowbreak{}BehavioralRewrite.\allowbreak{}existsUnique\_\allowbreak{}exactIndependentResidualTopology} {\tiny\ttfamily Paper1PL/\allowbreak BehavioralRewrite.lean}
\item \textbf{\nolinkurl{BWR22}}\hypertarget{lh:BWR22}{}\enspace{\ttfamily Ssot.\allowbreak{}BehavioralRewrite.\allowbreak{}boundedSemanticEditCapacity\_\allowbreak{}lower\_\allowbreak{}bound} {\tiny\ttfamily Paper1PL/\allowbreak BehavioralRewrite.lean}
\item \textbf{\nolinkurl{BWR23}}\hypertarget{lh:BWR23}{}\enspace{\ttfamily Ssot.\allowbreak{}BehavioralRewrite.\allowbreak{}minimumIndexedLaminarGapCost\_\allowbreak{}ceilDiv\_\allowbreak{}le\_\allowbreak{}semanticEdits} {\tiny\ttfamily Paper1PL/\allowbreak BehavioralRewrite.lean}
\item \textbf{\nolinkurl{BWR24}}\hypertarget{lh:BWR24}{}\enspace{\ttfamily Ssot.\allowbreak{}BehavioralRewrite.\allowbreak{}minimumSemanticEditPlan\_\allowbreak{}singleOmission\_\allowbreak{}failures} {\tiny\ttfamily Paper1PL/\allowbreak BehavioralRewrite.lean}
\item \textbf{\nolinkurl{BWR25}}\hypertarget{lh:BWR25}{}\enspace{\ttfamily Ssot.\allowbreak{}BehavioralRewrite.\allowbreak{}exactSingleParent\_\allowbreak{}architectureCost\_\allowbreak{}lower\_\allowbreak{}bound} {\tiny\ttfamily Paper1PL/\allowbreak BehavioralRewrite.lean}
\item \textbf{\nolinkurl{BWR26}}\hypertarget{lh:BWR26}{}\enspace{\ttfamily Ssot.\allowbreak{}BehavioralRewrite.\allowbreak{}singleParent\_\allowbreak{}choice\_\allowbreak{}includes\_\allowbreak{}pricedGap} {\tiny\ttfamily Paper1PL/\allowbreak BehavioralRewrite.lean}
\item \textbf{\nolinkurl{BWR27}}\hypertarget{lh:BWR27}{}\enspace{\ttfamily Ssot.\allowbreak{}BehavioralRewrite.\allowbreak{}equalOtherCost\_\allowbreak{}multiParent\_\allowbreak{}strictlyCheaper} {\tiny\ttfamily Paper1PL/\allowbreak BehavioralRewrite.lean}
\item \textbf{\nolinkurl{BWR28}}\hypertarget{lh:BWR28}{}\enspace{\ttfamily Ssot.\allowbreak{}BehavioralRewrite.\allowbreak{}pricedSemanticEdit\_\allowbreak{}lower\_\allowbreak{}bound} {\tiny\ttfamily Paper1PL/\allowbreak BehavioralRewrite.lean}
\item \textbf{\nolinkurl{BWR29}}\hypertarget{lh:BWR29}{}\enspace{\ttfamily Ssot.\allowbreak{}BehavioralRewrite.\allowbreak{}allIndependentResidualTopologies\_\allowbreak{}not\_\allowbreak{}zeroError\_\allowbreak{}of\_\allowbreak{}nonempty} {\tiny\ttfamily Paper1PL/\allowbreak BehavioralRewrite.lean}
\item \textbf{\nolinkurl{BWR30}}\hypertarget{lh:BWR30}{}\enspace{\ttfamily Ssot.\allowbreak{}BehavioralRewrite.\allowbreak{}checkedCompleteResidualTopology\_\allowbreak{}zeroError} {\tiny\ttfamily Paper1PL/\allowbreak BehavioralRewrite.lean}
\item \textbf{\nolinkurl{BWR31}}\hypertarget{lh:BWR31}{}\enspace{\ttfamily Ssot.\allowbreak{}BehavioralRewrite.\allowbreak{}unrestrictedDuplicateOwners\_\allowbreak{}not\_\allowbreak{}zeroError\_\allowbreak{}of\_\allowbreak{}distinct} {\tiny\ttfamily Paper1PL/\allowbreak BehavioralRewrite.lean}
\item \textbf{\nolinkurl{BWR32}}\hypertarget{lh:BWR32}{}\enspace{\ttfamily Ssot.\allowbreak{}BehavioralRewrite.\allowbreak{}checkedSynchronizedDuplicateOwners\_\allowbreak{}zeroError} {\tiny\ttfamily Paper1PL/\allowbreak BehavioralRewrite.lean}
\item \textbf{\nolinkurl{BWR33}}\hypertarget{lh:BWR33}{}\enspace{\ttfamily Ssot.\allowbreak{}BehavioralRewrite.\allowbreak{}universalNative\_\allowbreak{}laminar} {\tiny\ttfamily Paper1PL/\allowbreak BehavioralRewrite.lean}
\item \textbf{\nolinkurl{BWR34}}\hypertarget{lh:BWR34}{}\enspace{\ttfamily Ssot.\allowbreak{}BehavioralRewrite.\allowbreak{}universalNative\_\allowbreak{}misses\_\allowbreak{}no\_\allowbreak{}required} {\tiny\ttfamily Paper1PL/\allowbreak BehavioralRewrite.lean}
\item \textbf{\nolinkurl{BWR35}}\hypertarget{lh:BWR35}{}\enspace{\ttfamily Ssot.\allowbreak{}BehavioralRewrite.\allowbreak{}universalNative\_\allowbreak{}violates\_\allowbreak{}exactReach\_\allowbreak{}of\_\allowbreak{}unrequired} {\tiny\ttfamily Paper1PL/\allowbreak BehavioralRewrite.lean}
\item \textbf{\nolinkurl{BWR36}}\hypertarget{lh:BWR36}{}\enspace{\ttfamily Ssot.\allowbreak{}BehavioralRewrite.\allowbreak{}universalNative\_\allowbreak{}exactReach\_\allowbreak{}iff\_\allowbreak{}requiredUniversal} {\tiny\ttfamily Paper1PL/\allowbreak BehavioralRewrite.lean}
\item \textbf{\nolinkurl{BWR37}}\hypertarget{lh:BWR37}{}\enspace{\ttfamily Ssot.\allowbreak{}BehavioralRewrite.\allowbreak{}zeroErrorOverReachableResidualTopologies\_\allowbreak{}iff\_\allowbreak{}subset\_\allowbreak{}complete} {\tiny\ttfamily Paper1PL/\allowbreak BehavioralRewrite.lean}
\item \textbf{\nolinkurl{BWR38}}\hypertarget{lh:BWR38}{}\enspace{\ttfamily Ssot.\allowbreak{}BehavioralRewrite.\allowbreak{}exploit\_\allowbreak{}unreachable\_\allowbreak{}iff\_\allowbreak{}zeroError} {\tiny\ttfamily Paper1PL/\allowbreak BehavioralRewrite.lean}
\item \textbf{\nolinkurl{BWR39}}\hypertarget{lh:BWR39}{}\enspace{\ttfamily Ssot.\allowbreak{}BehavioralRewrite.\allowbreak{}residualOnly\_\allowbreak{}expressive\_\allowbreak{}and\_\allowbreak{}zeroError\_\allowbreak{}iff\_\allowbreak{}invalid\_\allowbreak{}unreachable} {\tiny\ttfamily Paper1PL/\allowbreak BehavioralRewrite.lean}
\item \textbf{\nolinkurl{BWR40}}\hypertarget{lh:BWR40}{}\enspace{\ttfamily Ssot.\allowbreak{}BehavioralRewrite.\allowbreak{}ZeroErrorPolicyEquivalent} {\tiny\ttfamily Paper1PL/\allowbreak BehavioralRewrite.lean}
\item \textbf{\nolinkurl{BWR41}}\hypertarget{lh:BWR41}{}\enspace{\ttfamily Ssot.\allowbreak{}BehavioralRewrite.\allowbreak{}ZeroErrorPolicyEquivalent.\allowbreak{}pointwiseAgreement} {\tiny\ttfamily Paper1PL/\allowbreak BehavioralRewrite.lean}
\item \textbf{\nolinkurl{BWR42}}\hypertarget{lh:BWR42}{}\enspace{\ttfamily Ssot.\allowbreak{}BehavioralRewrite.\allowbreak{}same\_\allowbreak{}empty\_\allowbreak{}native\_\allowbreak{}ancestry\_\allowbreak{}exact\_\allowbreak{}for\_\allowbreak{}laminar\_\allowbreak{}and\_\allowbreak{}crossing} {\tiny\ttfamily Paper1PL/\allowbreak BehavioralRewrite.lean}
\item \textbf{\nolinkurl{BWR43}}\hypertarget{lh:BWR43}{}\enspace{\ttfamily Ssot.\allowbreak{}BehavioralRewrite.\allowbreak{}representationState\_\allowbreak{}ne\_\allowbreak{}of\_\allowbreak{}requiredRelation\_\allowbreak{}ne} {\tiny\ttfamily Paper1PL/\allowbreak BehavioralRewrite.lean}
\item \textbf{\nolinkurl{BWR44}}\hypertarget{lh:BWR44}{}\enspace{\ttfamily Ssot.\allowbreak{}BehavioralRewrite.\allowbreak{}residual\_\allowbreak{}ne\_\allowbreak{}of\_\allowbreak{}scope\_\allowbreak{}ne\_\allowbreak{}of\_\allowbreak{}native\_\allowbreak{}eq} {\tiny\ttfamily Paper1PL/\allowbreak BehavioralRewrite.lean}
\item \textbf{\nolinkurl{BWR45}}\hypertarget{lh:BWR45}{}\enspace{\ttfamily Ssot.\allowbreak{}BehavioralRewrite.\allowbreak{}bundledResidualAnswer\_\allowbreak{}iff\_\allowbreak{}of\_\allowbreak{}same\_\allowbreak{}bundle} {\tiny\ttfamily Paper1PL/\allowbreak BehavioralRewrite.lean}
\item \textbf{\nolinkurl{BWR46}}\hypertarget{lh:BWR46}{}\enspace{\ttfamily Ssot.\allowbreak{}BehavioralRewrite.\allowbreak{}exactBundledResidualTopology\_\allowbreak{}iff\_\allowbreak{}eq\_\allowbreak{}univ\_\allowbreak{}of\_\allowbreak{}surjective} {\tiny\ttfamily Paper1PL/\allowbreak BehavioralRewrite.lean}
\item \textbf{\nolinkurl{BWR47}}\hypertarget{lh:BWR47}{}\enspace{\ttfamily Ssot.\allowbreak{}BehavioralRewrite.\allowbreak{}independentBundledResidualTopology\_\allowbreak{}card} {\tiny\ttfamily Paper1PL/\allowbreak BehavioralRewrite.lean}
\item \textbf{\nolinkurl{BWR48}}\hypertarget{lh:BWR48}{}\enspace{\ttfamily Ssot.\allowbreak{}BehavioralRewrite.\allowbreak{}existsUnique\_\allowbreak{}exactBundledResidualTopology\_\allowbreak{}of\_\allowbreak{}surjective} {\tiny\ttfamily Paper1PL/\allowbreak BehavioralRewrite.lean}
\item \textbf{\nolinkurl{BWR49}}\hypertarget{lh:BWR49}{}\enspace{\ttfamily Ssot.\allowbreak{}BehavioralRewrite.\allowbreak{}zeroErrorOverReachableBundledResidualTopologies\_\allowbreak{}iff\_\allowbreak{}subset\_\allowbreak{}complete} {\tiny\ttfamily Paper1PL/\allowbreak BehavioralRewrite.lean}
\item \textbf{\nolinkurl{BWR50}}\hypertarget{lh:BWR50}{}\enspace{\ttfamily Ssot.\allowbreak{}BehavioralRewrite.\allowbreak{}disjointAdditiveCost\_\allowbreak{}eq\_\allowbreak{}sum\_\allowbreak{}singletons} {\tiny\ttfamily Paper1PL/\allowbreak BehavioralRewrite.lean}
\item \textbf{\nolinkurl{BWR51}}\hypertarget{lh:BWR51}{}\enspace{\ttfamily Ssot.\allowbreak{}BehavioralRewrite.\allowbreak{}uniformDisjointAdditiveCost\_\allowbreak{}eq\_\allowbreak{}card\_\allowbreak{}mul\_\allowbreak{}unit} {\tiny\ttfamily Paper1PL/\allowbreak BehavioralRewrite.lean}
\item \textbf{\nolinkurl{BWR52}}\hypertarget{lh:BWR52}{}\enspace{\ttfamily Ssot.\allowbreak{}BehavioralRewrite.\allowbreak{}normalizedDisjointAdditiveCost\_\allowbreak{}eq\_\allowbreak{}card} {\tiny\ttfamily Paper1PL/\allowbreak BehavioralRewrite.lean}
\item \textbf{\nolinkurl{BWR53}}\hypertarget{lh:BWR53}{}\enspace{\ttfamily Ssot.\allowbreak{}BehavioralRewrite.\allowbreak{}positiveSingletonDisjointAdditiveCost\_\allowbreak{}pos\_\allowbreak{}of\_\allowbreak{}nonempty} {\tiny\ttfamily Paper1PL/\allowbreak BehavioralRewrite.lean}
\item \textbf{\nolinkurl{BWR54}}\hypertarget{lh:BWR54}{}\enspace{\ttfamily Ssot.\allowbreak{}BehavioralRewrite.\allowbreak{}semanticEditDistance\_\allowbreak{}eq\_\allowbreak{}zero\_\allowbreak{}iff} {\tiny\ttfamily Paper1PL/\allowbreak BehavioralRewrite.lean}
\item \textbf{\nolinkurl{BWR55}}\hypertarget{lh:BWR55}{}\enspace{\ttfamily Ssot.\allowbreak{}BehavioralRewrite.\allowbreak{}normalizedDisjointAdditiveCost\_\allowbreak{}changedAnswers\_\allowbreak{}eq\_\allowbreak{}semanticEditDistance} {\tiny\ttfamily Paper1PL/\allowbreak BehavioralRewrite.lean}
\item \textbf{\nolinkurl{BWR56}}\hypertarget{lh:BWR56}{}\enspace{\ttfamily Ssot.\allowbreak{}BehavioralRewrite.\allowbreak{}semanticEditDistance\_\allowbreak{}ceilDiv\_\allowbreak{}le\_\allowbreak{}edits} {\tiny\ttfamily Paper1PL/\allowbreak BehavioralRewrite.lean}
\item \textbf{\nolinkurl{BWR57}}\hypertarget{lh:BWR57}{}\enspace{\ttfamily Ssot.\allowbreak{}BehavioralRewrite.\allowbreak{}reachableInvalidResidualTopologyCount\_\allowbreak{}eq\_\allowbreak{}zero\_\allowbreak{}iff\_\allowbreak{}zeroError} {\tiny\ttfamily Paper1PL/\allowbreak BehavioralRewrite.lean}
\item \textbf{\nolinkurl{BWR58}}\hypertarget{lh:BWR58}{}\enspace{\ttfamily Ssot.\allowbreak{}BehavioralRewrite.\allowbreak{}ExactValidExpansion.\allowbreak{}valid\_\allowbreak{}iff\_\allowbreak{}mem\_\allowbreak{}range} {\tiny\ttfamily Paper1PL/\allowbreak BehavioralRewrite.lean}
\item \textbf{\nolinkurl{BWR59}}\hypertarget{lh:BWR59}{}\enspace{\ttfamily Ssot.\allowbreak{}BehavioralRewrite.\allowbreak{}ExactValidExpansion.\allowbreak{}range\_\allowbreak{}expand\_\allowbreak{}eq\_\allowbreak{}setOf\_\allowbreak{}valid} {\tiny\ttfamily Paper1PL/\allowbreak BehavioralRewrite.lean}
\item \textbf{\nolinkurl{BWR60}}\hypertarget{lh:BWR60}{}\enspace{\ttfamily Ssot.\allowbreak{}BehavioralRewrite.\allowbreak{}ExactValidExpansion.\allowbreak{}unrestricted\_\allowbreak{}excess\_\allowbreak{}eq\_\allowbreak{}invalid} {\tiny\ttfamily Paper1PL/\allowbreak BehavioralRewrite.lean}
\item \textbf{\nolinkurl{BWR61}}\hypertarget{lh:BWR61}{}\enspace{\ttfamily Ssot.\allowbreak{}BehavioralRewrite.\allowbreak{}replicatedExpansion\_\allowbreak{}range\_\allowbreak{}eq\_\allowbreak{}agreement} {\tiny\ttfamily Paper1PL/\allowbreak BehavioralRewrite.lean}
\item \textbf{\nolinkurl{BWR62}}\hypertarget{lh:BWR62}{}\enspace{\ttfamily Ssot.\allowbreak{}BehavioralRewrite.\allowbreak{}unrestrictedReplicatedExcess\_\allowbreak{}eq\_\allowbreak{}disagreement} {\tiny\ttfamily Paper1PL/\allowbreak BehavioralRewrite.lean}
\item \textbf{\nolinkurl{BWR63}}\hypertarget{lh:BWR63}{}\enspace{\ttfamily Ssot.\allowbreak{}BehavioralRewrite.\allowbreak{}pricedRealization\_\allowbreak{}cost\_\allowbreak{}floor} {\tiny\ttfamily Paper1PL/\allowbreak BehavioralRewrite.lean}
\item \textbf{\nolinkurl{CAA1}}\hypertarget{lh:CAA1}{}\enspace{\ttfamily Ssot.\allowbreak{}CompletionAudit.\allowbreak{}uninspected\_\allowbreak{}omission\_\allowbreak{}has\_\allowbreak{}same\_\allowbreak{}verdict} {\tiny\ttfamily Paper1PL/\allowbreak CompletionAudit.lean}
\item \textbf{\nolinkurl{CAA2}}\hypertarget{lh:CAA2}{}\enspace{\ttfamily Ssot.\allowbreak{}CompletionAudit.\allowbreak{}detects\_\allowbreak{}single\_\allowbreak{}omissions\_\allowbreak{}reads\_\allowbreak{}every\_\allowbreak{}entry} {\tiny\ttfamily Paper1PL/\allowbreak CompletionAudit.lean}
\item \textbf{\nolinkurl{CAA3}}\hypertarget{lh:CAA3}{}\enspace{\ttfamily Ssot.\allowbreak{}CompletionAudit.\allowbreak{}exact\_\allowbreak{}audit\_\allowbreak{}operation\_\allowbreak{}lower\_\allowbreak{}bound} {\tiny\ttfamily Paper1PL/\allowbreak CompletionAudit.lean}
\item \textbf{\nolinkurl{CAA4}}\hypertarget{lh:CAA4}{}\enspace{\ttfamily Ssot.\allowbreak{}CompletionAudit.\allowbreak{}independent\_\allowbreak{}releases\_\allowbreak{}operation\_\allowbreak{}lower\_\allowbreak{}bound} {\tiny\ttfamily Paper1PL/\allowbreak CompletionAudit.lean}
\item \textbf{\nolinkurl{CAA5}}\hypertarget{lh:CAA5}{}\enspace{\ttfamily Ssot.\allowbreak{}CompletionAudit.\allowbreak{}priced\_\allowbreak{}independent\_\allowbreak{}release\_\allowbreak{}audits} {\tiny\ttfamily Paper1PL/\allowbreak CompletionAudit.lean}
\item \textbf{\nolinkurl{CAA6}}\hypertarget{lh:CAA6}{}\enspace{\ttfamily Ssot.\allowbreak{}CompletionAudit.\allowbreak{}unbounded\_\allowbreak{}batch\_\allowbreak{}checks\_\allowbreak{}all\_\allowbreak{}in\_\allowbreak{}one\_\allowbreak{}probe} {\tiny\ttfamily Paper1PL/\allowbreak CompletionAudit.lean}
\item \textbf{\nolinkurl{CAA7}}\hypertarget{lh:CAA7}{}\enspace{\ttfamily Ssot.\allowbreak{}CompletionAudit.\allowbreak{}proved\_\allowbreak{}complete\_\allowbreak{}inputs\_\allowbreak{}need\_\allowbreak{}no\_\allowbreak{}completion\_\allowbreak{}probes} {\tiny\ttfamily Paper1PL/\allowbreak CompletionAudit.lean}
\item \textbf{\nolinkurl{CAP2}}\hypertarget{lh:CAP2}{}\enspace{\ttfamily Ssot.\allowbreak{}NativeCapacityLadder.\allowbreak{}oneContractHierarchy\_\allowbreak{}iff\_\allowbreak{}membershipLaminar} {\tiny\ttfamily Paper1PL/\allowbreak NativeCapacityLadder.lean}
\item \textbf{\nolinkurl{CAP3}}\hypertarget{lh:CAP3}{}\enspace{\ttfamily Ssot.\allowbreak{}NativeCapacityLadder.\allowbreak{}minimumOneContractCompletion\_\allowbreak{}iff\_\allowbreak{}membershipGap} {\tiny\ttfamily Paper1PL/\allowbreak NativeCapacityLadder.lean}
\item \textbf{\nolinkurl{CAP4}}\hypertarget{lh:CAP4}{}\enspace{\ttfamily Ssot.\allowbreak{}NativeCapacityLadder.\allowbreak{}oneContractNative\_\allowbreak{}has\_\allowbreak{}no\_\allowbreak{}unwantedMembership} {\tiny\ttfamily Paper1PL/\allowbreak NativeCapacityLadder.lean}
\item \textbf{\nolinkurl{CAP5}}\hypertarget{lh:CAP5}{}\enspace{\ttfamily Ssot.\allowbreak{}NativeCapacityLadder.\allowbreak{}addOnlyCompletion\_\allowbreak{}cannot\_\allowbreak{}cancel\_\allowbreak{}nativeMembership} {\tiny\ttfamily Paper1PL/\allowbreak NativeCapacityLadder.lean}
\item \textbf{\nolinkurl{CAP6}}\hypertarget{lh:CAP6}{}\enspace{\ttfamily Ssot.\allowbreak{}NativeCapacityLadder.\allowbreak{}multipleContracts\_\allowbreak{}complete} {\tiny\ttfamily Paper1PL/\allowbreak NativeCapacityLadder.lean}
\item \textbf{\nolinkurl{CAP7}}\hypertarget{lh:CAP7}{}\enspace{\ttfamily Ssot.\allowbreak{}NativeCapacityLadder.\allowbreak{}jointOutsideAnswers\_\allowbreak{}card} {\tiny\ttfamily Paper1PL/\allowbreak NativeCapacityLadder.lean}
\item \textbf{\nolinkurl{CND1}}\hypertarget{lh:CND1}{}\enspace{\ttfamily Ssot.\allowbreak{}CarrierNonDichotomy.\allowbreak{}implementation\_\allowbreak{}ancestry\_\allowbreak{}iff\_\allowbreak{}contract\_\allowbreak{}requires\_\allowbreak{}implementation} {\tiny\ttfamily Paper1PL/\allowbreak CarrierNonDichotomy.lean}
\item \textbf{\nolinkurl{CND2}}\hypertarget{lh:CND2}{}\enspace{\ttfamily Ssot.\allowbreak{}CarrierNonDichotomy.\allowbreak{}ancestry\_\allowbreak{}supplies\_\allowbreak{}required\_\allowbreak{}implementation} {\tiny\ttfamily Paper1PL/\allowbreak CarrierNonDichotomy.lean}
\item \textbf{\nolinkurl{CND3}}\hypertarget{lh:CND3}{}\enspace{\ttfamily Ssot.\allowbreak{}CarrierNonDichotomy.\allowbreak{}routing\_\allowbreak{}supplies\_\allowbreak{}required\_\allowbreak{}implementation} {\tiny\ttfamily Paper1PL/\allowbreak CarrierNonDichotomy.lean}
\item \textbf{\nolinkurl{CND4}}\hypertarget{lh:CND4}{}\enspace{\ttfamily Ssot.\allowbreak{}CarrierNonDichotomy.\allowbreak{}exact\_\allowbreak{}ancestry\_\allowbreak{}and\_\allowbreak{}routing\_\allowbreak{}agree\_\allowbreak{}on\_\allowbreak{}required\_\allowbreak{}implementation} {\tiny\ttfamily Paper1PL/\allowbreak CarrierNonDichotomy.lean}
\item \textbf{\nolinkurl{CND5}}\hypertarget{lh:CND5}{}\enspace{\ttfamily Ssot.\allowbreak{}CarrierNonDichotomy.\allowbreak{}ancestry\_\allowbreak{}and\_\allowbreak{}required\_\allowbreak{}possession\_\allowbreak{}cooccur} {\tiny\ttfamily Paper1PL/\allowbreak CarrierNonDichotomy.lean}
\item \textbf{\nolinkurl{CND6}}\hypertarget{lh:CND6}{}\enspace{\ttfamily Ssot.\allowbreak{}CarrierNonDichotomy.\allowbreak{}ancestry\_\allowbreak{}derives\_\allowbreak{}supply} {\tiny\ttfamily Paper1PL/\allowbreak CarrierNonDichotomy.lean}
\item \textbf{\nolinkurl{CND7}}\hypertarget{lh:CND7}{}\enspace{\ttfamily Ssot.\allowbreak{}CarrierNonDichotomy.\allowbreak{}ancestry\_\allowbreak{}and\_\allowbreak{}supply\_\allowbreak{}not\_\allowbreak{}mutually\_\allowbreak{}exclusive} {\tiny\ttfamily Paper1PL/\allowbreak CarrierNonDichotomy.lean}
\item \textbf{\nolinkurl{CPL1}}\hypertarget{lh:CPL1}{}\enspace{\ttfamily Ssot.\allowbreak{}CoupledEdges.\allowbreak{}independentGaps\_\allowbreak{}le\_\allowbreak{}coupledGap} {\tiny\ttfamily Paper1PL/\allowbreak CoupledEdges.lean}
\item \textbf{\nolinkurl{CPL2}}\hypertarget{lh:CPL2}{}\enspace{\ttfamily Ssot.\allowbreak{}CoupledEdges.\allowbreak{}coupledGap\_\allowbreak{}strict\_\allowbreak{}witness} {\tiny\ttfamily Paper1PL/\allowbreak CoupledEdges.lean}
\item \textbf{\nolinkurl{CPL3}}\hypertarget{lh:CPL3}{}\enspace{\ttfamily Ssot.\allowbreak{}CoupledEdges.\allowbreak{}symmDiff\_\allowbreak{}le\_\allowbreak{}dualRoleGap} {\tiny\ttfamily Paper1PL/\allowbreak CoupledEdges.lean}
\item \textbf{\nolinkurl{CPL4}}\hypertarget{lh:CPL4}{}\enspace{\ttfamily Ssot.\allowbreak{}CoupledEdges.\allowbreak{}dualRoleGap\_\allowbreak{}inter\_\allowbreak{}eq\_\allowbreak{}symmDiff} {\tiny\ttfamily Paper1PL/\allowbreak CoupledEdges.lean}
\item \textbf{\nolinkurl{CRS3}}\hypertarget{lh:CRS3}{}\enspace{\ttfamily Ssot.\allowbreak{}ContractRelativeSoundness.\allowbreak{}conformance\_\allowbreak{}does\_\allowbreak{}not\_\allowbreak{}determine\_\allowbreak{}inheritedBehavior} {\tiny\ttfamily Paper1PL/\allowbreak ContractRelativeSoundness.lean}
\item \textbf{\nolinkurl{CRS4}}\hypertarget{lh:CRS4}{}\enspace{\ttfamily Ssot.\allowbreak{}ContractRelativeSoundness.\allowbreak{}inheritedBehavior\_\allowbreak{}does\_\allowbreak{}not\_\allowbreak{}determine\_\allowbreak{}inheritedInvariant} {\tiny\ttfamily Paper1PL/\allowbreak ContractRelativeSoundness.lean}
\item \textbf{\nolinkurl{CSD1}}\hypertarget{lh:CSD1}{}\enspace{\ttfamily Ssot.\allowbreak{}CapacityDescent.\allowbreak{}outsideAnswers\_\allowbreak{}mono\_\allowbreak{}lowerOne} {\tiny\ttfamily Paper1PL/\allowbreak CapacityDescent.lean}
\item \textbf{\nolinkurl{CSD2}}\hypertarget{lh:CSD2}{}\enspace{\ttfamily Ssot.\allowbreak{}CapacityDescent.\allowbreak{}outsideAnswers\_\allowbreak{}strict\_\allowbreak{}lowerOne\_\allowbreak{}of\_\allowbreak{}used} {\tiny\ttfamily Paper1PL/\allowbreak CapacityDescent.lean}
\item \textbf{\nolinkurl{CSD3}}\hypertarget{lh:CSD3}{}\enspace{\ttfamily Ssot.\allowbreak{}CapacityDescent.\allowbreak{}simplifyingStep\_\allowbreak{}eq\_\allowbreak{}lowerOne\_\allowbreak{}of\_\allowbreak{}positive} {\tiny\ttfamily Paper1PL/\allowbreak CapacityDescent.lean}
\item \textbf{\nolinkurl{CSD4}}\hypertarget{lh:CSD4}{}\enspace{\ttfamily Ssot.\allowbreak{}CapacityDescent.\allowbreak{}simplifyingStep\_\allowbreak{}rearms} {\tiny\ttfamily Paper1PL/\allowbreak CapacityDescent.lean}
\item \textbf{\nolinkurl{CSD5}}\hypertarget{lh:CSD5}{}\enspace{\ttfamily Ssot.\allowbreak{}CapacityDescent.\allowbreak{}lowerOne\_\allowbreak{}four\_\allowbreak{}steps} {\tiny\ttfamily Paper1PL/\allowbreak CapacityDescent.lean}
\item \textbf{\nolinkurl{CSD6}}\hypertarget{lh:CSD6}{}\enspace{\ttfamily Ssot.\allowbreak{}CapacityDescent.\allowbreak{}simplifyingPolicy\_\allowbreak{}four\_\allowbreak{}steps} {\tiny\ttfamily Paper1PL/\allowbreak CapacityDescent.lean}
\item \textbf{\nolinkurl{CSD7}}\hypertarget{lh:CSD7}{}\enspace{\ttfamily Ssot.\allowbreak{}CapacityDescent.\allowbreak{}structuralOnly\_\allowbreak{}outsideAnswers} {\tiny\ttfamily Paper1PL/\allowbreak CapacityDescent.lean}
\item \textbf{\nolinkurl{CSD8}}\hypertarget{lh:CSD8}{}\enspace{\ttfamily Ssot.\allowbreak{}CapacityDescent.\allowbreak{}maintenanceCharge\_\allowbreak{}mono\_\allowbreak{}of\_\allowbreak{}unitCost\_\allowbreak{}le} {\tiny\ttfamily Paper1PL/\allowbreak CapacityDescent.lean}
\item \textbf{\nolinkurl{CWF1}}\hypertarget{lh:CWF1}{}\enspace{\ttfamily Ssot.\allowbreak{}RelationCarrier.\allowbreak{}chargedTranscript\_\allowbreak{}bits\_\allowbreak{}lower\_\allowbreak{}bound} {\tiny\ttfamily Paper1IT/\allowbreak RelationCarrier.lean}
\item \textbf{\nolinkurl{CWF2}}\hypertarget{lh:CWF2}{}\enspace{\ttfamily Ssot.\allowbreak{}RelationCarrier.\allowbreak{}chargedTranscript\_\allowbreak{}actions\_\allowbreak{}lower\_\allowbreak{}bound} {\tiny\ttfamily Paper1IT/\allowbreak RelationCarrier.lean}
\item \textbf{\nolinkurl{CWF3}}\hypertarget{lh:CWF3}{}\enspace{\ttfamily Ssot.\allowbreak{}RelationCarrier.\allowbreak{}chargedTranscript\_\allowbreak{}incremental\_\allowbreak{}cost\_\allowbreak{}lower\_\allowbreak{}bound} {\tiny\ttfamily Paper1IT/\allowbreak RelationCarrier.lean}
\item \textbf{\nolinkurl{CWF4}}\hypertarget{lh:CWF4}{}\enspace{\ttfamily Ssot.\allowbreak{}RelationCarrier.\allowbreak{}chargedTranscript\_\allowbreak{}actions\_\allowbreak{}pos} {\tiny\ttfamily Paper1IT/\allowbreak RelationCarrier.lean}
\item \textbf{\nolinkurl{CWF5}}\hypertarget{lh:CWF5}{}\enspace{\ttfamily Ssot.\allowbreak{}RelationCarrier.\allowbreak{}chargedTranscript\_\allowbreak{}incremental\_\allowbreak{}cost\_\allowbreak{}pos} {\tiny\ttfamily Paper1IT/\allowbreak RelationCarrier.lean}
\item \textbf{\nolinkurl{CWF6}}\hypertarget{lh:CWF6}{}\enspace{\ttfamily Ssot.\allowbreak{}RelationCarrier.\allowbreak{}independentRelation\_\allowbreak{}acquisition\_\allowbreak{}actions\_\allowbreak{}lower\_\allowbreak{}bound} {\tiny\ttfamily Paper1IT/\allowbreak RelationCarrier.lean}
\item \textbf{\nolinkurl{CWF7}}\hypertarget{lh:CWF7}{}\enspace{\ttfamily Ssot.\allowbreak{}RelationCarrier.\allowbreak{}sharedSpecification\_\allowbreak{}zero\_\allowbreak{}acquisition} {\tiny\ttfamily Paper1IT/\allowbreak RelationCarrier.lean}
\item \textbf{\nolinkurl{CWF8}}\hypertarget{lh:CWF8}{}\enspace{\ttfamily Ssot.\allowbreak{}RelationCarrier.\allowbreak{}unavailableBatcher\_\allowbreak{}setup\_\allowbreak{}pos} {\tiny\ttfamily Paper1IT/\allowbreak RelationCarrier.lean}
\item \textbf{\nolinkurl{CWF9}}\hypertarget{lh:CWF9}{}\enspace{\ttfamily Ssot.\allowbreak{}RelationCarrier.\allowbreak{}unavailableBatcher\_\allowbreak{}setup\_\allowbreak{}cost\_\allowbreak{}floor} {\tiny\ttfamily Paper1IT/\allowbreak RelationCarrier.lean}
\item \textbf{\nolinkurl{DEC1}}\hypertarget{lh:DEC1}{}\enspace{\ttfamily Ssot.\allowbreak{}Decidability.\allowbreak{}adequate\_\allowbreak{}decidable} {\tiny\ttfamily Paper1PL/\allowbreak Decidability.lean}
\item \textbf{\nolinkurl{DEC2}}\hypertarget{lh:DEC2}{}\enspace{\ttfamily Ssot.\allowbreak{}Decidability.\allowbreak{}finiteIndexedLaminar\_\allowbreak{}decidable} {\tiny\ttfamily Paper1PL/\allowbreak Decidability.lean}
\item \textbf{\nolinkurl{DEC3}}\hypertarget{lh:DEC3}{}\enspace{\ttfamily Ssot.\allowbreak{}Decidability.\allowbreak{}bruteForceGap\_\allowbreak{}spec} {\tiny\ttfamily Paper1PL/\allowbreak Decidability.lean}
\item \textbf{\nolinkurl{DEC4}}\hypertarget{lh:DEC4}{}\enspace{\ttfamily Ssot.\allowbreak{}Decidability.\allowbreak{}gap\_\allowbreak{}computable} {\tiny\ttfamily Paper1PL/\allowbreak Decidability.lean}
\item \textbf{\nolinkurl{DEC5}}\hypertarget{lh:DEC5}{}\enspace{\ttfamily Ssot.\allowbreak{}Decidability.\allowbreak{}contractOnly\_\allowbreak{}gap\_\allowbreak{}eq\_\allowbreak{}total} {\tiny\ttfamily Paper1PL/\allowbreak Decidability.lean}
\item \textbf{\nolinkurl{DEC6}}\hypertarget{lh:DEC6}{}\enspace{\ttfamily Ssot.\allowbreak{}Decidability.\allowbreak{}finite\_\allowbreak{}native\_\allowbreak{}price\_\allowbreak{}chain} {\tiny\ttfamily Paper1PL/\allowbreak Decidability.lean}
\item \textbf{\nolinkurl{DEC7}}\hypertarget{lh:DEC7}{}\enspace{\ttfamily Ssot.\allowbreak{}Decidability.\allowbreak{}dropped\_\allowbreak{}requirements\_\allowbreak{}gap\_\allowbreak{}incomparable} {\tiny\ttfamily Paper1PL/\allowbreak Decidability.lean}
\item \textbf{\nolinkurl{DRP1}}\hypertarget{lh:DRP1}{}\enspace{\ttfamily Ssot.\allowbreak{}DerivationRetention.\allowbreak{}Path.\allowbreak{}unique} {\tiny\ttfamily Paper1PL/\allowbreak DerivationRetention.lean}
\item \textbf{\nolinkurl{DRP2}}\hypertarget{lh:DRP2}{}\enspace{\ttfamily Ssot.\allowbreak{}DerivationRetention.\allowbreak{}derivations\_\allowbreak{}subsingleton} {\tiny\ttfamily Paper1PL/\allowbreak DerivationRetention.lean}
\item \textbf{\nolinkurl{DRP3}}\hypertarget{lh:DRP3}{}\enspace{\ttfamily Ssot.\allowbreak{}DerivationRetention.\allowbreak{}charged\_\allowbreak{}deletion\_\allowbreak{}iff\_\allowbreak{}unretained} {\tiny\ttfamily Paper1PL/\allowbreak DerivationRetention.lean}
\item \textbf{\nolinkurl{DRP4}}\hypertarget{lh:DRP4}{}\enspace{\ttfamily Ssot.\allowbreak{}DerivationRetention.\allowbreak{}direct\_\allowbreak{}growth\_\allowbreak{}retains\_\allowbreak{}derivations} {\tiny\ttfamily Paper1PL/\allowbreak DerivationRetention.lean}
\item \textbf{\nolinkurl{DRP5}}\hypertarget{lh:DRP5}{}\enspace{\ttfamily Ssot.\allowbreak{}DerivationRetention.\allowbreak{}derivations\_\allowbreak{}mono} {\tiny\ttfamily Paper1PL/\allowbreak DerivationRetention.lean}
\item \textbf{\nolinkurl{ENC1}}\hypertarget{lh:ENC1}{}\enspace{\ttfamily encoding\_\allowbreak{}location\_\allowbreak{}gap} {\tiny\ttfamily AbstractClassSystem/\allowbreak Extended.lean}
\item \textbf{\nolinkurl{EVO17}}\hypertarget{lh:EVO17}{}\enspace{\ttfamily Ssot.\allowbreak{}EvolvingObservation.\allowbreak{}behaviorAdequateThroughout\_\allowbreak{}iff\_\allowbreak{}properColoring} {\tiny\ttfamily Paper1IT/\allowbreak EvolvingObservation.lean}
\item \textbf{\nolinkurl{EVO18}}\hypertarget{lh:EVO18}{}\enspace{\ttfamily Ssot.\allowbreak{}EvolvingObservation.\allowbreak{}conflictGraph\_\allowbreak{}observedWhen\_\allowbreak{}adj\_\allowbreak{}iff} {\tiny\ttfamily Paper1IT/\allowbreak EvolvingObservation.lean}
\item \textbf{\nolinkurl{EVO23}}\hypertarget{lh:EVO23}{}\enspace{\ttfamily Ssot.\allowbreak{}EvolvingObservation.\allowbreak{}behaviorAdequateThroughout\_\allowbreak{}of\_\allowbreak{}observe\_\allowbreak{}injective} {\tiny\ttfamily Paper1IT/\allowbreak EvolvingObservation.lean}
\item \textbf{\nolinkurl{EVO24}}\hypertarget{lh:EVO24}{}\enspace{\ttfamily Ssot.\allowbreak{}EvolvingObservation.\allowbreak{}tag\_\allowbreak{}injective\_\allowbreak{}of\_\allowbreak{}constant\_\allowbreak{}observation\_\allowbreak{}of\_\allowbreak{}behavior\_\allowbreak{}injective} {\tiny\ttfamily Paper1IT/\allowbreak EvolvingObservation.lean}
\item \textbf{\nolinkurl{EVO25}}\hypertarget{lh:EVO25}{}\enspace{\ttfamily Ssot.\allowbreak{}EvolvingObservation.\allowbreak{}overlappingPairTrace\_\allowbreak{}union\_\allowbreak{}not\_\allowbreak{}cluster} {\tiny\ttfamily Paper1IT/\allowbreak EvolvingObservation.lean}
\item \textbf{\nolinkurl{EVO26}}\hypertarget{lh:EVO26}{}\enspace{\ttfamily Ssot.\allowbreak{}EvolvingObservation.\allowbreak{}behaviorAdequateThroughout\_\allowbreak{}of\_\allowbreak{}tag\_\allowbreak{}injective} {\tiny\ttfamily Paper1IT/\allowbreak EvolvingObservation.lean}
\item \textbf{\nolinkurl{EVO27}}\hypertarget{lh:EVO27}{}\enspace{\ttfamily Ssot.\allowbreak{}EvolvingObservation.\allowbreak{}exists\_\allowbreak{}behaviorAdequateThroughout\_\allowbreak{}card} {\tiny\ttfamily Paper1IT/\allowbreak EvolvingObservation.lean}
\item \textbf{\nolinkurl{EVO28}}\hypertarget{lh:EVO28}{}\enspace{\ttfamily Ssot.\allowbreak{}EvolvingObservation.\allowbreak{}behaviorKeyPrice\_\allowbreak{}range} {\tiny\ttfamily Paper1IT/\allowbreak EvolvingObservation.lean}
\item \textbf{\nolinkurl{EVO29}}\hypertarget{lh:EVO29}{}\enspace{\ttfamily Ssot.\allowbreak{}EvolvingObservation.\allowbreak{}card\_\allowbreak{}le\_\allowbreak{}palette\_\allowbreak{}of\_\allowbreak{}constant\_\allowbreak{}observation\_\allowbreak{}of\_\allowbreak{}behavior\_\allowbreak{}injective} {\tiny\ttfamily Paper1IT/\allowbreak EvolvingObservation.lean}
\item \textbf{\nolinkurl{EVO30}}\hypertarget{lh:EVO30}{}\enspace{\ttfamily Ssot.\allowbreak{}EvolvingObservation.\allowbreak{}exists\_\allowbreak{}behaviorAdequateThroughout\_\allowbreak{}one\_\allowbreak{}of\_\allowbreak{}observe\_\allowbreak{}injective} {\tiny\ttfamily Paper1IT/\allowbreak EvolvingObservation.lean}
\item \textbf{\nolinkurl{EVO38}}\hypertarget{lh:EVO38}{}\enspace{\ttfamily Ssot.\allowbreak{}EvolvingObservation.\allowbreak{}adequateThroughout\_\allowbreak{}precomp} {\tiny\ttfamily Paper1IT/\allowbreak EvolvingObservation.lean}
\item \textbf{\nolinkurl{EVO39}}\hypertarget{lh:EVO39}{}\enspace{\ttfamily Ssot.\allowbreak{}EvolvingObservation.\allowbreak{}conflictGraph\_\allowbreak{}precomp\_\allowbreak{}le} {\tiny\ttfamily Paper1IT/\allowbreak EvolvingObservation.lean}
\item \textbf{\nolinkurl{EVO40}}\hypertarget{lh:EVO40}{}\enspace{\ttfamily Ssot.\allowbreak{}EvolvingObservation.\allowbreak{}sum\_\allowbreak{}palette\_\allowbreak{}eq\_\allowbreak{}sum\_\allowbreak{}additionalNameCount\_\allowbreak{}add\_\allowbreak{}card} {\tiny\ttfamily Paper1IT/\allowbreak EvolvingObservation.lean}
\item \textbf{\nolinkurl{FDC1}}\hypertarget{lh:FDC1}{}\enspace{\ttfamily Ssot.\allowbreak{}FiniteRuleDescriptions.\allowbreak{}exact\_\allowbreak{}generator\_\allowbreak{}iff\_\allowbreak{}computable\_\allowbreak{}provider\_\allowbreak{}bound} {\tiny\ttfamily Paper1PL/\allowbreak FiniteRuleDescriptions.lean}
\item \textbf{\nolinkurl{FDC2}}\hypertarget{lh:FDC2}{}\enspace{\ttfamily Ssot.\allowbreak{}FiniteRuleDescriptions.\allowbreak{}fixed\_\allowbreak{}finite\_\allowbreak{}universe\_\allowbreak{}has\_\allowbreak{}exact\_\allowbreak{}generator} {\tiny\ttfamily Paper1PL/\allowbreak FiniteRuleDescriptions.lean}
\item \textbf{\nolinkurl{FDC3}}\hypertarget{lh:FDC3}{}\enspace{\ttfamily Ssot.\allowbreak{}FiniteRuleDescriptions.\allowbreak{}exact\_\allowbreak{}computable\_\allowbreak{}generator\_\allowbreak{}has\_\allowbreak{}computable\_\allowbreak{}lookup} {\tiny\ttfamily Paper1PL/\allowbreak FiniteRuleDescriptions.lean}
\item \textbf{\nolinkurl{FDC4}}\hypertarget{lh:FDC4}{}\enspace{\ttfamily Ssot.\allowbreak{}FiniteRuleDescriptions.\allowbreak{}firstSuccess\_\allowbreak{}primrec} {\tiny\ttfamily Paper1PL/\allowbreak FiniteRuleDescriptions.lean}
\item \textbf{\nolinkurl{FDC5}}\hypertarget{lh:FDC5}{}\enspace{\ttfamily Ssot.\allowbreak{}FiniteRuleDescriptions.\allowbreak{}firstSuccess\_\allowbreak{}unique} {\tiny\ttfamily Paper1PL/\allowbreak FiniteRuleDescriptions.lean}
\item \textbf{\nolinkurl{FDC6}}\hypertarget{lh:FDC6}{}\enspace{\ttfamily Ssot.\allowbreak{}FiniteRuleDescriptions.\allowbreak{}firstSuccess\_\allowbreak{}exists\_\allowbreak{}iff\_\allowbreak{}halting} {\tiny\ttfamily Paper1PL/\allowbreak FiniteRuleDescriptions.lean}
\item \textbf{\nolinkurl{FDC7}}\hypertarget{lh:FDC7}{}\enspace{\ttfamily Ssot.\allowbreak{}FiniteRuleDescriptions.\allowbreak{}singleton\_\allowbreak{}fibers\_\allowbreak{}do\_\allowbreak{}not\_\allowbreak{}suffice\_\allowbreak{}for\_\allowbreak{}generation} {\tiny\ttfamily Paper1PL/\allowbreak FiniteRuleDescriptions.lean}
\item \textbf{\nolinkurl{FDC8}}\hypertarget{lh:FDC8}{}\enspace{\ttfamily Ssot.\allowbreak{}FiniteRuleDescriptions.\allowbreak{}firstSuccess\_\allowbreak{}no\_\allowbreak{}computable\_\allowbreak{}provider\_\allowbreak{}bound} {\tiny\ttfamily Paper1PL/\allowbreak FiniteRuleDescriptions.lean}
\item \textbf{\nolinkurl{FLD1}}\hypertarget{lh:FLD1}{}\enspace{\ttfamily Ssot.\allowbreak{}FieldMenus.\allowbreak{}parityField\_\allowbreak{}card} {\tiny\ttfamily Paper1PL/\allowbreak FieldMenus.lean}
\item \textbf{\nolinkurl{FLD2}}\hypertarget{lh:FLD2}{}\enspace{\ttfamily Ssot.\allowbreak{}FieldMenus.\allowbreak{}full\_\allowbreak{}profile\_\allowbreak{}injective} {\tiny\ttfamily Paper1PL/\allowbreak FieldMenus.lean}
\item \textbf{\nolinkurl{FLD3}}\hypertarget{lh:FLD3}{}\enspace{\ttfamily Ssot.\allowbreak{}FieldMenus.\allowbreak{}profile\_\allowbreak{}tagFeasible\_\allowbreak{}two} {\tiny\ttfamily Paper1PL/\allowbreak FieldMenus.lean}
\item \textbf{\nolinkurl{FLD4}}\hypertarget{lh:FLD4}{}\enspace{\ttfamily Ssot.\allowbreak{}FieldMenus.\allowbreak{}profile\_\allowbreak{}eq\_\allowbreak{}of\_\allowbreak{}sub\_\allowbreak{}eq} {\tiny\ttfamily Paper1PL/\allowbreak FieldMenus.lean}
\item \textbf{\nolinkurl{FLD5}}\hypertarget{lh:FLD5}{}\enspace{\ttfamily Ssot.\allowbreak{}FieldMenus.\allowbreak{}parity\_\allowbreak{}menu\_\allowbreak{}conflictGraph\_\allowbreak{}eq\_\allowbreak{}top} {\tiny\ttfamily Paper1PL/\allowbreak FieldMenus.lean}
\item \textbf{\nolinkurl{FLD6}}\hypertarget{lh:FLD6}{}\enspace{\ttfamily Ssot.\allowbreak{}FieldMenus.\allowbreak{}parity\_\allowbreak{}menu\_\allowbreak{}each\_\allowbreak{}branch\_\allowbreak{}tagFeasible\_\allowbreak{}two} {\tiny\ttfamily Paper1PL/\allowbreak FieldMenus.lean}
\item \textbf{\nolinkurl{FLD7}}\hypertarget{lh:FLD7}{}\enspace{\ttfamily Ssot.\allowbreak{}FieldMenus.\allowbreak{}parity\_\allowbreak{}menu\_\allowbreak{}needs\_\allowbreak{}global\_\allowbreak{}uniqueness} {\tiny\ttfamily Paper1PL/\allowbreak FieldMenus.lean}
\item \textbf{\nolinkurl{FLD8}}\hypertarget{lh:FLD8}{}\enspace{\ttfamily Ssot.\allowbreak{}FieldMenus.\allowbreak{}profile\_\allowbreak{}eq\_\allowbreak{}iff\_\allowbreak{}eq\_\allowbreak{}or\_\allowbreak{}eq\_\allowbreak{}sub} {\tiny\ttfamily Paper1PL/\allowbreak FieldMenus.lean}
\item \textbf{\nolinkurl{GRI1}}\hypertarget{lh:GRI1}{}\enspace{\ttfamily Ssot.\allowbreak{}GapRatioIndependence.\allowbreak{}bruteForceGap\_\allowbreak{}appendDisjointRow} {\tiny\ttfamily Paper1PL/\allowbreak GapRatioIndependence.lean}
\item \textbf{\nolinkurl{GRI2}}\hypertarget{lh:GRI2}{}\enspace{\ttfamily Ssot.\allowbreak{}GapRatioIndependence.\allowbreak{}appendDisjointRow\_\allowbreak{}total} {\tiny\ttfamily Paper1PL/\allowbreak GapRatioIndependence.lean}
\item \textbf{\nolinkurl{GRI3}}\hypertarget{lh:GRI3}{}\enspace{\ttfamily Ssot.\allowbreak{}GapRatioIndependence.\allowbreak{}appendDisjointRow\_\allowbreak{}gap\_\allowbreak{}and\_\allowbreak{}total} {\tiny\ttfamily Paper1PL/\allowbreak GapRatioIndependence.lean}
\item \textbf{\nolinkurl{GRI4}}\hypertarget{lh:GRI4}{}\enspace{\ttfamily Ssot.\allowbreak{}GapRatioIndependence.\allowbreak{}appendFinRow\_\allowbreak{}gap\_\allowbreak{}and\_\allowbreak{}total} {\tiny\ttfamily Paper1PL/\allowbreak GapRatioIndependence.lean}
\item \textbf{\nolinkurl{HCC1}}\hypertarget{lh:HCC1}{}\enspace{\ttfamily Ssot.\allowbreak{}HolderCoverConfigurations.\allowbreak{}complete\_\allowbreak{}iff\_\allowbreak{}all\_\allowbreak{}holders\_\allowbreak{}enabled} {\tiny\ttfamily Paper1PL/\allowbreak HolderCoverConfigurations.lean}
\item \textbf{\nolinkurl{HCC2}}\hypertarget{lh:HCC2}{}\enspace{\ttfamily Ssot.\allowbreak{}HolderCoverConfigurations.\allowbreak{}distinct\_\allowbreak{}completion\_\allowbreak{}maps\_\allowbreak{}card} {\tiny\ttfamily Paper1PL/\allowbreak HolderCoverConfigurations.lean}
\item \textbf{\nolinkurl{HCC3}}\hypertarget{lh:HCC3}{}\enspace{\ttfamily Ssot.\allowbreak{}HolderCoverConfigurations.\allowbreak{}enabled\_\allowbreak{}subsets\_\allowbreak{}give\_\allowbreak{}distinct\_\allowbreak{}answers} {\tiny\ttfamily Paper1PL/\allowbreak HolderCoverConfigurations.lean}
\item \textbf{\nolinkurl{HCC4}}\hypertarget{lh:HCC4}{}\enspace{\ttfamily Ssot.\allowbreak{}HolderCoverConfigurations.\allowbreak{}minimum\_\allowbreak{}cover\_\allowbreak{}has\_\allowbreak{}private\_\allowbreak{}answers} {\tiny\ttfamily Paper1PL/\allowbreak HolderCoverConfigurations.lean}
\item \textbf{\nolinkurl{HVD1}}\hypertarget{lh:HVD1}{}\enspace{\ttfamily Ssot.\allowbreak{}HistoryViewDominance.\allowbreak{}adequateThroughout\_\allowbreak{}iff\_\allowbreak{}exact\_\allowbreak{}resolvers} {\tiny\ttfamily Paper1PL/\allowbreak HistoryViewDominance.lean}
\item \textbf{\nolinkurl{HVD3}}\hypertarget{lh:HVD3}{}\enspace{\ttfamily Ssot.\allowbreak{}HistoryViewDominance.\allowbreak{}retained\_\allowbreak{}nominal\_\allowbreak{}view\_\allowbreak{}weakly\_\allowbreak{}dominates} {\tiny\ttfamily Paper1PL/\allowbreak HistoryViewDominance.lean}
\item \textbf{\nolinkurl{HVD4}}\hypertarget{lh:HVD4}{}\enspace{\ttfamily Ssot.\allowbreak{}HistoryViewDominance.\allowbreak{}retained\_\allowbreak{}nominal\_\allowbreak{}view\_\allowbreak{}needs\_\allowbreak{}no\_\allowbreak{}residual\_\allowbreak{}names} {\tiny\ttfamily Paper1PL/\allowbreak HistoryViewDominance.lean}
\item \textbf{\nolinkurl{HVD7}}\hypertarget{lh:HVD7}{}\enspace{\ttfamily Ssot.\allowbreak{}HistoryViewDominance.\allowbreak{}collision\_\allowbreak{}gives\_\allowbreak{}strict\_\allowbreak{}history\_\allowbreak{}separation} {\tiny\ttfamily Paper1PL/\allowbreak HistoryViewDominance.lean}
\item \textbf{\nolinkurl{HVD8}}\hypertarget{lh:HVD8}{}\enspace{\ttfamily Ssot.\allowbreak{}HistoryViewDominance.\allowbreak{}derived\_\allowbreak{}helpers\_\allowbreak{}cannot\_\allowbreak{}remove\_\allowbreak{}collision} {\tiny\ttfamily Paper1PL/\allowbreak HistoryViewDominance.lean}
\item \textbf{\nolinkurl{HVD11}}\hypertarget{lh:HVD11}{}\enspace{\ttfamily Ssot.\allowbreak{}HistoryViewDominance.\allowbreak{}merging\_\allowbreak{}history\_\allowbreak{}strict\_\allowbreak{}separation} {\tiny\ttfamily Paper1PL/\allowbreak HistoryViewDominance.lean}
\item \textbf{\nolinkurl{HVD12}}\hypertarget{lh:HVD12}{}\enspace{\ttfamily Ssot.\allowbreak{}HistoryViewDominance.\allowbreak{}collision\_\allowbreak{}forces\_\allowbreak{}recourse\_\allowbreak{}but\_\allowbreak{}nominal\_\allowbreak{}stays\_\allowbreak{}unchanged} {\tiny\ttfamily Paper1PL/\allowbreak HistoryViewDominance.lean}
\item \textbf{\nolinkurl{IDD1}}\hypertarget{lh:IDD1}{}\enspace{\ttfamily Ssot.\allowbreak{}IdentityDiscipline.\allowbreak{}structural\_\allowbreak{}adequate\_\allowbreak{}iff\_\allowbreak{}profile\_\allowbreak{}injective} {\tiny\ttfamily Paper1IT/\allowbreak IdentityDiscipline.lean}
\item \textbf{\nolinkurl{IDD2}}\hypertarget{lh:IDD2}{}\enspace{\ttfamily Ssot.\allowbreak{}IdentityDiscipline.\allowbreak{}structural\_\allowbreak{}inadequate\_\allowbreak{}of\_\allowbreak{}clone} {\tiny\ttfamily Paper1IT/\allowbreak IdentityDiscipline.lean}
\item \textbf{\nolinkurl{IDD3}}\hypertarget{lh:IDD3}{}\enspace{\ttfamily Ssot.\allowbreak{}IdentityDiscipline.\allowbreak{}every\_\allowbreak{}structural\_\allowbreak{}discipline\_\allowbreak{}fails\_\allowbreak{}on\_\allowbreak{}clone\_\allowbreak{}extension} {\tiny\ttfamily Paper1IT/\allowbreak IdentityDiscipline.lean}
\item \textbf{\nolinkurl{IDD4}}\hypertarget{lh:IDD4}{}\enspace{\ttfamily Ssot.\allowbreak{}IdentityDiscipline.\allowbreak{}nominal\_\allowbreak{}extension\_\allowbreak{}stable} {\tiny\ttfamily Paper1IT/\allowbreak IdentityDiscipline.lean}
\item \textbf{\nolinkurl{IDD5}}\hypertarget{lh:IDD5}{}\enspace{\ttfamily Ssot.\allowbreak{}IdentityDiscipline.\allowbreak{}canonical\_\allowbreak{}adequate} {\tiny\ttfamily Paper1IT/\allowbreak IdentityDiscipline.lean}
\item \textbf{\nolinkurl{IDD6}}\hypertarget{lh:IDD6}{}\enspace{\ttfamily Ssot.\allowbreak{}IdentityDiscipline.\allowbreak{}structurallyEnforceable\_\allowbreak{}iff\_\allowbreak{}fiberSaturated} {\tiny\ttfamily Paper1IT/\allowbreak IdentityDiscipline.lean}
\item \textbf{\nolinkurl{IDD7}}\hypertarget{lh:IDD7}{}\enspace{\ttfamily Ssot.\allowbreak{}IdentityDiscipline.\allowbreak{}not\_\allowbreak{}structurallyEnforceable\_\allowbreak{}of\_\allowbreak{}boundary\_\allowbreak{}clone} {\tiny\ttfamily Paper1IT/\allowbreak IdentityDiscipline.lean}
\item \textbf{\nolinkurl{IMR1}}\hypertarget{lh:IMR1}{}\enspace{\ttfamily Ssot.\allowbreak{}IdentityMigration.\allowbreak{}clone\_\allowbreak{}extension\_\allowbreak{}forces\_\allowbreak{}disciplineChange\_\allowbreak{}eq\_\allowbreak{}one} {\tiny\ttfamily Paper1PL/\allowbreak IdentityMigration.lean}
\item \textbf{\nolinkurl{IMR2}}\hypertarget{lh:IMR2}{}\enspace{\ttfamily Ssot.\allowbreak{}IdentityMigration.\allowbreak{}nominal\_\allowbreak{}extension\_\allowbreak{}has\_\allowbreak{}zero\_\allowbreak{}disciplineChange} {\tiny\ttfamily Paper1PL/\allowbreak IdentityMigration.lean}
\item \textbf{\nolinkurl{IMR3}}\hypertarget{lh:IMR3}{}\enspace{\ttfamily Ssot.\allowbreak{}IdentityMigration.\allowbreak{}equal\_\allowbreak{}initial\_\allowbreak{}cost\_\allowbreak{}clone\_\allowbreak{}upgrade\_\allowbreak{}strict} {\tiny\ttfamily Paper1PL/\allowbreak IdentityMigration.lean}
\item \textbf{\nolinkurl{IMR4}}\hypertarget{lh:IMR4}{}\enspace{\ttfamily Ssot.\allowbreak{}IdentityMigration.\allowbreak{}retained\_\allowbreak{}structural\_\allowbreak{}upgrade\_\allowbreak{}costs\_\allowbreak{}twice\_\allowbreak{}direct} {\tiny\ttfamily Paper1PL/\allowbreak IdentityMigration.lean}
\item \textbf{\nolinkurl{IMR5}}\hypertarget{lh:IMR5}{}\enspace{\ttfamily Ssot.\allowbreak{}IdentityMigration.\allowbreak{}replaced\_\allowbreak{}structural\_\allowbreak{}upgrade\_\allowbreak{}costs\_\allowbreak{}three\_\allowbreak{}times\_\allowbreak{}direct} {\tiny\ttfamily Paper1PL/\allowbreak IdentityMigration.lean}
\item \textbf{\nolinkurl{INC1}}\hypertarget{lh:INC1}{}\enspace{\ttfamily Ssot.\allowbreak{}DeclaredIncentive.\allowbreak{}declaredLedger\_\allowbreak{}strictly\_\allowbreak{}prefers\_\allowbreak{}decline} {\tiny\ttfamily Paper1PL/\allowbreak DeclaredIncentive.lean}
\item \textbf{\nolinkurl{INC2}}\hypertarget{lh:INC2}{}\enspace{\ttfamily Ssot.\allowbreak{}DeclaredIncentive.\allowbreak{}hiddenCompletionIndicator\_\allowbreak{}laminar\_\allowbreak{}eq\_\allowbreak{}zero} {\tiny\ttfamily Paper1PL/\allowbreak DeclaredIncentive.lean}
\item \textbf{\nolinkurl{INC3}}\hypertarget{lh:INC3}{}\enspace{\ttfamily Ssot.\allowbreak{}DeclaredIncentive.\allowbreak{}hiddenCompletionIndicator\_\allowbreak{}crossing\_\allowbreak{}eq\_\allowbreak{}one} {\tiny\ttfamily Paper1PL/\allowbreak DeclaredIncentive.lean}
\item \textbf{\nolinkurl{INC4}}\hypertarget{lh:INC4}{}\enspace{\ttfamily Ssot.\allowbreak{}DeclaredIncentive.\allowbreak{}no\_\allowbreak{}successful\_\allowbreak{}native\_\allowbreak{}observation\_\allowbreak{}estimator\_\allowbreak{}for\_\allowbreak{}completion} {\tiny\ttfamily Paper1PL/\allowbreak DeclaredIncentive.lean}
\item \textbf{\nolinkurl{INC5}}\hypertarget{lh:INC5}{}\enspace{\ttfamily Ssot.\allowbreak{}DeclaredIncentive.\allowbreak{}maintenanceLedger\_\allowbreak{}eventually\_\allowbreak{}prefers\_\allowbreak{}acquire} {\tiny\ttfamily Paper1PL/\allowbreak DeclaredIncentive.lean}
\item \textbf{\nolinkurl{INC6}}\hypertarget{lh:INC6}{}\enspace{\ttfamily Ssot.\allowbreak{}DeclaredIncentive.\allowbreak{}declared\_\allowbreak{}and\_\allowbreak{}maintenance\_\allowbreak{}orders\_\allowbreak{}reverse\_\allowbreak{}after\_\allowbreak{}someHorizon} {\tiny\ttfamily Paper1PL/\allowbreak DeclaredIncentive.lean}
\item \textbf{\nolinkurl{INC7}}\hypertarget{lh:INC7}{}\enspace{\ttfamily Ssot.\allowbreak{}DeclaredIncentive.\allowbreak{}zeroValue\_\allowbreak{}decline\_\allowbreak{}cost\_\allowbreak{}independent\_\allowbreak{}of\_\allowbreak{}repairPrice} {\tiny\ttfamily Paper1PL/\allowbreak DeclaredIncentive.lean}
\item \textbf{\nolinkurl{INC8}}\hypertarget{lh:INC8}{}\enspace{\ttfamily Ssot.\allowbreak{}DeclaredIncentive.\allowbreak{}reversalThreshold\_\allowbreak{}antitone\_\allowbreak{}in\_\allowbreak{}repairPrice} {\tiny\ttfamily Paper1PL/\allowbreak DeclaredIncentive.lean}
\item \textbf{\nolinkurl{INC9}}\hypertarget{lh:INC9}{}\enspace{\ttfamily Ssot.\allowbreak{}DeclaredIncentive.\allowbreak{}observationHidden\_\allowbreak{}completion\_\allowbreak{}reverses\_\allowbreak{}outputOnly\_\allowbreak{}order} {\tiny\ttfamily Paper1PL/\allowbreak DeclaredIncentive.lean}
\item \textbf{\nolinkurl{INT1}}\hypertarget{lh:INT1}{}\enspace{\ttfamily Ssot.\allowbreak{}BoundaryIntent.\allowbreak{}Intent.\allowbreak{}sameAnswer\_\allowbreak{}equivalence} {\tiny\ttfamily Paper1IT/\allowbreak BoundaryIntent.lean}
\item \textbf{\nolinkurl{INT2}}\hypertarget{lh:INT2}{}\enspace{\ttfamily Ssot.\allowbreak{}BoundaryIntent.\allowbreak{}Intent.\allowbreak{}Exact.\allowbreak{}adequate} {\tiny\ttfamily Paper1IT/\allowbreak BoundaryIntent.lean}
\item \textbf{\nolinkurl{INT3}}\hypertarget{lh:INT3}{}\enspace{\ttfamily Ssot.\allowbreak{}BoundaryIntent.\allowbreak{}Intent.\allowbreak{}adequate\_\allowbreak{}iff\_\allowbreak{}exists\_\allowbreak{}reachable\_\allowbreak{}resolver} {\tiny\ttfamily Paper1IT/\allowbreak BoundaryIntent.lean}
\item \textbf{\nolinkurl{INT4}}\hypertarget{lh:INT4}{}\enspace{\ttfamily Ssot.\allowbreak{}BoundaryIntent.\allowbreak{}Intent.\allowbreak{}exact\_\allowbreak{}requiredRelation\_\allowbreak{}iff} {\tiny\ttfamily Paper1IT/\allowbreak BoundaryIntent.lean}
\item \textbf{\nolinkurl{INT5}}\hypertarget{lh:INT5}{}\enspace{\ttfamily Ssot.\allowbreak{}BoundaryIntent.\allowbreak{}Intent.\allowbreak{}no\_\allowbreak{}intent\_\allowbreak{}free\_\allowbreak{}defect} {\tiny\ttfamily Paper1IT/\allowbreak BoundaryIntent.lean}
\item \textbf{\nolinkurl{INT6}}\hypertarget{lh:INT6}{}\enspace{\ttfamily Ssot.\allowbreak{}EvolvingObservation.\allowbreak{}behaviorAdequateThroughout\_\allowbreak{}iff\_\allowbreak{}boundaryIntent} {\tiny\ttfamily Paper1IT/\allowbreak EvolvingObservation.lean}
\item \textbf{\nolinkurl{INT7}}\hypertarget{lh:INT7}{}\enspace{\ttfamily Ssot.\allowbreak{}NominalFactorization.\allowbreak{}IndexedDerivationRepresentation.\allowbreak{}realizes\_\allowbreak{}boundaryIntent} {\tiny\ttfamily Paper1PL/\allowbreak NominalFactorization.lean}
\item \textbf{\nolinkurl{INT8}}\hypertarget{lh:INT8}{}\enspace{\ttfamily Ssot.\allowbreak{}BoundaryIntent.\allowbreak{}Intent.\allowbreak{}Exact.\allowbreak{}alternativeDispatch\_\allowbreak{}eq\_\allowbreak{}contractCall} {\tiny\ttfamily Paper1IT/\allowbreak BoundaryIntent.lean}
\item \textbf{\nolinkurl{INT9}}\hypertarget{lh:INT9}{}\enspace{\ttfamily Ssot.\allowbreak{}BoundaryIntent.\allowbreak{}Intent.\allowbreak{}no\_\allowbreak{}exact\_\allowbreak{}resolver\_\allowbreak{}of\_\allowbreak{}visible\_\allowbreak{}collision} {\tiny\ttfamily Paper1IT/\allowbreak BoundaryIntent.lean}
\item \textbf{\nolinkurl{IPE1}}\hypertarget{lh:IPE1}{}\enspace{\ttfamily Ssot.\allowbreak{}ImmediateParentEvolution.\allowbreak{}between\_\allowbreak{}preserved\_\allowbreak{}path} {\tiny\ttfamily Paper1PL/\allowbreak ImmediateParentEvolution.lean}
\item \textbf{\nolinkurl{IPE2}}\hypertarget{lh:IPE2}{}\enspace{\ttfamily Ssot.\allowbreak{}ImmediateParentEvolution.\allowbreak{}nested\_\allowbreak{}growth\_\allowbreak{}below\_\allowbreak{}stable\_\allowbreak{}base\_\allowbreak{}forces\_\allowbreak{}parent\_\allowbreak{}edge\_\allowbreak{}removal} {\tiny\ttfamily Paper1PL/\allowbreak ImmediateParentEvolution.lean}
\item \textbf{\nolinkurl{IPE3}}\hypertarget{lh:IPE3}{}\enspace{\ttfamily Ssot.\allowbreak{}ImmediateParentEvolution.\allowbreak{}insertionScope\_\allowbreak{}each\_\allowbreak{}state\_\allowbreak{}has\_\allowbreak{}exact\_\allowbreak{}singleParent} {\tiny\ttfamily Paper1PL/\allowbreak ImmediateParentEvolution.lean}
\item \textbf{\nolinkurl{IPE4}}\hypertarget{lh:IPE4}{}\enspace{\ttfamily Ssot.\allowbreak{}ImmediateParentEvolution.\allowbreak{}insertionSingleParentPolicy\_\allowbreak{}removedParentEdgeCount\_\allowbreak{}eq\_\allowbreak{}one} {\tiny\ttfamily Paper1PL/\allowbreak ImmediateParentEvolution.lean}
\item \textbf{\nolinkurl{IPE5}}\hypertarget{lh:IPE5}{}\enspace{\ttfamily Ssot.\allowbreak{}ImmediateParentEvolution.\allowbreak{}insertionScope\_\allowbreak{}multiParent\_\allowbreak{}edges\_\allowbreak{}monotone} {\tiny\ttfamily Paper1PL/\allowbreak ImmediateParentEvolution.lean}
\item \textbf{\nolinkurl{IPE6}}\hypertarget{lh:IPE6}{}\enspace{\ttfamily Ssot.\allowbreak{}ImmediateParentEvolution.\allowbreak{}stagedScope\_\allowbreak{}laminar} {\tiny\ttfamily Paper1PL/\allowbreak ImmediateParentEvolution.lean}
\item \textbf{\nolinkurl{IPE7}}\hypertarget{lh:IPE7}{}\enspace{\ttfamily Ssot.\allowbreak{}ImmediateParentEvolution.\allowbreak{}stagedScope\_\allowbreak{}cumulative\_\allowbreak{}parent\_\allowbreak{}edge\_\allowbreak{}removal\_\allowbreak{}lower\_\allowbreak{}bound} {\tiny\ttfamily Paper1PL/\allowbreak ImmediateParentEvolution.lean}
\item \textbf{\nolinkurl{IPE8}}\hypertarget{lh:IPE8}{}\enspace{\ttfamily Ssot.\allowbreak{}ImmediateParentEvolution.\allowbreak{}stagedScope\_\allowbreak{}multiParent\_\allowbreak{}edges\_\allowbreak{}monotone} {\tiny\ttfamily Paper1PL/\allowbreak ImmediateParentEvolution.lean}
\item \textbf{\nolinkurl{IPE9}}\hypertarget{lh:IPE9}{}\enspace{\ttfamily Ssot.\allowbreak{}ImmediateParentEvolution.\allowbreak{}stagedScope\_\allowbreak{}unbounded\_\allowbreak{}parent\_\allowbreak{}edge\_\allowbreak{}separation} {\tiny\ttfamily Paper1PL/\allowbreak ImmediateParentEvolution.lean}
\item \textbf{\nolinkurl{IPE10}}\hypertarget{lh:IPE10}{}\enspace{\ttfamily Ssot.\allowbreak{}ImmediateParentEvolution.\allowbreak{}ImmediateSingleParentPolicy.\allowbreak{}exists\_\allowbreak{}addedParentEdge\_\allowbreak{}of\_\allowbreak{}new\_\allowbreak{}ancestor} {\tiny\ttfamily Paper1PL/\allowbreak ImmediateParentEvolution.lean}
\item \textbf{\nolinkurl{IPE11}}\hypertarget{lh:IPE11}{}\enspace{\ttfamily Ssot.\allowbreak{}ImmediateParentEvolution.\allowbreak{}insertionScope\_\allowbreak{}oneParent\_\allowbreak{}two\_\allowbreak{}le\_\allowbreak{}parentEdgeEditCount} {\tiny\ttfamily Paper1PL/\allowbreak ImmediateParentEvolution.lean}
\item \textbf{\nolinkurl{IPE12}}\hypertarget{lh:IPE12}{}\enspace{\ttfamily Ssot.\allowbreak{}ImmediateParentEvolution.\allowbreak{}insertionSingleParentPolicy\_\allowbreak{}parentEdgeEditCount\_\allowbreak{}eq\_\allowbreak{}two} {\tiny\ttfamily Paper1PL/\allowbreak ImmediateParentEvolution.lean}
\item \textbf{\nolinkurl{IPE13}}\hypertarget{lh:IPE13}{}\enspace{\ttfamily Ssot.\allowbreak{}ImmediateParentEvolution.\allowbreak{}insertionMultiParent\_\allowbreak{}exact} {\tiny\ttfamily Paper1PL/\allowbreak ImmediateParentEvolution.lean}
\item \textbf{\nolinkurl{IPE14}}\hypertarget{lh:IPE14}{}\enspace{\ttfamily Ssot.\allowbreak{}ImmediateParentEvolution.\allowbreak{}insertionMultiParent\_\allowbreak{}edges\_\allowbreak{}monotone} {\tiny\ttfamily Paper1PL/\allowbreak ImmediateParentEvolution.lean}
\item \textbf{\nolinkurl{IPE15}}\hypertarget{lh:IPE15}{}\enspace{\ttfamily Ssot.\allowbreak{}ImmediateParentEvolution.\allowbreak{}insertionMultiParentAddedEdgeCount\_\allowbreak{}eq\_\allowbreak{}one} {\tiny\ttfamily Paper1PL/\allowbreak ImmediateParentEvolution.lean}
\item \textbf{\nolinkurl{IPE16}}\hypertarget{lh:IPE16}{}\enspace{\ttfamily Ssot.\allowbreak{}ImmediateParentEvolution.\allowbreak{}stagedScope\_\allowbreak{}cumulative\_\allowbreak{}parent\_\allowbreak{}edge\_\allowbreak{}edit\_\allowbreak{}lower\_\allowbreak{}bound} {\tiny\ttfamily Paper1PL/\allowbreak ImmediateParentEvolution.lean}
\item \textbf{\nolinkurl{IPE17}}\hypertarget{lh:IPE17}{}\enspace{\ttfamily Ssot.\allowbreak{}ImmediateParentEvolution.\allowbreak{}stagedMultiParent\_\allowbreak{}exact} {\tiny\ttfamily Paper1PL/\allowbreak ImmediateParentEvolution.lean}
\item \textbf{\nolinkurl{IPE18}}\hypertarget{lh:IPE18}{}\enspace{\ttfamily Ssot.\allowbreak{}ImmediateParentEvolution.\allowbreak{}stagedMultiParentAddedEdgeCount\_\allowbreak{}eq\_\allowbreak{}one} {\tiny\ttfamily Paper1PL/\allowbreak ImmediateParentEvolution.lean}
\item \textbf{\nolinkurl{IPE19}}\hypertarget{lh:IPE19}{}\enspace{\ttfamily Ssot.\allowbreak{}ImmediateParentEvolution.\allowbreak{}stagedMultiParent\_\allowbreak{}cumulative\_\allowbreak{}edge\_\allowbreak{}edit\_\allowbreak{}count\_\allowbreak{}eq\_\allowbreak{}n} {\tiny\ttfamily Paper1PL/\allowbreak ImmediateParentEvolution.lean}
\item \textbf{\nolinkurl{IPE20}}\hypertarget{lh:IPE20}{}\enspace{\ttfamily Ssot.\allowbreak{}ImmediateParentEvolution.\allowbreak{}stagedMultiParent\_\allowbreak{}edges\_\allowbreak{}monotone} {\tiny\ttfamily Paper1PL/\allowbreak ImmediateParentEvolution.lean}
\item \textbf{\nolinkurl{IPE21}}\hypertarget{lh:IPE21}{}\enspace{\ttfamily Ssot.\allowbreak{}ImmediateParentEvolution.\allowbreak{}stagedScope\_\allowbreak{}step\_\allowbreak{}two\_\allowbreak{}edits\_\allowbreak{}or\_\allowbreak{}external\_\allowbreak{}pair} {\tiny\ttfamily Paper1PL/\allowbreak ImmediateParentEvolution.lean}
\item \textbf{\nolinkurl{IPE22}}\hypertarget{lh:IPE22}{}\enspace{\ttfamily Ssot.\allowbreak{}ImmediateParentEvolution.\allowbreak{}stagedFrozenWitness\_\allowbreak{}card} {\tiny\ttfamily Paper1PL/\allowbreak ImmediateParentEvolution.lean}
\item \textbf{\nolinkurl{IPE23}}\hypertarget{lh:IPE23}{}\enspace{\ttfamily Ssot.\allowbreak{}ImmediateParentEvolution.\allowbreak{}stagedScope\_\allowbreak{}frozen\_\allowbreak{}parent\_\allowbreak{}quadratic\_\allowbreak{}external\_\allowbreak{}lower\_\allowbreak{}bound} {\tiny\ttfamily Paper1PL/\allowbreak ImmediateParentEvolution.lean}
\item \textbf{\nolinkurl{IPE24}}\hypertarget{lh:IPE24}{}\enspace{\ttfamily Ssot.\allowbreak{}ImmediateParentEvolution.\allowbreak{}stagedScope\_\allowbreak{}monotone} {\tiny\ttfamily Paper1PL/\allowbreak ImmediateParentEvolution.lean}
\item \textbf{\nolinkurl{IPE25}}\hypertarget{lh:IPE25}{}\enspace{\ttfamily Ssot.\allowbreak{}ImmediateParentEvolution.\allowbreak{}stagedScope\_\allowbreak{}step\_\allowbreak{}retraction\_\allowbreak{}or\_\allowbreak{}external\_\allowbreak{}pair} {\tiny\ttfamily Paper1PL/\allowbreak ImmediateParentEvolution.lean}
\item \textbf{\nolinkurl{IPE26}}\hypertarget{lh:IPE26}{}\enspace{\ttfamily Ssot.\allowbreak{}ImmediateParentEvolution.\allowbreak{}ImmediateDelegatingSingleParentPolicy.\allowbreak{}retracted\_\allowbreak{}edge\_\allowbreak{}pair\_\allowbreak{}remains\_\allowbreak{}required} {\tiny\ttfamily Paper1PL/\allowbreak ImmediateParentEvolution.lean}
\item \textbf{\nolinkurl{IPE27}}\hypertarget{lh:IPE27}{}\enspace{\ttfamily Ssot.\allowbreak{}ImmediateParentEvolution.\allowbreak{}stagedScope\_\allowbreak{}step\_\allowbreak{}edgeMonotone\_\allowbreak{}implies\_\allowbreak{}external\_\allowbreak{}endpoint} {\tiny\ttfamily Paper1PL/\allowbreak ImmediateParentEvolution.lean}
\item \textbf{\nolinkurl{IPE28}}\hypertarget{lh:IPE28}{}\enspace{\ttfamily Ssot.\allowbreak{}ImmediateParentEvolution.\allowbreak{}directMultiParent\_\allowbreak{}edges\_\allowbreak{}monotone} {\tiny\ttfamily Paper1PL/\allowbreak ImmediateParentEvolution.lean}
\item \textbf{\nolinkurl{IPE29}}\hypertarget{lh:IPE29}{}\enspace{\ttfamily Ssot.\allowbreak{}ImmediateParentEvolution.\allowbreak{}stagedGapToFrozenWitness} {\tiny\ttfamily Paper1PL/\allowbreak ImmediateParentEvolution.lean}
\item \textbf{\nolinkurl{IPE30}}\hypertarget{lh:IPE30}{}\enspace{\ttfamily Ssot.\allowbreak{}ImmediateParentEvolution.\allowbreak{}stagedFrozenWitnessToGap\_\allowbreak{}gapToWitness} {\tiny\ttfamily Paper1PL/\allowbreak ImmediateParentEvolution.lean}
\item \textbf{\nolinkurl{IPE31}}\hypertarget{lh:IPE31}{}\enspace{\ttfamily Ssot.\allowbreak{}ImmediateParentEvolution.\allowbreak{}stagedGapToFrozenWitness\_\allowbreak{}injective} {\tiny\ttfamily Paper1PL/\allowbreak ImmediateParentEvolution.lean}
\item \textbf{\nolinkurl{IPE32}}\hypertarget{lh:IPE32}{}\enspace{\ttfamily Ssot.\allowbreak{}ImmediateParentEvolution.\allowbreak{}stagedScope\_\allowbreak{}frozen\_\allowbreak{}initially\_\allowbreak{}exact\_\allowbreak{}external\_\allowbreak{}eq} {\tiny\ttfamily Paper1PL/\allowbreak ImmediateParentEvolution.lean}
\item \textbf{\nolinkurl{IPE33}}\hypertarget{lh:IPE33}{}\enspace{\ttfamily Ssot.\allowbreak{}ImmediateParentEvolution.\allowbreak{}ownerChain\_\allowbreak{}exact} {\tiny\ttfamily Paper1PL/\allowbreak ImmediateParentEvolution.lean}
\item \textbf{\nolinkurl{IPE34}}\hypertarget{lh:IPE34}{}\enspace{\ttfamily Ssot.\allowbreak{}ImmediateParentEvolution.\allowbreak{}ownerChain\_\allowbreak{}acyclic} {\tiny\ttfamily Paper1PL/\allowbreak ImmediateParentEvolution.lean}
\item \textbf{\nolinkurl{IPE35}}\hypertarget{lh:IPE35}{}\enspace{\ttfamily Ssot.\allowbreak{}ImmediateParentEvolution.\allowbreak{}ownerChain\_\allowbreak{}edges\_\allowbreak{}monotone} {\tiny\ttfamily Paper1PL/\allowbreak ImmediateParentEvolution.lean}
\item \textbf{\nolinkurl{IPE36}}\hypertarget{lh:IPE36}{}\enspace{\ttfamily Ssot.\allowbreak{}ImmediateParentEvolution.\allowbreak{}ownerChain\_\allowbreak{}update\_\allowbreak{}iff} {\tiny\ttfamily Paper1PL/\allowbreak ImmediateParentEvolution.lean}
\item \textbf{\nolinkurl{IPE37}}\hypertarget{lh:IPE37}{}\enspace{\ttfamily Ssot.\allowbreak{}ImmediateParentEvolution.\allowbreak{}ownerChainAddedEdgeCount\_\allowbreak{}eq\_\allowbreak{}one} {\tiny\ttfamily Paper1PL/\allowbreak ImmediateParentEvolution.lean}
\item \textbf{\nolinkurl{IPE38}}\hypertarget{lh:IPE38}{}\enspace{\ttfamily Ssot.\allowbreak{}ImmediateParentEvolution.\allowbreak{}ownerChain\_\allowbreak{}cumulative\_\allowbreak{}edge\_\allowbreak{}edit\_\allowbreak{}count\_\allowbreak{}eq\_\allowbreak{}n} {\tiny\ttfamily Paper1PL/\allowbreak ImmediateParentEvolution.lean}
\item \textbf{\nolinkurl{L10}}\hypertarget{lh:L10}{}\enspace{\ttfamily finset\_\allowbreak{}matroid\_\allowbreak{}basis\_\allowbreak{}equicardinality} {\tiny\ttfamily axis\_framework.lean}
\item \textbf{\nolinkurl{L11}}\hypertarget{lh:L11}{}\enspace{\ttfamily inclusionMinimal\_\allowbreak{}spanning\_\allowbreak{}iff\_\allowbreak{}basis} {\tiny\ttfamily axis\_framework.lean}
\item \textbf{\nolinkurl{L12}}\hypertarget{lh:L12}{}\enspace{\ttfamily matroidal\_\allowbreak{}distinguishing\_\allowbreak{}minimal\_\allowbreak{}iff\_\allowbreak{}basis} {\tiny\ttfamily axis\_framework.lean}
\item \textbf{\nolinkurl{L13}}\hypertarget{lh:L13}{}\enspace{\ttfamily exists\_\allowbreak{}inclusionMinimalFamily\_\allowbreak{}subset} {\tiny\ttfamily axis\_framework.lean}
\item \textbf{\nolinkurl{L14}}\hypertarget{lh:L14}{}\enspace{\ttfamily matroidal\_\allowbreak{}distinguishing\_\allowbreak{}criterion\_\allowbreak{}iff} {\tiny\ttfamily axis\_framework.lean}
\item \textbf{\nolinkurl{L15}}\hypertarget{lh:L15}{}\enspace{\ttfamily freeAxisMatroid\_\allowbreak{}basis\_\allowbreak{}eq\_\allowbreak{}ground} {\tiny\ttfamily axis\_framework.lean}
\item \textbf{\nolinkurl{L16}}\hypertarget{lh:L16}{}\enspace{\ttfamily closureInducedAxisMatroid} {\tiny\ttfamily axis\_framework.lean}
\item \textbf{\nolinkurl{L17}}\hypertarget{lh:L17}{}\enspace{\ttfamily closureInducedBasisEquicardinality} {\tiny\ttfamily axis\_framework.lean}
\item \textbf{\nolinkurl{L18}}\hypertarget{lh:L18}{}\enspace{\ttfamily not\_\allowbreak{}spanning\_\allowbreak{}of\_\allowbreak{}card\_\allowbreak{}lt\_\allowbreak{}basis} {\tiny\ttfamily axis\_framework.lean}
\item \textbf{\nolinkurl{L19}}\hypertarget{lh:L19}{}\enspace{\ttfamily spanning\_\allowbreak{}of\_\allowbreak{}basis\_\allowbreak{}card\_\allowbreak{}lt\_\allowbreak{}has\_\allowbreak{}redundant\_\allowbreak{}subset} {\tiny\ttfamily axis\_framework.lean}
\item \textbf{\nolinkurl{LRR1}}\hypertarget{lh:LRR1}{}\enspace{\ttfamily Ssot.\allowbreak{}LocalRepairRatchet.\allowbreak{}finiteGapCost\_\allowbreak{}eq\_\allowbreak{}sum\_\allowbreak{}sdiff} {\tiny\ttfamily Paper1PL/\allowbreak LocalRepairRatchet.lean}
\item \textbf{\nolinkurl{LRR2}}\hypertarget{lh:LRR2}{}\enspace{\ttfamily Ssot.\allowbreak{}LocalRepairRatchet.\allowbreak{}fixedCarrier\_\allowbreak{}completion\_\allowbreak{}growth\_\allowbreak{}eq} {\tiny\ttfamily Paper1PL/\allowbreak LocalRepairRatchet.lean}
\item \textbf{\nolinkurl{LRR3}}\hypertarget{lh:LRR3}{}\enspace{\ttfamily Ssot.\allowbreak{}LocalRepairRatchet.\allowbreak{}fixedCarrier\_\allowbreak{}admissible\_\allowbreak{}after\_\allowbreak{}growth} {\tiny\ttfamily Paper1PL/\allowbreak LocalRepairRatchet.lean}
\item \textbf{\nolinkurl{LRR4}}\hypertarget{lh:LRR4}{}\enspace{\ttfamily Ssot.\allowbreak{}LocalRepairRatchet.\allowbreak{}bruteForceGap\_\allowbreak{}growth\_\allowbreak{}le} {\tiny\ttfamily Paper1PL/\allowbreak LocalRepairRatchet.lean}
\item \textbf{\nolinkurl{LRR5}}\hypertarget{lh:LRR5}{}\enspace{\ttfamily Ssot.\allowbreak{}LocalRepairRatchet.\allowbreak{}arrangementExcess\_\allowbreak{}mono\_\allowbreak{}fixedCarrier} {\tiny\ttfamily Paper1PL/\allowbreak LocalRepairRatchet.lean}
\item \textbf{\nolinkurl{LRR6}}\hypertarget{lh:LRR6}{}\enspace{\ttfamily Ssot.\allowbreak{}LocalRepairRatchet.\allowbreak{}arrangementExcess\_\allowbreak{}strict\_\allowbreak{}fixedCarrier} {\tiny\ttfamily Paper1PL/\allowbreak LocalRepairRatchet.lean}
\item \textbf{\nolinkurl{LRR7}}\hypertarget{lh:LRR7}{}\enspace{\ttfamily Ssot.\allowbreak{}LocalRepairRatchet.\allowbreak{}arrangementExcess\_\allowbreak{}growth\_\allowbreak{}eq\_\allowbreak{}of\_\allowbreak{}gap\_\allowbreak{}eq} {\tiny\ttfamily Paper1PL/\allowbreak LocalRepairRatchet.lean}
\item \textbf{\nolinkurl{MGH1}}\hypertarget{lh:MGH1}{}\enspace{\ttfamily Ssot.\allowbreak{}MonotoneMultiParentHistory.\allowbreak{}path\_\allowbreak{}eq\_\allowbreak{}or\_\allowbreak{}step} {\tiny\ttfamily Paper1PL/\allowbreak MonotoneMultiParentHistory.lean}
\item \textbf{\nolinkurl{MGH2}}\hypertarget{lh:MGH2}{}\enspace{\ttfamily Ssot.\allowbreak{}MonotoneMultiParentHistory.\allowbreak{}exact\_\allowbreak{}supply} {\tiny\ttfamily Paper1PL/\allowbreak MonotoneMultiParentHistory.lean}
\item \textbf{\nolinkurl{MGH3}}\hypertarget{lh:MGH3}{}\enspace{\ttfamily Ssot.\allowbreak{}MonotoneMultiParentHistory.\allowbreak{}acyclic} {\tiny\ttfamily Paper1PL/\allowbreak MonotoneMultiParentHistory.lean}
\item \textbf{\nolinkurl{MGH9}}\hypertarget{lh:MGH9}{}\enspace{\ttfamily Ssot.\allowbreak{}MonotoneMultiParentHistory.\allowbreak{}repairEdges\_\allowbreak{}empty} {\tiny\ttfamily Paper1PL/\allowbreak MonotoneMultiParentHistory.lean}
\item \textbf{\nolinkurl{MGH10}}\hypertarget{lh:MGH10}{}\enspace{\ttfamily Ssot.\allowbreak{}MonotoneMultiParentHistory.\allowbreak{}normalizedStepRepair\_\allowbreak{}zero} {\tiny\ttfamily Paper1PL/\allowbreak MonotoneMultiParentHistory.lean}
\item \textbf{\nolinkurl{MGH11}}\hypertarget{lh:MGH11}{}\enspace{\ttfamily Ssot.\allowbreak{}MonotoneMultiParentHistory.\allowbreak{}arbitrary\_\allowbreak{}history\_\allowbreak{}joint\_\allowbreak{}zero} {\tiny\ttfamily Paper1PL/\allowbreak MonotoneMultiParentHistory.lean}
\item \textbf{\nolinkurl{MRC1}}\hypertarget{lh:MRC1}{}\enspace{\ttfamily Ssot.\allowbreak{}MatchingRecourse.\allowbreak{}population\_\allowbreak{}le\_\allowbreak{}twice\_\allowbreak{}pairCapacity\_\allowbreak{}add\_\allowbreak{}colors} {\tiny\ttfamily Paper1PL/\allowbreak MatchingRecourse.lean}
\item \textbf{\nolinkurl{MRC2}}\hypertarget{lh:MRC2}{}\enspace{\ttfamily Ssot.\allowbreak{}MatchingRecourse.\allowbreak{}population\_\allowbreak{}le\_\allowbreak{}colors\_\allowbreak{}add\_\allowbreak{}twice\_\allowbreak{}recourse} {\tiny\ttfamily Paper1PL/\allowbreak MatchingRecourse.lean}
\item \textbf{\nolinkurl{MRC3}}\hypertarget{lh:MRC3}{}\enspace{\ttfamily Ssot.\allowbreak{}MatchingRecourse.\allowbreak{}exact\_\allowbreak{}price\_\allowbreak{}lower\_\allowbreak{}bound} {\tiny\ttfamily Paper1PL/\allowbreak MatchingRecourse.lean}
\item \textbf{\nolinkurl{MRC4}}\hypertarget{lh:MRC4}{}\enspace{\ttfamily Ssot.\allowbreak{}MatchingRecourse.\allowbreak{}extremal\_\allowbreak{}class\_\allowbreak{}capacity\_\allowbreak{}le\_\allowbreak{}recourse} {\tiny\ttfamily Paper1PL/\allowbreak MatchingRecourse.lean}
\item \textbf{\nolinkurl{MRC5}}\hypertarget{lh:MRC5}{}\enspace{\ttfamily Ssot.\allowbreak{}MatchingRecourse.\allowbreak{}exact\_\allowbreak{}price\_\allowbreak{}upper\_\allowbreak{}parameters} {\tiny\ttfamily Paper1PL/\allowbreak MatchingRecourse.lean}
\item \textbf{\nolinkurl{NHF1}}\hypertarget{lh:NHF1}{}\enspace{\ttfamily Ssot.\allowbreak{}NominalFactorization.\allowbreak{}nominal\_\allowbreak{}not\_\allowbreak{}determined\_\allowbreak{}by\_\allowbreak{}observation} {\tiny\ttfamily Paper1PL/\allowbreak NominalFactorization.lean}
\item \textbf{\nolinkurl{NHF3}}\hypertarget{lh:NHF3}{}\enspace{\ttfamily Ssot.\allowbreak{}NominalFactorization.\allowbreak{}structural\_\allowbreak{}helper\_\allowbreak{}preserves\_\allowbreak{}has\_\allowbreak{}collision} {\tiny\ttfamily Paper1PL/\allowbreak NominalFactorization.lean}
\item \textbf{\nolinkurl{NHF4}}\hypertarget{lh:NHF4}{}\enspace{\ttfamily Ssot.\allowbreak{}NominalFactorization.\allowbreak{}every\_\allowbreak{}behavior\_\allowbreak{}factors\_\allowbreak{}through\_\allowbreak{}nominal\_\allowbreak{}baseline} {\tiny\ttfamily Paper1PL/\allowbreak NominalFactorization.lean}
\item \textbf{\nolinkurl{NHF5}}\hypertarget{lh:NHF5}{}\enspace{\ttfamily Ssot.\allowbreak{}NominalFactorization.\allowbreak{}retaining\_\allowbreak{}nominal\_\allowbreak{}identity\_\allowbreak{}has\_\allowbreak{}no\_\allowbreak{}marginal\_\allowbreak{}penalty} {\tiny\ttfamily Paper1PL/\allowbreak NominalFactorization.lean}
\item \textbf{\nolinkurl{NHF6}}\hypertarget{lh:NHF6}{}\enspace{\ttfamily Ssot.\allowbreak{}NominalFactorization.\allowbreak{}laminar\_\allowbreak{}of\_\allowbreak{}singleInheritanceRepresentation} {\tiny\ttfamily Paper1PL/\allowbreak NominalFactorization.lean}
\item \textbf{\nolinkurl{NHF7}}\hypertarget{lh:NHF7}{}\enspace{\ttfamily Ssot.\allowbreak{}NominalFactorization.\allowbreak{}singleInheritance\_\allowbreak{}iff\_\allowbreak{}laminar} {\tiny\ttfamily Paper1PL/\allowbreak NominalFactorization.lean}
\item \textbf{\nolinkurl{NHF8}}\hypertarget{lh:NHF8}{}\enspace{\ttfamily Ssot.\allowbreak{}NominalFactorization.\allowbreak{}nonlaminar\_\allowbreak{}requires\_\allowbreak{}multiParent} {\tiny\ttfamily Paper1PL/\allowbreak NominalFactorization.lean}
\item \textbf{\nolinkurl{NHF9}}\hypertarget{lh:NHF9}{}\enspace{\ttfamily Ssot.\allowbreak{}NominalFactorization.\allowbreak{}nonlaminar\_\allowbreak{}has\_\allowbreak{}incomparable\_\allowbreak{}applicable\_\allowbreak{}owners} {\tiny\ttfamily Paper1PL/\allowbreak NominalFactorization.lean}
\item \textbf{\nolinkurl{NHF10}}\hypertarget{lh:NHF10}{}\enspace{\ttfamily Ssot.\allowbreak{}NominalFactorization.\allowbreak{}contraction\_\allowbreak{}consistent\_\allowbreak{}resolution\_\allowbreak{}requires\_\allowbreak{}linearization} {\tiny\ttfamily Paper1PL/\allowbreak NominalFactorization.lean}
\item \textbf{\nolinkurl{NHF11}}\hypertarget{lh:NHF11}{}\enspace{\ttfamily Ssot.\allowbreak{}NominalFactorization.\allowbreak{}identity\_\allowbreak{}dependent\_\allowbreak{}behavior\_\allowbreak{}not\_\allowbreak{}structurally\_\allowbreak{}decodable} {\tiny\ttfamily Paper1PL/\allowbreak NominalFactorization.lean}
\item \textbf{\nolinkurl{NHF12}}\hypertarget{lh:NHF12}{}\enspace{\ttfamily Ssot.\allowbreak{}NominalFactorization.\allowbreak{}nominal\_\allowbreak{}hierarchical\_\allowbreak{}factorization} {\tiny\ttfamily Paper1PL/\allowbreak NominalFactorization.lean}
\item \textbf{\nolinkurl{NHF13}}\hypertarget{lh:NHF13}{}\enspace{\ttfamily Ssot.\allowbreak{}NominalFactorization.\allowbreak{}nonlaminar\_\allowbreak{}forces\_\allowbreak{}externalized\_\allowbreak{}derivation} {\tiny\ttfamily Paper1PL/\allowbreak NominalFactorization.lean}
\item \textbf{\nolinkurl{NHF14}}\hypertarget{lh:NHF14}{}\enspace{\ttfamily Ssot.\allowbreak{}NominalFactorization.\allowbreak{}nonlaminar\_\allowbreak{}multiParent\_\allowbreak{}avoids\_\allowbreak{}externalization} {\tiny\ttfamily Paper1PL/\allowbreak NominalFactorization.lean}
\item \textbf{\nolinkurl{NHF15}}\hypertarget{lh:NHF15}{}\enspace{\ttfamily Ssot.\allowbreak{}NominalFactorization.\allowbreak{}external\_\allowbreak{}repair\_\allowbreak{}provider\_\allowbreak{}lookup\_\allowbreak{}coloring} {\tiny\ttfamily Paper1PL/\allowbreak NominalFactorization.lean}
\item \textbf{\nolinkurl{NHF16}}\hypertarget{lh:NHF16}{}\enspace{\ttfamily Ssot.\allowbreak{}ProductResolution.\allowbreak{}typeFirst\_\allowbreak{}eq\_\allowbreak{}scopeFirst\_\allowbreak{}iff\_\allowbreak{}coordinateMinimum\_\allowbreak{}mem} {\tiny\ttfamily Paper1PL/\allowbreak ProductResolution.lean}
\item \textbf{\nolinkurl{NHF17}}\hypertarget{lh:NHF17}{}\enspace{\ttfamily Ssot.\allowbreak{}ProductResolution.\allowbreak{}typeFirstResolution\_\allowbreak{}eq\_\allowbreak{}of\_\allowbreak{}columnEq} {\tiny\ttfamily Paper1PL/\allowbreak ProductResolution.lean}
\item \textbf{\nolinkurl{NHF18}}\hypertarget{lh:NHF18}{}\enspace{\ttfamily Ssot.\allowbreak{}ProductResolution.\allowbreak{}typeFirstResolution\_\allowbreak{}insert\_\allowbreak{}empty\_\allowbreak{}provider} {\tiny\ttfamily Paper1PL/\allowbreak ProductResolution.lean}
\item \textbf{\nolinkurl{NHF19}}\hypertarget{lh:NHF19}{}\enspace{\ttfamily Ssot.\allowbreak{}NominalFactorization.\allowbreak{}FiniteResolver.\allowbreak{}toLinearization\_\allowbreak{}extends\_\allowbreak{}reverseAncestry} {\tiny\ttfamily Paper1PL/\allowbreak NominalFactorization.lean}
\item \textbf{\nolinkurl{NHF20}}\hypertarget{lh:NHF20}{}\enspace{\ttfamily Ssot.\allowbreak{}NominalFactorization.\allowbreak{}DerivationRepresentation.\allowbreak{}owner\_\allowbreak{}injective} {\tiny\ttfamily Paper1PL/\allowbreak NominalFactorization.lean}
\item \textbf{\nolinkurl{NHF21}}\hypertarget{lh:NHF21}{}\enspace{\ttfamily Ssot.\allowbreak{}NominalFactorization.\allowbreak{}ancestryOnly\_\allowbreak{}singleInheritance\_\allowbreak{}iff\_\allowbreak{}laminar} {\tiny\ttfamily Paper1PL/\allowbreak NominalFactorization.lean}
\item \textbf{\nolinkurl{NHF22}}\hypertarget{lh:NHF22}{}\enspace{\ttfamily Ssot.\allowbreak{}NominalFactorization.\allowbreak{}ancestry\_\allowbreak{}not\_\allowbreak{}determined\_\allowbreak{}by\_\allowbreak{}observation\_\allowbreak{}and\_\allowbreak{}nominal} {\tiny\ttfamily Paper1PL/\allowbreak NominalFactorization.lean}
\item \textbf{\nolinkurl{NHF23}}\hypertarget{lh:NHF23}{}\enspace{\ttfamily Ssot.\allowbreak{}NominalFactorization.\allowbreak{}ExternalizedSingleInheritanceRepresentation.\allowbreak{}external\_\allowbreak{}iff\_\allowbreak{}applicable\_\allowbreak{}and\_\allowbreak{}gap} {\tiny\ttfamily Paper1PL/\allowbreak NominalFactorization.lean}
\item \textbf{\nolinkurl{NHF24}}\hypertarget{lh:NHF24}{}\enspace{\ttfamily Ssot.\allowbreak{}NominalFactorization.\allowbreak{}ExternalizedSingleInheritanceRepresentation.\allowbreak{}ownerGrowthNeedsNoRepair\_\allowbreak{}iff} {\tiny\ttfamily Paper1PL/\allowbreak NominalFactorization.lean}
\item \textbf{\nolinkurl{NHF25}}\hypertarget{lh:NHF25}{}\enspace{\ttfamily Ssot.\allowbreak{}NominalFactorization.\allowbreak{}DelegatingSingleInheritanceRepresentation.\allowbreak{}gap\_\allowbreak{}forces\_\allowbreak{}forward\_\allowbreak{}target} {\tiny\ttfamily Paper1PL/\allowbreak NominalFactorization.lean}
\item \textbf{\nolinkurl{NHF26}}\hypertarget{lh:NHF26}{}\enspace{\ttfamily Ssot.\allowbreak{}NominalFactorization.\allowbreak{}DelegatingSingleInheritanceRepresentation.\allowbreak{}ancestryGap\_\allowbreak{}card\_\allowbreak{}le\_\allowbreak{}forwardingDecision\_\allowbreak{}card} {\tiny\ttfamily Paper1PL/\allowbreak NominalFactorization.lean}
\item \textbf{\nolinkurl{NHF27}}\hypertarget{lh:NHF27}{}\enspace{\ttfamily Ssot.\allowbreak{}NominalFactorization.\allowbreak{}DelegatingSingleInheritanceRepresentation.\allowbreak{}ancestryGap\_\allowbreak{}card\_\allowbreak{}eq\_\allowbreak{}sum} {\tiny\ttfamily Paper1PL/\allowbreak NominalFactorization.lean}
\item \textbf{\nolinkurl{NHF28}}\hypertarget{lh:NHF28}{}\enspace{\ttfamily Ssot.\allowbreak{}NominalFactorization.\allowbreak{}minimumDelegationCost\_\allowbreak{}iff\_\allowbreak{}minimumLaminarGapCost} {\tiny\ttfamily Paper1PL/\allowbreak NominalFactorization.lean}
\item \textbf{\nolinkurl{NHF29}}\hypertarget{lh:NHF29}{}\enspace{\ttfamily Ssot.\allowbreak{}DelegationCompression.\allowbreak{}FactObliviousDelegation.\allowbreak{}gap\_\allowbreak{}of\_\allowbreak{}provides\_\allowbreak{}of\_\allowbreak{}reaches} {\tiny\ttfamily Paper1PL/\allowbreak DelegationCompression.lean}
\item \textbf{\nolinkurl{NHF30}}\hypertarget{lh:NHF30}{}\enspace{\ttfamily Ssot.\allowbreak{}DelegationCompression.\allowbreak{}FoolingFamily.\allowbreak{}card\_\allowbreak{}le\_\allowbreak{}activeHolder\_\allowbreak{}card} {\tiny\ttfamily Paper1PL/\allowbreak DelegationCompression.lean}
\item \textbf{\nolinkurl{NHF31}}\hypertarget{lh:NHF31}{}\enspace{\ttfamily Ssot.\allowbreak{}LaminarRepairFPT.\allowbreak{}local\_\allowbreak{}conflict\_\allowbreak{}forces\_\allowbreak{}branch} {\tiny\ttfamily Paper1PL/\allowbreak LaminarRepairFPT.lean}
\item \textbf{\nolinkurl{NHF34}}\hypertarget{lh:NHF34}{}\enspace{\ttfamily Ssot.\allowbreak{}LaminarRepairFPT.\allowbreak{}laminarOn\_\allowbreak{}union\_\allowbreak{}of\_\allowbreak{}cross\_\allowbreak{}disjoint} {\tiny\ttfamily Paper1PL/\allowbreak LaminarRepairFPT.lean}
\item \textbf{\nolinkurl{NHF35}}\hypertarget{lh:NHF35}{}\enspace{\ttfamily Ssot.\allowbreak{}FixedHierarchyScopes.\allowbreak{}replaceRow\_\allowbreak{}laminar} {\tiny\ttfamily Paper1PL/\allowbreak FixedHierarchyScopes.lean}
\item \textbf{\nolinkurl{NHF36}}\hypertarget{lh:NHF36}{}\enspace{\ttfamily Ssot.\allowbreak{}FixedHierarchyScopes.\allowbreak{}gapCost\_\allowbreak{}replace\_\allowbreak{}equal\_\allowbreak{}scope\_\allowbreak{}le} {\tiny\ttfamily Paper1PL/\allowbreak FixedHierarchyScopes.lean}
\item \textbf{\nolinkurl{NHF37}}\hypertarget{lh:NHF37}{}\enspace{\ttfamily Ssot.\allowbreak{}FixedHierarchyScopes.\allowbreak{}fixedHierarchy\_\allowbreak{}pointwise\_\allowbreak{}minimizes} {\tiny\ttfamily Paper1PL/\allowbreak FixedHierarchyScopes.lean}
\item \textbf{\nolinkurl{NHF38}}\hypertarget{lh:NHF38}{}\enspace{\ttfamily Ssot.\allowbreak{}FixedHierarchyScopes.\allowbreak{}normalizeRows\_\allowbreak{}laminar} {\tiny\ttfamily Paper1PL/\allowbreak FixedHierarchyScopes.lean}
\item \textbf{\nolinkurl{NHF39}}\hypertarget{lh:NHF39}{}\enspace{\ttfamily Ssot.\allowbreak{}FixedHierarchyScopes.\allowbreak{}gapCost\_\allowbreak{}normalize\_\allowbreak{}equal\_\allowbreak{}scope\_\allowbreak{}le} {\tiny\ttfamily Paper1PL/\allowbreak FixedHierarchyScopes.lean}
\item \textbf{\nolinkurl{NHF40}}\hypertarget{lh:NHF40}{}\enspace{\ttfamily Ssot.\allowbreak{}FixedHierarchyScopes.\allowbreak{}normalizeRows\_\allowbreak{}eq\_\allowbreak{}of\_\allowbreak{}representative\_\allowbreak{}eq} {\tiny\ttfamily Paper1PL/\allowbreak FixedHierarchyScopes.lean}
\item \textbf{\nolinkurl{NHF41}}\hypertarget{lh:NHF41}{}\enspace{\ttfamily Ssot.\allowbreak{}FixedHierarchyScopes.\allowbreak{}fixedHierarchy\_\allowbreak{}pointwise\_\allowbreak{}minimizes\_\allowbreak{}with\_\allowbreak{}empty} {\tiny\ttfamily Paper1PL/\allowbreak FixedHierarchyScopes.lean}
\item \textbf{\nolinkurl{NHF44}}\hypertarget{lh:NHF44}{}\enspace{\ttfamily Ssot.\allowbreak{}NominalFactorization.\allowbreak{}canonicalDerivation\_\allowbreak{}applies\_\allowbreak{}iff} {\tiny\ttfamily Paper1PL/\allowbreak NominalFactorization.lean}
\item \textbf{\nolinkurl{NHF45}}\hypertarget{lh:NHF45}{}\enspace{\ttfamily Ssot.\allowbreak{}NominalFactorization.\allowbreak{}every\_\allowbreak{}applicability\_\allowbreak{}family\_\allowbreak{}has\_\allowbreak{}exact\_\allowbreak{}multiParent} {\tiny\ttfamily Paper1PL/\allowbreak NominalFactorization.lean}
\item \textbf{\nolinkurl{NHF47}}\hypertarget{lh:NHF47}{}\enspace{\ttfamily Ssot.\allowbreak{}DelegationCompression.\allowbreak{}RectangleCover.\allowbreak{}toFactObliviousDelegation} {\tiny\ttfamily Paper1PL/\allowbreak DelegationCompression.lean}
\item \textbf{\nolinkurl{NHF48}}\hypertarget{lh:NHF48}{}\enspace{\ttfamily Ssot.\allowbreak{}DelegationCompression.\allowbreak{}FactObliviousDelegation.\allowbreak{}toActiveRectangleCover} {\tiny\ttfamily Paper1PL/\allowbreak DelegationCompression.lean}
\item \textbf{\nolinkurl{NHF50}}\hypertarget{lh:NHF50}{}\enspace{\ttfamily Ssot.\allowbreak{}NominalFactorization.\allowbreak{}every\_\allowbreak{}indexed\_\allowbreak{}applicability\_\allowbreak{}family\_\allowbreak{}has\_\allowbreak{}owner\_\allowbreak{}preserving\_\allowbreak{}multiParent} {\tiny\ttfamily Paper1PL/\allowbreak NominalFactorization.lean}
\item \textbf{\nolinkurl{NHF51}}\hypertarget{lh:NHF51}{}\enspace{\ttfamily Ssot.\allowbreak{}NominalFactorization.\allowbreak{}indexed\_\allowbreak{}singleInheritance\_\allowbreak{}iff\_\allowbreak{}laminar} {\tiny\ttfamily Paper1PL/\allowbreak NominalFactorization.lean}
\item \textbf{\nolinkurl{NHF52}}\hypertarget{lh:NHF52}{}\enspace{\ttfamily Ssot.\allowbreak{}NominalFactorization.\allowbreak{}workflowScopes\_\allowbreak{}nonlaminar} {\tiny\ttfamily Paper1PL/\allowbreak NominalFactorization.lean}
\item \textbf{\nolinkurl{NHF53}}\hypertarget{lh:NHF53}{}\enspace{\ttfamily Ssot.\allowbreak{}NominalFactorization.\allowbreak{}DelegatingSingleInheritanceRepresentation.\allowbreak{}coverage\_\allowbreak{}resolution\_\allowbreak{}lower\_\allowbreak{}bound} {\tiny\ttfamily Paper1PL/\allowbreak CoverageResolution.lean}
\item \textbf{\nolinkurl{NHF54}}\hypertarget{lh:NHF54}{}\enspace{\ttfamily Ssot.\allowbreak{}NominalFactorization.\allowbreak{}DelegatingSingleInheritanceRepresentation.\allowbreak{}coverage\_\allowbreak{}resolution\_\allowbreak{}strict} {\tiny\ttfamily Paper1PL/\allowbreak CoverageResolution.lean}
\item \textbf{\nolinkurl{NHF55}}\hypertarget{lh:NHF55}{}\enspace{\ttfamily Ssot.\allowbreak{}NominalFactorization.\allowbreak{}DelegatingSingleInheritanceRepresentation.\allowbreak{}zero\_\allowbreak{}conflict\_\allowbreak{}strict} {\tiny\ttfamily Paper1PL/\allowbreak CoverageResolution.lean}
\item \textbf{\nolinkurl{NHF56}}\hypertarget{lh:NHF56}{}\enspace{\ttfamily Ssot.\allowbreak{}NominalFactorization.\allowbreak{}minimum\_\allowbreak{}coverage\_\allowbreak{}resolution\_\allowbreak{}cost} {\tiny\ttfamily Paper1PL/\allowbreak CoverageResolution.lean}
\item \textbf{\nolinkurl{NHF57}}\hypertarget{lh:NHF57}{}\enspace{\ttfamily Ssot.\allowbreak{}NominalFactorization.\allowbreak{}minimum\_\allowbreak{}coverage\_\allowbreak{}resolution\_\allowbreak{}cost\_\allowbreak{}strict} {\tiny\ttfamily Paper1PL/\allowbreak CoverageResolution.lean}
\item \textbf{\nolinkurl{NHF58}}\hypertarget{lh:NHF58}{}\enspace{\ttfamily Ssot.\allowbreak{}NominalFactorization.\allowbreak{}DelegatingSingleInheritanceRepresentation.\allowbreak{}coverageResolutionLedger\_\allowbreak{}strict} {\tiny\ttfamily Paper1PL/\allowbreak CoverageResolution.lean}
\item \textbf{\nolinkurl{NHF59}}\hypertarget{lh:NHF59}{}\enspace{\ttfamily Ssot.\allowbreak{}NominalFactorization.\allowbreak{}every\_\allowbreak{}policy\_\allowbreak{}state\_\allowbreak{}has\_\allowbreak{}owner\_\allowbreak{}preserving\_\allowbreak{}multiParent} {\tiny\ttfamily Paper1PL/\allowbreak NominalFactorization.lean}
\item \textbf{\nolinkurl{NHF60}}\hypertarget{lh:NHF60}{}\enspace{\ttfamily Ssot.\allowbreak{}NominalFactorization.\allowbreak{}policy\_\allowbreak{}singleInheritance\_\allowbreak{}iff\_\allowbreak{}every\_\allowbreak{}state\_\allowbreak{}laminar} {\tiny\ttfamily Paper1PL/\allowbreak NominalFactorization.lean}
\item \textbf{\nolinkurl{NHF61}}\hypertarget{lh:NHF61}{}\enspace{\ttfamily Ssot.\allowbreak{}NominalFactorization.\allowbreak{}policy\_\allowbreak{}pointwise\_\allowbreak{}minimumDelegationCost} {\tiny\ttfamily Paper1PL/\allowbreak NominalFactorization.lean}
\item \textbf{\nolinkurl{NHF62}}\hypertarget{lh:NHF62}{}\enspace{\ttfamily Ssot.\allowbreak{}NominalFactorization.\allowbreak{}policy\_\allowbreak{}crossing\_\allowbreak{}state\_\allowbreak{}has\_\allowbreak{}positive\_\allowbreak{}minimumDelegationCost} {\tiny\ttfamily Paper1PL/\allowbreak NominalFactorization.lean}
\item \textbf{\nolinkurl{NHF63}}\hypertarget{lh:NHF63}{}\enspace{\ttfamily Ssot.\allowbreak{}NominalFactorization.\allowbreak{}minimumDelegationCost\_\allowbreak{}isLeast\_\allowbreak{}exactRepresentations} {\tiny\ttfamily Paper1PL/\allowbreak NominalFactorization.lean}
\item \textbf{\nolinkurl{NHF64}}\hypertarget{lh:NHF64}{}\enspace{\ttfamily Ssot.\allowbreak{}NominalFactorization.\allowbreak{}SelectorCarrier.\allowbreak{}conflictSite\_\allowbreak{}iff\_\allowbreak{}of\_\allowbreak{}exact} {\tiny\ttfamily Paper1PL/\allowbreak CoverageResolution.lean}
\item \textbf{\nolinkurl{NHF65}}\hypertarget{lh:NHF65}{}\enspace{\ttfamily Ssot.\allowbreak{}NominalFactorization.\allowbreak{}SelectorCarrier.\allowbreak{}exact\_\allowbreak{}carriers\_\allowbreak{}conflictSite\_\allowbreak{}iff} {\tiny\ttfamily Paper1PL/\allowbreak CoverageResolution.lean}
\item \textbf{\nolinkurl{NHF66}}\hypertarget{lh:NHF66}{}\enspace{\ttfamily Ssot.\allowbreak{}NominalFactorization.\allowbreak{}SelectorCarrier.\allowbreak{}exact\_\allowbreak{}carriers\_\allowbreak{}winner\_\allowbreak{}eq} {\tiny\ttfamily Paper1PL/\allowbreak CoverageResolution.lean}
\item \textbf{\nolinkurl{NHF67}}\hypertarget{lh:NHF67}{}\enspace{\ttfamily Ssot.\allowbreak{}NominalFactorization.\allowbreak{}SelectorCarrier.\allowbreak{}exact\_\allowbreak{}carriers\_\allowbreak{}conflictSet\_\allowbreak{}eq} {\tiny\ttfamily Paper1PL/\allowbreak CoverageResolution.lean}
\item \textbf{\nolinkurl{NHF68}}\hypertarget{lh:NHF68}{}\enspace{\ttfamily Ssot.\allowbreak{}NominalFactorization.\allowbreak{}SelectorCarrier.\allowbreak{}conflictSet\_\allowbreak{}eq\_\allowbreak{}contract\_\allowbreak{}of\_\allowbreak{}exact} {\tiny\ttfamily Paper1PL/\allowbreak CoverageResolution.lean}
\item \textbf{\nolinkurl{NOM1}}\hypertarget{lh:NOM1}{}\enspace{\ttfamily nominal\_\allowbreak{}centralized} {\tiny\ttfamily AbstractClassSystem/\allowbreak Extended.lean}
\item \textbf{\nolinkurl{NPS1}}\hypertarget{lh:NPS1}{}\enspace{\ttfamily Ssot.\allowbreak{}NativeProofSystem.\allowbreak{}derives\_\allowbreak{}iff\_\allowbreak{}declared} {\tiny\ttfamily Paper1PL/\allowbreak NativeProofSystem.lean}
\item \textbf{\nolinkurl{NPS2}}\hypertarget{lh:NPS2}{}\enspace{\ttfamily Ssot.\allowbreak{}NativeProofSystem.\allowbreak{}derivation\_\allowbreak{}sound} {\tiny\ttfamily Paper1PL/\allowbreak NativeProofSystem.lean}
\item \textbf{\nolinkurl{NPS3}}\hypertarget{lh:NPS3}{}\enspace{\ttfamily Ssot.\allowbreak{}NativeProofSystem.\allowbreak{}unrestrictedNative\_\allowbreak{}sound} {\tiny\ttfamily Paper1PL/\allowbreak NativeProofSystem.lean}
\item \textbf{\nolinkurl{NPS4}}\hypertarget{lh:NPS4}{}\enspace{\ttfamily Ssot.\allowbreak{}NativeProofSystem.\allowbreak{}unrestrictedNative\_\allowbreak{}complete} {\tiny\ttfamily Paper1PL/\allowbreak NativeProofSystem.lean}
\item \textbf{\nolinkurl{NPS5}}\hypertarget{lh:NPS5}{}\enspace{\ttfamily Ssot.\allowbreak{}NativeProofSystem.\allowbreak{}unrestrictedNative\_\allowbreak{}has\_\allowbreak{}exact\_\allowbreak{}multiParent} {\tiny\ttfamily Paper1PL/\allowbreak NativeProofSystem.lean}
\item \textbf{\nolinkurl{NPS6}}\hypertarget{lh:NPS6}{}\enspace{\ttfamily Ssot.\allowbreak{}NativeProofSystem.\allowbreak{}oneParent\_\allowbreak{}sound\_\allowbreak{}complete\_\allowbreak{}implies\_\allowbreak{}required\_\allowbreak{}laminar} {\tiny\ttfamily Paper1PL/\allowbreak NativeProofSystem.lean}
\item \textbf{\nolinkurl{NPS7}}\hypertarget{lh:NPS7}{}\enspace{\ttfamily Ssot.\allowbreak{}NativeProofSystem.\allowbreak{}crossingScope\_\allowbreak{}not\_\allowbreak{}laminar} {\tiny\ttfamily Paper1PL/\allowbreak NativeProofSystem.lean}
\item \textbf{\nolinkurl{NPS8}}\hypertarget{lh:NPS8}{}\enspace{\ttfamily Ssot.\allowbreak{}NativeProofSystem.\allowbreak{}crossing\_\allowbreak{}no\_\allowbreak{}sound\_\allowbreak{}complete\_\allowbreak{}oneParent} {\tiny\ttfamily Paper1PL/\allowbreak NativeProofSystem.lean}
\item \textbf{\nolinkurl{NPS9}}\hypertarget{lh:NPS9}{}\enspace{\ttfamily Ssot.\allowbreak{}NativeProofSystem.\allowbreak{}crossing\_\allowbreak{}exists\_\allowbreak{}valid\_\allowbreak{}not\_\allowbreak{}derivable} {\tiny\ttfamily Paper1PL/\allowbreak NativeProofSystem.lean}
\item \textbf{\nolinkurl{NRH1}}\hypertarget{lh:NRH1}{}\enspace{\ttfamily Ssot.\allowbreak{}SemanticRepairHistory.\allowbreak{}ancestor\_\allowbreak{}preserved\_\allowbreak{}on\_\allowbreak{}path} {\tiny\ttfamily Paper1PL/\allowbreak SemanticRepairHistory.lean}
\item \textbf{\nolinkurl{NRH2}}\hypertarget{lh:NRH2}{}\enspace{\ttfamily Ssot.\allowbreak{}SemanticRepairHistory.\allowbreak{}between\_\allowbreak{}preserved\_\allowbreak{}supported\_\allowbreak{}path} {\tiny\ttfamily Paper1PL/\allowbreak SemanticRepairHistory.lean}
\item \textbf{\nolinkurl{NRH3}}\hypertarget{lh:NRH3}{}\enspace{\ttfamily Ssot.\allowbreak{}SemanticRepairHistory.\allowbreak{}strict\_\allowbreak{}insertion\_\allowbreak{}exact\_\allowbreak{}forces\_\allowbreak{}repairDeletion} {\tiny\ttfamily Paper1PL/\allowbreak SemanticRepairHistory.lean}
\item \textbf{\nolinkurl{NRH4}}\hypertarget{lh:NRH4}{}\enspace{\ttfamily Ssot.\allowbreak{}SemanticRepairHistory.\allowbreak{}strict\_\allowbreak{}insertion\_\allowbreak{}forces\_\allowbreak{}normalized\_\allowbreak{}repair} {\tiny\ttfamily Paper1PL/\allowbreak SemanticRepairHistory.lean}
\item \textbf{\nolinkurl{NRH5}}\hypertarget{lh:NRH5}{}\enspace{\ttfamily Ssot.\allowbreak{}SemanticRepairHistory.\allowbreak{}strict\_\allowbreak{}insertion\_\allowbreak{}normalized\_\allowbreak{}cost\_\allowbreak{}pos} {\tiny\ttfamily Paper1PL/\allowbreak SemanticRepairHistory.lean}
\item \textbf{\nolinkurl{NRH6}}\hypertarget{lh:NRH6}{}\enspace{\ttfamily Ssot.\allowbreak{}SemanticRepairHistory.\allowbreak{}staged\_\allowbreak{}normalized\_\allowbreak{}cost\_\allowbreak{}lower\_\allowbreak{}bound} {\tiny\ttfamily Paper1PL/\allowbreak SemanticRepairHistory.lean}
\item \textbf{\nolinkurl{NRH7}}\hypertarget{lh:NRH7}{}\enspace{\ttfamily Ssot.\allowbreak{}SemanticRepairHistory.\allowbreak{}withdrawn\_\allowbreak{}path\_\allowbreak{}requires\_\allowbreak{}raw\_\allowbreak{}deletion} {\tiny\ttfamily Paper1PL/\allowbreak SemanticRepairHistory.lean}
\item \textbf{\nolinkurl{NRH8}}\hypertarget{lh:NRH8}{}\enspace{\ttfamily Ssot.\allowbreak{}SemanticRepairHistory.\allowbreak{}staged\_\allowbreak{}normalized\_\allowbreak{}repair} {\tiny\ttfamily Paper1PL/\allowbreak SemanticRepairHistory.lean}
\item \textbf{\nolinkurl{NRH9}}\hypertarget{lh:NRH9}{}\enspace{\ttfamily Ssot.\allowbreak{}SemanticRepairHistory.\allowbreak{}staged\_\allowbreak{}deletions\_\allowbreak{}add\_\allowbreak{}outside\_\allowbreak{}updates\_\allowbreak{}lower\_\allowbreak{}bound} {\tiny\ttfamily Paper1PL/\allowbreak SemanticRepairHistory.lean}
\item \textbf{\nolinkurl{OLC1}}\hypertarget{lh:OLC1}{}\enspace{\ttfamily Ssot.\allowbreak{}ObservationLocalCapacity.\allowbreak{}no\_\allowbreak{}observation\_\allowbreak{}classifier\_\allowbreak{}of\_\allowbreak{}collision} {\tiny\ttfamily Paper1PL/\allowbreak ObservationLocalCapacity.lean}
\item \textbf{\nolinkurl{OLC2}}\hypertarget{lh:OLC2}{}\enspace{\ttfamily Ssot.\allowbreak{}ObservationLocalCapacity.\allowbreak{}hiddenWorkflowSuccessfulNativeObservation\_\allowbreak{}eq} {\tiny\ttfamily Paper1PL/\allowbreak ObservationLocalCapacity.lean}
\item \textbf{\nolinkurl{OLC3}}\hypertarget{lh:OLC3}{}\enspace{\ttfamily Ssot.\allowbreak{}ObservationLocalCapacity.\allowbreak{}no\_\allowbreak{}successful\_\allowbreak{}native\_\allowbreak{}observation\_\allowbreak{}classifier\_\allowbreak{}for\_\allowbreak{}treeShape} {\tiny\ttfamily Paper1PL/\allowbreak ObservationLocalCapacity.lean}
\item \textbf{\nolinkurl{OLC4}}\hypertarget{lh:OLC4}{}\enspace{\ttfamily Ssot.\allowbreak{}ObservationLocalCapacity.\allowbreak{}successful\_\allowbreak{}native\_\allowbreak{}observation\_\allowbreak{}policy\_\allowbreak{}same\_\allowbreak{}decision} {\tiny\ttfamily Paper1PL/\allowbreak ObservationLocalCapacity.lean}
\item \textbf{\nolinkurl{OLC5}}\hypertarget{lh:OLC5}{}\enspace{\ttfamily Ssot.\allowbreak{}ObservationLocalCapacity.\allowbreak{}no\_\allowbreak{}successful\_\allowbreak{}native\_\allowbreak{}observation\_\allowbreak{}policy\_\allowbreak{}fires\_\allowbreak{}exactly\_\allowbreak{}on\_\allowbreak{}crossing} {\tiny\ttfamily Paper1PL/\allowbreak ObservationLocalCapacity.lean}
\item \textbf{\nolinkurl{OLC7}}\hypertarget{lh:OLC7}{}\enspace{\ttfamily Ssot.\allowbreak{}ObservationLocalCapacity.\allowbreak{}total\_\allowbreak{}successful\_\allowbreak{}native\_\allowbreak{}attributor\_\allowbreak{}misclassifies\_\allowbreak{}one} {\tiny\ttfamily Paper1PL/\allowbreak ObservationLocalCapacity.lean}
\item \textbf{\nolinkurl{OLC8}}\hypertarget{lh:OLC8}{}\enspace{\ttfamily Ssot.\allowbreak{}ObservationLocalCapacity.\allowbreak{}sound\_\allowbreak{}partial\_\allowbreak{}successful\_\allowbreak{}native\_\allowbreak{}attributor\_\allowbreak{}abstains} {\tiny\ttfamily Paper1PL/\allowbreak ObservationLocalCapacity.lean}
\item \textbf{\nolinkurl{OLC9}}\hypertarget{lh:OLC9}{}\enspace{\ttfamily Ssot.\allowbreak{}ObservationLocalCapacity.\allowbreak{}correct\_\allowbreak{}supply\_\allowbreak{}routes\_\allowbreak{}recover\_\allowbreak{}requirements} {\tiny\ttfamily Paper1PL/\allowbreak ObservationLocalCapacity.lean}
\item \textbf{\nolinkurl{OPT1}}\hypertarget{lh:OPT1}{}\enspace{\ttfamily Ssot.\allowbreak{}OptionRegret.\allowbreak{}restriction\_\allowbreak{}regret} {\tiny\ttfamily Paper1PL/\allowbreak OptionRegret.lean}
\item \textbf{\nolinkurl{OPT3}}\hypertarget{lh:OPT3}{}\enspace{\ttfamily Ssot.\allowbreak{}OptionRegret.\allowbreak{}stagedStepBurden\_\allowbreak{}pos} {\tiny\ttfamily Paper1PL/\allowbreak OptionRegret.lean}
\item \textbf{\nolinkurl{OPT4}}\hypertarget{lh:OPT4}{}\enspace{\ttfamily Ssot.\allowbreak{}OptionRegret.\allowbreak{}stagedCumulativeBurden\_\allowbreak{}ge\_\allowbreak{}horizon} {\tiny\ttfamily Paper1PL/\allowbreak OptionRegret.lean}
\item \textbf{\nolinkurl{OPT5}}\hypertarget{lh:OPT5}{}\enspace{\ttfamily Ssot.\allowbreak{}OptionRegret.\allowbreak{}exists\_\allowbreak{}horizon\_\allowbreak{}recurringBurden\_\allowbreak{}exceeds\_\allowbreak{}finiteAcquisition} {\tiny\ttfamily Paper1PL/\allowbreak OptionRegret.lean}
\item \textbf{\nolinkurl{OPT6}}\hypertarget{lh:OPT6}{}\enspace{\ttfamily Ssot.\allowbreak{}OptionRegret.\allowbreak{}finitePreventionCost\_\allowbreak{}lt\_\allowbreak{}recurringBurden\_\allowbreak{}after\_\allowbreak{}someHorizon} {\tiny\ttfamily Paper1PL/\allowbreak OptionRegret.lean}
\item \textbf{\nolinkurl{OPT7}}\hypertarget{lh:OPT7}{}\enspace{\ttfamily Ssot.\allowbreak{}OptionRegret.\allowbreak{}stagedRestriction\_\allowbreak{}includes\_\allowbreak{}cumulativeBurden} {\tiny\ttfamily Paper1PL/\allowbreak OptionRegret.lean}
\item \textbf{\nolinkurl{OPT8}}\hypertarget{lh:OPT8}{}\enspace{\ttfamily Ssot.\allowbreak{}OptionRegret.\allowbreak{}acquisitionExcess\_\allowbreak{}lt\_\allowbreak{}recurringBurden\_\allowbreak{}after\_\allowbreak{}someHorizon} {\tiny\ttfamily Paper1PL/\allowbreak OptionRegret.lean}
\item \textbf{\nolinkurl{OPT9}}\hypertarget{lh:OPT9}{}\enspace{\ttfamily Ssot.\allowbreak{}OptionRegret.\allowbreak{}acquisitionExcess\_\allowbreak{}lt\_\allowbreak{}linearRepair\_\allowbreak{}from\_\allowbreak{}affineThreshold} {\tiny\ttfamily Paper1PL/\allowbreak OptionRegret.lean}
\item \textbf{\nolinkurl{OPT10}}\hypertarget{lh:OPT10}{}\enspace{\ttfamily Ssot.\allowbreak{}OptionRegret.\allowbreak{}acquisitionExcess\_\allowbreak{}lt\_\allowbreak{}stagedBurden\_\allowbreak{}from\_\allowbreak{}affineThreshold} {\tiny\ttfamily Paper1PL/\allowbreak OptionRegret.lean}
\item \textbf{\nolinkurl{OPT11}}\hypertarget{lh:OPT11}{}\enspace{\ttfamily Ssot.\allowbreak{}OptionRegret.\allowbreak{}lowerRateAcquisition\_\allowbreak{}eventually\_\allowbreak{}beats\_\allowbreak{}permanentRestriction} {\tiny\ttfamily Paper1PL/\allowbreak OptionRegret.lean}
\item \textbf{\nolinkurl{OPT12}}\hypertarget{lh:OPT12}{}\enspace{\ttfamily Ssot.\allowbreak{}OptionRegret.\allowbreak{}lowerRateAcquisition\_\allowbreak{}refutes\_\allowbreak{}permanentRestrictionBound} {\tiny\ttfamily Paper1PL/\allowbreak OptionRegret.lean}
\item \textbf{\nolinkurl{OPT13}}\hypertarget{lh:OPT13}{}\enspace{\ttfamily Ssot.\allowbreak{}OptionRegret.\allowbreak{}policySet\_\allowbreak{}infimum\_\allowbreak{}antitone} {\tiny\ttfamily Paper1PL/\allowbreak OptionRegret.lean}
\item \textbf{\nolinkurl{OPT14}}\hypertarget{lh:OPT14}{}\enspace{\ttfamily Ssot.\allowbreak{}OptionRegret.\allowbreak{}acquisitionUnion\_\allowbreak{}infimum\_\allowbreak{}le\_\allowbreak{}staying} {\tiny\ttfamily Paper1PL/\allowbreak OptionRegret.lean}
\item \textbf{\nolinkurl{OPT15}}\hypertarget{lh:OPT15}{}\enspace{\ttfamily Ssot.\allowbreak{}OptionRegret.\allowbreak{}cheaper\_\allowbreak{}feasible\_\allowbreak{}excludes\_\allowbreak{}staying} {\tiny\ttfamily Paper1PL/\allowbreak OptionRegret.lean}
\item \textbf{\nolinkurl{OPT16}}\hypertarget{lh:OPT16}{}\enspace{\ttfamily Ssot.\allowbreak{}OptionRegret.\allowbreak{}lowerRateAcquisition\_\allowbreak{}excludes\_\allowbreak{}stagedStayingPolicies} {\tiny\ttfamily Paper1PL/\allowbreak OptionRegret.lean}
\item \textbf{\nolinkurl{OPT17}}\hypertarget{lh:OPT17}{}\enspace{\ttfamily Ssot.\allowbreak{}OptionRegret.\allowbreak{}attained\_\allowbreak{}minimum\_\allowbreak{}eq\_\allowbreak{}iff\_\allowbreak{}retained\_\allowbreak{}optimum} {\tiny\ttfamily Paper1PL/\allowbreak OptionRegret.lean}
\item \textbf{\nolinkurl{OPT18}}\hypertarget{lh:OPT18}{}\enspace{\ttfamily Ssot.\allowbreak{}OptionRegret.\allowbreak{}attained\_\allowbreak{}minimum\_\allowbreak{}strict\_\allowbreak{}of\_\allowbreak{}no\_\allowbreak{}retained\_\allowbreak{}optimum} {\tiny\ttfamily Paper1PL/\allowbreak OptionRegret.lean}
\item \textbf{\nolinkurl{OPT19}}\hypertarget{lh:OPT19}{}\enspace{\ttfamily Ssot.\allowbreak{}OptionRegret.\allowbreak{}dominated\_\allowbreak{}removal\_\allowbreak{}preserves\_\allowbreak{}attained\_\allowbreak{}minimum} {\tiny\ttfamily Paper1PL/\allowbreak OptionRegret.lean}
\item \textbf{\nolinkurl{OPT20}}\hypertarget{lh:OPT20}{}\enspace{\ttfamily Ssot.\allowbreak{}OptionRegret.\allowbreak{}conservativeExtension\_\allowbreak{}preserves\_\allowbreak{}old\_\allowbreak{}answers} {\tiny\ttfamily Paper1PL/\allowbreak OptionRegret.lean}
\item \textbf{\nolinkurl{OPT21}}\hypertarget{lh:OPT21}{}\enspace{\ttfamily Ssot.\allowbreak{}OptionRegret.\allowbreak{}rungUnion\_\allowbreak{}eq\_\allowbreak{}top} {\tiny\ttfamily Paper1PL/\allowbreak OptionRegret.lean}
\item \textbf{\nolinkurl{OPT22}}\hypertarget{lh:OPT22}{}\enspace{\ttfamily Ssot.\allowbreak{}OptionRegret.\allowbreak{}rungUnion\_\allowbreak{}infimum\_\allowbreak{}eq\_\allowbreak{}top} {\tiny\ttfamily Paper1PL/\allowbreak OptionRegret.lean}
\item \textbf{\nolinkurl{OWM1}}\hypertarget{lh:OWM1}{}\enspace{\ttfamily Ssot.\allowbreak{}NominalFactorization.\allowbreak{}structural\_\allowbreak{}current\_\allowbreak{}exact\_\allowbreak{}but\_\allowbreak{}nominal\_\allowbreak{}extension\_\allowbreak{}stable} {\tiny\ttfamily Paper1PL/\allowbreak NominalFactorization.lean}
\item \textbf{\nolinkurl{OWM2}}\hypertarget{lh:OWM2}{}\enspace{\ttfamily Ssot.\allowbreak{}NominalFactorization.\allowbreak{}workflow\_\allowbreak{}singleParent\_\allowbreak{}not\_\allowbreak{}extension\_\allowbreak{}stable} {\tiny\ttfamily Paper1PL/\allowbreak NominalFactorization.lean}
\item \textbf{\nolinkurl{PD1}}\hypertarget{lh:PD1}{}\enspace{\ttfamily nominal\_\allowbreak{}pareto\_\allowbreak{}dominates\_\allowbreak{}shape} {\tiny\ttfamily discipline\_migration.lean}
\item \textbf{\nolinkurl{PIS1}}\hypertarget{lh:PIS1}{}\enspace{\ttfamily Ssot.\allowbreak{}PriceInvariantSeparation.\allowbreak{}nonempty\_\allowbreak{}cost\_\allowbreak{}pos} {\tiny\ttfamily Paper1PL/\allowbreak PriceInvariantSeparation.lean}
\item \textbf{\nolinkurl{PIS2}}\hypertarget{lh:PIS2}{}\enspace{\ttfamily Ssot.\allowbreak{}PriceInvariantSeparation.\allowbreak{}zero\_\allowbreak{}cost\_\allowbreak{}iff\_\allowbreak{}empty} {\tiny\ttfamily Paper1PL/\allowbreak PriceInvariantSeparation.lean}
\item \textbf{\nolinkurl{PIS3}}\hypertarget{lh:PIS3}{}\enspace{\ttfamily Ssot.\allowbreak{}PriceInvariantSeparation.\allowbreak{}finite\_\allowbreak{}universe\_\allowbreak{}uniform\_\allowbreak{}floor} {\tiny\ttfamily Paper1PL/\allowbreak PriceInvariantSeparation.lean}
\item \textbf{\nolinkurl{PIS4}}\hypertarget{lh:PIS4}{}\enspace{\ttfamily Ssot.\allowbreak{}PriceInvariantSeparation.\allowbreak{}every\_\allowbreak{}price\_\allowbreak{}preserves\_\allowbreak{}witness} {\tiny\ttfamily Paper1PL/\allowbreak PriceInvariantSeparation.lean}
\item \textbf{\nolinkurl{PIS5}}\hypertarget{lh:PIS5}{}\enspace{\ttfamily Ssot.\allowbreak{}PriceInvariantSeparation.\allowbreak{}crossing\_\allowbreak{}forces\_\allowbreak{}nonempty\_\allowbreak{}residual} {\tiny\ttfamily Paper1PL/\allowbreak PriceInvariantSeparation.lean}
\item \textbf{\nolinkurl{PIS6}}\hypertarget{lh:PIS6}{}\enspace{\ttfamily Ssot.\allowbreak{}PriceInvariantSeparation.\allowbreak{}crossing\_\allowbreak{}cost\_\allowbreak{}positive\_\allowbreak{}under\_\allowbreak{}every\_\allowbreak{}price} {\tiny\ttfamily Paper1PL/\allowbreak PriceInvariantSeparation.lean}
\item \textbf{\nolinkurl{PJD1}}\hypertarget{lh:PJD1}{}\enspace{\ttfamily Ssot.\allowbreak{}ProjectionDuality.\allowbreak{}exactRecovery\_\allowbreak{}iff\_\allowbreak{}no\_\allowbreak{}fiberConflict} {\tiny\ttfamily Paper1IT/\allowbreak ProjectionDuality.lean}
\item \textbf{\nolinkurl{PJD2}}\hypertarget{lh:PJD2}{}\enspace{\ttfamily Ssot.\allowbreak{}ProjectionDuality.\allowbreak{}everyDecoderFails\_\allowbreak{}iff\_\allowbreak{}hasFiberConflict} {\tiny\ttfamily Paper1IT/\allowbreak ProjectionDuality.lean}
\item \textbf{\nolinkurl{PJD3}}\hypertarget{lh:PJD3}{}\enspace{\ttfamily Ssot.\allowbreak{}ProjectionDuality.\allowbreak{}dual\_\allowbreak{}certification} {\tiny\ttfamily Paper1IT/\allowbreak ProjectionDuality.lean}
\item \textbf{\nolinkurl{PJD4}}\hypertarget{lh:PJD4}{}\enspace{\ttfamily Ssot.\allowbreak{}ProjectionDuality.\allowbreak{}channel\_\allowbreak{}separation} {\tiny\ttfamily Paper1IT/\allowbreak ProjectionDuality.lean}
\item \textbf{\nolinkurl{PRT2}}\hypertarget{lh:PRT2}{}\enspace{\ttfamily AbstractClassSystem.\allowbreak{}ProtocolRuntime.\allowbreak{}protoCheck\_\allowbreak{}in\_\allowbreak{}shapeQuerySet} {\tiny\ttfamily AbstractClassSystem/\allowbreak Extended.lean}
\item \textbf{\nolinkurl{PST1}}\hypertarget{lh:PST1}{}\enspace{\ttfamily Ssot.\allowbreak{}PartitionStateThreshold.\allowbreak{}zeroErrorTagging\_\allowbreak{}iff\_\allowbreak{}colorClasses\_\allowbreak{}profileIndependent} {\tiny\ttfamily Paper1PL/\allowbreak PartitionStateThreshold.lean}
\item \textbf{\nolinkurl{PST2}}\hypertarget{lh:PST2}{}\enspace{\ttfamily Ssot.\allowbreak{}PartitionStateThreshold.\allowbreak{}twoStateTagging\_\allowbreak{}iff\_\allowbreak{}commonPartitionIndependent} {\tiny\ttfamily Paper1PL/\allowbreak PartitionStateThreshold.lean}
\item \textbf{\nolinkurl{PST3}}\hypertarget{lh:PST3}{}\enspace{\ttfamily Ssot.\allowbreak{}PartitionStateThreshold.\allowbreak{}commonPartitionIndependent\_\allowbreak{}iff\_\allowbreak{}bipartiteMatchingClass} {\tiny\ttfamily Paper1PL/\allowbreak PartitionStateThreshold.lean}
\item \textbf{\nolinkurl{PST4}}\hypertarget{lh:PST4}{}\enspace{\ttfamily Ssot.\allowbreak{}PartitionStateThreshold.\allowbreak{}three\_\allowbreak{}states\_\allowbreak{}double\_\allowbreak{}snapshot\_\allowbreak{}price} {\tiny\ttfamily Paper1PL/\allowbreak PartitionStateThreshold.lean}
\item \textbf{\nolinkurl{RBF1}}\hypertarget{lh:RBF1}{}\enspace{\ttfamily Ssot.\allowbreak{}RuleBasedFactoring.\allowbreak{}compiled\_\allowbreak{}ancestry\_\allowbreak{}exact} {\tiny\ttfamily Paper1PL/\allowbreak RuleBasedFactoring.lean}
\item \textbf{\nolinkurl{RBF2}}\hypertarget{lh:RBF2}{}\enspace{\ttfamily Ssot.\allowbreak{}RuleBasedFactoring.\allowbreak{}compileRule\_\allowbreak{}preserves\_\allowbreak{}revision\_\allowbreak{}count} {\tiny\ttfamily Paper1PL/\allowbreak RuleBasedFactoring.lean}
\item \textbf{\nolinkurl{RBF3}}\hypertarget{lh:RBF3}{}\enspace{\ttfamily Ssot.\allowbreak{}RuleBasedFactoring.\allowbreak{}unchanged\_\allowbreak{}rule\_\allowbreak{}iff\_\allowbreak{}input\_\allowbreak{}sufficient} {\tiny\ttfamily Paper1PL/\allowbreak RuleBasedFactoring.lean}
\item \textbf{\nolinkurl{RBF4}}\hypertarget{lh:RBF4}{}\enspace{\ttfamily Ssot.\allowbreak{}RuleBasedFactoring.\allowbreak{}exact\_\allowbreak{}rules\_\allowbreak{}with\_\allowbreak{}allowed\_\allowbreak{}changes\_\allowbreak{}iff\_\allowbreak{}hits\_\allowbreak{}conflicts} {\tiny\ttfamily Paper1PL/\allowbreak RuleBasedFactoring.lean}
\item \textbf{\nolinkurl{RBF5}}\hypertarget{lh:RBF5}{}\enspace{\ttfamily Ssot.\allowbreak{}RuleBasedFactoring.\allowbreak{}rule\_\allowbreak{}revision\_\allowbreak{}budget\_\allowbreak{}iff\_\allowbreak{}conflict\_\allowbreak{}hitting\_\allowbreak{}budget} {\tiny\ttfamily Paper1PL/\allowbreak RuleBasedFactoring.lean}
\item \textbf{\nolinkurl{RBF6}}\hypertarget{lh:RBF6}{}\enspace{\ttfamily Ssot.\allowbreak{}RuleBasedFactoring.\allowbreak{}conflicting\_\allowbreak{}updates\_\allowbreak{}revision\_\allowbreak{}lower\_\allowbreak{}bound} {\tiny\ttfamily Paper1PL/\allowbreak RuleBasedFactoring.lean}
\item \textbf{\nolinkurl{RCN1}}\hypertarget{lh:RCN1}{}\enspace{\ttfamily Ssot.\allowbreak{}RepresentationCanonicity.\allowbreak{}exactEncoding\_\allowbreak{}refines\_\allowbreak{}canonicalQuotient} {\tiny\ttfamily Paper1PL/\allowbreak RepresentationCanonicity.lean}
\item \textbf{\nolinkurl{RCN2}}\hypertarget{lh:RCN2}{}\enspace{\ttfamily Ssot.\allowbreak{}RepresentationCanonicity.\allowbreak{}minimalExact\_\allowbreak{}iff\_\allowbreak{}samePartition\_\allowbreak{}canonicalQuotient} {\tiny\ttfamily Paper1PL/\allowbreak RepresentationCanonicity.lean}
\item \textbf{\nolinkurl{RCN3}}\hypertarget{lh:RCN3}{}\enspace{\ttfamily Ssot.\allowbreak{}RepresentationCanonicity.\allowbreak{}canonicalDecisionEncoding\_\allowbreak{}minimalExact} {\tiny\ttfamily Paper1PL/\allowbreak RepresentationCanonicity.lean}
\item \textbf{\nolinkurl{RCN4}}\hypertarget{lh:RCN4}{}\enspace{\ttfamily Ssot.\allowbreak{}RepresentationCanonicity.\allowbreak{}minimalExact\_\allowbreak{}sameRepresentationPartition} {\tiny\ttfamily Paper1PL/\allowbreak RepresentationCanonicity.lean}
\item \textbf{\nolinkurl{RCN5}}\hypertarget{lh:RCN5}{}\enspace{\ttfamily Ssot.\allowbreak{}RepresentationCanonicity.\allowbreak{}answerCapacityEquivalent\_\allowbreak{}iff\_\allowbreak{}sameRepresentationPartition} {\tiny\ttfamily Paper1PL/\allowbreak RepresentationCanonicity.lean}
\item \textbf{\nolinkurl{RCN6}}\hypertarget{lh:RCN6}{}\enspace{\ttfamily Ssot.\allowbreak{}RepresentationCanonicity.\allowbreak{}minimalExact\_\allowbreak{}answerCapacityEquivalent} {\tiny\ttfamily Paper1PL/\allowbreak RepresentationCanonicity.lean}
\item \textbf{\nolinkurl{RCN7}}\hypertarget{lh:RCN7}{}\enspace{\ttfamily Ssot.\allowbreak{}RepresentationCanonicity.\allowbreak{}minimalExact\_\allowbreak{}exists\_\allowbreak{}presentationEquivalence} {\tiny\ttfamily Paper1PL/\allowbreak RepresentationCanonicity.lean}
\item \textbf{\nolinkurl{RCN8}}\hypertarget{lh:RCN8}{}\enspace{\ttfamily Ssot.\allowbreak{}RepresentationCanonicity.\allowbreak{}IndexedDerivationRepresentation.\allowbreak{}toExactDecisionEncoding} {\tiny\ttfamily Paper1PL/\allowbreak RepresentationCanonicity.lean}
\item \textbf{\nolinkurl{RCN9}}\hypertarget{lh:RCN9}{}\enspace{\ttfamily Ssot.\allowbreak{}RepresentationCanonicity.\allowbreak{}DelegatingSingleInheritanceRepresentation.\allowbreak{}toNominalDecisionEncoding} {\tiny\ttfamily Paper1PL/\allowbreak RepresentationCanonicity.lean}
\item \textbf{\nolinkurl{RCN10}}\hypertarget{lh:RCN10}{}\enspace{\ttfamily Ssot.\allowbreak{}RepresentationCanonicity.\allowbreak{}optimizerQuotientMap\_\allowbreak{}eq\_\allowbreak{}iff} {\tiny\ttfamily Paper1PL/\allowbreak RepresentationCanonicity.lean}
\item \textbf{\nolinkurl{RCN11}}\hypertarget{lh:RCN11}{}\enspace{\ttfamily Ssot.\allowbreak{}RepresentationCanonicity.\allowbreak{}exactValidActionEncoding\_\allowbreak{}refines\_\allowbreak{}optimizerQuotient} {\tiny\ttfamily Paper1PL/\allowbreak RepresentationCanonicity.lean}
\item \textbf{\nolinkurl{RLC3}}\hypertarget{lh:RLC3}{}\enspace{\ttfamily Ssot.\allowbreak{}RelationCarrier.\allowbreak{}independentRelation\_\allowbreak{}carrier\_\allowbreak{}card\_\allowbreak{}lower\_\allowbreak{}bound} {\tiny\ttfamily Paper1IT/\allowbreak RelationCarrier.lean}
\item \textbf{\nolinkurl{RLC4}}\hypertarget{lh:RLC4}{}\enspace{\ttfamily Ssot.\allowbreak{}RelationCarrier.\allowbreak{}independentRelation\_\allowbreak{}requiredBits\_\allowbreak{}lower\_\allowbreak{}bound} {\tiny\ttfamily Paper1IT/\allowbreak RelationCarrier.lean}
\item \textbf{\nolinkurl{RLC6}}\hypertarget{lh:RLC6}{}\enspace{\ttfamily Ssot.\allowbreak{}RelationCarrier.\allowbreak{}zeroError\_\allowbreak{}iff\_\allowbreak{}of\_\allowbreak{}same\_\allowbreak{}realizedRelations} {\tiny\ttfamily Paper1IT/\allowbreak RelationCarrier.lean}
\item \textbf{\nolinkurl{RLC7}}\hypertarget{lh:RLC7}{}\enspace{\ttfamily Ssot.\allowbreak{}RelationCarrier.\allowbreak{}reachableRelation\_\allowbreak{}input\_\allowbreak{}injective} {\tiny\ttfamily Paper1IT/\allowbreak RelationCarrier.lean}
\item \textbf{\nolinkurl{RLC8}}\hypertarget{lh:RLC8}{}\enspace{\ttfamily Ssot.\allowbreak{}RelationCarrier.\allowbreak{}reachableRelation\_\allowbreak{}input\_\allowbreak{}card\_\allowbreak{}lower\_\allowbreak{}bound} {\tiny\ttfamily Paper1IT/\allowbreak RelationCarrier.lean}
\item \textbf{\nolinkurl{RLC9}}\hypertarget{lh:RLC9}{}\enspace{\ttfamily Ssot.\allowbreak{}RelationCarrier.\allowbreak{}reachableRelation\_\allowbreak{}input\_\allowbreak{}bits\_\allowbreak{}lower\_\allowbreak{}bound} {\tiny\ttfamily Paper1IT/\allowbreak RelationCarrier.lean}
\item \textbf{\nolinkurl{RLC10}}\hypertarget{lh:RLC10}{}\enspace{\ttfamily Ssot.\allowbreak{}RelationCarrier.\allowbreak{}independentRelation\_\allowbreak{}fixed\_\allowbreak{}generator\_\allowbreak{}bits\_\allowbreak{}lower\_\allowbreak{}bound} {\tiny\ttfamily Paper1IT/\allowbreak RelationCarrier.lean}
\item \textbf{\nolinkurl{SCC1}}\hypertarget{lh:SCC1}{}\enspace{\ttfamily Ssot.\allowbreak{}SemanticChangeCost.\allowbreak{}faithful\_\allowbreak{}endpoints\_\allowbreak{}differ} {\tiny\ttfamily Paper1PL/\allowbreak SemanticChangeCost.lean}
\item \textbf{\nolinkurl{SCC2}}\hypertarget{lh:SCC2}{}\enspace{\ttfamily Ssot.\allowbreak{}SemanticChangeCost.\allowbreak{}faithful\_\allowbreak{}reencoding\_\allowbreak{}preserves\_\allowbreak{}change} {\tiny\ttfamily Paper1PL/\allowbreak SemanticChangeCost.lean}
\item \textbf{\nolinkurl{SCC3}}\hypertarget{lh:SCC3}{}\enspace{\ttfamily Ssot.\allowbreak{}SemanticChangeCost.\allowbreak{}semantic\_\allowbreak{}change\_\allowbreak{}requires\_\allowbreak{}operation} {\tiny\ttfamily Paper1PL/\allowbreak SemanticChangeCost.lean}
\item \textbf{\nolinkurl{SCC4}}\hypertarget{lh:SCC4}{}\enspace{\ttfamily Ssot.\allowbreak{}SemanticChangeCost.\allowbreak{}semantic\_\allowbreak{}change\_\allowbreak{}has\_\allowbreak{}changing\_\allowbreak{}step} {\tiny\ttfamily Paper1PL/\allowbreak SemanticChangeCost.lean}
\item \textbf{\nolinkurl{SCC5}}\hypertarget{lh:SCC5}{}\enspace{\ttfamily Ssot.\allowbreak{}SemanticChangeCost.\allowbreak{}preserving\_\allowbreak{}operations\_\allowbreak{}preserve\_\allowbreak{}meaning} {\tiny\ttfamily Paper1PL/\allowbreak SemanticChangeCost.lean}
\item \textbf{\nolinkurl{SCC6}}\hypertarget{lh:SCC6}{}\enspace{\ttfamily Ssot.\allowbreak{}SemanticChangeCost.\allowbreak{}nonempty\_\allowbreak{}trace\_\allowbreak{}cost\_\allowbreak{}pos} {\tiny\ttfamily Paper1PL/\allowbreak SemanticChangeCost.lean}
\item \textbf{\nolinkurl{SCC7}}\hypertarget{lh:SCC7}{}\enspace{\ttfamily Ssot.\allowbreak{}SemanticChangeCost.\allowbreak{}faithful\_\allowbreak{}semantic\_\allowbreak{}change\_\allowbreak{}cost\_\allowbreak{}pos} {\tiny\ttfamily Paper1PL/\allowbreak SemanticChangeCost.lean}
\item \textbf{\nolinkurl{SCC8}}\hypertarget{lh:SCC8}{}\enspace{\ttfamily Ssot.\allowbreak{}SemanticChangeCost.\allowbreak{}operation\_\allowbreak{}floor\_\allowbreak{}bounds\_\allowbreak{}trace} {\tiny\ttfamily Paper1PL/\allowbreak SemanticChangeCost.lean}
\item \textbf{\nolinkurl{SCC9}}\hypertarget{lh:SCC9}{}\enspace{\ttfamily Ssot.\allowbreak{}SemanticChangeCost.\allowbreak{}finite\_\allowbreak{}positive\_\allowbreak{}trace\_\allowbreak{}has\_\allowbreak{}floor} {\tiny\ttfamily Paper1PL/\allowbreak SemanticChangeCost.lean}
\item \textbf{\nolinkurl{SCI1}}\hypertarget{lh:SCI1}{}\enspace{\ttfamily Ssot.\allowbreak{}NominalFactorization.\allowbreak{}CoordinationCarrier.\allowbreak{}correct\_\allowbreak{}answers\_\allowbreak{}invariant} {\tiny\ttfamily Paper1PL/\allowbreak CoverageResolution.lean}
\item \textbf{\nolinkurl{SCI2}}\hypertarget{lh:SCI2}{}\enspace{\ttfamily Ssot.\allowbreak{}NominalFactorization.\allowbreak{}CoordinationCarrier.\allowbreak{}correct\_\allowbreak{}answers\_\allowbreak{}all} {\tiny\ttfamily Paper1PL/\allowbreak CoverageResolution.lean}
\item \textbf{\nolinkurl{SCI3}}\hypertarget{lh:SCI3}{}\enspace{\ttfamily Ssot.\allowbreak{}NominalFactorization.\allowbreak{}CoordinationCarrier.\allowbreak{}correct\_\allowbreak{}answer\_\allowbreak{}count} {\tiny\ttfamily Paper1PL/\allowbreak CoverageResolution.lean}
\item \textbf{\nolinkurl{SCI4}}\hypertarget{lh:SCI4}{}\enspace{\ttfamily Ssot.\allowbreak{}NominalFactorization.\allowbreak{}cooperativeTrace\_\allowbreak{}eq} {\tiny\ttfamily Paper1PL/\allowbreak CoverageResolution.lean}
\item \textbf{\nolinkurl{SCI5}}\hypertarget{lh:SCI5}{}\enspace{\ttfamily Ssot.\allowbreak{}NominalFactorization.\allowbreak{}initializationRank\_\allowbreak{}lt\_\allowbreak{}iff} {\tiny\ttfamily Paper1PL/\allowbreak CoverageResolution.lean}
\item \textbf{\nolinkurl{SCI6}}\hypertarget{lh:SCI6}{}\enspace{\ttfamily Ssot.\allowbreak{}NominalFactorization.\allowbreak{}distinct\_\allowbreak{}sharing\_\allowbreak{}requires\_\allowbreak{}outside\_\allowbreak{}access} {\tiny\ttfamily Paper1PL/\allowbreak CoverageResolution.lean}
\item \textbf{\nolinkurl{SCI7}}\hypertarget{lh:SCI7}{}\enspace{\ttfamily Ssot.\allowbreak{}NominalFactorization.\allowbreak{}distinct\_\allowbreak{}sharing\_\allowbreak{}count\_\allowbreak{}lower\_\allowbreak{}bound} {\tiny\ttfamily Paper1PL/\allowbreak CoverageResolution.lean}
\item \textbf{\nolinkurl{SCI8}}\hypertarget{lh:SCI8}{}\enspace{\ttfamily Ssot.\allowbreak{}NominalFactorization.\allowbreak{}no\_\allowbreak{}native\_\allowbreak{}sharing\_\allowbreak{}count} {\tiny\ttfamily Paper1PL/\allowbreak CoverageResolution.lean}
\item \textbf{\nolinkurl{SCI9}}\hypertarget{lh:SCI9}{}\enspace{\ttfamily Ssot.\allowbreak{}NominalFactorization.\allowbreak{}coupling\_\allowbreak{}witness\_\allowbreak{}lower\_\allowbreak{}bound} {\tiny\ttfamily Paper1PL/\allowbreak CoverageResolution.lean}
\item \textbf{\nolinkurl{SCI10}}\hypertarget{lh:SCI10}{}\enspace{\ttfamily Ssot.\allowbreak{}NominalFactorization.\allowbreak{}coupling\_\allowbreak{}witness\_\allowbreak{}attained} {\tiny\ttfamily Paper1PL/\allowbreak CoverageResolution.lean}
\item \textbf{\nolinkurl{SCI11}}\hypertarget{lh:SCI11}{}\enspace{\ttfamily Ssot.\allowbreak{}NominalFactorization.\allowbreak{}CoordinationCarrier.\allowbreak{}lookup\_\allowbreak{}and\_\allowbreak{}state\_\allowbreak{}answer\_\allowbreak{}count} {\tiny\ttfamily Paper1PL/\allowbreak CoverageResolution.lean}
\item \textbf{\nolinkurl{SCM1}}\hypertarget{lh:SCM1}{}\enspace{\ttfamily Ssot.\allowbreak{}HistoryViewDominance.\allowbreak{}structural\_\allowbreak{}membership\_\allowbreak{}exact\_\allowbreak{}iff} {\tiny\ttfamily Paper1PL/\allowbreak HistoryViewDominance.lean}
\item \textbf{\nolinkurl{SCM2}}\hypertarget{lh:SCM2}{}\enspace{\ttfamily Ssot.\allowbreak{}HistoryViewDominance.\allowbreak{}unwanted\_\allowbreak{}match\_\allowbreak{}refutes\_\allowbreak{}exact\_\allowbreak{}membership} {\tiny\ttfamily Paper1PL/\allowbreak HistoryViewDominance.lean}
\item \textbf{\nolinkurl{SCM3}}\hypertarget{lh:SCM3}{}\enspace{\ttfamily Ssot.\allowbreak{}HistoryViewDominance.\allowbreak{}nominal\_\allowbreak{}admission\_\allowbreak{}exact} {\tiny\ttfamily Paper1PL/\allowbreak HistoryViewDominance.lean}
\item \textbf{\nolinkurl{SCM4}}\hypertarget{lh:SCM4}{}\enspace{\ttfamily Ssot.\allowbreak{}HistoryViewDominance.\allowbreak{}nominal\_\allowbreak{}exact\_\allowbreak{}structural\_\allowbreak{}inexact} {\tiny\ttfamily Paper1PL/\allowbreak HistoryViewDominance.lean}
\item \textbf{\nolinkurl{SCT1}}\hypertarget{lh:SCT1}{}\enspace{\ttfamily Ssot.\allowbreak{}SemanticCompletionTax.\allowbreak{}semanticConsequence\_\allowbreak{}iff\_\allowbreak{}semanticallyValid} {\tiny\ttfamily Paper1PL/\allowbreak SemanticCompletionTax.lean}
\item \textbf{\nolinkurl{SCT2}}\hypertarget{lh:SCT2}{}\enspace{\ttfamily Ssot.\allowbreak{}SemanticCompletionTax.\allowbreak{}zeroErrorFor\_\allowbreak{}iff\_\allowbreak{}soundFor\_\allowbreak{}and\_\allowbreak{}completeFor} {\tiny\ttfamily Paper1PL/\allowbreak SemanticCompletionTax.lean}
\item \textbf{\nolinkurl{SCT3}}\hypertarget{lh:SCT3}{}\enspace{\ttfamily Ssot.\allowbreak{}SemanticCompletionTax.\allowbreak{}AuxiliaryCompletion.\allowbreak{}deficitToDecision\_\allowbreak{}injective} {\tiny\ttfamily Paper1PL/\allowbreak SemanticCompletionTax.lean}
\item \textbf{\nolinkurl{SCT4}}\hypertarget{lh:SCT4}{}\enspace{\ttfamily Ssot.\allowbreak{}SemanticCompletionTax.\allowbreak{}exists\_\allowbreak{}auxiliaryCompletion\_\allowbreak{}iff\_\allowbreak{}soundFor} {\tiny\ttfamily Paper1PL/\allowbreak SemanticCompletionTax.lean}
\item \textbf{\nolinkurl{SCT5}}\hypertarget{lh:SCT5}{}\enspace{\ttfamily Ssot.\allowbreak{}SemanticCompletionTax.\allowbreak{}nativeDeficitEquivGap} {\tiny\ttfamily Paper1PL/\allowbreak SemanticCompletionTax.lean}
\item \textbf{\nolinkurl{SCT6}}\hypertarget{lh:SCT6}{}\enspace{\ttfamily Ssot.\allowbreak{}SemanticCompletionTax.\allowbreak{}gap\_\allowbreak{}card\_\allowbreak{}le\_\allowbreak{}auxiliaryDecision\_\allowbreak{}card} {\tiny\ttfamily Paper1PL/\allowbreak SemanticCompletionTax.lean}
\item \textbf{\nolinkurl{SCT7}}\hypertarget{lh:SCT7}{}\enspace{\ttfamily Ssot.\allowbreak{}SemanticCompletionTax.\allowbreak{}minimumSemanticCompletionTax\_\allowbreak{}iff\_\allowbreak{}minimumLaminarGapCost} {\tiny\ttfamily Paper1PL/\allowbreak SemanticCompletionTax.lean}
\item \textbf{\nolinkurl{SCT8}}\hypertarget{lh:SCT8}{}\enspace{\ttfamily Ssot.\allowbreak{}SemanticCompletionTax.\allowbreak{}minimumSemanticCompletionTax\_\allowbreak{}pos\_\allowbreak{}of\_\allowbreak{}not\_\allowbreak{}laminar} {\tiny\ttfamily Paper1PL/\allowbreak SemanticCompletionTax.lean}
\item \textbf{\nolinkurl{SCT9}}\hypertarget{lh:SCT9}{}\enspace{\ttfamily Ssot.\allowbreak{}SemanticCompletionTax.\allowbreak{}multiParent\_\allowbreak{}realizes\_\allowbreak{}zeroError\_\allowbreak{}with\_\allowbreak{}zero\_\allowbreak{}completionTax} {\tiny\ttfamily Paper1PL/\allowbreak SemanticCompletionTax.lean}
\item \textbf{\nolinkurl{SCT10}}\hypertarget{lh:SCT10}{}\enspace{\ttfamily Ssot.\allowbreak{}SemanticCompletionTax.\allowbreak{}nativeDeficitEquivPositiveGapAt} {\tiny\ttfamily Paper1PL/\allowbreak SemanticCompletionTax.lean}
\item \textbf{\nolinkurl{SCT11}}\hypertarget{lh:SCT11}{}\enspace{\ttfamily Ssot.\allowbreak{}SemanticCompletionTax.\allowbreak{}nativeExcessEquivEffectiveBlockAt} {\tiny\ttfamily Paper1PL/\allowbreak SemanticCompletionTax.lean}
\item \textbf{\nolinkurl{SCT12}}\hypertarget{lh:SCT12}{}\enspace{\ttfamily Ssot.\allowbreak{}SemanticCompletionTax.\allowbreak{}stagedScope\_\allowbreak{}step\_\allowbreak{}retraction\_\allowbreak{}or\_\allowbreak{}semanticCompletionBurden} {\tiny\ttfamily Paper1PL/\allowbreak SemanticCompletionTax.lean}
\item \textbf{\nolinkurl{SCT13}}\hypertarget{lh:SCT13}{}\enspace{\ttfamily Ssot.\allowbreak{}SemanticCompletionTax.\allowbreak{}stagedScope\_\allowbreak{}recurrent\_\allowbreak{}parent\_\allowbreak{}quadratic\_\allowbreak{}semanticCompletion\_\allowbreak{}lower\_\allowbreak{}bound} {\tiny\ttfamily Paper1PL/\allowbreak SemanticCompletionTax.lean}
\item \textbf{\nolinkurl{SCT14}}\hypertarget{lh:SCT14}{}\enspace{\ttfamily Ssot.\allowbreak{}SemanticCompletionTax.\allowbreak{}minimumSemanticCompletionTax\_\allowbreak{}iff\_\allowbreak{}minimumDelegationCost} {\tiny\ttfamily Paper1PL/\allowbreak SemanticCompletionTax.lean}
\item \textbf{\nolinkurl{SCT15}}\hypertarget{lh:SCT15}{}\enspace{\ttfamily Ssot.\allowbreak{}SemanticCompletionTax.\allowbreak{}contractOnly\_\allowbreak{}canonicalCompletion\_\allowbreak{}card\_\allowbreak{}eq\_\allowbreak{}requiredRelation} {\tiny\ttfamily Paper1PL/\allowbreak SemanticCompletionTax.lean}
\item \textbf{\nolinkurl{SCT16}}\hypertarget{lh:SCT16}{}\enspace{\ttfamily Ssot.\allowbreak{}SemanticCompletionTax.\allowbreak{}minimumSemanticCompletionTax\_\allowbreak{}le\_\allowbreak{}contractOnly} {\tiny\ttfamily Paper1PL/\allowbreak SemanticCompletionTax.lean}
\item \textbf{\nolinkurl{SCT17}}\hypertarget{lh:SCT17}{}\enspace{\ttfamily Ssot.\allowbreak{}SemanticCompletionTax.\allowbreak{}contractOnly\_\allowbreak{}positive\_\allowbreak{}of\_\allowbreak{}requiredIncidence} {\tiny\ttfamily Paper1PL/\allowbreak SemanticCompletionTax.lean}
\item \textbf{\nolinkurl{SCT18}}\hypertarget{lh:SCT18}{}\enspace{\ttfamily Ssot.\allowbreak{}SemanticCompletionTax.\allowbreak{}family\_\allowbreak{}optimizerQuotient\_\allowbreak{}eq\_\allowbreak{}iff\_\allowbreak{}same\_\allowbreak{}required\_\allowbreak{}owners} {\tiny\ttfamily Paper1PL/\allowbreak SemanticCompletionTax.lean}
\item \textbf{\nolinkurl{SCT19}}\hypertarget{lh:SCT19}{}\enspace{\ttfamily Ssot.\allowbreak{}SemanticCompletionTax.\allowbreak{}minimumSemanticCompletionTax\_\allowbreak{}lt\_\allowbreak{}contractOnly\_\allowbreak{}of\_\allowbreak{}requiredIncidence} {\tiny\ttfamily Paper1PL/\allowbreak SemanticCompletionTax.lean}
\item \textbf{\nolinkurl{SCT20}}\hypertarget{lh:SCT20}{}\enspace{\ttfamily Ssot.\allowbreak{}SemanticCompletionTax.\allowbreak{}indexedNativeDeficitEquivGap} {\tiny\ttfamily Paper1PL/\allowbreak SemanticCompletionTax.lean}
\item \textbf{\nolinkurl{SCT21}}\hypertarget{lh:SCT21}{}\enspace{\ttfamily Ssot.\allowbreak{}SemanticCompletionTax.\allowbreak{}minimumIndexedSemanticCompletionTax\_\allowbreak{}iff\_\allowbreak{}minimumIndexedLaminarGapCost} {\tiny\ttfamily Paper1PL/\allowbreak SemanticCompletionTax.lean}
\item \textbf{\nolinkurl{SCT22}}\hypertarget{lh:SCT22}{}\enspace{\ttfamily Ssot.\allowbreak{}SemanticCompletionTax.\allowbreak{}indexedContractOnly\_\allowbreak{}canonicalCompletion\_\allowbreak{}card\_\allowbreak{}eq\_\allowbreak{}requiredRelation} {\tiny\ttfamily Paper1PL/\allowbreak SemanticCompletionTax.lean}
\item \textbf{\nolinkurl{SCT23}}\hypertarget{lh:SCT23}{}\enspace{\ttfamily Ssot.\allowbreak{}SemanticCompletionTax.\allowbreak{}minimumIndexedSemanticCompletionTax\_\allowbreak{}lt\_\allowbreak{}contractOnly\_\allowbreak{}of\_\allowbreak{}requiredIncidence} {\tiny\ttfamily Paper1PL/\allowbreak SemanticCompletionTax.lean}
\item \textbf{\nolinkurl{SCT24}}\hypertarget{lh:SCT24}{}\enspace{\ttfamily Ssot.\allowbreak{}SemanticCompletionTax.\allowbreak{}minimumIndexedDelegationCost\_\allowbreak{}iff\_\allowbreak{}minimumIndexedLaminarGapCost} {\tiny\ttfamily Paper1PL/\allowbreak SemanticCompletionTax.lean}
\item \textbf{\nolinkurl{SCT25}}\hypertarget{lh:SCT25}{}\enspace{\ttfamily Ssot.\allowbreak{}SemanticCompletionTax.\allowbreak{}minimumIndexedSemanticCompletionTax\_\allowbreak{}iff\_\allowbreak{}minimumIndexedDelegationCost} {\tiny\ttfamily Paper1PL/\allowbreak SemanticCompletionTax.lean}
\item \textbf{\nolinkurl{SCT26}}\hypertarget{lh:SCT26}{}\enspace{\ttfamily Ssot.\allowbreak{}SemanticCompletionTax.\allowbreak{}indexedMultiParent\_\allowbreak{}realizes\_\allowbreak{}zeroError\_\allowbreak{}with\_\allowbreak{}zero\_\allowbreak{}completionTax} {\tiny\ttfamily Paper1PL/\allowbreak SemanticCompletionTax.lean}
\item \textbf{\nolinkurl{SCT27}}\hypertarget{lh:SCT27}{}\enspace{\ttfamily Ssot.\allowbreak{}SemanticCompletionTax.\allowbreak{}IndexedDelegatingSingleInheritanceRepresentation.\allowbreak{}gap\_\allowbreak{}forces\_\allowbreak{}forward\_\allowbreak{}target} {\tiny\ttfamily Paper1PL/\allowbreak SemanticCompletionTax.lean}
\item \textbf{\nolinkurl{SCT28}}\hypertarget{lh:SCT28}{}\enspace{\ttfamily Ssot.\allowbreak{}SemanticCompletionTax.\allowbreak{}IndexedDelegatingSingleInheritanceRepresentation.\allowbreak{}gapToForwardingDecision\_\allowbreak{}injective} {\tiny\ttfamily Paper1PL/\allowbreak SemanticCompletionTax.lean}
\item \textbf{\nolinkurl{SCT29}}\hypertarget{lh:SCT29}{}\enspace{\ttfamily Ssot.\allowbreak{}SemanticCompletionTax.\allowbreak{}IndexedDelegatingSingleInheritanceRepresentation.\allowbreak{}gap\_\allowbreak{}card\_\allowbreak{}le\_\allowbreak{}forwardingDecision\_\allowbreak{}card} {\tiny\ttfamily Paper1PL/\allowbreak SemanticCompletionTax.lean}
\item \textbf{\nolinkurl{SCT30}}\hypertarget{lh:SCT30}{}\enspace{\ttfamily Ssot.\allowbreak{}SemanticCompletionTax.\allowbreak{}minimumIndexedSemanticCompletionTax\_\allowbreak{}pos\_\allowbreak{}of\_\allowbreak{}not\_\allowbreak{}laminar} {\tiny\ttfamily Paper1PL/\allowbreak SemanticCompletionTax.lean}
\item \textbf{\nolinkurl{SCT31}}\hypertarget{lh:SCT31}{}\enspace{\ttfamily Ssot.\allowbreak{}SemanticCompletionTax.\allowbreak{}minimumIndexedSemanticCompletionTax\_\allowbreak{}eq\_\allowbreak{}zero\_\allowbreak{}iff\_\allowbreak{}laminar} {\tiny\ttfamily Paper1PL/\allowbreak SemanticCompletionTax.lean}
\item \textbf{\nolinkurl{SCT32}}\hypertarget{lh:SCT32}{}\enspace{\ttfamily Ssot.\allowbreak{}SemanticCompletionTax.\allowbreak{}indexedZeroErrorMaintenanceCostLadder} {\tiny\ttfamily Paper1PL/\allowbreak SemanticCompletionTax.lean}
\item \textbf{\nolinkurl{SCT33}}\hypertarget{lh:SCT33}{}\enspace{\ttfamily Ssot.\allowbreak{}SemanticCompletionTax.\allowbreak{}indexedExactCarrierTrichotomy} {\tiny\ttfamily Paper1PL/\allowbreak SemanticCompletionTax.lean}
\item \textbf{\nolinkurl{SCT34}}\hypertarget{lh:SCT34}{}\enspace{\ttfamily Ssot.\allowbreak{}SemanticCompletionTax.\allowbreak{}indexedExactOneParentDichotomy} {\tiny\ttfamily Paper1PL/\allowbreak SemanticCompletionTax.lean}
\item \textbf{\nolinkurl{SCT35}}\hypertarget{lh:SCT35}{}\enspace{\ttfamily Ssot.\allowbreak{}SemanticCompletionTax.\allowbreak{}exactStructuralDebtBalance\_\allowbreak{}and\_\allowbreak{}cases} {\tiny\ttfamily Paper1PL/\allowbreak SemanticCompletionTax.lean}
\item \textbf{\nolinkurl{SCT36}}\hypertarget{lh:SCT36}{}\enspace{\ttfamily Ssot.\allowbreak{}SemanticCompletionTax.\allowbreak{}languageOwnedCompleteness\_\allowbreak{}eliminates\_\allowbreak{}programSpecificRepair} {\tiny\ttfamily Paper1PL/\allowbreak SemanticCompletionTax.lean}
\item \textbf{\nolinkurl{SCT37}}\hypertarget{lh:SCT37}{}\enspace{\ttfamily Ssot.\allowbreak{}SemanticCompletionTax.\allowbreak{}zeroDeficit\_\allowbreak{}sound\_\allowbreak{}nativeProvenance\_\allowbreak{}eq\_\allowbreak{}required\_\allowbreak{}and\_\allowbreak{}has\_\allowbreak{}multiParent} {\tiny\ttfamily Paper1PL/\allowbreak SemanticCompletionTax.lean}
\item \textbf{\nolinkurl{SCT38}}\hypertarget{lh:SCT38}{}\enspace{\ttfamily Ssot.\allowbreak{}SemanticCompletionTax.\allowbreak{}stableExactViewErrorEquivSemanticChange} {\tiny\ttfamily Paper1PL/\allowbreak SemanticCompletionTax.lean}
\item \textbf{\nolinkurl{SCT39}}\hypertarget{lh:SCT39}{}\enspace{\ttfamily Ssot.\allowbreak{}SemanticCompletionTax.\allowbreak{}stableExactView\_\allowbreak{}errorCount\_\allowbreak{}eq\_\allowbreak{}semanticChangeCount} {\tiny\ttfamily Paper1PL/\allowbreak SemanticCompletionTax.lean}
\item \textbf{\nolinkurl{SCT40}}\hypertarget{lh:SCT40}{}\enspace{\ttfamily Ssot.\allowbreak{}SemanticCompletionTax.\allowbreak{}stableExactViewCoverageEquivRequiredBefore} {\tiny\ttfamily Paper1PL/\allowbreak SemanticCompletionTax.lean}
\item \textbf{\nolinkurl{SCT41}}\hypertarget{lh:SCT41}{}\enspace{\ttfamily Ssot.\allowbreak{}SemanticCompletionTax.\allowbreak{}stableExactView\_\allowbreak{}coverageCount\_\allowbreak{}eq\_\allowbreak{}initialRequiredCount} {\tiny\ttfamily Paper1PL/\allowbreak SemanticCompletionTax.lean}
\item \textbf{\nolinkurl{SCT42}}\hypertarget{lh:SCT42}{}\enspace{\ttfamily Ssot.\allowbreak{}SemanticCompletionTax.\allowbreak{}stableExactView\_\allowbreak{}coverageFraction\_\allowbreak{}eq} {\tiny\ttfamily Paper1PL/\allowbreak SemanticCompletionTax.lean}
\item \textbf{\nolinkurl{SCT43}}\hypertarget{lh:SCT43}{}\enspace{\ttfamily Ssot.\allowbreak{}SemanticCompletionTax.\allowbreak{}stableExactView\_\allowbreak{}coverageFraction\_\allowbreak{}strictly\_\allowbreak{}decreases} {\tiny\ttfamily Paper1PL/\allowbreak SemanticCompletionTax.lean}
\item \textbf{\nolinkurl{SCT44}}\hypertarget{lh:SCT44}{}\enspace{\ttfamily Ssot.\allowbreak{}SemanticCompletionTax.\allowbreak{}boundedView\_\allowbreak{}coverageFraction\_\allowbreak{}le} {\tiny\ttfamily Paper1PL/\allowbreak SemanticCompletionTax.lean}
\item \textbf{\nolinkurl{SCT45}}\hypertarget{lh:SCT45}{}\enspace{\ttfamily Ssot.\allowbreak{}SemanticCompletionTax.\allowbreak{}boundedView\_\allowbreak{}coverageCeiling\_\allowbreak{}strictly\_\allowbreak{}decreases} {\tiny\ttfamily Paper1PL/\allowbreak SemanticCompletionTax.lean}
\item \textbf{\nolinkurl{SCT46}}\hypertarget{lh:SCT46}{}\enspace{\ttfamily Ssot.\allowbreak{}SemanticCompletionTax.\allowbreak{}zeroErrorFor\_\allowbreak{}iff\_\allowbreak{}exactCoupling} {\tiny\ttfamily Paper1PL/\allowbreak SemanticCompletionTax.lean}
\item \textbf{\nolinkurl{SCT47}}\hypertarget{lh:SCT47}{}\enspace{\ttfamily Ssot.\allowbreak{}SemanticCompletionTax.\allowbreak{}zeroErrorFor\_\allowbreak{}iff\_\allowbreak{}correctMaintenanceDebt\_\allowbreak{}empty} {\tiny\ttfamily Paper1PL/\allowbreak SemanticCompletionTax.lean}
\item \textbf{\nolinkurl{SCT48}}\hypertarget{lh:SCT48}{}\enspace{\ttfamily Ssot.\allowbreak{}SemanticCompletionTax.\allowbreak{}correctMaintenanceDebt\_\allowbreak{}card\_\allowbreak{}eq\_\allowbreak{}deficit\_\allowbreak{}add\_\allowbreak{}excess} {\tiny\ttfamily Paper1PL/\allowbreak SemanticCompletionTax.lean}
\item \textbf{\nolinkurl{SCT49}}\hypertarget{lh:SCT49}{}\enspace{\ttfamily Ssot.\allowbreak{}SemanticCompletionTax.\allowbreak{}correctMaintenanceDebt\_\allowbreak{}card\_\allowbreak{}pos\_\allowbreak{}iff\_\allowbreak{}not\_\allowbreak{}zeroErrorFor} {\tiny\ttfamily Paper1PL/\allowbreak SemanticCompletionTax.lean}
\item \textbf{\nolinkurl{SCT50}}\hypertarget{lh:SCT50}{}\enspace{\ttfamily Ssot.\allowbreak{}SemanticCompletionTax.\allowbreak{}not\_\allowbreak{}zeroErrorFor\_\allowbreak{}iff\_\allowbreak{}underCoupled\_\allowbreak{}or\_\allowbreak{}overCoupled} {\tiny\ttfamily Paper1PL/\allowbreak SemanticCompletionTax.lean}
\item \textbf{\nolinkurl{SCT51}}\hypertarget{lh:SCT51}{}\enspace{\ttfamily Ssot.\allowbreak{}SemanticCompletionTax.\allowbreak{}exists\_\allowbreak{}auxiliaryCompletion\_\allowbreak{}iff\_\allowbreak{}nativeExcess\_\allowbreak{}empty} {\tiny\ttfamily Paper1PL/\allowbreak SemanticCompletionTax.lean}
\item \textbf{\nolinkurl{SCT52}}\hypertarget{lh:SCT52}{}\enspace{\ttfamily Ssot.\allowbreak{}SemanticCompletionTax.\allowbreak{}indexedOneParentDebtEquivGap} {\tiny\ttfamily Paper1PL/\allowbreak SemanticCompletionTax.lean}
\item \textbf{\nolinkurl{SCT53}}\hypertarget{lh:SCT53}{}\enspace{\ttfamily Ssot.\allowbreak{}SemanticCompletionTax.\allowbreak{}indexedOneParentDebt\_\allowbreak{}card\_\allowbreak{}eq\_\allowbreak{}gap} {\tiny\ttfamily Paper1PL/\allowbreak SemanticCompletionTax.lean}
\item \textbf{\nolinkurl{SCT54}}\hypertarget{lh:SCT54}{}\enspace{\ttfamily Ssot.\allowbreak{}SemanticCompletionTax.\allowbreak{}oneParentDebt\_\allowbreak{}eq\_\allowbreak{}forcedFloor\_\allowbreak{}add\_\allowbreak{}avoidable} {\tiny\ttfamily Paper1PL/\allowbreak SemanticCompletionTax.lean}
\item \textbf{\nolinkurl{SCT55}}\hypertarget{lh:SCT55}{}\enspace{\ttfamily Ssot.\allowbreak{}SemanticCompletionTax.\allowbreak{}oneParentAvoidableDebt\_\allowbreak{}eq\_\allowbreak{}zero\_\allowbreak{}iff\_\allowbreak{}optimal} {\tiny\ttfamily Paper1PL/\allowbreak SemanticCompletionTax.lean}
\item \textbf{\nolinkurl{SCT56}}\hypertarget{lh:SCT56}{}\enspace{\ttfamily Ssot.\allowbreak{}SemanticCompletionTax.\allowbreak{}AuxiliaryCompletion.\allowbreak{}outsideDecision\_\allowbreak{}iff\_\allowbreak{}nativeDeficit} {\tiny\ttfamily Paper1PL/\allowbreak SemanticCompletionTax.lean}
\item \textbf{\nolinkurl{SCT57}}\hypertarget{lh:SCT57}{}\enspace{\ttfamily Ssot.\allowbreak{}SemanticCompletionTax.\allowbreak{}AuxiliaryCompletion.\allowbreak{}nativeDeficitEquivOutsideDecision} {\tiny\ttfamily Paper1PL/\allowbreak SemanticCompletionTax.lean}
\item \textbf{\nolinkurl{SCT58}}\hypertarget{lh:SCT58}{}\enspace{\ttfamily Ssot.\allowbreak{}SemanticCompletionTax.\allowbreak{}indexedReplicaCompletion\_\allowbreak{}outsideDecision\_\allowbreak{}card\_\allowbreak{}eq\_\allowbreak{}gap} {\tiny\ttfamily Paper1PL/\allowbreak SemanticCompletionTax.lean}
\item \textbf{\nolinkurl{SCT59}}\hypertarget{lh:SCT59}{}\enspace{\ttfamily Ssot.\allowbreak{}SemanticCompletionTax.\allowbreak{}minimumIndexedLaminarGapCost\_\allowbreak{}le\_\allowbreak{}replicaCompletion\_\allowbreak{}outsideDecision\_\allowbreak{}card} {\tiny\ttfamily Paper1PL/\allowbreak SemanticCompletionTax.lean}
\item \textbf{\nolinkurl{SDN1}}\hypertarget{lh:SDN1}{}\enspace{\ttfamily Ssot.\allowbreak{}SemanticDenotation.\allowbreak{}Specification.\allowbreak{}exact\_\allowbreak{}iff\_\allowbreak{}complete\_\allowbreak{}and\_\allowbreak{}sound} {\tiny\ttfamily Paper1PL/\allowbreak SemanticDenotation.lean}
\item \textbf{\nolinkurl{SDN2}}\hypertarget{lh:SDN2}{}\enspace{\ttfamily Ssot.\allowbreak{}SemanticDenotation.\allowbreak{}Specification.\allowbreak{}exact\_\allowbreak{}of\_\allowbreak{}equivalent} {\tiny\ttfamily Paper1PL/\allowbreak SemanticDenotation.lean}
\item \textbf{\nolinkurl{SDN3}}\hypertarget{lh:SDN3}{}\enspace{\ttfamily Ssot.\allowbreak{}SemanticDenotation.\allowbreak{}Specification.\allowbreak{}equivalent\_\allowbreak{}erase\_\allowbreak{}of\_\allowbreak{}denotation\_\allowbreak{}covered} {\tiny\ttfamily Paper1PL/\allowbreak SemanticDenotation.lean}
\item \textbf{\nolinkurl{SDN4}}\hypertarget{lh:SDN4}{}\enspace{\ttfamily Ssot.\allowbreak{}SemanticDenotation.\allowbreak{}Specification.\allowbreak{}exact\_\allowbreak{}erase\_\allowbreak{}of\_\allowbreak{}denotation\_\allowbreak{}covered} {\tiny\ttfamily Paper1PL/\allowbreak SemanticDenotation.lean}
\item \textbf{\nolinkurl{SDN5}}\hypertarget{lh:SDN5}{}\enspace{\ttfamily Ssot.\allowbreak{}SemanticDenotation.\allowbreak{}AtomicAuthoritySystem.\allowbreak{}toRoleSystem\_\allowbreak{}role\_\allowbreak{}eq\_\allowbreak{}iff} {\tiny\ttfamily Paper1PL/\allowbreak SemanticDenotation.lean}
\item \textbf{\nolinkurl{SDN6}}\hypertarget{lh:SDN6}{}\enspace{\ttfamily Ssot.\allowbreak{}SemanticDenotation.\allowbreak{}AtomicAuthoritySystem.\allowbreak{}basis\_\allowbreak{}denotations\_\allowbreak{}eq\_\allowbreak{}required} {\tiny\ttfamily Paper1PL/\allowbreak SemanticDenotation.lean}
\item \textbf{\nolinkurl{SDN7}}\hypertarget{lh:SDN7}{}\enspace{\ttfamily Ssot.\allowbreak{}SemanticDenotation.\allowbreak{}AtomicAuthoritySystem.\allowbreak{}basis\_\allowbreak{}card\_\allowbreak{}eq\_\allowbreak{}required\_\allowbreak{}card} {\tiny\ttfamily Paper1PL/\allowbreak SemanticDenotation.lean}
\item \textbf{\nolinkurl{SDN8}}\hypertarget{lh:SDN8}{}\enspace{\ttfamily Ssot.\allowbreak{}SemanticDenotation.\allowbreak{}AtomicAuthoritySystem.\allowbreak{}role\_\allowbreak{}invariant\_\allowbreak{}quantity\_\allowbreak{}eq} {\tiny\ttfamily Paper1PL/\allowbreak SemanticDenotation.lean}
\item \textbf{\nolinkurl{SDN9}}\hypertarget{lh:SDN9}{}\enspace{\ttfamily Ssot.\allowbreak{}SemanticDenotation.\allowbreak{}DenotationalRoleSystem.\allowbreak{}existsUnique\_\allowbreak{}realizes} {\tiny\ttfamily Paper1PL/\allowbreak SemanticDenotation.lean}
\item \textbf{\nolinkurl{SDN10}}\hypertarget{lh:SDN10}{}\enspace{\ttfamily Ssot.\allowbreak{}SemanticDenotation.\allowbreak{}DenotationalRoleSystem.\allowbreak{}realizes\_\allowbreak{}iff\_\allowbreak{}roleOf\_\allowbreak{}eq} {\tiny\ttfamily Paper1PL/\allowbreak SemanticDenotation.lean}
\item \textbf{\nolinkurl{SDN11}}\hypertarget{lh:SDN11}{}\enspace{\ttfamily Ssot.\allowbreak{}SemanticDenotation.\allowbreak{}DenotationalRoleSystem.\allowbreak{}roleOf\_\allowbreak{}surjective} {\tiny\ttfamily Paper1PL/\allowbreak SemanticDenotation.lean}
\item \textbf{\nolinkurl{SDN12}}\hypertarget{lh:SDN12}{}\enspace{\ttfamily Ssot.\allowbreak{}SemanticDenotation.\allowbreak{}DenotationalRoleSystem.\allowbreak{}toRoleSystem} {\tiny\ttfamily Paper1PL/\allowbreak SemanticDenotation.lean}
\item \textbf{\nolinkurl{SDN13}}\hypertarget{lh:SDN13}{}\enspace{\ttfamily Ssot.\allowbreak{}SemanticDenotation.\allowbreak{}DenotationalRoleSystem.\allowbreak{}exact\_\allowbreak{}iff\_\allowbreak{}complete} {\tiny\ttfamily Paper1PL/\allowbreak SemanticDenotation.lean}
\item \textbf{\nolinkurl{SDN14}}\hypertarget{lh:SDN14}{}\enspace{\ttfamily Ssot.\allowbreak{}SemanticDenotation.\allowbreak{}DenotationalRoleSystem.\allowbreak{}irredundantExact\_\allowbreak{}iff\_\allowbreak{}isBasis} {\tiny\ttfamily Paper1PL/\allowbreak SemanticDenotation.lean}
\item \textbf{\nolinkurl{SDN15}}\hypertarget{lh:SDN15}{}\enspace{\ttfamily Ssot.\allowbreak{}SemanticDenotation.\allowbreak{}DenotationalRoleSystem.\allowbreak{}encodedScopes\_\allowbreak{}eq\_\allowbreak{}intendedScopes\_\allowbreak{}of\_\allowbreak{}exact} {\tiny\ttfamily Paper1PL/\allowbreak SemanticDenotation.lean}
\item \textbf{\nolinkurl{SDN16}}\hypertarget{lh:SDN16}{}\enspace{\ttfamily Ssot.\allowbreak{}SemanticDenotation.\allowbreak{}DenotationalRoleSystem.\allowbreak{}encodedScopes\_\allowbreak{}eq\_\allowbreak{}intendedScopes\_\allowbreak{}of\_\allowbreak{}basis} {\tiny\ttfamily Paper1PL/\allowbreak SemanticDenotation.lean}
\item \textbf{\nolinkurl{SDN17}}\hypertarget{lh:SDN17}{}\enspace{\ttfamily Ssot.\allowbreak{}SemanticDenotation.\allowbreak{}DenotationalRoleSystem.\allowbreak{}scopeQuantity\_\allowbreak{}basis\_\allowbreak{}invariant} {\tiny\ttfamily Paper1PL/\allowbreak SemanticDenotation.lean}
\item \textbf{\nolinkurl{SDN18}}\hypertarget{lh:SDN18}{}\enspace{\ttfamily Ssot.\allowbreak{}SemanticDenotation.\allowbreak{}DenotationalRoleSystem.\allowbreak{}scopeQuantity\_\allowbreak{}same\_\allowbreak{}or\_\allowbreak{}changedAnswer} {\tiny\ttfamily Paper1PL/\allowbreak SemanticDenotation.lean}
\item \textbf{\nolinkurl{SDN19}}\hypertarget{lh:SDN19}{}\enspace{\ttfamily Ssot.\allowbreak{}SemanticDenotation.\allowbreak{}DenotationalRoleSystem.\allowbreak{}roleOf\_\allowbreak{}eq\_\allowbreak{}iff\_\allowbreak{}declDenote\_\allowbreak{}eq} {\tiny\ttfamily Paper1PL/\allowbreak SemanticDenotation.lean}
\item \textbf{\nolinkurl{SDN20}}\hypertarget{lh:SDN20}{}\enspace{\ttfamily Ssot.\allowbreak{}SemanticDenotation.\allowbreak{}DenotationalRoleSystem.\allowbreak{}coarseRoles\_\allowbreak{}agree\_\allowbreak{}or\_\allowbreak{}semanticRefinementWitness} {\tiny\ttfamily Paper1PL/\allowbreak SemanticDenotation.lean}
\item \textbf{\nolinkurl{SDN21}}\hypertarget{lh:SDN21}{}\enspace{\ttfamily Ssot.\allowbreak{}SemanticDenotation.\allowbreak{}DenotationalRoleSystem.\allowbreak{}scopeQuantity\_\allowbreak{}eq\_\allowbreak{}across\_\allowbreak{}bases} {\tiny\ttfamily Paper1PL/\allowbreak SemanticDenotation.lean}
\item \textbf{\nolinkurl{SEF1}}\hypertarget{lh:SEF1}{}\enspace{\ttfamily Ssot.\allowbreak{}SemanticFactoring.\allowbreak{}sameDecisionColumn\_\allowbreak{}historyDecision\_\allowbreak{}iff} {\tiny\ttfamily Paper1PL/\allowbreak SemanticFactoring.lean}
\item \textbf{\nolinkurl{SEF2}}\hypertarget{lh:SEF2}{}\enspace{\ttfamily Ssot.\allowbreak{}SemanticFactoring.\allowbreak{}exactHistoryEncoding\_\allowbreak{}code\_\allowbreak{}eq\_\allowbreak{}implies\_\allowbreak{}sameMaintenanceAnswers} {\tiny\ttfamily Paper1PL/\allowbreak SemanticFactoring.lean}
\item \textbf{\nolinkurl{SEF3}}\hypertarget{lh:SEF3}{}\enspace{\ttfamily Ssot.\allowbreak{}SemanticFactoring.\allowbreak{}canonicalHistoryCode\_\allowbreak{}eq\_\allowbreak{}iff} {\tiny\ttfamily Paper1PL/\allowbreak SemanticFactoring.lean}
\item \textbf{\nolinkurl{SEF4}}\hypertarget{lh:SEF4}{}\enspace{\ttfamily Ssot.\allowbreak{}SemanticFactoring.\allowbreak{}minimalExactHistoryEncoding\_\allowbreak{}code\_\allowbreak{}eq\_\allowbreak{}iff} {\tiny\ttfamily Paper1PL/\allowbreak SemanticFactoring.lean}
\item \textbf{\nolinkurl{SEF5}}\hypertarget{lh:SEF5}{}\enspace{\ttfamily Ssot.\allowbreak{}SemanticFactoring.\allowbreak{}sameCompleteDecisionColumn\_\allowbreak{}iff} {\tiny\ttfamily Paper1PL/\allowbreak SemanticFactoring.lean}
\item \textbf{\nolinkurl{SEF6}}\hypertarget{lh:SEF6}{}\enspace{\ttfamily Ssot.\allowbreak{}SemanticFactoring.\allowbreak{}merge\_\allowbreak{}share\_\allowbreak{}or\_\allowbreak{}separate} {\tiny\ttfamily Paper1PL/\allowbreak SemanticFactoring.lean}
\item \textbf{\nolinkurl{SEF7}}\hypertarget{lh:SEF7}{}\enspace{\ttfamily Ssot.\allowbreak{}SemanticFactoring.\allowbreak{}sameImplementation\_\allowbreak{}distinctMeaning\_\allowbreak{}shareFactor\_\allowbreak{}notIdentity} {\tiny\ttfamily Paper1PL/\allowbreak SemanticFactoring.lean}
\item \textbf{\nolinkurl{SEF8}}\hypertarget{lh:SEF8}{}\enspace{\ttfamily Ssot.\allowbreak{}SemanticFactoring.\allowbreak{}differentImplementation\_\allowbreak{}requires\_\allowbreak{}separateCodes} {\tiny\ttfamily Paper1PL/\allowbreak SemanticFactoring.lean}
\item \textbf{\nolinkurl{SEF9}}\hypertarget{lh:SEF9}{}\enspace{\ttfamily Ssot.\allowbreak{}SemanticFactoring.\allowbreak{}sameImplementationPattern\_\allowbreak{}merge\_\allowbreak{}and\_\allowbreak{}separate\_\allowbreak{}witness} {\tiny\ttfamily Paper1PL/\allowbreak SemanticFactoring.lean}
\item \textbf{\nolinkurl{SEF10}}\hypertarget{lh:SEF10}{}\enspace{\ttfamily Ssot.\allowbreak{}SemanticFactoring.\allowbreak{}noImplementationOnlyRule\_\allowbreak{}decidesSemanticIdentity} {\tiny\ttfamily Paper1PL/\allowbreak SemanticFactoring.lean}
\item \textbf{\nolinkurl{SEF11}}\hypertarget{lh:SEF11}{}\enspace{\ttfamily Ssot.\allowbreak{}SemanticFactoring.\allowbreak{}semanticEditDistance\_\allowbreak{}le\_\allowbreak{}sharedAnswerError} {\tiny\ttfamily Paper1PL/\allowbreak SemanticFactoring.lean}
\item \textbf{\nolinkurl{SEF12}}\hypertarget{lh:SEF12}{}\enspace{\ttfamily Ssot.\allowbreak{}SemanticFactoring.\allowbreak{}semanticEditDistance\_\allowbreak{}is\_\allowbreak{}exact\_\allowbreak{}sharedAnswerFloor} {\tiny\ttfamily Paper1PL/\allowbreak SemanticFactoring.lean}
\item \textbf{\nolinkurl{SEF13}}\hypertarget{lh:SEF13}{}\enspace{\ttfamily Ssot.\allowbreak{}SemanticFactoring.\allowbreak{}sharedAnswerError\_\allowbreak{}pos\_\allowbreak{}of\_\allowbreak{}distinct} {\tiny\ttfamily Paper1PL/\allowbreak SemanticFactoring.lean}
\item \textbf{\nolinkurl{SEF14}}\hypertarget{lh:SEF14}{}\enspace{\ttfamily Ssot.\allowbreak{}SemanticFactoring.\allowbreak{}oneAsymmetricAnswer\_\allowbreak{}forces\_\allowbreak{}positive\_\allowbreak{}sharedAnswerError} {\tiny\ttfamily Paper1PL/\allowbreak SemanticFactoring.lean}
\item \textbf{\nolinkurl{SEF15}}\hypertarget{lh:SEF15}{}\enspace{\ttfamily Ssot.\allowbreak{}SemanticFactoring.\allowbreak{}mergedRepresentation\_\allowbreak{}to\_\allowbreak{}exactDistinct\_\allowbreak{}requires\_\allowbreak{}recode} {\tiny\ttfamily Paper1PL/\allowbreak SemanticFactoring.lean}
\item \textbf{\nolinkurl{SEF16}}\hypertarget{lh:SEF16}{}\enspace{\ttfamily Ssot.\allowbreak{}SemanticFactoring.\allowbreak{}expandSharedParentFactor\_\allowbreak{}eq\_\allowbreak{}commonParentRequirements} {\tiny\ttfamily Paper1PL/\allowbreak SemanticFactoring.lean}
\item \textbf{\nolinkurl{SEF17}}\hypertarget{lh:SEF17}{}\enspace{\ttfamily Ssot.\allowbreak{}SemanticFactoring.\allowbreak{}repeatedCommonParentEditCount} {\tiny\ttfamily Paper1PL/\allowbreak SemanticFactoring.lean}
\item \textbf{\nolinkurl{SEF18}}\hypertarget{lh:SEF18}{}\enspace{\ttfamily Ssot.\allowbreak{}SemanticFactoring.\allowbreak{}sharedParentFactor\_\allowbreak{}exactSavings} {\tiny\ttfamily Paper1PL/\allowbreak SemanticFactoring.lean}
\item \textbf{\nolinkurl{SEF19}}\hypertarget{lh:SEF19}{}\enspace{\ttfamily Ssot.\allowbreak{}SemanticFactoring.\allowbreak{}sharedParentFactor\_\allowbreak{}strict\_\allowbreak{}iff} {\tiny\ttfamily Paper1PL/\allowbreak SemanticFactoring.lean}
\item \textbf{\nolinkurl{SEF20}}\hypertarget{lh:SEF20}{}\enspace{\ttfamily Ssot.\allowbreak{}SemanticFactoring.\allowbreak{}oneCommonParent\_\allowbreak{}editCounts} {\tiny\ttfamily Paper1PL/\allowbreak SemanticFactoring.lean}
\item \textbf{\nolinkurl{SEF21}}\hypertarget{lh:SEF21}{}\enspace{\ttfamily Ssot.\allowbreak{}SemanticFactoring.\allowbreak{}oneCommonParent\_\allowbreak{}factoring\_\allowbreak{}strictly\_\allowbreak{}reduces\_\allowbreak{}edits} {\tiny\ttfamily Paper1PL/\allowbreak SemanticFactoring.lean}
\item \textbf{\nolinkurl{SEF22}}\hypertarget{lh:SEF22}{}\enspace{\ttfamily Ssot.\allowbreak{}SemanticFactoring.\allowbreak{}commonParentFactor\_\allowbreak{}exact\_\allowbreak{}edit\_\allowbreak{}law} {\tiny\ttfamily Paper1PL/\allowbreak SemanticFactoring.lean}
\item \textbf{\nolinkurl{SEF23}}\hypertarget{lh:SEF23}{}\enspace{\ttfamily Ssot.\allowbreak{}SemanticFactoring.\allowbreak{}sharedParentFactor\_\allowbreak{}finiteMigrationCrossover} {\tiny\ttfamily Paper1PL/\allowbreak SemanticFactoring.lean}
\item \textbf{\nolinkurl{SEF24}}\hypertarget{lh:SEF24}{}\enspace{\ttfamily Ssot.\allowbreak{}SemanticFactoring.\allowbreak{}sameCurrentParents\_\allowbreak{}do\_\allowbreak{}not\_\allowbreak{}decide\_\allowbreak{}sharedFuture} {\tiny\ttfamily Paper1PL/\allowbreak SemanticFactoring.lean}
\item \textbf{\nolinkurl{SEF25}}\hypertarget{lh:SEF25}{}\enspace{\ttfamily Ssot.\allowbreak{}SemanticFactoring.\allowbreak{}noExactDecoderThroughEqualStructuralProfile} {\tiny\ttfamily Paper1PL/\allowbreak SemanticFactoring.lean}
\item \textbf{\nolinkurl{SIN1}}\hypertarget{lh:SIN1}{}\enspace{\ttfamily Ssot.\allowbreak{}ImmediateParentEvolution.\allowbreak{}strict\_\allowbreak{}scope\_\allowbreak{}containment\_\allowbreak{}forces\_\allowbreak{}owner\_\allowbreak{}ancestor} {\tiny\ttfamily Paper1PL/\allowbreak SignedImmediateParentEvolution.lean}
\item \textbf{\nolinkurl{SIN2}}\hypertarget{lh:SIN2}{}\enspace{\ttfamily Ssot.\allowbreak{}ImmediateParentEvolution.\allowbreak{}strict\_\allowbreak{}insertion\_\allowbreak{}signedFree\_\allowbreak{}forces\_\allowbreak{}parent\_\allowbreak{}edge\_\allowbreak{}removal} {\tiny\ttfamily Paper1PL/\allowbreak SignedImmediateParentEvolution.lean}
\item \textbf{\nolinkurl{SIN3}}\hypertarget{lh:SIN3}{}\enspace{\ttfamily Ssot.\allowbreak{}ImmediateParentEvolution.\allowbreak{}strict\_\allowbreak{}insertion\_\allowbreak{}forces\_\allowbreak{}repair} {\tiny\ttfamily Paper1PL/\allowbreak SignedImmediateParentEvolution.lean}
\item \textbf{\nolinkurl{SIN4}}\hypertarget{lh:SIN4}{}\enspace{\ttfamily Ssot.\allowbreak{}ImmediateParentEvolution.\allowbreak{}strict\_\allowbreak{}insertion\_\allowbreak{}signedFree\_\allowbreak{}forces\_\allowbreak{}parent\_\allowbreak{}edge\_\allowbreak{}removal\_\allowbreak{}of\_\allowbreak{}nonempty} {\tiny\ttfamily Paper1PL/\allowbreak SignedImmediateParentEvolution.lean}
\item \textbf{\nolinkurl{SIN5}}\hypertarget{lh:SIN5}{}\enspace{\ttfamily Ssot.\allowbreak{}ImmediateParentEvolution.\allowbreak{}strict\_\allowbreak{}insertion\_\allowbreak{}forces\_\allowbreak{}repair\_\allowbreak{}of\_\allowbreak{}nonempty} {\tiny\ttfamily Paper1PL/\allowbreak SignedImmediateParentEvolution.lean}
\item \textbf{\nolinkurl{SIP1}}\hypertarget{lh:SIP1}{}\enspace{\ttfamily Ssot.\allowbreak{}ImmediateParentEvolution.\allowbreak{}ImmediateSignedSingleParentPolicy.\allowbreak{}positiveGap\_\allowbreak{}is\_\allowbreak{}forwarded} {\tiny\ttfamily Paper1PL/\allowbreak SignedImmediateParentEvolution.lean}
\item \textbf{\nolinkurl{SIP2}}\hypertarget{lh:SIP2}{}\enspace{\ttfamily Ssot.\allowbreak{}ImmediateParentEvolution.\allowbreak{}ImmediateSignedSingleParentPolicy.\allowbreak{}effectiveBlock\_\allowbreak{}is\_\allowbreak{}blocked} {\tiny\ttfamily Paper1PL/\allowbreak SignedImmediateParentEvolution.lean}
\item \textbf{\nolinkurl{SIP3}}\hypertarget{lh:SIP3}{}\enspace{\ttfamily Ssot.\allowbreak{}ImmediateParentEvolution.\allowbreak{}ImmediateSignedSingleParentPolicy.\allowbreak{}one\_\allowbreak{}le\_\allowbreak{}signedExternalCount} {\tiny\ttfamily Paper1PL/\allowbreak SignedImmediateParentEvolution.lean}
\item \textbf{\nolinkurl{SIP4}}\hypertarget{lh:SIP4}{}\enspace{\ttfamily Ssot.\allowbreak{}ImmediateParentEvolution.\allowbreak{}nested\_\allowbreak{}growth\_\allowbreak{}signedFree\_\allowbreak{}endpoints\_\allowbreak{}force\_\allowbreak{}parent\_\allowbreak{}edge\_\allowbreak{}removal} {\tiny\ttfamily Paper1PL/\allowbreak SignedImmediateParentEvolution.lean}
\item \textbf{\nolinkurl{SIP5}}\hypertarget{lh:SIP5}{}\enspace{\ttfamily Ssot.\allowbreak{}ImmediateParentEvolution.\allowbreak{}stagedScope\_\allowbreak{}step\_\allowbreak{}retraction\_\allowbreak{}or\_\allowbreak{}signed\_\allowbreak{}external} {\tiny\ttfamily Paper1PL/\allowbreak SignedImmediateParentEvolution.lean}
\item \textbf{\nolinkurl{SIP6}}\hypertarget{lh:SIP6}{}\enspace{\ttfamily Ssot.\allowbreak{}ImmediateParentEvolution.\allowbreak{}stagedFrozenWitnessToSignedExternal\_\allowbreak{}injective} {\tiny\ttfamily Paper1PL/\allowbreak SignedImmediateParentEvolution.lean}
\item \textbf{\nolinkurl{SIP7}}\hypertarget{lh:SIP7}{}\enspace{\ttfamily Ssot.\allowbreak{}ImmediateParentEvolution.\allowbreak{}stagedScope\_\allowbreak{}recurrent\_\allowbreak{}parent\_\allowbreak{}quadratic\_\allowbreak{}signed\_\allowbreak{}external\_\allowbreak{}lower\_\allowbreak{}bound} {\tiny\ttfamily Paper1PL/\allowbreak SignedImmediateParentEvolution.lean}
\item \textbf{\nolinkurl{SLF1}}\hypertarget{lh:SLF1}{}\enspace{\ttfamily Ssot.\allowbreak{}SelfCalls.\allowbreak{}Completion.\allowbreak{}requiredExternal\_\allowbreak{}card\_\allowbreak{}le\_\allowbreak{}materialized} {\tiny\ttfamily Paper1PL/\allowbreak SelfCalls.lean}
\item \textbf{\nolinkurl{SLF2}}\hypertarget{lh:SLF2}{}\enspace{\ttfamily Ssot.\allowbreak{}SelfCalls.\allowbreak{}canonicalCompletion\_\allowbreak{}materializedSites} {\tiny\ttfamily Paper1PL/\allowbreak SelfCalls.lean}
\item \textbf{\nolinkurl{SLF3}}\hypertarget{lh:SLF3}{}\enspace{\ttfamily Ssot.\allowbreak{}SelfCalls.\allowbreak{}subclass\_\allowbreak{}dominates\_\allowbreak{}wrapper\_\allowbreak{}iff} {\tiny\ttfamily Paper1PL/\allowbreak SelfCalls.lean}
\item \textbf{\nolinkurl{SLF4}}\hypertarget{lh:SLF4}{}\enspace{\ttfamily Ssot.\allowbreak{}SelfCalls.\allowbreak{}wrapper\_\allowbreak{}not\_\allowbreak{}dominates\_\allowbreak{}subclass} {\tiny\ttfamily Paper1PL/\allowbreak SelfCalls.lean}
\item \textbf{\nolinkurl{SLF5}}\hypertarget{lh:SLF5}{}\enspace{\ttfamily Ssot.\allowbreak{}SelfCalls.\allowbreak{}incomparable\_\allowbreak{}of\_\allowbreak{}base\_\allowbreak{}sites\_\allowbreak{}gt\_\allowbreak{}derived\_\allowbreak{}sites} {\tiny\ttfamily Paper1PL/\allowbreak SelfCalls.lean}
\item \textbf{\nolinkurl{SLF6}}\hypertarget{lh:SLF6}{}\enspace{\ttfamily Ssot.\allowbreak{}SelfCalls.\allowbreak{}derivedWinner\_\allowbreak{}subclass\_\allowbreak{}strict\_\allowbreak{}witness} {\tiny\ttfamily Paper1PL/\allowbreak SelfCalls.lean}
\item \textbf{\nolinkurl{SLF7}}\hypertarget{lh:SLF7}{}\enspace{\ttfamily Ssot.\allowbreak{}SelfCalls.\allowbreak{}baseWinner\_\allowbreak{}incomparable\_\allowbreak{}witness} {\tiny\ttfamily Paper1PL/\allowbreak SelfCalls.lean}
\item \textbf{\nolinkurl{SPN1}}\hypertarget{lh:SPN1}{}\enspace{\ttfamily INT4} {\tiny\ttfamily HandleAliases.lean}
\item \textbf{\nolinkurl{SPN2}}\hypertarget{lh:SPN2}{}\enspace{\ttfamily RCN1} {\tiny\ttfamily HandleAliases.lean}
\item \textbf{\nolinkurl{SPN3}}\hypertarget{lh:SPN3}{}\enspace{\ttfamily SCT5} {\tiny\ttfamily HandleAliases.lean}
\item \textbf{\nolinkurl{SPN4}}\hypertarget{lh:SPN4}{}\enspace{\ttfamily SCT14} {\tiny\ttfamily HandleAliases.lean}
\item \textbf{\nolinkurl{SPN5}}\hypertarget{lh:SPN5}{}\enspace{\ttfamily SCT32} {\tiny\ttfamily HandleAliases.lean}
\item \textbf{\nolinkurl{SPN6}}\hypertarget{lh:SPN6}{}\enspace{\ttfamily NHF63} {\tiny\ttfamily HandleAliases.lean}
\item \textbf{\nolinkurl{SPN7}}\hypertarget{lh:SPN7}{}\enspace{\ttfamily BLK8} {\tiny\ttfamily HandleAliases.lean}
\item \textbf{\nolinkurl{SPN8}}\hypertarget{lh:SPN8}{}\enspace{\ttfamily NHF68} {\tiny\ttfamily HandleAliases.lean}
\item \textbf{\nolinkurl{SPN9}}\hypertarget{lh:SPN9}{}\enspace{\ttfamily BWR48} {\tiny\ttfamily HandleAliases.lean}
\item \textbf{\nolinkurl{SPN10}}\hypertarget{lh:SPN10}{}\enspace{\ttfamily BWR49} {\tiny\ttfamily HandleAliases.lean}
\item \textbf{\nolinkurl{SPN11}}\hypertarget{lh:SPN11}{}\enspace{\ttfamily IPE28} {\tiny\ttfamily HandleAliases.lean}
\item \textbf{\nolinkurl{SPN12}}\hypertarget{lh:SPN12}{}\enspace{\ttfamily SIP7} {\tiny\ttfamily HandleAliases.lean}
\item \textbf{\nolinkurl{SPN13}}\hypertarget{lh:SPN13}{}\enspace{\ttfamily UND37} {\tiny\ttfamily HandleAliases.lean}
\item \textbf{\nolinkurl{SPN14}}\hypertarget{lh:SPN14}{}\enspace{\ttfamily NPS9} {\tiny\ttfamily HandleAliases.lean}
\item \textbf{\nolinkurl{SPN15}}\hypertarget{lh:SPN15}{}\enspace{\ttfamily PJD3} {\tiny\ttfamily HandleAliases.lean}
\item \textbf{\nolinkurl{SPN16}}\hypertarget{lh:SPN16}{}\enspace{\ttfamily OPT6} {\tiny\ttfamily HandleAliases.lean}
\item \textbf{\nolinkurl{SPN17}}\hypertarget{lh:SPN17}{}\enspace{\ttfamily SCT35} {\tiny\ttfamily HandleAliases.lean}
\item \textbf{\nolinkurl{SPN18}}\hypertarget{lh:SPN18}{}\enspace{\ttfamily BWR38} {\tiny\ttfamily HandleAliases.lean}
\item \textbf{\nolinkurl{SPN19}}\hypertarget{lh:SPN19}{}\enspace{\ttfamily BWR10} {\tiny\ttfamily HandleAliases.lean}
\item \textbf{\nolinkurl{SPN20}}\hypertarget{lh:SPN20}{}\enspace{\ttfamily NHF45} {\tiny\ttfamily HandleAliases.lean}
\item \textbf{\nolinkurl{SRA1}}\hypertarget{lh:SRA1}{}\enspace{\ttfamily Ssot.\allowbreak{}LaminarRepairFPT.\allowbreak{}laminarAfter\_\allowbreak{}restrictFacts} {\tiny\ttfamily Paper1PL/\allowbreak LaminarRepairFPT.lean}
\item \textbf{\nolinkurl{SRA2}}\hypertarget{lh:SRA2}{}\enspace{\ttfamily Ssot.\allowbreak{}LaminarRepairFPT.\allowbreak{}restrictedMinimum\_\allowbreak{}le\_\allowbreak{}filteredRepair} {\tiny\ttfamily Paper1PL/\allowbreak LaminarRepairFPT.lean}
\item \textbf{\nolinkurl{TOL1}}\hypertarget{lh:TOL1}{}\enspace{\ttfamily Ssot.\allowbreak{}DeclaredTolerance.\allowbreak{}gap\_\allowbreak{}attained} {\tiny\ttfamily Paper1PL/\allowbreak DeclaredTolerance.lean}
\item \textbf{\nolinkurl{TOL2}}\hypertarget{lh:TOL2}{}\enspace{\ttfamily Ssot.\allowbreak{}DeclaredTolerance.\allowbreak{}gap\_\allowbreak{}le\_\allowbreak{}candidate} {\tiny\ttfamily Paper1PL/\allowbreak DeclaredTolerance.lean}
\item \textbf{\nolinkurl{TOL3}}\hypertarget{lh:TOL3}{}\enspace{\ttfamily Ssot.\allowbreak{}DeclaredTolerance.\allowbreak{}omitted\_\allowbreak{}card\_\allowbreak{}le\_\allowbreak{}charged\_\allowbreak{}add\_\allowbreak{}tolerated} {\tiny\ttfamily Paper1PL/\allowbreak DeclaredTolerance.lean}
\item \textbf{\nolinkurl{TOL4}}\hypertarget{lh:TOL4}{}\enspace{\ttfamily Ssot.\allowbreak{}DeclaredTolerance.\allowbreak{}omission\_\allowbreak{}tolerance\_\allowbreak{}bounds} {\tiny\ttfamily Paper1PL/\allowbreak DeclaredTolerance.lean}
\item \textbf{\nolinkurl{TOL5}}\hypertarget{lh:TOL5}{}\enspace{\ttfamily Ssot.\allowbreak{}DeclaredTolerance.\allowbreak{}declared\_\allowbreak{}overreach\_\allowbreak{}within\_\allowbreak{}budget} {\tiny\ttfamily Paper1PL/\allowbreak DeclaredTolerance.lean}
\item \textbf{\nolinkurl{TOL6}}\hypertarget{lh:TOL6}{}\enspace{\ttfamily Ssot.\allowbreak{}DeclaredTolerance.\allowbreak{}budget\_\allowbreak{}lower\_\allowbreak{}bound\_\allowbreak{}applies\_\allowbreak{}to\_\allowbreak{}every\_\allowbreak{}declaration} {\tiny\ttfamily Paper1PL/\allowbreak DeclaredTolerance.lean}
\item \textbf{\nolinkurl{TOL7}}\hypertarget{lh:TOL7}{}\enspace{\ttfamily Ssot.\allowbreak{}DeclaredTolerance.\allowbreak{}saving\_\allowbreak{}requires\_\allowbreak{}tolerated\_\allowbreak{}pairs} {\tiny\ttfamily Paper1PL/\allowbreak DeclaredTolerance.lean}
\item \textbf{\nolinkurl{TST1}}\hypertarget{lh:TST1}{}\enspace{\ttfamily Ssot.\allowbreak{}FiniteTesting.\allowbreak{}catchesEveryViolation\_\allowbreak{}iff\_\allowbreak{}missedViolations\_\allowbreak{}empty} {\tiny\ttfamily Paper1PL/\allowbreak FiniteTesting.lean}
\item \textbf{\nolinkurl{TST2}}\hypertarget{lh:TST2}{}\enspace{\ttfamily Ssot.\allowbreak{}FiniteTesting.\allowbreak{}testDeficit\_\allowbreak{}eq\_\allowbreak{}zero\_\allowbreak{}iff\_\allowbreak{}subset} {\tiny\ttfamily Paper1PL/\allowbreak FiniteTesting.lean}
\item \textbf{\nolinkurl{TST3}}\hypertarget{lh:TST3}{}\enspace{\ttfamily Ssot.\allowbreak{}FiniteTesting.\allowbreak{}zeroError\_\allowbreak{}iff\_\allowbreak{}passes\_\allowbreak{}and\_\allowbreak{}catches} {\tiny\ttfamily Paper1PL/\allowbreak FiniteTesting.lean}
\item \textbf{\nolinkurl{TST4}}\hypertarget{lh:TST4}{}\enspace{\ttfamily Ssot.\allowbreak{}FiniteTesting.\allowbreak{}finite\_\allowbreak{}suite\_\allowbreak{}omits} {\tiny\ttfamily Paper1PL/\allowbreak FiniteTesting.lean}
\item \textbf{\nolinkurl{TST5}}\hypertarget{lh:TST5}{}\enspace{\ttfamily Ssot.\allowbreak{}FiniteTesting.\allowbreak{}finite\_\allowbreak{}suite\_\allowbreak{}has\_\allowbreak{}positive\_\allowbreak{}testDeficit} {\tiny\ttfamily Paper1PL/\allowbreak FiniteTesting.lean}
\item \textbf{\nolinkurl{TST6}}\hypertarget{lh:TST6}{}\enspace{\ttfamily Ssot.\allowbreak{}FiniteTesting.\allowbreak{}finite\_\allowbreak{}suite\_\allowbreak{}admits\_\allowbreak{}undetected\_\allowbreak{}violation} {\tiny\ttfamily Paper1PL/\allowbreak FiniteTesting.lean}
\item \textbf{\nolinkurl{TST7}}\hypertarget{lh:TST7}{}\enspace{\ttfamily Ssot.\allowbreak{}FiniteTesting.\allowbreak{}shiftedRequirementClockTrace\_\allowbreak{}all\_\allowbreak{}laminar\_\allowbreak{}iff\_\allowbreak{}nonhalting} {\tiny\ttfamily Paper1PL/\allowbreak FiniteTesting.lean}
\item \textbf{\nolinkurl{TST8}}\hypertarget{lh:TST8}{}\enspace{\ttfamily Ssot.\allowbreak{}FiniteTesting.\allowbreak{}singletonRequirementSuite\_\allowbreak{}complete\_\allowbreak{}iff\_\allowbreak{}nonhalting} {\tiny\ttfamily Paper1PL/\allowbreak FiniteTesting.lean}
\item \textbf{\nolinkurl{TST9}}\hypertarget{lh:TST9}{}\enspace{\ttfamily Ssot.\allowbreak{}FiniteTesting.\allowbreak{}singletonRequirementSuite\_\allowbreak{}completeness\_\allowbreak{}not\_\allowbreak{}re} {\tiny\ttfamily Paper1PL/\allowbreak FiniteTesting.lean}
\item \textbf{\nolinkurl{TST10}}\hypertarget{lh:TST10}{}\enspace{\ttfamily Ssot.\allowbreak{}FiniteTesting.\allowbreak{}singletonRequirementSuite\_\allowbreak{}incompleteness\_\allowbreak{}re} {\tiny\ttfamily Paper1PL/\allowbreak FiniteTesting.lean}
\item \textbf{\nolinkurl{UMS1}}\hypertarget{lh:UMS1}{}\enspace{\ttfamily Ssot.\allowbreak{}SupplyMaintenanceCharacterization.\allowbreak{}one\_\allowbreak{}crossing\_\allowbreak{}impossible} {\tiny\ttfamily Paper1PL/\allowbreak SupplyMaintenanceCharacterization.lean}
\item \textbf{\nolinkurl{UMS2}}\hypertarget{lh:UMS2}{}\enspace{\ttfamily Ssot.\allowbreak{}SupplyMaintenanceCharacterization.\allowbreak{}no\_\allowbreak{}parent\_\allowbreak{}nonempty\_\allowbreak{}impossible} {\tiny\ttfamily Paper1PL/\allowbreak SupplyMaintenanceCharacterization.lean}
\item \textbf{\nolinkurl{UMS3}}\hypertarget{lh:UMS3}{}\enspace{\ttfamily Ssot.\allowbreak{}SupplyMaintenanceCharacterization.\allowbreak{}universal\_\allowbreak{}attainment\_\allowbreak{}iff\_\allowbreak{}multiple} {\tiny\ttfamily Paper1PL/\allowbreak SupplyMaintenanceCharacterization.lean}
\item \textbf{\nolinkurl{UMS4}}\hypertarget{lh:UMS4}{}\enspace{\ttfamily Ssot.\allowbreak{}SupplyMaintenanceCharacterization.\allowbreak{}both\_\allowbreak{}supply\_\allowbreak{}rungs\_\allowbreak{}iff} {\tiny\ttfamily Paper1PL/\allowbreak SupplyMaintenanceCharacterization.lean}
\item \textbf{\nolinkurl{UMS5}}\hypertarget{lh:UMS5}{}\enspace{\ttfamily Ssot.\allowbreak{}SupplyMaintenanceCharacterization.\allowbreak{}directHistory} {\tiny\ttfamily Paper1PL/\allowbreak SupplyMaintenanceCharacterization.lean}
\item \textbf{\nolinkurl{UMS6}}\hypertarget{lh:UMS6}{}\enspace{\ttfamily Ssot.\allowbreak{}SupplyMaintenanceCharacterization.\allowbreak{}separate\_\allowbreak{}supply\_\allowbreak{}histories\_\allowbreak{}jointly\_\allowbreak{}zero} {\tiny\ttfamily Paper1PL/\allowbreak SupplyMaintenanceCharacterization.lean}
\item \textbf{\nolinkurl{UND5}}\hypertarget{lh:UND5}{}\enspace{\ttfamily AbstractClassSystem.\allowbreak{}OpenWorld.\allowbreak{}finiteBarrierWitness\_\allowbreak{}primrec} {\tiny\ttfamily AbstractClassSystem/\allowbreak BarrierComplexity.lean}
\item \textbf{\nolinkurl{UND6}}\hypertarget{lh:UND6}{}\enspace{\ttfamily AbstractClassSystem.\allowbreak{}OpenWorld.\allowbreak{}hasBarrier\_\allowbreak{}re} {\tiny\ttfamily AbstractClassSystem/\allowbreak BarrierComplexity.lean}
\item \textbf{\nolinkurl{UND7}}\hypertarget{lh:UND7}{}\enspace{\ttfamily AbstractClassSystem.\allowbreak{}OpenWorld.\allowbreak{}halting\_\allowbreak{}manyOneReducible\_\allowbreak{}hasBarrier} {\tiny\ttfamily AbstractClassSystem/\allowbreak BarrierComplexity.lean}
\item \textbf{\nolinkurl{UND8}}\hypertarget{lh:UND8}{}\enspace{\ttfamily AbstractClassSystem.\allowbreak{}OpenWorld.\allowbreak{}nonhalting\_\allowbreak{}manyOneReducible\_\allowbreak{}barrierFree} {\tiny\ttfamily AbstractClassSystem/\allowbreak BarrierComplexity.lean}
\item \textbf{\nolinkurl{UND9}}\hypertarget{lh:UND9}{}\enspace{\ttfamily AbstractClassSystem.\allowbreak{}OpenWorld.\allowbreak{}barrierFree\_\allowbreak{}not\_\allowbreak{}re} {\tiny\ttfamily AbstractClassSystem/\allowbreak BarrierComplexity.lean}
\item \textbf{\nolinkurl{UND10}}\hypertarget{lh:UND10}{}\enspace{\ttfamily AbstractClassSystem.\allowbreak{}OpenWorld.\allowbreak{}barrier\_\allowbreak{}effective\_\allowbreak{}classification} {\tiny\ttfamily AbstractClassSystem/\allowbreak BarrierComplexity.lean}
\item \textbf{\nolinkurl{UND11}}\hypertarget{lh:UND11}{}\enspace{\ttfamily AbstractClassSystem.\allowbreak{}OpenWorld.\allowbreak{}HasMultiplicityBarrier} {\tiny\ttfamily AbstractClassSystem/\allowbreak BarrierComplexity.lean}
\item \textbf{\nolinkurl{UND12}}\hypertarget{lh:UND12}{}\enspace{\ttfamily AbstractClassSystem.\allowbreak{}OpenWorld.\allowbreak{}hasMultiplicityBarrier\_\allowbreak{}one\_\allowbreak{}iff} {\tiny\ttfamily AbstractClassSystem/\allowbreak BarrierComplexity.lean}
\item \textbf{\nolinkurl{UND13}}\hypertarget{lh:UND13}{}\enspace{\ttfamily AbstractClassSystem.\allowbreak{}OpenWorld.\allowbreak{}totalBarrierClock\_\allowbreak{}hasMultiplicity\_\allowbreak{}of\_\allowbreak{}halting} {\tiny\ttfamily AbstractClassSystem/\allowbreak BarrierComplexity.lean}
\item \textbf{\nolinkurl{UND14}}\hypertarget{lh:UND14}{}\enspace{\ttfamily AbstractClassSystem.\allowbreak{}OpenWorld.\allowbreak{}haltToTotalBarrier\_\allowbreak{}is\_\allowbreak{}total} {\tiny\ttfamily AbstractClassSystem/\allowbreak BarrierComplexity.lean}
\item \textbf{\nolinkurl{UND15}}\hypertarget{lh:UND15}{}\enspace{\ttfamily AbstractClassSystem.\allowbreak{}OpenWorld.\allowbreak{}halting\_\allowbreak{}iff\_\allowbreak{}transformed\_\allowbreak{}total\_\allowbreak{}hasBarrier} {\tiny\ttfamily AbstractClassSystem/\allowbreak BarrierComplexity.lean}
\item \textbf{\nolinkurl{UND16}}\hypertarget{lh:UND16}{}\enspace{\ttfamily AbstractClassSystem.\allowbreak{}OpenWorld.\allowbreak{}hasBarrier\_\allowbreak{}of\_\allowbreak{}hasMultiplicity} {\tiny\ttfamily AbstractClassSystem/\allowbreak BarrierComplexity.lean}
\item \textbf{\nolinkurl{UND17}}\hypertarget{lh:UND17}{}\enspace{\ttfamily AbstractClassSystem.\allowbreak{}OpenWorld.\allowbreak{}halting\_\allowbreak{}iff\_\allowbreak{}transformed\_\allowbreak{}total\_\allowbreak{}hasMultiplicity} {\tiny\ttfamily AbstractClassSystem/\allowbreak BarrierComplexity.lean}
\item \textbf{\nolinkurl{UND18}}\hypertarget{lh:UND18}{}\enspace{\ttfamily AbstractClassSystem.\allowbreak{}OpenWorld.\allowbreak{}halting\_\allowbreak{}manyOneReducible\_\allowbreak{}total\_\allowbreak{}hasMultiplicity} {\tiny\ttfamily AbstractClassSystem/\allowbreak BarrierComplexity.lean}
\item \textbf{\nolinkurl{UND19}}\hypertarget{lh:UND19}{}\enspace{\ttfamily AbstractClassSystem.\allowbreak{}OpenWorld.\allowbreak{}MultiplicityBound} {\tiny\ttfamily AbstractClassSystem/\allowbreak BarrierComplexity.lean}
\item \textbf{\nolinkurl{UND20}}\hypertarget{lh:UND20}{}\enspace{\ttfamily AbstractClassSystem.\allowbreak{}OpenWorld.\allowbreak{}nonhalting\_\allowbreak{}manyOneReducible\_\allowbreak{}total\_\allowbreak{}multiplicityBound} {\tiny\ttfamily AbstractClassSystem/\allowbreak BarrierComplexity.lean}
\item \textbf{\nolinkurl{UND21}}\hypertarget{lh:UND21}{}\enspace{\ttfamily AbstractClassSystem.\allowbreak{}OpenWorld.\allowbreak{}multiplicityBound\_\allowbreak{}not\_\allowbreak{}re} {\tiny\ttfamily AbstractClassSystem/\allowbreak BarrierComplexity.lean}
\item \textbf{\nolinkurl{UND22}}\hypertarget{lh:UND22}{}\enspace{\ttfamily AbstractClassSystem.\allowbreak{}OpenWorld.\allowbreak{}totalObserverMultiplicityBound\_\allowbreak{}not\_\allowbreak{}re} {\tiny\ttfamily AbstractClassSystem/\allowbreak BarrierComplexity.lean}
\item \textbf{\nolinkurl{UND23}}\hypertarget{lh:UND23}{}\enspace{\ttfamily AbstractClassSystem.\allowbreak{}OpenWorld.\allowbreak{}finiteClockTrace\_\allowbreak{}eq\_\allowbreak{}id\_\allowbreak{}of\_\allowbreak{}nonhalting} {\tiny\ttfamily AbstractClassSystem/\allowbreak FiniteTraceUndecidability.lean}
\item \textbf{\nolinkurl{UND24}}\hypertarget{lh:UND24}{}\enspace{\ttfamily AbstractClassSystem.\allowbreak{}OpenWorld.\allowbreak{}finiteClockTrace\_\allowbreak{}constant\_\allowbreak{}of\_\allowbreak{}evaln} {\tiny\ttfamily AbstractClassSystem/\allowbreak FiniteTraceUndecidability.lean}
\item \textbf{\nolinkurl{UND25}}\hypertarget{lh:UND25}{}\enspace{\ttfamily AbstractClassSystem.\allowbreak{}OpenWorld.\allowbreak{}halting\_\allowbreak{}iff\_\allowbreak{}finiteClockTrace\_\allowbreak{}hasMultiplicity} {\tiny\ttfamily AbstractClassSystem/\allowbreak FiniteTraceUndecidability.lean}
\item \textbf{\nolinkurl{UND26}}\hypertarget{lh:UND26}{}\enspace{\ttfamily AbstractClassSystem.\allowbreak{}OpenWorld.\allowbreak{}finiteClockTrace\_\allowbreak{}safe\_\allowbreak{}iff\_\allowbreak{}nonhalting} {\tiny\ttfamily AbstractClassSystem/\allowbreak FiniteTraceUndecidability.lean}
\item \textbf{\nolinkurl{UND27}}\hypertarget{lh:UND27}{}\enspace{\ttfamily AbstractClassSystem.\allowbreak{}OpenWorld.\allowbreak{}finiteClockTrace\_\allowbreak{}safety\_\allowbreak{}not\_\allowbreak{}re} {\tiny\ttfamily AbstractClassSystem/\allowbreak FiniteTraceUndecidability.lean}
\item \textbf{\nolinkurl{UND28}}\hypertarget{lh:UND28}{}\enspace{\ttfamily AbstractClassSystem.\allowbreak{}OpenWorld.\allowbreak{}finiteTableMultiplicityBound\_\allowbreak{}decidable} {\tiny\ttfamily AbstractClassSystem/\allowbreak FiniteTraceUndecidability.lean}
\item \textbf{\nolinkurl{UND29}}\hypertarget{lh:UND29}{}\enspace{\ttfamily AbstractClassSystem.\allowbreak{}OpenWorld.\allowbreak{}finiteClockTrace\_\allowbreak{}eq\_\allowbreak{}id\_\allowbreak{}or\_\allowbreak{}constant} {\tiny\ttfamily AbstractClassSystem/\allowbreak FiniteTraceUndecidability.lean}
\item \textbf{\nolinkurl{UND30}}\hypertarget{lh:UND30}{}\enspace{\ttfamily AbstractClassSystem.\allowbreak{}OpenWorld.\allowbreak{}finiteBinaryTrace\_\allowbreak{}safety\_\allowbreak{}not\_\allowbreak{}re} {\tiny\ttfamily AbstractClassSystem/\allowbreak FiniteTraceUndecidability.lean}
\item \textbf{\nolinkurl{UND31}}\hypertarget{lh:UND31}{}\enspace{\ttfamily Ssot.\allowbreak{}OpenWorldLaminarity.\allowbreak{}laminarRequirementTable\_\allowbreak{}laminar} {\tiny\ttfamily Paper1PL/\allowbreak OpenWorldLaminarity.lean}
\item \textbf{\nolinkurl{UND32}}\hypertarget{lh:UND32}{}\enspace{\ttfamily Ssot.\allowbreak{}OpenWorldLaminarity.\allowbreak{}crossingRequirementTable\_\allowbreak{}not\_\allowbreak{}laminar} {\tiny\ttfamily Paper1PL/\allowbreak OpenWorldLaminarity.lean}
\item \textbf{\nolinkurl{UND33}}\hypertarget{lh:UND33}{}\enspace{\ttfamily Ssot.\allowbreak{}OpenWorldLaminarity.\allowbreak{}finiteRequirementTable\_\allowbreak{}laminar\_\allowbreak{}decidable} {\tiny\ttfamily Paper1PL/\allowbreak OpenWorldLaminarity.lean}
\item \textbf{\nolinkurl{UND34}}\hypertarget{lh:UND34}{}\enspace{\ttfamily Ssot.\allowbreak{}OpenWorldLaminarity.\allowbreak{}finiteRequirementClockTrace\_\allowbreak{}eq\_\allowbreak{}laminar\_\allowbreak{}or\_\allowbreak{}crossing} {\tiny\ttfamily Paper1PL/\allowbreak OpenWorldLaminarity.lean}
\item \textbf{\nolinkurl{UND35}}\hypertarget{lh:UND35}{}\enspace{\ttfamily Ssot.\allowbreak{}OpenWorldLaminarity.\allowbreak{}halting\_\allowbreak{}iff\_\allowbreak{}finiteRequirementClockTrace\_\allowbreak{}hasNonlaminar} {\tiny\ttfamily Paper1PL/\allowbreak OpenWorldLaminarity.lean}
\item \textbf{\nolinkurl{UND36}}\hypertarget{lh:UND36}{}\enspace{\ttfamily Ssot.\allowbreak{}OpenWorldLaminarity.\allowbreak{}finiteRequirementClockTrace\_\allowbreak{}all\_\allowbreak{}laminar\_\allowbreak{}iff\_\allowbreak{}nonhalting} {\tiny\ttfamily Paper1PL/\allowbreak OpenWorldLaminarity.lean}
\item \textbf{\nolinkurl{UND37}}\hypertarget{lh:UND37}{}\enspace{\ttfamily Ssot.\allowbreak{}OpenWorldLaminarity.\allowbreak{}finiteRequirementClockTrace\_\allowbreak{}all\_\allowbreak{}laminar\_\allowbreak{}not\_\allowbreak{}re} {\tiny\ttfamily Paper1PL/\allowbreak OpenWorldLaminarity.lean}
\item \textbf{\nolinkurl{UND38}}\hypertarget{lh:UND38}{}\enspace{\ttfamily Ssot.\allowbreak{}OpenWorldLaminarity.\allowbreak{}finiteRequirementClockTrace\_\allowbreak{}snapshot\_\allowbreak{}has\_\allowbreak{}multiParent} {\tiny\ttfamily Paper1PL/\allowbreak OpenWorldLaminarity.lean}
\item \textbf{\nolinkurl{UND39}}\hypertarget{lh:UND39}{}\enspace{\ttfamily Ssot.\allowbreak{}OpenWorldLaminarity.\allowbreak{}sound\_\allowbreak{}restriction\_\allowbreak{}omits\_\allowbreak{}safe\_\allowbreak{}generator} {\tiny\ttfamily Paper1PL/\allowbreak OpenWorldLaminarity.lean}
\item \textbf{\nolinkurl{UND40}}\hypertarget{lh:UND40}{}\enspace{\ttfamily AbstractClassSystem.\allowbreak{}OpenWorld.\allowbreak{}finiteClockTrace\_\allowbreak{}hasMultiplicityBarrier\_\allowbreak{}re} {\tiny\ttfamily AbstractClassSystem/\allowbreak FiniteTraceUndecidability.lean}
\item \textbf{\nolinkurl{UND41}}\hypertarget{lh:UND41}{}\enspace{\ttfamily Ssot.\allowbreak{}OpenWorldLaminarity.\allowbreak{}finiteRequirementClockTrace\_\allowbreak{}hasNonlaminar\_\allowbreak{}re} {\tiny\ttfamily Paper1PL/\allowbreak OpenWorldLaminarity.lean}
\item \textbf{\nolinkurl{WAL1}}\hypertarget{lh:WAL1}{}\enspace{\ttfamily Ssot.\allowbreak{}AssignmentExpressions.\allowbreak{}assignmentFree\_\allowbreak{}preserves\_\allowbreak{}outer} {\tiny\ttfamily Paper1PL/\allowbreak AssignmentExpressions.lean}
\item \textbf{\nolinkurl{WAL2}}\hypertarget{lh:WAL2}{}\enspace{\ttfamily Ssot.\allowbreak{}AssignmentExpressions.\allowbreak{}no\_\allowbreak{}termination\_\allowbreak{}preserving\_\allowbreak{}local\_\allowbreak{}replacement} {\tiny\ttfamily Paper1PL/\allowbreak AssignmentExpressions.lean}
\item \textbf{\nolinkurl{WAL3}}\hypertarget{lh:WAL3}{}\enspace{\ttfamily Ssot.\allowbreak{}AssignmentExpressions.\allowbreak{}lambda\_\allowbreak{}binding\_\allowbreak{}reuses\_\allowbreak{}argument} {\tiny\ttfamily Paper1PL/\allowbreak AssignmentExpressions.lean}
\item \textbf{\nolinkurl{WAL4}}\hypertarget{lh:WAL4}{}\enspace{\ttfamily Ssot.\allowbreak{}AssignmentExpressions.\allowbreak{}comprehension\_\allowbreak{}binding\_\allowbreak{}reuses\_\allowbreak{}argument} {\tiny\ttfamily Paper1PL/\allowbreak AssignmentExpressions.lean}
\item \textbf{\nolinkurl{WAL5}}\hypertarget{lh:WAL5}{}\enspace{\ttfamily Ssot.\allowbreak{}AssignmentExpressions.\allowbreak{}minimum\_\allowbreak{}monotone} {\tiny\ttfamily Paper1PL/\allowbreak AssignmentExpressions.lean}
\item \textbf{\nolinkurl{WAL6}}\hypertarget{lh:WAL6}{}\enspace{\ttfamily Ssot.\allowbreak{}AssignmentExpressions.\allowbreak{}no\_\allowbreak{}local\_\allowbreak{}replacement} {\tiny\ttfamily Paper1PL/\allowbreak AssignmentExpressions.lean}
\item \textbf{\nolinkurl{WAL7}}\hypertarget{lh:WAL7}{}\enspace{\ttfamily Ssot.\allowbreak{}AssignmentExpressions.\allowbreak{}escaping\_\allowbreak{}binding\_\allowbreak{}observed} {\tiny\ttfamily Paper1PL/\allowbreak AssignmentExpressions.lean}
\end{list}
\else
\begin{longtable}{@{}>{\raggedright\arraybackslash}m{0.10\linewidth}>{\raggedright\arraybackslash}m{0.86\linewidth}@{}}
\caption{Lean handle index.}\label{tab:lean-handle-index}\\
\toprule
\textbf{ID} & \textbf{Lean Handle / Source} \\
\midrule
\endfirsthead
\toprule
\textbf{ID} & \textbf{Lean Handle / Source} \\
\midrule
\endhead
\endfoot
\bottomrule
\endlastfoot
\hypertarget{lh:ACF1}{\textbf{\nolinkurl{ACF1}}} & {\fontsize{8}{9}\selectfont\ttfamily Ssot.\allowbreak{}AccountingCharacterization.\allowbreak{}positive\_\allowbreak{}sets\_\allowbreak{}preserve\_\allowbreak{}crossings}\par\vspace{0.1ex}{\fontsize{8}{9}\selectfont\ttfamily Paper1PL/\allowbreak AccountingCharacterization.lean} \\
\midrule
\hypertarget{lh:ACF2}{\textbf{\nolinkurl{ACF2}}} & {\fontsize{8}{9}\selectfont\ttfamily Ssot.\allowbreak{}AccountingCharacterization.\allowbreak{}finite\_\allowbreak{}declaration\_\allowbreak{}positive\_\allowbreak{}floor}\par\vspace{0.1ex}{\fontsize{8}{9}\selectfont\ttfamily Paper1PL/\allowbreak AccountingCharacterization.lean} \\
\midrule
\hypertarget{lh:ACF3}{\textbf{\nolinkurl{ACF3}}} & {\fontsize{8}{9}\selectfont\ttfamily Ssot.\allowbreak{}AccountingCharacterization.\allowbreak{}every\_\allowbreak{}nonempty\_\allowbreak{}set\_\allowbreak{}is\_\allowbreak{}crossing\_\allowbreak{}residual}\par\vspace{0.1ex}{\fontsize{8}{9}\selectfont\ttfamily Paper1PL/\allowbreak AccountingCharacterization.lean} \\
\midrule
\hypertarget{lh:ACF4}{\textbf{\nolinkurl{ACF4}}} & {\fontsize{8}{9}\selectfont\ttfamily Ssot.\allowbreak{}AccountingCharacterization.\allowbreak{}preserves\_\allowbreak{}all\_\allowbreak{}crossings\_\allowbreak{}iff\_\allowbreak{}positive}\par\vspace{0.1ex}{\fontsize{8}{9}\selectfont\ttfamily Paper1PL/\allowbreak AccountingCharacterization.lean} \\
\midrule
\hypertarget{lh:AGP1}{\textbf{\nolinkurl{AGP1}}} & {\fontsize{8}{9}\selectfont\ttfamily Ssot.\allowbreak{}AncestryGapProbability.\allowbreak{}zeroGap\_\allowbreak{}iff\_\allowbreak{}regionMinimum\_\allowbreak{}zero}\par\vspace{0.1ex}{\fontsize{8}{9}\selectfont\ttfamily Paper1PL/\allowbreak AncestryGapProbability.lean} \\
\end{longtable}
\fi
\makeatother
\endgroup

}{%
  \textbf{Error:} \texttt{content/lean\_handle\_ids\_auto.tex} not found.
}

\section{Claim-to-Lean Handle Mapping}
\label{claim-handle-mapping}

This section maps each paper claim to its corresponding Lean formalization.
The generated table is checked against the final manuscript labels and cited
handle set.  PST1--PST3 formalize the two-state reduction to bipartite edge
coloring.  Equality with maximum degree is the cited K\H{o}nig theorem and is
not imported as a Lean axiom.
The proof index contains two alias labels used by other cross-references.  The
table covers the labeled formal claims, including unnumbered prose claims.

\IfFileExists{content/claim_mapping_auto.tex}{%
\noindent Prose rows quote the local annotated passage; reference numbers come from the compiled source documents. Handles support the tagged mathematical claim, not surrounding external observations or recorded executions.\par
\begingroup
\footnotesize
\setlength{\tabcolsep}{3pt}
\renewcommand{\arraystretch}{1.08}
\ifdefined\PaperSupplementTableSetup\PaperSupplementTableSetup\fi
\setlength{\LTpre}{0pt}
\setlength{\LTpost}{0pt}
\ifdefined\PaperGenericLayout
\fontsize{10}{12}\selectfont
\setlength{\LTpre}{\baselineskip}
\setlength{\LTpost}{\baselineskip}
\fi
\begin{longtable}{@{}>{\raggedright\arraybackslash}m{0.68\linewidth}>{\raggedleft\arraybackslash}m{0.27\linewidth}@{}}
\caption{Claim-to-Lean handle mapping.}\label{tab:claim-lean-handle-mapping}\\
\toprule
\textbf{Manuscript claim} & \textbf{Lean handle} \\
\midrule
\endfirsthead
\toprule
\textbf{Manuscript claim} & \textbf{Lean handle} \\
\midrule
\endhead
\endfoot
\bottomrule
\endlastfoot
Proposition 3.1: Same structure can hide different meanings & \LHrng{EVO}{23}{30}, \LHrng{IDD}{1}{5}, \LHrng{IMR}{1}{5}, \LHrng{NHF}{3}{5}, \LH{NHF11}, \LH{OWM1}, \LH{PD1} \\
\midrule
Theorem 3.2: Fewest stable key values & \LHrng{EVO}{17}{18} \\
\midrule
Proposition 3.3: Structural contracts admit unwanted members & \LHrng{SCM}{1}{4} \\
\midrule
Proposition 4.2: One parent suffices exactly for tree-shaped requirements & \LH{NHF7}, \LH{NHF20}, \LHrng{NHF}{44}{45}, \LHrng{NHF}{50}{51} \\
\midrule
Missing and unwanted supply & \LH{BWR29}, \LHrng{BWR}{37}{39}, \LH{BWR52}, \LH{BWR57}, \LHrng{SCT}{46}{51} \\
\midrule
Theorem 4.3: The exact price of inheritance incompleteness & \LH{BWR42}, \LHrng{SCT}{1}{4}, \LHrng{SCT}{20}{26}, \LHrng{SCT}{30}{34}, \LHrng{SCT}{52}{55} \\
\midrule
Theorem 4.4: Contract membership obeys the same gap law & \LHrng{CAP}{2}{7}, \LHrng{CPL}{1}{4} \\
\midrule
Theorem 4.5: Exact gaps for two implementations and a row--column grid & \LHrng{AVG}{1}{7} \\
\midrule
Theorem 4.6: The ancestry gap is the minimum outside supply & \LH{SCT24}, \LHrng{SCT}{27}{34} \\
\midrule
Theorem 4.7: Copying cannot lower the ancestry gap & \LHrng{BWR}{58}{62}, \LHrng{SCT}{56}{59}, \LHrng{SDN}{17}{21} \\
\midrule
Theorem 4.8: Merge, share, or separate & \LHrng{SEF}{1}{25} \\
\midrule
Proposition 4.9: Generator inputs distinguish outside answers & \LHrng{BWR}{45}{49}, \LH{RBF3}, \LHrng{RLC}{7}{10}, \LHrng{SCT}{3}{7}, \LHrng{SCT}{20}{26} \\
\midrule
Corollary 4.10: Passing runs and ancestry hide requirement shape; supply routes reveal it & \LH{BWR42}, \LHrng{OLC}{1}{5}, \LH{OLC9}, \LH{SCT24}, \LH{SCT34}, \LH{SCT59} \\
\midrule
Theorem 4.11: Even relaxed supply keeps crossings costly & \LHrng{BLK}{1}{10} \\
\midrule
Corollary 4.12: Complete supply characterizes correct maintenance & \LHrng{BWR}{9}{11}, \LHrng{MGH}{9}{11}, \LHrng{NHF}{13}{14}, \LHrng{UMS}{1}{6} \\
\midrule
Theorem 4.13: Reachable omissions violate required supply & \LHrng{BWR}{19}{21}, \LH{BWR24}, \LHrng{BWR}{29}{30}, \LHrng{BWR}{37}{38}, \LHrng{BWR}{45}{49}, \LH{SCT23}, \LHrng{SCT}{30}{34} \\
\midrule
Theorem 4.14: Strict insertion forces one-parent repair & \LHrng{BWR}{50}{56}, \LHrng{DRP}{3}{4}, \LHrng{IPE}{1}{28}, \LHrng{IPE}{33}{38}, \LHrng{NRH}{1}{6}, \LHrng{NRH}{8}{9}, \LHrng{OPT}{3}{4}, \LHrng{SIN}{1}{5}, \LHrng{SIP}{1}{7} \\
\midrule
Corollary 4.15: An unchanged hierarchy leaves new requirements outside & \LHrng{IPE}{29}{32}, \LHrng{LRR}{1}{7}, \LHrng{SCT}{38}{39} \\
\midrule
Proposition 4.16: Batching shares operations across facts & \LHrng{BWR}{22}{23}, \LH{BWR28}, \LHrng{BWR}{43}{44}, \LHrng{BWR}{50}{56}, \LHrng{GRI}{1}{4}, \LHrng{SCC}{1}{9} \\
\midrule
Theorem 4.17: Conflicts survive answer-preserving refactoring & \LHrng{BLK}{1}{10}, \LHrng{NHF}{53}{58}, \LHrng{NHF}{64}{68} \\
\midrule
Proposition 4.18: Self-calls move answers between routing and precedence & \LHrng{SLF}{1}{7} \\
\midrule
Corollary 4.19: Changing required implementations changes the comparison & \LHrng{NHF}{35}{41}, \LH{SDN18} \\
\midrule
Theorem 4.20: No algorithm certifies every permanently tree-shaped future & \LHrng{UND}{31}{38}, \LH{UND41} \\
\midrule
Proposition 5.1: Omitting required pairs can raise or lower the gap & \LH{DEC7} \\
\midrule
Theorem 6.1: Per-pair behavioral checks cannot choose the required pairs & \LHrng{BAP}{1}{25}, \LHrng{INT}{8}{9} \\
\midrule
Is-a and has-a are not opposites & \LHrng{CND}{1}{7} \\
\midrule
Corollary 6.2: Finite adoption cost cannot erase recurring workaround cost & \LHrng{BWR}{37}{39}, \LHrng{BWR}{45}{49}, \LH{BWR63}, \LHrng{IPE}{23}{27}, \LH{NRH6}, \LHrng{OPT}{3}{16} \\
\midrule
Corollary 6.3: Conservative capacity additions & \LHrng{CSD}{1}{8}, \LHrng{IPE}{21}{28}, \LH{OPT13}, \LH{OPT20}, \LHrng{SIN}{1}{5} \\
\midrule
Tagged prose (main text, p.~3): Workarounds can preserve required facts and behavioral correctness. At a crossing, sound one-parent workarounds still supply at least $\PaperNotationLink{PaperGap}{d_\downarrow}>0$ outside pairs and fail correct maintenance (Theorem~4.6). We count required facts: extra identity names, class--contract memberships, and implementation--class pairs. Each \emph{required answer} must be supplied. Lines measure encoding length; the gap counts facts even the best one-parent hierarchy leaves to code or tooling. Price outside work however you like. Multiple parents remain strictly cheaper on every crossing exactly when doing any outside work costs more than doing none. Bundling and bulk discounts do not change that test (Supplement~Theorem~4.2). Unit charges give exact counts (Proposition~4.16). Batching shares an operation's positive cost across facts still requiring supply (Corollary~6.2). & \LHrng{ACF}{1}{4} \\
\midrule
Tagged prose (main text, p.~3): Code or generators must encode the missing facts. Packing four values into Figure~4's three parameters still leaves a fourth binding. The compiler may derive it, or the body must unpack it. With ancestry and generator fixed, $M$ possible sets of missing required facts need $\lceil\log_2 M\rceil$ input bits regardless of source length (Proposition~4.9). & \LH{ARI1} \\
\midrule
Tagged prose (main text, p.~4): No rule reading only successful outputs and ancestry can distinguish zero from positive gap. Reading ancestry, removals, and outside routes recovers $\PaperNotationLink{PaperRequired}{R^{\mathrm{req}}}$ exactly (Corollary~4.10). & \LHrng{IPE}{21}{28}, \LH{NHF63}, \LHrng{OLC}{1}{5}, \LHrng{PJD}{1}{4}, \LHrng{RCN}{1}{11}, \LHrng{SIP}{1}{7} \\
\midrule
Tagged prose (main text, p.~4): If a ban removes choices without changing each remaining design's cost, it cannot make the cheapest choice cheaper, even when adding costs across projects. Savings require people otherwise to choose a more expensive design; the cited rationales do not measure those choices (Section~6.5). & \LH{OPT13}, \LHrng{OPT}{17}{19} \\
\midrule
Tagged prose (main text, p.~5): Each left-to-right step in Figure~1 derives more required facts; in reverse, it is a language or project-policy ban. Adjacent rungs can tie today; an allowed change can leave more facts for code or tooling under the lower rung. & \LH{BWR43}, \LH{BWR54}, \LH{CAA7}, \LH{ENC1}, \LH{FDC2}, \LH{NOM1}, \LH{PD1}, \LHrng{RBF}{1}{2}, \LHrng{SCT}{1}{4}, \LHrng{SCT}{20}{26}, \LHrng{SCT}{30}{36} \\
\midrule
Tagged prose (main text, p.~5): Let $\PaperInput$ contain every interface input, $\PaperAnswerSpace$ the answers (including output sets), and $a:X\to\PaperAnswerSpace$ the specification. Encoding $e:X\to\PaperEncodingSpace$ with decoder $r:\PaperEncodingSpace\to\PaperAnswerSpace$ preserves required answers iff $r(e(x))=\PaperNotationLink{PaperAnswer}{a(x)}$ for every $x$. The program presents $e(x)$; the decoder must recover $\PaperNotationLink{PaperAnswer}{a(x)}$. Refinement decides whether such a decoder exists. Swapping required and forbidden cases keeps their groups but changes permitted behavior. & \LHrng{INT}{1}{7} \\
\midrule
Tagged prose (main text, p.~6): For schemas, field deletion can make separately declared types look identical while requiring different interpretations. $X=\PaperNotationLink{PaperClasses}{C}$ tracks meanings across edits; $e$ is the visible record plus any retained identity key. Write $p:\PaperNotationLink{PaperClasses}{C}\to\mathcal V$ for the current record. A decoder from $p$ exists on every reachable value exactly when same-looking meanings need the same answer: $p(x)=p(y)\Longrightarrow \PaperNotationLink{PaperAnswer}{a(x)}=a(y)$. This characterizes safe erasure. & \LHrng{BWR}{3}{4} \\
\midrule
Tagged prose (main text, p.~6): A helper computed from $p$ cannot distinguish same-looking meanings; a nominal key can. & \LHrng{IDD}{1}{7}, \LHrng{NHF}{3}{4}, \LH{NHF11} \\
\midrule
Tagged prose (main text, p.~6): Every rung satisfies $\mathcal D_m(\mathcal H)\subseteq\mathcal D_{\mathrm{complete}}(\mathcal H)$: the complete mechanism retains forwarding, registration, and generation. Thus the ideal compares designs across the whole hierarchy (Lemma~1.1). Repair obligations form one sum, $\PaperNotationLink{PaperBurden}{B_{\mathrm{rep}}}$: parent-edge deletions carrying still-required pairs, plus missing or removed pairs before and after each update. The history fixes changes to required facts. Complete supply attains zero on every finite declared supply history, including removals; lower supply rungs fail on single-version examples (Corollary~4.12; Supplement~\S4.5). & \LH{BPI1}, \LHrng{BWR}{37}{39}, \LH{BWR57}, \LHrng{INT}{1}{7}, \LHrng{OPT}{21}{22} \\
\midrule
Tagged prose (main text, p.~6): Every rung satisfies $\mathcal D_m(\mathcal H)\subseteq\mathcal D_{\mathrm{complete}}(\mathcal H)$: the complete mechanism retains forwarding, registration, and generation. Thus the ideal compares designs across the whole hierarchy (Lemma~1.1). Repair obligations form one sum, $\PaperNotationLink{PaperBurden}{B_{\mathrm{rep}}}$: parent-edge deletions carrying still-required pairs, plus missing or removed pairs before and after each update. The history fixes changes to required facts. Complete supply attains zero on every finite declared supply history, including removals; lower supply rungs fail on single-version examples (Corollary~4.12; Supplement~\S4.5). & \LHrng{MGH}{9}{11}, \LH{SCT9}, \LHrng{UMS}{1}{5} \\
\midrule
Tagged prose (main text, p.~6): Every representation bound fixes an arbitrary finite declared relation before choosing a mechanism. Reshaping requirements changes the specification (Corollary~4.19); no algorithm certifies all and only tree-shaped futures (Theorem~4.20), and strict insertion between tree-shaped versions forces one-parent repair (Theorem~4.14(0)). Name each required fact's declaration, required and forbidden cases, and independent changes. Unknown or external cases stay unscored; including them requires recomputing the relation and gap. Section~5 recovers pinned-source relations. & \LHrng{BWR}{43}{44}, \LH{BWR54}, \LHrng{IPE}{21}{28}, \LH{SCT2}, \LH{SCT32}, \LH{SDN1} \\
\midrule
Tagged prose (main text, p.~6): Refactorings preserve every required fact on every reachable input. Identical fields or base lists today do not justify merging declarations. Merge only when no required fact distinguishes them through any supported change. Otherwise, share agreeing implementations and keep identities distinct (Theorem~4.8); changing a required fact changes the specification (Corollary~4.19). Ruling out missing connections needs the complete map. Untested reachable states can still fail. Naming costs and ancestry gaps are independent (Section~6.3). & \LHrng{BWR}{29}{30}, \LHrng{BWR}{37}{38}, \LH{INT5}, \LHrng{SDN}{15}{18} \\
\midrule
Tagged prose (main text, p.~8): For a nonempty relation, one constant key value requires no maintained name. Additional names therefore number $\PaperNotationLink{PaperKeys}{\chi(G^{\rm req})}-1$. When each package has its own key namespace, add each package's $\chi-1$ extra names; this can differ from $\chi-1$ of the union graph. & \LH{EVO40} \\
\midrule
Tagged prose (main text, p.~8): Finite policies can be checked. For arbitrary generators, unsafe cases have finite counterexamples, but no checker certifies exactly all safe futures, even with two meanings and finite views (co-r.e.-complete)~. Supplement~\S\S5--7 give fixed-field constructions and known-schema bounds. & \LH{DEC1}, \LHrng{UND}{23}{30}, \LH{UND40} \\
\midrule
Tagged prose (main text, p.~8): Nominal identity never needs more extra keys across supported changes; look-alikes needing different handling make the advantage strict. & \LHrng{HVD}{3}{4}, \LH{HVD7} \\
\midrule
Tagged prose (main text, p.~9): In the declared table $\PaperNotationLink{PaperRequired}{R^{\mathrm{req}}}$, rows are independently changing shared implementations and columns are classes; $(f,c)\in \PaperNotationLink{PaperRequired}{R^{\mathrm{req}}}$ means $c$ needs $f$. Contract parents supply no implementation. One implementation parent places each class's code providers on one ancestor chain; multiple parents allow incomparable providers. Code supplies implementations lost when an inheritance path is removed. We count two kinds of compensation: forwarding a missing pair and removing an unwanted one. & \LHrng{DEC}{5}{6} \\
\midrule
Tagged prose (main text, p.~10): If class $c$ does not inherit implementation $f$, $\PaperNotationLink{PaperRoutes}{\varphi}(f,c)$ chooses a forwarding destination $v$. In routing graph $\PaperNotationLink{PaperRoutingGraph}{\Gamma}$, $c\to v$ is one forwarding edge; $c\to^*v$ follows zero or more. Read $f@c$ as ``class $c$ can reach an implementation of $f$'': [displayed formula] Per-pair forwarding, nonpropagating registries, adapters, and generated routes supply missing pairs outside ancestry. Native delegation has its own capacity: Kotlin's \texttt{by} forwards an interface with one declaration; Rust's \texttt{Deref} lets member access follow one target chain~. Stein models delegation as inheritance~; per-pair delegation here is outside supply included in the outside-supply minimum. Copies supply pairs the original provider's ancestry misses; copying its body does not make the consumer inherit it (Theorem~4.7). & \LHrng{BWR}{31}{32} \\
\midrule
Tagged prose (main text, p.~10): An unoverridden method added to a code provider reaches every required class through ancestry exactly when that provider has no outside pair; otherwise, forwarding code or its shared rule must supply the missing routes. In Figure~2, a new \texttt{ByteReceiveStream} method reaches the file and buffered readers. After changing to one parent, a new \texttt{\_BaseFileStream} method reaches the writer directly, but the file reader only if its forwarding code or rule covers it. & \LH{NHF1}, \LH{NHF20}, \LH{NHF22}, \LHrng{NHF}{25}{28} \\
\midrule
Tagged prose (main text, p.~10): An unoverridden method added to a code provider reaches every required class through ancestry exactly when that provider has no outside pair; otherwise, forwarding code or its shared rule must supply the missing routes. In Figure~2, a new \texttt{ByteReceiveStream} method reaches the file and buffered readers. After changing to one parent, a new \texttt{\_BaseFileStream} method reaches the writer directly, but the file reader only if its forwarding code or rule covers it. & \LHrng{NHF}{23}{24} \\
\midrule
Tagged prose (main text, p.~11): The \emph{ancestry gap} is the fewest required pairs that a sound one-parent hierarchy leaves outside. Retain as many required pairs as possible while their class groups fit one tree: [displayed formula] A deleted $1$ leaves code or tooling to supply an implementation the class still needs but no longer inherits. A crossing describes required groups; the gap counts pairs that sound one-parent ancestry cannot derive. The definition optimizes tables; Theorem~4.6 shows that every program design in its scope reduces to this optimization. & \LHrng{NHF}{6}{9}, \LHrng{NHF}{13}{14}, \LHrng{NHF}{20}{21} \\
\midrule
Tagged prose (main text, p.~12): Contract and implementation incompleteness obey the same one-parent law. Registries or generated tables still determine omitted memberships. The hierarchy's four strict steps use Figure~5's identity collision, AnyIO's crossing declared as memberships, any nonempty $\PaperNotationLink{PaperRequired}{R^{\mathrm{req}}}$, and Figure~2's implementation crossing. & \LH{CAP2}, \LH{NHF52}, \LH{OWM2}, \LH{PD1}, \LHrng{SCT}{31}{32} \\
\midrule
Tagged prose (main text, p.~12): A tree-shaped map has zero gap but its updates can force deletion, forwarding, or removal (Theorem~4.14); Corollary~6.2 prices the recurring obligation. & \LH{BWR9}, \LHrng{NHF}{8}{14} \\
\midrule
Tagged prose (main text, p.~12): Specified lookup winners, state sharing, and initialization order are representation-invariant, outside the supply triples (Theorem~4.17; Supplement Theorem~8.1). With the same state requirements, multiple parents still permit every one-parent and composition design (Corollary~8.5). A shared ancestor creates a diamond; conflicts require competing definitions. C++ virtual bases and Python's instance namespace decide sharing. Zero outside supply while meeting state requirements needs a compatibility proof; Python's lookup/initialization coupling costs exactly one outside answer in the two-class witness (Supplement~\S8.1). The specification fixes these answers for every representation; layout has a separate runtime price. & \LH{NHF10}, \LH{NHF16}, \LH{NHF19}, \LHrng{SCI}{1}{3} \\
\midrule
Tagged prose (main text, p.~13): For independent classes, cell probabilities $q_{00},q_{10},q_{01},q_{11}$ mean neither, only $A$, only $B$, or both. With $q_{\min}=\min\{q_{10},q_{01},q_{11}\}>0$, zero-gap probability lies between $(1-q_{\min})^n$ and $3(1-q_{\min})^n$ (exact inclusion--exclusion proof: Supplement~\S4.9). & \LHrng{AGP}{1}{7} \\
\midrule
Tagged prose (main text, p.~13): Tree shape decides whether one parent derives every pair in a finite requirement table. Deleting at most $\ell$ pairs to reach it is NP-complete: Chen et al.'s deletion-flip problem on complete $0/1$ tables~. Their four-way search deletes one of a crossing's four pairs ~. Lean checks this step; the audit replays the search. Supplement~\S9.9 gives both finder runtimes and distinguishes deletion-only from minimum-flip search. & \LHrng{DEC}{2}{4}, \LH{NHF31} \\
\midrule
Tagged prose (main text, p.~15): Overrides must preserve superclass contracts; typechecking alone is insufficient (Section~6.1). Removing unwanted implementation supply cannot cancel a superclass promise. The preceding results count missing pairs; here we also charge removal. Allowing unwanted pairs instead changes requirements (Section~5). & \LHrng{BLK}{1}{8}, \LHrng{BWR}{33}{36} \\
\midrule
Tagged prose (main text, p.~16): Among behaviorally correct designs, multiple parents leave zero outside pairs against exactly $\PaperNotationLink{PaperGap}{d_\downarrow}>0$ in the best sound one-parent design. Price that work by the fact, by the batch, or with bulk discounts. Multiple parents remain strictly cheaper on every crossing exactly when doing any outside work costs more than doing none (Supplement~Theorem~4.2). & \LHrng{ACF}{1}{4}, \LHrng{BWR}{25}{27}, \LH{OPT1} \\
\midrule
Tagged prose (main text, p.~16): Even zero-gap updates can force an old parent-edge deletion, forwarding, or removal (Theorem~4.14). External routing creates choices that ancestry no longer fixes, so required supply must be preserved in every reachable configuration. & \LHrng{NHF}{59}{62} \\
\midrule
Tagged prose (main text, p.~16): Two representations \emph{preserve the same required supply} iff, at every reachable state $t$, every reachable setup supplies every required implementation and no forbidden one: $\PaperNotationLink{PaperSupplied}{R_D}=\PaperNotationLink{PaperRequired}{R_t^{\mathrm{req}}}$. One working setup does not establish this. & \LHrng{BWR}{40}{41} \\
\midrule
Tagged prose (main text, p.~19): New behavior is defined at its source. Slop restates facts native mechanisms would derive from that source. Metaclasses, descriptors, \texttt{\_\_getattribute\_\_}, and generated ancestry can reuse a rule iff equal inputs always require equal results. A different required result for the same input forces a rule change (exact minimum: Supplement~\S13.3). & \LHrng{RBF}{3}{6} \\
\midrule
Tagged prose (main text, p.~19): Asking whether class $c$ gets implementation $f$ does not mean one can list everything $c$ gets (Supplement~\S13.2). Grouping or generating routes does not lower the ancestry gap. & \LHrng{BWR}{1}{2}, \LHrng{BWR}{8}{9}, \LH{BWR12}, \LHrng{SCT}{24}{25}, \LHrng{SDN}{13}{17} \\
\midrule
Tagged prose (main text, p.~20): A helper can bundle outside pairs~; Theorem~4.6 fixes their minimum. & \LHrng{BWR}{3}{8}, \LHrng{BWR}{12}{18}, \LHrng{BWR}{31}{32}, \LHrng{NHF}{29}{30}, \LHrng{NHF}{47}{48}, \LHrng{SDN}{13}{17} \\
\midrule
Tagged prose (main text, p.~20): A computable checker that never accepts an unsafe generator must leave some permanently tree-shaped generator uncertified. A different generator producing the same history may still be certified. & \LH{UND39} \\
\midrule
Tagged prose (main text, p.~21): Passing tests can miss reachable cases; coverage is complete iff $|V\setminus T|=0$, which a separate proof can establish. One-test completeness in a fixed computable trace family is co-r.e.-complete (proof: Supplement~\S4.1). & \LHrng{TST}{1}{10} \\
\midrule
Tagged prose (main text, p.~21): Each selected public method must still resolve to its original concrete provider, including dunders. This public-name contract forces \PublicApiPairs{} provider--class pairs with certified exact gap \PublicApiGap{} under sound ancestry (Table~2; Supplement~\S9.8.1). & \LHrng{APC}{1}{2} \\
\midrule
Tagged prose (main text, p.~22): Table~3 totals preserve distinct providers and replicas (Proposition~4.1; Theorems~4.6, 4.7). In recorded optima, \ProviderStatePairs{}/\ProviderStateTotal{} outside pairs (\ProviderStatePercent{}\%) have state-bearing providers: they or resolved ancestors declare nonempty \texttt{\_\_slots\_\_} or write instance attributes. Stateless defaults cannot supply this state. Re-optimizing those providers' requirements alone forces \StatefulGap{} outside pairs, certified exact on one sound stateful chain, granting other providers defaults with zero outside supply, keeping each implementation and its state together in its provider (Supplement~\S9.8.2). & \LHrng{SRA}{1}{2} \\
\midrule
Tagged prose (main text, p.~23): \textbf{Primary component optimum.} SQLAlchemy's largest component has 223 code providers, 4,831 required pairs, and gap \SQLAlchemyComponentTax{} of the project's \PythonLongitudinalLastGap{}. A relation-level hierarchy and its outside routes attain \SQLAlchemyComponentTax{}; executable integration is separate. A Lean-checked LRAT certificate rules out one fewer outside pair under sound one-parent ancestry. \textbf{Bundles for the recorded design.} A bundle is a rectangle of missing pairs: implementations grouped with classes needing all of them. The \emph{minimum biclique cover} of $\PaperNotationLink{PaperRequired}{R^{\mathrm{req}}}\setminus \PaperNotationLink{PaperAncestry}{H}$ for this design is \SQLAlchemyComponentBooleanRank{} bundles~; other gap-optimal hierarchies can need different counts. Lean-checked LRAT excludes \IndependentHolderExcludedBound{} bundles; Supplement~\S9.1 gives the maximal-rectangle construction. Each bundle adds uncovered pairs, so independent enabling makes all $2^{\SQLAlchemyComponentBooleanRank{}}$ subsets distinct, exactly one complete (Theorem~4.13; Supplement~\S9.1). & \LHrng{HCC}{1}{2} \\
\midrule
Tagged prose (main text, p.~24): With unit-cost forwarding and removal, the source-class minimum is \PythonPaidBlockingAggregate{} (solver lower; checked design upper), against the sound minimum \PythonProviderOptimumTotal{}. Each provider--class pair is declared required, forbidden, or tolerated: tolerated pairs may be inherited or omitted without charge. Tolerating omission of $t$ recovered pairs lowers the gap by at most $t$. The reported lower bounds hold whichever unrequired classes are tolerated, within the per-implementation allowance (Supplement~\S9.9). & \LH{TOL4}, \LH{TOL6} \\
\midrule
Tagged prose (main text, p.~24): For implementation requirements alone, classes needing the same providers can share a supply node; a class needing a subset of those providers can be the supply parent. Grouping implementation supply does not decide contract membership or erase distinct identities. & \LH{BWR14}, \LH{BWR17}, \LHrng{RCN}{1}{7}, \LH{SCT18} \\
\midrule
Tagged prose (main text, p.~25): Even with all code inherited, look-alike records needing different handling need extra identity keys (Theorem~3.2): $m$ pairwise confusable meanings need $m-1$ even at zero gap. Independently, distinct identities do not remove crossings: the $h\times h$ row--column family needs $h(h-1)$ outside pairs but no extra name. & \LHrng{AVG}{1}{7}, \LH{EVO17}, \LHrng{EVO}{23}{30}, \LH{EVO40}, \LH{NHF15}, \LH{NHF28} \\
\midrule
Tagged prose (main text, p.~25): Successful outputs and ancestry cannot identify the gap; ancestry, removals, and outside routes recover $\PaperNotationLink{PaperRequired}{R^{\mathrm{req}}}$ (Corollary~4.10). & \LH{BWR12}, \LHrng{BWR}{29}{30}, \LHrng{BWR}{37}{39}, \LH{BWR42}, \LHrng{NPS}{1}{9}, \LHrng{OLC}{1}{5}, \LHrng{OLC}{7}{8}, \LHrng{SCT}{1}{9}, \LHrng{SCT}{33}{34} \\
\end{longtable}
\endgroup

}{%
  \textbf{Error:} \texttt{content/claim\_mapping\_auto.tex} not found.
}

\begingroup
\small
\bibliographystyle{plain}
\bibliography{references}
\endgroup